\documentclass[a4paper, 11pt, oneside, onecolumn]{article}

\usepackage[english]{babel} % set language
\usepackage{fontawesome}
\usepackage{microtype} % micro-typographical features and ligatures
\usepackage{xspace} % \xspace: if next character is punctuation -> no-space, else -> space
\usepackage[utf8]{inputenc}	% Allows for writing special characters in the tex-file
\usepackage[T1]{fontenc} % 8-bit font encoding
\usepackage{float}
\usepackage{dsfont}
\usepackage{soul}
\usepackage{setspace} 

\usepackage{amsmath,amssymb,bbm}
\usepackage{sansmath}

\DeclareMathVersion{sans}
\SetSymbolFont{operators}{sans}{OT1}{cmbr}{m}{n}
\SetSymbolFont{letters}{sans}{OML}{cmbrm}{m}{it}
\SetSymbolFont{symbols}{sans}{OMS}{cmbrs}{m}{n}
\SetMathAlphabet{\mathit}{sans}{OT1}{cmbr}{m}{sl}
\SetMathAlphabet{\mathbf}{sans}{OT1}{cmbr}{bx}{n}
\SetMathAlphabet{\mathtt}{sans}{OT1}{cmtl}{m}{n}
\SetSymbolFont{largesymbols}{sans}{OMX}{iwona}{m}{n}

\DeclareMathVersion{boldsans}
\SetSymbolFont{operators}{boldsans}{OT1}{cmbr}{b}{n}
\SetSymbolFont{letters}{boldsans}{OML}{cmbrm}{b}{it}
\SetSymbolFont{symbols}{boldsans}{OMS}{cmbrs}{b}{n}
\SetMathAlphabet{\mathit}{boldsans}{OT1}{cmbr}{b}{sl}
\SetMathAlphabet{\mathbf}{boldsans}{OT1}{cmbr}{bx}{n}
\SetMathAlphabet{\mathtt}{boldsans}{OT1}{cmtl}{b}{n}
\SetSymbolFont{largesymbols}{boldsans}{OMX}{iwona}{bx}{n}

\usepackage{afterpage}

\usepackage{amsfonts,amsmath,amssymb,bm,mathtools,slashed} % Math-tools
\usepackage{mathrsfs}

\usepackage{multicol} % text in multiple columns
\usepackage[dvipsnames,table]{xcolor}	% colors

\definecolor{indigo}{rgb}{0.0, 0.25, 0.42}

\definecolor{rwth}{RGB}{0,84,159}
\definecolor{rwth2}{RGB}{64,127,183}

\definecolor{rmp}{RGB}{41, 43, 133}
\newcommand{\LinkColor}[0]{rwth}

\colorlet{darkBlue}{blue!45!black}
\colorlet{linkColor}{blue!80!black}
\usepackage{hyperref} % references
\hypersetup{
	colorlinks, 
	bookmarksopen, 
	bookmarksnumbered,
	pdftoolbar=false, % show Acrobat toolbar
	pdfmenubar=false, % show Acrobat menu?
	pdffitwindow=false, % window fit to page when opened
	pdfstartview={FitH},	 % fits the width of the page to the window
	citecolor=\LinkColor, %NavyBlue, 	%color of links to bibliography
	linkcolor=\LinkColor,	%color of internal links
	urlcolor=\LinkColor, %color of external links 
	pdftitle={Systematic study of flavored WIMP DM}, % title
	pdfsubject={Flavored-WIMP-DM}, % subject
	pdfauthor={Felix Wilsch}, % author
	pdfnewwindow=true, % links in new window
	linktoc=all
}

\usepackage{lipsum}
\setlipsum{%
  par-before = \begingroup\color{gray},
  par-after = \endgroup
}

\usepackage{geometry}
\usepackage{pdflscape}

\usepackage{fancyhdr}
\fancypagestyle{main}{%
    \fancyhf{}
    \setcounter{page}{1}
    \pagenumbering{arabic}
	\cfoot{--~\thepage~--}
	\renewcommand{\headrulewidth}{0.0pt}
}

\fancypagestyle{intro}{%
    \fancyhf{}
    \setcounter{page}{1}
    \pagenumbering{roman}
	\cfoot{--~\thepage~--}
}

\usepackage{titlesec}
\titleformat{\section}{\LARGE\sffamily\bfseries}{\thesection}{1.0em}{}
\titleformat{\subsection}{\Large\sffamily\bfseries}{\thesubsection}{1.0em}{}
\titleformat{\subsubsection}{\large\sffamily\bfseries}{\thesubsubsection}{1.0em}{}
\titleformat{\paragraph}[runin]{\sffamily\bfseries}{}{1.0em}{}

\makeatletter
\g@addto@macro\bfseries{\boldmath} % use boldmath in titles
\makeatother

\usepackage[sort&compress,numbers,merge]{natbib}
\addto\captionsenglish{}

\usepackage{graphicx}
\usepackage{epstopdf}
\graphicspath{{./Figures/}{./figures/}}
\usepackage{caption,subcaption}
\DeclareCaptionFormat{custom}
{%
    \textbf{#1#2} #3
}
\DeclareCaptionLabelSeparator{custom}{ }
\numberwithin{equation}{section}

\usepackage{booktabs} % enhanced table quality
\usepackage{longtable} % long tables with page break
\usepackage{multirow} % multirows in tables
\renewcommand{\arraystretch}{1.2}

\usepackage[shortlabels]{enumitem}
\setlist{itemsep=.1em,topsep=.5em}
\SetEnumerateShortLabel{i}{\textit{\roman*}}

\allowdisplaybreaks

\usepackage{bbold,bbm}
\usepackage{slashed}

\newcommand{\unit}{\mathbbm{1}}

\newcommand{\cG}{\mathcal{G}}

\newcommand{\cL}{\mathcal{L}}

\newcommand{\cO}{\mathcal{O}}

\DeclareUnicodeCharacter{202F}{\,} % Due to witchcraft, removing this line causes an error...

\definecolor{blue-violet}{rgb}{0.54, 0.17, 0.89}

\renewcommand{\author}[3]{%
    {\large \sffamily \mathversion{sans} #1${}^{\,#3\,}$\footnote{\href{mailto:#2\!}{\sffamily #2}}}%
}

\newcommand{\affiliation}[2]{%
    {\small\it \sffamily \mathversion{sans} ${}^{#1}$~#2 \\[1.8pt]}%
}

\renewcommand*{\thefootnote}{\fnsymbol{footnote}}

\newcommand\blfootnote[1]{%
  \begingroup
  \renewcommand\thefootnote{}\footnote{#1}%
  \addtocounter{footnote}{-1}%
  \endgroup
}

\begin{document}

% - - - - TITLE PAGE - - - - %
\newgeometry{bindingoffset=0cm, inner=3. cm, 
	outer=3. cm, 
	top=3.35cm}
	
\newgeometry{bindingoffset=0cm, inner=3. cm, 
	outer=3. cm, 
	top=3.35cm}
	
%\begin{titlepage}
\renewcommand*{\thefootnote}{\fnsymbol{footnote}}
\begin{center} 
\begin{minipage}{15.5cm}
\vspace{-0.7cm}
\begin{flushright}
{\footnotesize \ttfamily
P3H-25-091\\[-.8pt]
TTP25-043\\[-3pt]
TTK-25-37
}
\end{flushright}

\vspace{0.5cm}

\end{minipage}
\end{center}

\begin{center}
\vspace*{10pt}
{\bfseries  \sffamily \fontsize{20}{29}\selectfont \mathversion{boldsans} \baselineskip=30pt
Toward a Comprehensive Exploration of \\
\vspace{0.5ex}
Flavored Dark Matter Models
}
\vspace{26 pt} 
{\color{rwth}
\rule{\textwidth}{1.5pt}
}
% Authors
\begin{spacing}{1.2}
\author{Benedetta Belfatto}{benedetta.belfatto@kit.edu}{1}\!,\!
\author{Monika Blanke}{monika.blanke@kit.edu}{1,2}\!,\!
\author{Jan Heisig}{heisig@physik.rwth-aachen.de}{3}\!,\!
\author{Michael Kr\"amer}{mkraemer@physik.rwth-aachen.de}{3}\!,\!
\author{Lena~Rathmann}{lena.rathmann@kit.edu}{2,3}\!,\!
\author{and Felix Wilsch}{felix.wilsch@physik.rwth-aachen.de}{3}
\end{spacing}

\vspace{1.5ex}

% Affiliations
\affiliation{1}{Institut f\"ur Theoretische Teilchenphysik,
  Karlsruhe Institute of Technology, 
Engesserstra\ss e~7,
  D-76128~Karlsruhe, Germany}
\affiliation{2}{Institut f\"ur Astroteilchenphysik, Karlsruhe Institute of Technology,
  Hermann-von-Helmholtz-Platz 1,
  D-76344 Eggenstein-Leopoldshafen, Germany}
\affiliation{3}{Institute for Theoretical Particle Physics and Cosmology, RWTH Aachen University, Sommerfeldstr.~16, D-52056 Aachen, Germany}

\setcounter{footnote}{0}

\thispagestyle{empty}
\end{center}
\renewcommand*{\thefootnote}{\arabic{footnote}}%
\suppressfloats

\renewcommand{\abstractname}{\sffamily \bfseries \LARGE Abstract}
\linespread{1.05}
\vspace*{1cm}

\begin{abstract}
\vspace{0.5cm}
\noindent
We present a comprehensive framework for the study of flavored dark matter models, combining relic density calculations with direct and indirect detection limits, collider constraints, and a global analysis of flavor observables based on SMEFT matching and renormalization-group evolution. The framework applies to scalar or fermionic dark matter, including both self-conjugate and non-self-conjugate cases. As a proof of principle, we analyze two scenarios with Majorana dark matter coupling to right-handed charged leptons and to right-handed down-type quarks, assuming a thermal freeze-out. In the leptophilic case, flavor-violating decays such as $\mu \to e \gamma$ dominate the constraints, while LHC searches still leave sizable parameter space. For quark couplings, direct detection bounds and meson mixing severely restrict the allowed couplings, favoring hierarchical flavor structures. The toolchain presented in this paper is publicly available on GitHub~\href{https://github.com/lena-ra/Flavored-Dark-Matter}{\faicon{github}}\blfootnote{\href{https://github.com/lena-ra/Flavored-Dark-Matter}{\faicon{github}} \url{https://github.com/lena-ra/Flavored-Dark-Matter}}\setcounter{footnote}{0}.
\end{abstract}

\restoregeometry

% - - - - TABLE OF CONTENTS - - - - %
\newpage
{
    \sffamily
    \hypersetup{linkcolor=black}
	\addtocontents{toc}{\protect\hypertarget{toc}{}} % add label to toc
    % \begin{spacing}{1.035}
	\tableofcontents
    % \end{spacing}
}

% - - - - INTRODUCTION - - - - %
\section{Introduction}

The Standard Model (SM) of particle physics has proven to be extremely successful in describing laboratory-based experiments. However, it fails to address some fundamental theoretical questions like the origin of its gauge symmetry structure, the dynamics behind electroweak symmetry breaking (EWSB), and the origin of flavor and its hierarchies. Notably, the majority of free SM parameters appear in the latter sector, calling for an extended theory of flavor. Additionally, and perhaps even more pressingly, the SM also falls short of explaining several fundamental astrophysical and cosmological observations such as the origin of the matter-antimatter asymmetry and the existence of Dark Matter (DM) in our Universe. It is hence expected that a more complete theory addressing some or ideally all of these issues will ultimately replace the SM. 

A common approach toward achieving this goal is the study of minimal renormalizable extensions of the SM -- so-called simplified models -- that address one or more of the aforementioned issues. Analyzing their phenomenology provides insight into the necessary ingredients of a more complete theory.

A popular class of simplified models of DM introduces a $t$-channel mediator that couples the DM particle to the SM fermion sector. A recent review of this class of models can be found in~\cite{Arina:2025zpi}. While the simplest $t$-channel models introduce only a single DM and mediator field each, a non-trivial flavor structure can also be present in the dark sector. For example, in simplified models of the Minimal Supersymmetric Standard Model, the lightest neutralino -- a Majorana fermion -- is the non-flavored DM candidate while the squarks or sleptons are flavored scalars that serve as the $t$-channel mediators. Models with multiple mediator flavors have also been considered in~\cite{Bai:2013iqa,DiFranzo:2013vra,Biondini:2025gpg}. Flavored DM models, on the other hand, introduce the DM candidate as the lightest state of a flavor multiplet, while the mediator field remains a singlet under the flavor group~\cite{Kile:2011mn,Batell:2011tc,Agrawal:2011ze}. Lastly, in skew-flavored DM models both mediator and DM are flavored~\cite{Agrawal:2015kje}.

Flavored DM, in particular, comes along with significant phenomenological advantages. On the one hand, the tension between the observed relic abundance and the non-observation of DM in direct detection experiments can be relieved by weakening the coupling of the DM candidate to the detector nuclei~\cite{Agrawal:2011ze,Agrawal:2014aoa}. On the other hand, the models' phenomenology becomes much richer, including signatures of more complicated decay chains at the LHC, and the effects of flavor-violating new interactions in both flavor and collider physics experiments~\cite{Agrawal:2014aoa,Blanke:2017tnb,Blanke:2020bsf}. Additionally, flavored DM offers exciting opportunities for nonstandard freeze-out dynamics, such as conversion-driven freeze-out~\cite{Garny:2017rxs,DAgnolo:2017dbv}, whose realization in this framework~\cite{Acaroglu:2023phy} opens up further viable regions of parameter space and can simultaneously account for baryogenesis~\cite{Heisig:2024mwr}.

While a number of previous papers have analyzed the phenomenology of flavored DM models -- see Tab.~\ref{tab:overview-of-DM-mediator-combinations} for an overview -- significant parts of the flavored DM model space remain completely untouched. To improve our understanding of flavored DM, in this paper we therefore provide a toolchain that facilitates future phenomenological studies. The toolchain connects various publicly available tools for constraining models of DM and other new physics (NP) models, namely \textsc{FeynRules}~\cite{Alloul:2013bka}, \textsc{micrOMEGAs}~\cite{Belanger:2010gh,Belanger:2020gnr,Alguero:2023zol,Barducci:2016pcb}, \textsc{SModelS}~\cite{Alguero:2021dig,MahdiAltakach:2023bdn,Altakach:2024jwk}, \textsc{Matchete}~\cite{Fuentes-Martin:2022jrf} and \textsc{smelli}~\cite{Aebischer:2018iyb} {(based upon \textsc{flavio}~\cite{Straub:2018kue} and \textsc{Wilson}~\cite{Aebischer:2018bkb})}. 

In Sec.~\ref{sec:classific}, we begin with a classification of flavored DM models. Sec.~\ref{sec:toolchain} describes the setup of our toolchain, discussing in turn the handling of constraints from the DM relic abundance, direct and indirect detection, LHC and flavor physics. Then in Sec.~\ref{sec:application}, we apply the toolchain to two different flavored DM models that have not been studied in detail in the literature before, namely flavored Majorana DM coupled to either the right-handed charged leptons or to the right-handed down-type quarks, and discuss the results.
Appendix~\ref{app:lagran} summarizes the naming conventions used in the model files of the provided toolchain, while Appendix~\ref{app:lambda} provides a detailed discussion of the flavor structure and flavor constraints for the down-type quark scenario.

% - - - - CLASSIFICATION - - - - %
\section{Classification of Flavored Dark Matter Models}\label{sec:classific}

The flavored DM framework, broadly defined, comprises models in which the DM field is introduced as part of a flavor multiplet. In addition, it is commonly assumed that DM transforms as a singlet under the SM gauge symmetry and couples to the SM fermion sector via a $t$-channel mediator. In what follows, we adopt these assumptions. We further restrict our attention to simplified models in which only a single mediator is introduced in addition to the DM flavor multiplet. Models with multiple mediator flavors have been considered e.\,g.\ in~\cite{Bai:2013iqa,DiFranzo:2013vra,Biondini:2025gpg}. A~scenario in which both DM and the mediator carry flavor has been investigated in~\cite{Agrawal:2015kje}. Finally, a model with two mediators of different gauge representations coupling DM to both left- and right-handed leptons has been considered in~\cite{Acaroglu:2022boc}.

Massive spin-1 fields in a simplified model introduce additional complications and should therefore be considered with caution. 
In a possible UV completion such a spin-1 state arises either as a massive vector after spontaneous symmetry breaking of a bigger gauge group down to the SM gauge group $\cG_\mathrm{SM}$, or as a composite vector resonance of a strongly coupled sector. In both cases we generally expect further states of similar mass present in the theory that can alter conclusions. Therefore, while phenomenologically viable models with new vector states can be constructed, we do not include this case in our classification and restrict ourselves to dark sectors with only new spin-0 and spin-1/2 fields, i.\,e.\ scalars and fermions.

\begin{figure}[tbp]
    \centering
    \includegraphics[width=0.22\textwidth]{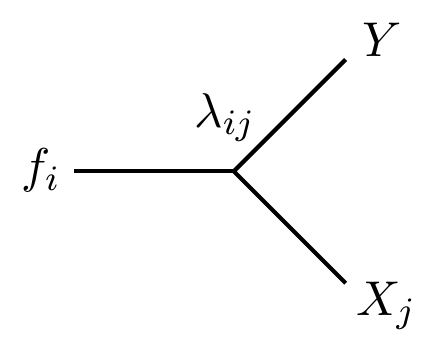}
    \caption{Schematic visualization of the interaction of a DM~field~$X$ with a SM~fermion~$f$ and a mediator field~$Y$. Here $i,j$ are (SM and dark) flavor indices.}
    \label{fig:generic-DM-diagram}
\end{figure}

Under these assumptions, the field content of flavored DM models generally contains the following particles:
\begin{itemize}
    \item $f={(f_1,f_2,f_3)}^\intercal$: 3-generation SM fermion field
    \item $X={(X_1, X_2, \dots X_n)}^\intercal$: $n$ generations of DM field
    \item $Y$: $t$-channel mediator field
\end{itemize}
where we adopt the notation of~\cite{Arina:2025zpi}. These states feature interactions of the form
\begin{equation}\label{eq:Lint}
    \mathcal{L}_\text{int} = -\lambda_{ij} \bar{f}_i X_j Y + {\rm h.c.}\,,
\end{equation}
schematically depicted in Fig.~\ref{fig:generic-DM-diagram}.

Since $X$ is assumed to be a gauge singlet, the mediator $Y$ has to carry the same gauge quantum numbers as the fermion $f$ in the interaction~\eqref{eq:Lint}. This leads us to five different classes of flavored DM models depending on the fermion $f$ coupled to the dark sector. In each case $Y$ can be chosen to be either a scalar or a fermion field,\footnote{Recall that we avoid the case of vector mediators for the reasons mentioned above.} which in turn fixes $X$ to be fermionic or scalar, respectively. Finally, $X$ can be either real or complex, resulting in four possible combinations for each choice of SM fermion $f$. In total, this yields 20 distinct scenarios involving a single flavored DM multiplet field and a single mediator field, which we summarize in Tab.~\ref{tab:overview-of-DM-mediator-combinations} along with references to previous studies where applicable.

To ensure DM stability, we assume a $\mathbb{Z}_2$ symmetry under which both the mediator $Y$ and the DM field $X$ are charged. The lightest flavor $X_1$ is thus stable and serves as DM candidate as long as $M_{X_1} < M_Y$. Note that in some of the simplified models introduced below this $\mathbb{Z}_2$ symmetry is accidental.  In order to ensure DM stability in UV-complete models of flavored DM, flavor symmetries can be used to stabilize the DM candidate~\cite{Batell:2011tc,Batell:2013zwa,Agrawal:2014aoa}.

\begin{table}[th]
    \centering
    \begin{tabular}{c | c | c | c | c }
        $f$ & $Y$: $\cG_\mathrm{SM} \times s$ & $X \sim (\boldsymbol{1},\boldsymbol{1})_0$ & MFV & beyond MFV
        \\\hline\hline
        \multirow{4}{*}{$q$} & \multirow{2}{*}{$(\boldsymbol{3},\boldsymbol{2})_{1/6} \times 0$} & {Dirac} fermion & ---  & \cite{Blanke:2017fum,Blanke:2020bsf}
        \\\cline{3-5}
        & & Majorana fermion & --- & ---
        \\\cline{2-5}
        & \multirow{2}{*}{$(\boldsymbol{3},\boldsymbol{2})_{1/6} \times \frac{1}{2}$} & complex scalar & \cite{Lopez-Honorez:2013wla,Mescia:2024rki} & ---
        \\\cline{3-5}
        & & real scalar & --- & ---
        \\\cline{2-5}
        \hline\hline
%%%%%%%%
        \multirow{4}{*}{$u$} & \multirow{2}{*}{$(\boldsymbol{3},\boldsymbol{1})_{2/3} \times 0$} & {Dirac} fermion & \cite{ Kilic:2015vka, Agrawal:2011ze,Batell:2013zwa,Kumar:2013hfa} & \cite{Blanke:2017tnb,Jubb:2017rhm,Blanke:2020bsf}
        \\\cline{3-5}
        & & Majorana fermion & --- & \cite{Acaroglu:2023phy,Acaroglu:2021qae}
        \\\cline{2-5}
        & \multirow{2}{*}{$(\boldsymbol{3},\boldsymbol{1})_{2/3} \times \frac{1}{2}$} & complex scalar & \cite{Lopez-Honorez:2013wla,Mescia:2024rki} & ---
        \\\cline{3-5}
        & & real scalar & --- & ---
        \\\cline{2-5}
        \hline\hline
%%%%%%%%
        \multirow{4}{*}{$d$} & \multirow{2}{*}{$(\boldsymbol{3},\boldsymbol{1})_{-1/3} \times 0$} & {Dirac} fermion & \cite{ Agrawal:2011ze,Agrawal:2014una}  & \cite{Agrawal:2014aoa, Bensalem:2021qtj,Kile:2011mn}
        \\\cline{3-5}
        & & Majorana fermion & --- & ---
        \\\cline{2-5}
        & \multirow{2}{*}{$(\boldsymbol{3},\boldsymbol{1})_{-1/3} \times \frac{1}{2}$} & complex scalar & \cite{Lopez-Honorez:2013wla,Mescia:2024rki} & ---
        \\\cline{3-5}
        & & real scalar & --- & ---
        \\\cline{2-5}
        \hline\hline
%%%%%%%%
        \multirow{4}{*}{$\ell$} & \multirow{2}{*}{$(\boldsymbol{1},\boldsymbol{2})_{-1/2} \times 0$} & {Dirac} fermion & \cite{Agrawal:2011ze} & ---
        \\\cline{3-5}
        & & Majorana fermion & --- & \cite{Ma:2006km,Zurek:2008qg}
        \\\cline{2-5}
        & \multirow{2}{*}{$(\boldsymbol{1},\boldsymbol{2})_{-1/2} \times \frac{1}{2}$} & complex scalar & \cite{Lee:2014rba} & \cite{Acaroglu:2022boc}
        \\\cline{3-5}
        & & real scalar & --- & ---
        \\\cline{2-5}
        \hline\hline
%%%%%%%%
        \multirow{4}{*}{$e$} & \multirow{2}{*}{$(\boldsymbol{1},\boldsymbol{1})_{-1} \times 0$} & {Dirac} fermion & \cite{ Agrawal:2011ze,Agrawal:2015tfa,Hamze:2014wca} & \cite{Chen:2015jkt,Desai:2020rwz}
        \\\cline{3-5}
        & & Majorana fermion & --- & \cite{Herms:2021fql,Heisig:2024mwr}
        \\\cline{2-5}
        & \multirow{2}{*}{$(\boldsymbol{1},\boldsymbol{1})_{-1} \times \frac{1}{2}$} & complex scalar & \cite{Lee:2014rba,Hamze:2014wca} & \cite{Acaroglu:2022hrm,Acaroglu:2022boc,Acaroglu:2023cza}
        \\\cline{3-5}
        & & real scalar & --- & ---
        \\\cline{2-5}
        \hline
    \end{tabular}%}
    \caption{%
    Overview of all possible combination of DM candidates and mediators that can couple to the various SM fermions. The second column shows the gauge quantum numbers of the mediator and its spin~$s$, the third column shows the nature of the DM fields, and the fourth and fifth column list the literature where such a combination was considered within the context of Minimal Flavor Violation~(MFV) or beyond, respectively (if available). Note that in Refs.~\cite{Mescia:2024rki,Lopez-Honorez:2013wla,Lee:2014rba,Hamze:2014wca,Kile:2011mn} considered flavored DM in the EFT limit without explicitly introducing the $t$-channel mediator. 
    }
    \label{tab:overview-of-DM-mediator-combinations}
\end{table}

\afterpage{\clearpage}

The form of the Lagrangian depends on the choice of spin representation for $X$ and $Y$. Below we collect the four possible cases, in which we adopt the more common notation $\tilde S$ and $S$ for real and complex scalar DM, $\tilde\chi$ and $\chi$ for Majorana and Dirac DM, and $\psi$ and $\varphi$ for a fermionic and scalar mediator, respectively. Furthermore, notice that in our notation, scalar mass parameters before EWSB are denoted by lower-case letters~($m$), whereas physical scalar masses after EWSB are labeled by capital letters~($M$). Fermion masses do not receive corrections from EWSB in our setup, and we label them by capital letters~($M$). Accordingly, for the Dirac and Majorana DM case, fermion masses are denoted by $M_\chi$ and~$M_{\tilde{\chi}}$, respectively, while for the mediator one has $m_\varphi \neq M_\varphi$. For the real and complex scalar DM case, this distinction similarly leads to $m_S \neq M_S$ and $m_{\tilde{S}} \neq M_{\tilde{S}}$, whereas the mediator mass is denoted by $M_\psi$.
\begin{itemize}
    \item {\bf Real scalar DM}

    In this case, we introduce a real scalar DM field $X \equiv \tilde{S}$ which is flavored and a mediator $Y \equiv \psi$ which is a Dirac fermion and carries the gauge quantum numbers of the SM fermion~$f$. The Lagrangian before EWSB describing all possible NP interactions is given by
\begin{align}
\begin{split}\label{eq:LrealSDM}
    \cL \supset & \frac{1}{2} (\partial_\mu \Tilde{S}) (\partial^\mu \Tilde{S}) - \frac{1}{2} m_{\Tilde{S}}^2 \Tilde{S} \Tilde{S} + \bar{\psi} (i \slashed{D} - M_\psi) \psi + (\lambda_{ij} \bar{f}_i \psi \Tilde{S}_j + \mathrm{h.c.}) \\
    & + \lambda_{\Tilde{S} H, ij} (\Tilde{S}_i \Tilde{S}_j)(H^\dagger H) + \lambda_{\Tilde{S} \Tilde{S}, ijkl} \Tilde{S}_i \Tilde{S}_j \Tilde{S}_k \Tilde{S}_l \, ,
\end{split}
\end{align}
where $m_{\tilde{S}}$ is the flavored DM mass matrix before EWSB. The complex matrix $\lambda_{ij}$ describes the coupling of the mediator to the SM fermions $f_i$ and the DM flavor multiplet. The hermitian coupling matrix $\lambda_{\Tilde{S} H}$ describes the interactions of $\tilde{S}$ with the Higgs doublet $H$, given by $H = (-i\,G^+, 1/\sqrt{2}\,(v + h + i\,G^0))^\intercal$ after EWSB. This Higgs portal interaction induces an additional contribution to the DM mass term after 
{\begin{equation}
    M_{\tilde{S}}^2 = m_{\tilde{S}}^2 - v^2 \lambda_{\Tilde{S} H}\, ,
    \label{eq:Higgs_mass_contribution}
\end{equation}}
where $v=246\,\text{GeV}$ is the vacuum expectation value of the Higgs. Finally, the coupling  $\lambda_{\Tilde{S} \Tilde{S}, ijkl}$ describes the flavored DM self-interactions.
    \item {\bf Complex scalar DM}

    The Lagrangian for complex scalar flavored DM, $S$, is very similar to the real DM case in Eq.~\eqref{eq:LrealSDM}. It reads 
\begin{align}
\begin{split}
    \cL \supset & (\partial_\mu S)^\dagger (\partial^\mu S) - m_S^2 S^\dagger S + \bar{\psi} (i \slashed{D} - M_\psi) \psi + (\lambda_{ij} \bar{f}_{i}\psi S_j + \mathrm{h.c.}) \\
    & + \lambda_{S H, ij} (S_i^\dagger S_j)(H^\dagger H) + \lambda_{S S, ijkl} (S_i^\dagger S_j) (S_k^\dagger S_l) \,.
\end{split}
\end{align}
Again, the Higgs portal interaction induces a contribution from EWSB to the DM mass matrix 
{\begin{equation}
    M_{{S}}^2 = m_{{S}}^2 - \frac{v^2}{2} \lambda_{{S} H} \,.
\end{equation}}
    \item {\bf Majorana DM}

    We write the Majorana DM field $X \equiv \tilde\chi$ as a four-component Majorana spinor $\tilde{\chi} = (\tilde{\chi}_L, i \sigma_2 \tilde{\chi}_L^*)^\intercal$ where $\tilde{\chi}_L$ is a two-component Weyl spinor. The mediator $Y \equiv \varphi$ is a complex scalar field that carries the same gauge quantum numbers  as the SM fermion $f$. The general Lagrangian before EWSB is given by
\begin{align}
\begin{split}
    \cL \supset &  \frac{1}{2} \bar{\Tilde{\chi}} (i \slashed{\partial} - M_{\Tilde{\chi}}) \Tilde{\chi} + (D_\mu \varphi)^\dagger (D^\mu \varphi) - m_\varphi^2 \varphi^\dagger \varphi + (\lambda_{ij} \bar{f}_{i} \Tilde{\chi}_j \varphi + \mathrm{h.c.}) \\
    & + \lambda_{\varphi H,1} (\varphi^\dagger \varphi)(H^\dagger H) + \lambda_{\varphi H,2} (H^\dagger \varphi)(\varphi^\dagger H) + \frac{1}{2}\lambda_{\varphi H,3} [(H_i \varepsilon^{ij} \varphi_j)^2 + \mathrm{h.c.}]  + \lambda_{\varphi \varphi} (\varphi^\dagger \varphi)^2 \,.
\end{split}
\end{align}
In contrast to the scalar DM models introduced above, the Majorana fermion DM field does not couple to the Higgs doublet at tree level. Instead, the $\lambda_{\varphi H}$ coupling terms couple the mediator $\varphi$ to the SM Higgs field. Therefore, the mass term $m^2_\varphi$ receives a correction from EWSB. Note that not all $\lambda_{\varphi H}$ terms in the Lagrangian are present in all Majorana DM models. The charge of the mediator  therefore determines the correction to the mass term after EWSB. For an $\mathrm{SU}(2)_L$-singlet mediator only the term $\lambda_{\varphi H,1} (\varphi^\dagger \varphi)(H^\dagger H)$ is present, resulting in the physical squared mass of the mediator after EWSB 
{\begin{equation}
    \label{eq:phi_mass_singlet}
    M_{\varphi}^2 = m_\varphi^2 - \frac{v^2}{2} \lambda_{\varphi H,1} \, .
\end{equation}}
An $\mathrm{SU}(2)_L$-doublet mediator coupled to the left-handed quark doublet additionally couples to the Higgs doublet through the $\lambda_{\varphi H,2} (H^\dagger \varphi)(\varphi^\dagger H)$ term. The components of the mediator doublet $\varphi = (\varphi_u, \varphi_d)^\intercal$ thus receive different contributions to their squared masses after EWSB, given by
{\begin{equation}
    \label{eq:phi_mass_q}
    M_{\varphi_u}^2 = m_\varphi^2 - \frac{v^2}{2} \lambda_{\varphi H,1} \, , \quad M_{\varphi_d}^2 = m_\varphi^2 - \frac{v^2}{2} (\lambda_{\varphi H,1} + \lambda_{\varphi H,2}) \, .
\end{equation}}
Lastly,  a doublet mediator coupled to the left-handed lepton doublet can be parameterized as $\varphi = (1/\sqrt{2}(\varphi_R + i \varphi_I), \varphi^-)^\intercal$, where we decomposed the neutral component of the doublet into its real and imaginary parts $\varphi_R$ and $\varphi_I$ as done in scotogenic models~\cite{Ma:2006km}. In this case all $\lambda_{\varphi H}$ terms are present, and the
squared masses of the physical neutral scalar bosons are given by
{\begin{equation}
    \label{eq:phi_mass_l1}
    M_{\varphi_R}^2 = m_\varphi^2 - \frac{v^2}{2} (\lambda_{\varphi H,1} + \lambda_{\varphi H,3}) \, , \quad M_{\varphi_I}^2 = m_\varphi^2 - \frac{v^2}{2} (\lambda_{\varphi H,1} - \lambda_{\varphi H,3}) \, .
\end{equation}}
The squared mass of the charged scalar boson is given by
{\begin{equation}
    \label{eq:phi_mass_l2}
    M_{\varphi^-}^2 = m_\varphi^2 - \frac{v^2}{2} (\lambda_{\varphi H,1} + \lambda_{\varphi H,2}) \, .
\end{equation}}
    \item{\bf Dirac DM}

For Dirac DM the flavor triplet $X \equiv \chi$ is a Dirac fermion while the scalar mediator $Y \equiv \varphi$ can be described analogously to the Majorana fermion DM models. The Lagrangian closely resembles the one of the Majorana fermion DM models and is given by
\begin{align}
\begin{split}
    \cL \supset &  \bar{\chi} (i \slashed{\partial} - M_\chi) \chi + (D_\mu \varphi)^\dagger (D^\mu \varphi) - m_\varphi^2 \varphi^\dagger \varphi + (\lambda_{ij} \bar{f}_{i} \chi_j \varphi + \mathrm{h.c.}) \\
    & + \lambda_{\varphi H,1} (\varphi^\dagger \varphi)(H^\dagger H) + \lambda_{\varphi H,2} (H^\dagger \varphi)(\varphi^\dagger H) + \frac{1}{2}\lambda_{\varphi H,3} [(H_i \varepsilon^{ij} \varphi_j)^2 + \mathrm{h.c.}]  + \lambda_{\varphi \varphi} (\varphi^\dagger \varphi)^2
\end{split}
\end{align}
Since the coupling of the mediator to the Higgs doublet is completely analogous to the Majorana fermion DM models, the mediator masses are again given by Eqs.~\eqref{eq:phi_mass_singlet},~\eqref{eq:phi_mass_q},~\eqref{eq:phi_mass_l1}, and~\eqref{eq:phi_mass_l2} depending on the gauge quantum numbers of $\varphi$.

\end{itemize}

Beyond the classification in Tab.~\ref{tab:overview-of-DM-mediator-combinations}, flavored DM models are distinguished by the dark sector's flavor structure. While a priori any number $n>1$ of generations is possible, the most common assumption in the literature is $n=3$ in analogy to the SM flavor structure. Furthermore, different possible scenarios for flavor symmetries in the dark sector have been discussed. The most common choices are Minimal Flavor Violation (MFV) or Dark Minimal Flavor Violation (DMFV).

MFV~\cite{Buras:2000dm,DAmbrosio:2002vsn,Chivukula:1987py,Hall:1990ac} is based on the assumption that the approximate flavor symmetry
\begin{equation}
G_\text{MFV} = U(3)_q \times U(3)_u \times U(3)_d \times U(3)_\ell \times U(3)_e
\end{equation}
is broken only by the SM Yukawa couplings $Y_u, Y_d, Y_e$ even in the presence of new particles and interactions. The DM flavor multiplet $X$ then has to transform under a non-trivial representation of one of the $U(3)$ factors in $G_\text{MFV}$. Therefore, only certain values for the number $n$ of dark generations are possible in MFV\@.   The structure of the coupling matrix $\lambda$ between the dark sector and the SM sector is also restricted by MFV and can be determined using the usual MFV spurion expansion. In the same way, also the DM mass matrix and the Higgs-portal coupling matrix for scalar DM are determined by the MFV expansion.

The DMFV hypothesis~\cite{Agrawal:2014aoa} extends the MFV framework in a minimal way. The flavor symmetry group $G_\text{MFV}$ is extended by a factor $G_X$ under which the DM field $X$ transforms non-trivially,
\begin{equation}
    G_\text{DMFV} = G_\text{MFV} \times G_X \, .
\end{equation} 
In analogy to MFV, in DMFV models the only sources of flavor breaking are the new flavor-violating coupling matrix~$\lambda$ and the SM Yukawa couplings $Y_u,~Y_d,~Y_e$. The DM mass matrix and Higgs-portal interactions can therefore be written in terms of a DMFV spurion expansion in~$\lambda$ (and in principle $Y_u,~Y_d,~Y_e$, but those enter only at higher order). In this way, the DMFV assumption reduces the number of new parameters in the model while still allowing for the phenomenological implications of a new source of flavor violation.

% - - - - TOOLCHAIN - - - - %
\section{Numerical Toolchain}\label{sec:toolchain}

\begin{figure}
   \centering
    \includegraphics[height=0.4\textheight]{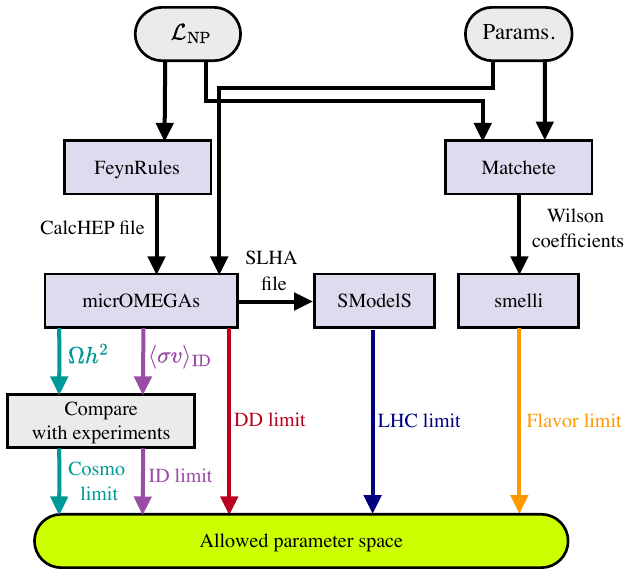}
    \caption{Illustration of the numerical toolchain used in our analysis of flavored DM models. The gray rounded rectangles show the inputs to the toolchain (i.e. the Lagrangian of the DM model and the values for all parameters) and light purple boxes show the tools used.
    }
    \label{fig:toolchain}
\end{figure}

For our phenomenological study, we built a numerical toolchain, illustrated in Fig.~\ref{fig:toolchain} and described in this section. For all models summarized in Tab.\ref{tab:overview-of-DM-mediator-combinations}, we derived the Feynman rules and generated \textsc{CalcHEP}~\cite{Belyaev_2013} and \textsc{UFO}~\cite{Darm__2023} model files using the \textsc{Mathematica}-based package \textsc{FeynRules}~\cite{Alloul:2013bka}. The toolchain as well as all model files are publicly available on GitHub~\href{https://github.com/lena-ra/Flavored-Dark-Matter}{\faicon{github}},\footnote{\label{fn:github}\href{https://github.com/lena-ra/Flavored-Dark-Matter}{https://github.com/lena-ra/Flavored-Dark-Matter}.} where the naming conventions used in the model files are described in Appendix \ref{app:lagran}.

\subsection{Dark Matter Observables}

For the automated computation of the DM relic density as well as the direct and indirect detection cross sections, we use \textsc{micrOMEGAs}~\cite{Belanger:2010gh,Belanger:2020gnr,Alguero:2023zol}, based on our \textsc{CalcHEP} model implementation. Further details are provided in the following subsections.

\subsubsection{Relic Abundance}

We use \textsc{micrOMEGAs} to numerically solve the Boltzmann equation to compute the DM relic density, taking into account coannihilation effects between all relevant particles. While the tool can also accommodate non-standard production mechanisms, such as freeze-in or conversion-driven freeze-out, we focus on the standard scenario of canonical thermal freeze-out in this work.

In this context, we assume that the DM particle is a thermal relic, and that its present-day abundance is set by the freeze-out mechanism and a standard cosmological history. {We consider both the case of the computed relic density being consistent with the value measured by the \textsc{Planck} satellite, $\Omega h^2 = 0.120 \pm 0.001$~\cite{Planck:2018}, within an assumed theoretical uncertainty of 10\%, as well as the case of underabundant DM\@. In the latter case the observed relic abundance is assumed to be accounted for by additional DM components. Parameter points that predict overabundant DM are considered experimentally excluded.}

\subsubsection{Direct Detection}

Direct detection of DM particles involves observing their rare scattering off a nucleus in a low-background environment. Constraints on the DM scattering cross section have been provided in terms of spin-independent (SI) and spin-dependent (SD) cross sections. Models where DM interacts with quarks are subject to stronger constraints compared to those coupling to leptons. {The former give rise to tree-level interactions as well as loop-induced Higgs and gluon interactions relevant in parts of the parameter space.} These loop-induced interactions, however, are suppressed for fermionic flavored DM due to the chiral coupling structure,  so that one-loop photon and $Z$-boson penguin and box diagrams become competitive~\cite{Agrawal:2011ze,Kumar:2013hfa,Agrawal:2014aoa,Blanke:2017tnb}. Leptophilic DM interacts via electroweak loop-diagrams only -- typically via photon or $Z$-boson exchange -- which are highly suppressed, in particular, for self-conjugate DM, see e.g.~\cite{Kawamura:2020qxo}.

We calculate both the SI and SD DM-nucleon scattering cross sections using \textsc{micrOMEGAs}.\footnote{{The current version of \textsc{micrOMEGAs} does not include DM-nucleon scattering via photon or $Z$-boson penguin or box diagrams potentially relevant for the leptophilic models}. Note that the photon exchange gives rise to operators beyond contact interactions.} We exclude parameter points that are ruled out at 90\% confidence level by the current direct detection experiments that are implemented in \textsc{micrOMEGAs}. This means that for the SI cross sections, we consider constraints from CRESST-III~\cite{Abdelhameed_2019}, LZ~\cite{Aalbers_2023}, XENON1T~\cite{Aprile_2018, Aprile_2019}, and DarkSide-50~\cite{Agnes_2018}. For SD cross sections, we include limits from PICO-60~\cite{Amole_2019}.

\subsubsection{Indirect Detection}\label{sec
:IDgen}

Indirect detection experiments constrain the velocity averaged DM annihilation cross section~$\langle \sigma v \rangle$ into SM final states by measuring the flux of cosmic messengers such as gamma rays, cosmic-ray antiprotons and positrons, and neutrinos. In our analysis, we use published constraints on annihilation into SM particle-antiparticle pairs derived from experimental data. These results are typically obtained either by the experimental collaborations themselves or by independent phenomenological studies based on publicly available data.

An overview of the indirect detection bounds included in our framework is provided in Table~\ref{tab:ID_summary}, grouped by cosmic messenger and SM final state.

\begin{table}[ht]
\centering
\renewcommand{\arraystretch}{1.2}
\begin{tabular}{lcccc}
\toprule
\textbf{Final state} & \textbf{Gamma rays} & \textbf{Antiprotons} & \textbf{Positrons} & \textbf{Neutrinos} \\
\midrule
$e^+ e^-$     & \cite{Archambault_2017, acharyya2024} & \cite{Cuoco:2017iax}                       & \cite{John_2021}         & -- \\
$\mu^+ \mu^-$ & \cite{Cirelli:2024ssz}          & \cite{Cuoco:2017iax}                       & \cite{John_2021}         & \cite{Cirelli:2024ssz} \\
$\tau^+ \tau^-$ & \cite{Albert_2017, Magic_2016}      & \cite{Cuoco:2017iax}                        & \cite{John_2021}         & -- \\
$q \bar{q}$ (light) & --                              & \cite{Calore:2022stf,Cuoco:2017iax}     & --                       & -- \\
$c \bar{c}$   & \cite{Archambault_2017}               & \cite{Cuoco:2017iax}                       & --                       & -- \\
$b \bar{b}$   & \cite{armand2021}                     & \cite{Calore:2022stf,Cuoco:2017iax}     & --                       & -- \\
$t \bar{t}$   & \cite{Abdallah_2016, Archambault_2017} & \cite{Cuoco:2017iax}                      & --                       & -- \\
$W^+ W^-$     & \cite{Ackermann_2015, Cirelli:2024ssz} & \cite{Calore:2022stf,Cuoco:2017iax} & --                       & -- \\
$Z Z$         & \cite{Aleksi__2014, Archambault_2017} & \cite{Cuoco:2017iax}                       & --                       & -- \\
$h h$         & --                                   & \cite{Calore:2022stf,Cuoco:2017iax}     & --                       & -- \\
$\nu \bar{\nu}$ (any flavor) & --                    & \cite{Cuoco:2017iax}                        & --                       & \cite{2023IceCubeDM, ANTARES_DM} \\
\bottomrule
\end{tabular}
\caption{Overview of indirect detection constraints used in this analysis, grouped by cosmic messenger and SM final state. References indicate the sources from which the limits were taken. ``--'' indicates that no constraint is included for that channel. All bounds assume a Navarro-Frenk-White %(NFW) 
halo profile, except for~\cite{Abdallah_2016, Aleksi__2014}, which use an Einasto profile.
}

\label{tab:ID_summary}
\end{table}

We use \textsc{micrOMEGAs} to compute the total annihilation cross section~$\langle \sigma v \rangle$ and the branching fractions into different primary final states,~$\mathrm{BR}_\mathrm{prim}$. To apply experimental limits, we compute the partial cross section for each final state as
\begin{equation}
\langle \sigma v \rangle_\mathrm{prim} = \mathrm{BR}_\mathrm{prim} \times \langle \sigma v \rangle \,,
\end{equation}
and compare it to the corresponding upper limits. Parameter points that yield a cross section above the bound for at least one channel are excluded.

{There are two situations in which final states beyond the standard channels listed in Table~\ref{tab:ID_summary} become relevant. In these cases, we apply approximate mappings that allow us to use the available experimental limits in a controlled way. First,} in our models, the DM particle couples to SM flavor triplets, and annihilation through $t$- or $u$-channel mediator exchange can result in mixed final states with different SM flavors. Since these mixed channels are not considered in the available constraints, we approximate the  cross section limit by using the average from the two corresponding single-flavor pair-annihilation channels:
\begin{equation}
    \langle \sigma v \rangle_{AB} = \frac{1}{2} \big( \langle \sigma v \rangle_{AA} + \langle \sigma v \rangle_{BB} \big) \,.
\end{equation}
This is equivalent to the prescription used in \textsc{micrOMEGAs} and can be applied if the DM mass satisfies $M_X > \max(M_A, M_B)$. In reality, mixed final states can occur if $2 M_X > M_A + M_B$, but such configurations are not included in this analysis. Simulating both the single-flavor and mixed-flavor annihilation processes with \textsc{MadDM}~\cite{Ambrogi:2018jqj,Arina:2021gfn}, we have verified that the spectrum from the averaged single-flavor channels approximates the mixed-flavor spectrum with sufficient accuracy to justify this approach.
(For neutrino final states, we neglect flavor-specific effects and sum over all neutrino flavors, consistent with the assumptions in the experimental bounds.)

{Secondly, radiative three-body final states can become relevant. If the two-body final state annihilation is $p$-wave suppressed, as is the case in the scenarios considered in Sec.~\ref{sec:application}, it is inefficient in the present Universe and the processes $X X \to \gamma f \bar f$ and, for quark-philic models, $X X \to g q \bar q$ can provide an important additional contribution, as they lift the $p$-wave suppression~\cite{Bringmann:2007nk,Garny:2011ii}. The process $X X \to \gamma f \bar f$ is automatically computed by \textsc{micrOMEGAs}. For quark final states, we additionally include $X X \to g q \bar q$ using
\begin{equation}
    \frac{\sigma_{X X \rightarrow q\bar q g}}{\sigma_{X X \rightarrow q\bar q \gamma}} = \frac{C_F}{Q_q^2} \frac{\alpha_s}{\alpha_{em}} \, ,
\end{equation}
where $C_F = 4/3$, $Q_q = 2/3$ for up-type quarks and $Q_q = -1/3$ for down-type quarks. Since experimental limits are generally not provided for these three-body final states, we incorporate them by summing the corresponding two- and three-body cross sections and applying the available bound for the associated two-body channel, following Ref.~\cite{Acaroglu:2022hrm}. This amounts to assuming that the continuum part of the spectra from $f\bar f \gamma$ and $q\bar q g$ is approximated sufficiently well by the corresponding $f\bar f$ and $q\bar q$ channels for the purpose of applying the published limits. This is also consistent with the analysis underlying Ref.~\cite{Arina:2023msd}, which indicates that radiative three-body spectra in related $t$-channel mediator models can be approximated reasonably well by standard two-body templates. Possible sharp spectral features from the radiated photon in $f\bar f\gamma$ are not included in this treatment and are beyond the scope of the present analysis.}

\subsection{Collider Constraints}

Collider searches for new physics can constrain regions of the parameter space where DM and/or its mediator are produced at the LHC\@. We use the \textsc{micrOMEGAs} interface~\cite{Barducci:2016pcb} to \textsc{SModelS}~v3~\cite{Alguero:2021dig,MahdiAltakach:2023bdn,Altakach:2024jwk} to evaluate these constraints. \textsc{SModelS} decomposes the full model into simplified model topologies, which are then tested against an extensive database of ATLAS and CMS results. The tool returns $r$-values, defined as the ratio of the predicted signal cross section to the experimental upper limit, and parameter points with $r \geq 1$ are considered excluded. For searches that provide additional statistical information beyond observed limits, \textsc{SModelS} can also compute an approximate likelihood for a given parameter point. However, to avoid potential biases from inconsistent likelihood availability across searches, we base our constraints solely on the $r$-value.

Constraints are typically strongest when the mediator carries QCD charge, as this enhances the production cross section. We also use the \textsc{SModelS} functionality to identify missing topologies, which are categorized as displaced, prompt, or outside the experimental results grid. This helps reveal gaps in the current experimental coverage that are particularly relevant for the scenario under study.

\subsection{Flavor Physics}

\subsubsection{Effective Field Theory Framework for Dark Matter Models}
\label{sec:DM-EFT}
Constraints from flavor observables are best explored by mapping the DM models onto their corresponding Effective Field Theory (EFT) description.\footnote{A similar approach was also followed in~\cite{Biondini:2025gpg} to analyze a $t$-channel DM model with a flavored mediator. However, in~\cite{Biondini:2025gpg} only LEP data was considered in the EFT analysis to constrain the model, whereas our focus here is on flavor observables, while LEP data can be trivially included in our setup.}
To this end, we have implemented the different DM models, with their Lagrangians given in Sec.~\ref{sec:classific}, in the \textsc{Matchete} package~\cite{Fuentes-Martin:2022jrf}, which performs the complete one-loop matching\footnote{See~\cite{Carmona:2021xtq,DasBakshi:2018vni} for other one-loop matching tools with a similar scope.} of the DM theories onto the Standard Model Effective Field Theory~(SMEFT)~\cite{Buchmuller:1985jz} in the \textit{Warsaw basis}~\cite{Grzadkowski:2010es}, for reviews see~\cite{Brivio:2017vri,Isidori:2023pyp,Aebischer:2025qhh}.

Subsequently, we use the Renormalization Group~(RG) equations of the SMEFT~\cite{Jenkins:2013zja,Jenkins:2013wua,Alonso:2013hga} to evolve the SMEFT Wilson coefficients from the high scale, where the NP parameters are defined, down to the electroweak scale. In a next step, the SMEFT is matched onto the Low Energy Effective Theory~(LEFT)~\cite{Jenkins:2017jig,Dekens:2019ept} and subsequently the coefficients of the LEFT are further run down to the energies of the flavor observables using the LEFT RG~equations~\cite{Jenkins:2017dyc}. The RG~evolution and SMEFT-LEFT matching is performed with the help of the \textsc{Wilson} package~\cite{Aebischer:2018bkb} (see also~\cite{Celis:2017hod}), and the flavor observables are computed using \textsc{flavio}~\cite{Straub:2018kue}.

The validity of the EFT description depends, of course, on the masses considered for the DM flavor multiplet~($X$) and the mediator~($Y$).  
The EFT constitutes a power series in~${E_\mathrm{obs}^2}/{M_\mathrm{NP}^2}$, where $E_\mathrm{obs}$ is the typical energy scale of the observables considered and $M_\mathrm{NP}$ is the mass scale of the NP states ($M_{X}$ and~$M_Y$). For flavor observables, we typically have $E_\mathrm{obs} \leq 5\,\mathrm{GeV}$.
Therefore, we expect the EFT to provide an accurate description even for DM masses in the range of a few times $10\,\mathrm{GeV}$.\footnote{For DM masses below the weak scale ($M_\mathrm{NP}<v$), the SMEFT expansion in $v^2/M_\mathrm{NP}^2$ suffers an apparent breakdown. However, since we subsequently match onto the LEFT and evolve to energies at or below the $B$-meson scale, the SMEFT expansion is always supplemented by the LEFT expansion in $E_\mathrm{obs}^2/v^2$. Hence, in combination, we find that the relevant observables have an expansion in $(E_\mathrm{obs}^2/v^2) \times (v^2/M_\mathrm{NP}^2) = E_\mathrm{obs}^2/M_\mathrm{NP}^2$ such that neglected higher-order SMEFT operators are still sufficiently suppressed as long as $E_\mathrm{obs} \ll M_\mathrm{NP}$, which validates our SMEFT expansion a posteriori. }
Notice, that such low masses are also strongly constrained by collider searches, for which we do not employ an EFT approximation.
Thus, first applying the LHC limits before determining the flavor constraints can improve the validity of the EFT description.
When also Higgs observables or electroweak precision data are included, $E_\mathrm{obs}$ is at the electroweak scale and the EFT cannot be expected to yield reliable results for such low DM masses. 
Thus, these constraints should be applied only for DM masses above a few times $100\,\mathrm{GeV}$, as done, e.g., in~\cite{Biondini:2025gpg}.
Ultimately, the mass scale up to which the EFT is deemed reliable must be decided individually for each analysis performed with the setup described here.
Therefore, this choice is left to the user of the toolchain we provide.

\subsubsection{Flavor Likelihood}
To construct a global flavor likelihood to obtain constraints on the DM models from all relevant flavor observables, we employ a variety of publicly available codes, partially mentioned before already.
At the core of the flavor toolchain is the \textsc{smelli} package\footnote{For the present analyses we use \textsc{v2.4.2} of the \textsc{smelli} code.}~\cite{Aebischer:2018iyb}, which allows one to compute a global likelihood in the SMEFT parameter space, including not only flavor observables, but also Higgs and electroweak precision measurements. 
Depending on the concrete scenario one is investigating, it can be convenient to only consider certain subsets of these measurements in the analysis.
A complete list of all available observables can be found in Appendix~D of~\cite{Aebischer:2018iyb}.
The \textsc{smelli} code is based on the \textsc{flavio} software~\cite{Straub:2018kue} to compute flavor observables and on the \textsc{Wilson} package~\cite{Aebischer:2018bkb} to perform the RG evolution in the SMEFT and LEFT, and the matching between the two EFTs.

The input to \textsc{smelli} is provided in terms of \texttt{WCxf} files~\cite{Aebischer:2017ugx} containing the numerical values of all SMEFT Wilson coefficients, corresponding to a specific point in the NP parameter space, at the high UV scale.
This input scale is set to 1\,TeV for the analyses presented in this article, but can be modified in the toolchain as desired.
These \texttt{WCxf} files are generated through a new interface to the \textsc{Matchete} package~\cite{Fuentes-Martin:2022jrf}, which was developed specifically for this project, but can also be used for any other NP theory.
This interface will be included and documented in the \textsc{Matchete} code from \textsc{v0.4.2} onward.
For a given point in NP parameter space, the interface numerically evaluates the one-loop matching conditions determined by \textsc{Matchete}, thus obtaining a list of numerical values for all Warsaw basis Wilson coefficients.
The results are subsequently written into \texttt{WCxf} files (one file per parameter point), which is then used as input to \textsc{smelli} to compute the global likelihood for the specified point.
The details of the parameter scan are, of course, specific to the scenario that is being investigated. 
Therefore, generating the NP parameter points used as input by the interface is left external to the toolchain.

Detailed instructions and examples on how to use the \textsc{Matchete}--\texttt{WCxf} interface\footnote{{This interface is also documented in detail within the \textsc{Matchete} package.}} for the DM models considered here, and for how to subsequently analyze these files with \textsc{smelli}, are provided on GitHub~\href{https://github.com/lena-ra/Flavored-Dark-Matter}{\faicon{github}}.\textsuperscript{\ref{fn:github}}

For every parameter point, \textsc{smelli} computes and returns the global log-likelihood ratio, which is defined as
\begin{align}
    \Delta \cL = \log \left( \frac{\cL_\mathrm{NP}}{\cL_\mathrm{SM}} \right)
    \,,
\end{align}
where $\cL_\mathrm{SM}$ is the likelihood of the SM and $\cL_\mathrm{NP}$ is the likelihood of the NP model at the specified parameter point.
For the analysis presented in this article, we reject all points with $\Delta \cL < -2$, that is, the parameter points that fit the data worse than the SM with a pull of approximately~$2\,\sigma$ with respect to the SM fit.
The threshold for rejecting or accepting points based on the value of the log-likelihood ratio can be easily modified if required.
While additional information on all computed observables is available through \textsc{smelli} in principle, our toolchain also provides a summary of the most relevant observables for every analysis.

\subsubsection{Flavor Observables}
The \textsc{smelli} package allows to include a wide range of observables, which are implemented in \textsc{flavio}, in the global likelihood analysis. 
The different types of available measurements include quark flavor observables, anomalous magnetic moments of leptons, neutrino trident production, charged- and neutral-current lepton-flavor-universality tests, and lepton-flavor-violation~(LFV) tests, but also electroweak precision observables~(EWPO) and Higgs decays.
A~list of all available observables can be found in appendix~D of~\cite{Aebischer:2018iyb} with an up-to-date version on GitHub~\href{https://github.com/smelli/smelli/tree/master/smelli/data/yaml}{\faicon{github}}.\footnote{\url{https://github.com/smelli/smelli/tree/master/smelli/data/yaml}}

Which of the observables are the most relevant depends heavily not only on the model considered but also on the specifics of the parameter scan. 
For example, if the DM couples only to the quark fields~($q,u,d$) the strongest constraints on the parameter space are presumably derived from meson--anti-meson mixing observables ($D^0$--$\bar{D}^0$ mixing in the up sector or $B^0_{d,s}$--$\bar{B}^0_{d,s}$ and $K^0$--$\bar{K}^0$ in the down sector, cf. Sec.~\ref{sec:down-quarks}), at least when the parameter scan allows sufficiently large contributions to these $\Delta F =2$ processes.
For leptophilic scenarios, on the other hand, the most stringently constrained charged-lepton flavor-violating observables such as $\mu \to e \gamma$ or $\mu \to e e e$ are likely to provide the strongest limits on the parameter space, if the scan allows for LFV, cf. Sec.~\ref{sec:rh-leptons}.
If this is not the case, EWPO can potentially give the leading constraints on the models (see, e.g.,~\cite{Biondini:2025gpg}).

Ultimately, the relevant observables have to be determined model-by-model and scan-by-scan.
Notice that once this is done, it can be useful to restrict \textsc{smelli} to compute only the relevant parts of the likelihood, in order to improve performance.

% - - - - APPLICATOONS - - - - %
\section{Applications and Results: Majorana Dark Matter}\label{sec:application}

In this section, we apply our toolchain to two models of Majorana flavored DM which have not yet been studied in the literature. In Sec.~\ref{sec:rh-leptons} we examine the coupling to right-handed charged leptons,\footnote{{While Majorana flavored DM coupled to right-handed charged leptons has been considered before, the analysis of~\cite{Herms:2021fql} was restricted to the coupling to only a single SM generation, and~\cite{Heisig:2024mwr} dealt exclusively with the possibility of obtaining conversion-driven leptogenesis.}} and in Sec.~\ref{sec:down-quarks} we consider the case of a coupling to the right-handed down-type quarks. In both cases, we assume DMFV, i.e.,~we assume $\lambda$ to be the only new source of flavor and CP violation. First, we describe the parametrization framework of $\lambda$ in Sec.~\ref{sec:parametrization}.

\subsection{General Parameterization of the Coupling Matrix and Masses} \label{sec:parametrization}

In this study, we make use of the parameterization framework for the matrix $\lambda$ introduced in~\cite{Acaroglu:2021qae}, which we summarize here for completeness. All expressions relevant for our analysis are presented below, and the reader is referred to~\cite{Acaroglu:2021qae} for additional context and discussion. Note that this parameterization can be used in all DMFV models with real DM. A~useful parameterization for complex DM in DMFV has been derived in~\cite{Agrawal:2014aoa}. It can be obtained from the one for real DM by removing the  orthogonal matrix $O$ and the diagonal matrix $d$ which become unphysical in the complex DM case.

The coupling matrix $\lambda$ is expressed in terms of two diagonal matrices $D$ and $d$, an orthogonal matrix $O$, and a unitary matrix $U$, respecting the flavor symmetry group $O(3)$:
\begin{equation}
\lambda = U \, \text{diag}(D_1, D_2, D_3) \, O \, \text{diag}(d_1, d_2, d_3) \, .
\label{eq:lambda-param}
\end{equation}
The unitary matrix $U$ is constructed as a product of three unitary rotations:
\begin{align}
    \label{eq:U_parameterization}
    U &= U_{23} \, U_{13} \, U_{12} \nonumber \\
    &= \begin{pmatrix}
    1 & 0 & 0 \\
    0 & c_{23}^\theta & s_{23}^\theta e^{-i \delta_{23}} \\
    0 & -s_{23}^\theta e^{i \delta_{23}} & c_{23}^\theta
    \end{pmatrix}
    \begin{pmatrix}
    c_{13}^\theta & 0 & s_{13}^\theta e^{-i \delta_{13}} \\
    0 & 1 & 0 \\
    -s_{13}^\theta e^{i \delta_{13}} & 0 & c_{13}^\theta
    \end{pmatrix}
    \begin{pmatrix}
    c_{12}^\theta & s_{12}^\theta e^{-i \delta_{12}} & 0 \\
    -s_{12}^\theta e^{i \delta_{12}} & c_{12}^\theta & 0 \\
    0 & 0 & 1
    \end{pmatrix} \, ,
\end{align}
where $c_{ij}^\theta = \cos(\theta_{ij})$ and $s_{ij}^\theta = \sin(\theta_{ij})$.
Similarly, the orthogonal matrix $O$ is defined as
\begin{equation}
    \label{eq:O_parameterization}
    O = O_{23} \, O_{13} \, O_{12}
    = \begin{pmatrix}
    1 & 0 & 0 \\
    0 & c_{23}^\phi & s_{23}^\phi \\
    0 & -s_{23}^\phi & c_{23}^\phi
    \end{pmatrix}
    \begin{pmatrix}
    c_{13}^\phi & 0 & s_{13}^\phi \\
    0 & 1 & 0 \\
    -s_{13}^\phi & 0 & c_{13}^\phi
    \end{pmatrix}
    \begin{pmatrix}
    c_{12}^\phi & s_{12}^\phi & 0 \\
    -s_{12}^\phi & c_{12}^\phi & 0 \\
    0 & 0 & 1
    \end{pmatrix} \, ,
\end{equation}
with $c_{ij}^\phi = \cos(\phi_{ij})$ and $s_{ij}^\phi = \sin(\phi_{ij})$. The diagonal unitary matrix $d$ is given by
\begin{equation}
d = \text{diag}(e^{i\gamma_1}, e^{i\gamma_2}, e^{i\gamma_3}) \, .
\end{equation}
In total, the coupling matrix $\lambda$ is characterized by the following 15 parameters:
\begin{equation}
\label{eq:params_coupling_matrix}
\theta_{23}, \theta_{13}, \theta_{12}, \phi_{23}, \phi_{13}, \phi_{12}, \delta_{23}, \delta_{13}, \delta_{12}, \gamma_1, \gamma_2, \gamma_3, D_1, D_2, D_3 \, .
\end{equation}

The DM mass matrix is parameterized in terms of $\lambda$ via the DMFV spurion expansion:
\begin{equation}
\tilde{M}_{\tilde{\chi}} = M_{\tilde{\chi}} \left[ \unit + \frac{\eta}{2} (\lambda^\dagger \lambda + \lambda^\intercal \lambda^*) + \mathcal{O}(\lambda^4)\right] \, .
\label{eq:mass-matrix}
\end{equation}
To diagonalize $\tilde{M}_{\tilde{\chi}}$, we apply the Autonne–Takagi factorization, yielding an orthogonal matrix $W$ such that
\begin{equation}\label{eq:W}
\tilde{M}_{\tilde{\chi}} = W^\intercal \tilde{M}_{\tilde{\chi}}^D W \, ,
\end{equation}
with $\tilde{M}_{\tilde{\chi}}^D = \text{diag}(M_{\tilde{\chi}_1}, M_{\tilde{\chi}_2}, M_{\tilde{\chi}_3})$ ordered such that $M_{\tilde{\chi}_1} < M_{\tilde{\chi}_2} < M_{\tilde{\chi}_3}$. The coupling matrix is accordingly rotated to the mass eigenbasis via $\tilde{\lambda} = \lambda W^\intercal$. Note that for vanishing phases $\gamma_i$ ($d=\unit$) the orthogonal transformation~$O$  simply amounts to a dark flavor symmetry transformation and is hence redundant such that $W=O$; see Appendix~\ref{app:DFBT} for further details on the dark flavor basis transformation.

Unlike previous analyses of Majorana flavored DM with DMFV in~\cite{Acaroglu:2021qae, Acaroglu:2023phy}, we allow for interactions of the mediator with the Higgs doublet, which modify the mediator’s physical mass after EWSB. The nature of this correction depends on the mediator’s gauge quantum numbers as discussed in Sec.~\ref{sec:classific}.

\subsection{Coupling to Right-Handed Charged Leptons}
\label{sec:rh-leptons}

\subsubsection{Scan Setup}
We consider a Majorana fermion DM flavor-triplet field and scalar mediator coupling to right-handed charged leptons. We perform a random scan of the model parameters as presented in Tab.~\ref{tab:scan_ranges-leptophilic} where the prior indicates whether the parameters are sampled linearly (lin) or logarithmically (log) in their scan range.\footnote{We note that to cover the full physical parameter space, one of the angles $\phi_{ij}$ should cover both positive and negative values between $0$ and $\pm \pi/4$. We have checked that restricting our scans to include only positive values does not impact our results.} To suppress LFV decays like $\mu \to e\gamma$, the mixing angles $\theta_{12}$ and $\phi_{12}$ are scanned logarithmically over $[10^{-6}, 10^{-2}]$. Because the mass matrix depends on the sign of $\eta$, which can be negative, we discard parameter points where the lightest mass eigenstate has a negative mass, guaranteeing $M_{\tilde{\chi}_1} > 0$.

\begin{table}[htb]
	\centering
	\begin{tabular}{c|c|c} 
    \textbf{Parameter}& \textbf{Range}& \textbf{Prior} \\ 
    \hline
	$D_i$ & $[0,2]$ & lin \\ 
    $\theta_{12}$& $[10^{-6},10^{-2}]$& log\\
    $\theta_{13}$& $[10^{-4},\pi/4]$& log\\
    $\theta_{23}$& $[10^{-4},\pi/4]$& log\\
    $\phi_{12}$& $[10^{-6},10^{-2}]$& log\\
    $\phi_{13}$& $[10^{-4},\pi/4]$& log\\
    $\phi_{23}$& $[10^{-4},\pi/4]$& log\\
    $\delta_{ij}$& $[0,2\pi)$& lin\\
    $\gamma_i$& $[0,2\pi)$& lin\\
    $\lambda_{\varphi H,1}$ & $[-1,-0.001]\cup [0.001, 1]$& log\\
    \hline
    $\eta$& $[-1,-0.001]\cup [0.001, 1]$& log\\ 
    $M_{\tilde{\chi}}\,[\mathrm{GeV}]$& $[100,1500]$& lin\\
    $M_\varphi\,[\mathrm{GeV}]$& $[M_{\tilde{\chi}_1},M_{\tilde{\chi}_1}+2000]$& lin\\
    \hline
	\end{tabular}
    \caption{Ranges and priors for the parameters for the scan with couplings to right-handed leptons. 
    The priors indicate whether the scan range is sampled linearly~(lin) or logarithmically~(log). 
    The upper part of the table shows the parameters needed for the couplings and the lower part shows the mass parameters.} 
	\label{tab:scan_ranges-leptophilic}
\end{table}

\subsubsection{Constraints and Results}

\begin{figure}
    \centering
    \begin{subfigure}[t]{0.49\textwidth}
        \includegraphics[width=\textwidth]{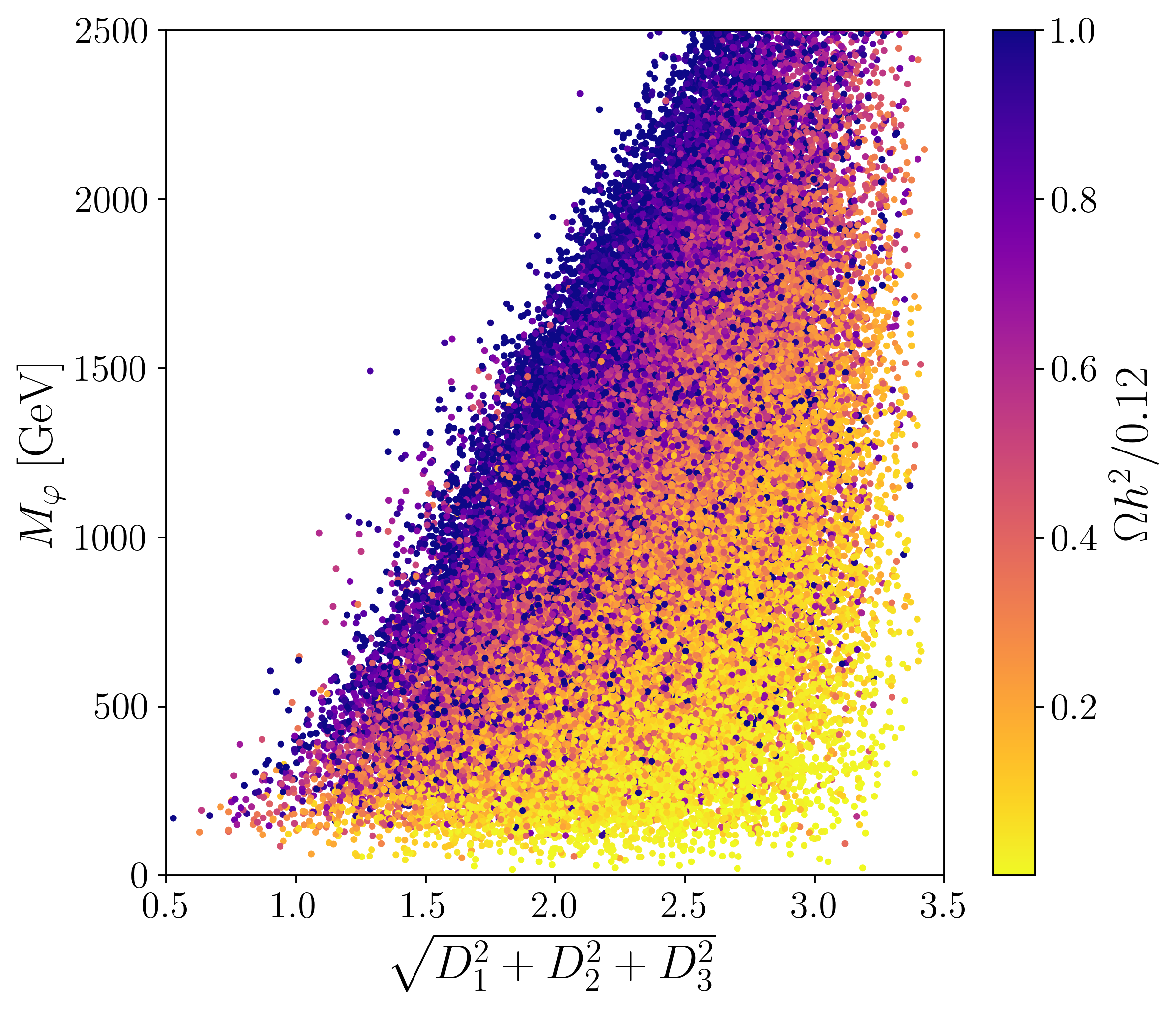}
    \end{subfigure}
    \begin{subfigure}[t]{0.49\textwidth}
        \includegraphics[width=\textwidth]{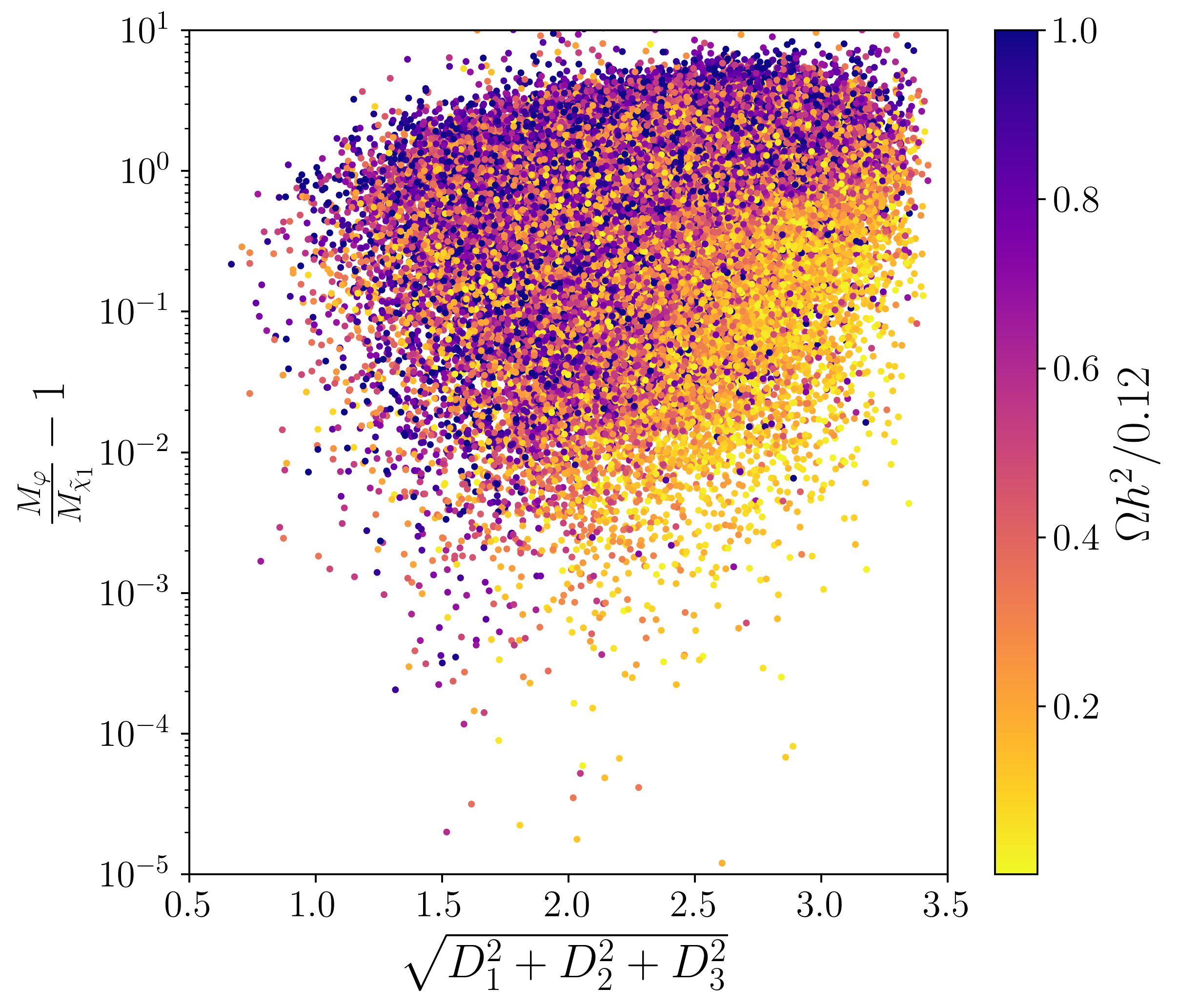}
    \end{subfigure}
    \caption{{Calculated relic abundance shown in color as indicated by the color scale. The left plot shows the parameter space in the $\sqrt{D_1^2+D_2^2+D_3^2}$--$M_\varphi$ plane, while the right plot shows the plane of $\sqrt{D_1^2+D_2^2+D_3^2}$ and $M_\varphi/M_{\tilde\chi_1}-1$.}}
    \label{fig:relic_abundance_leptons}
\end{figure}

\begin{figure}
    \centering
    \begin{subfigure}[t]{0.49\textwidth}
        \includegraphics[width=\textwidth]{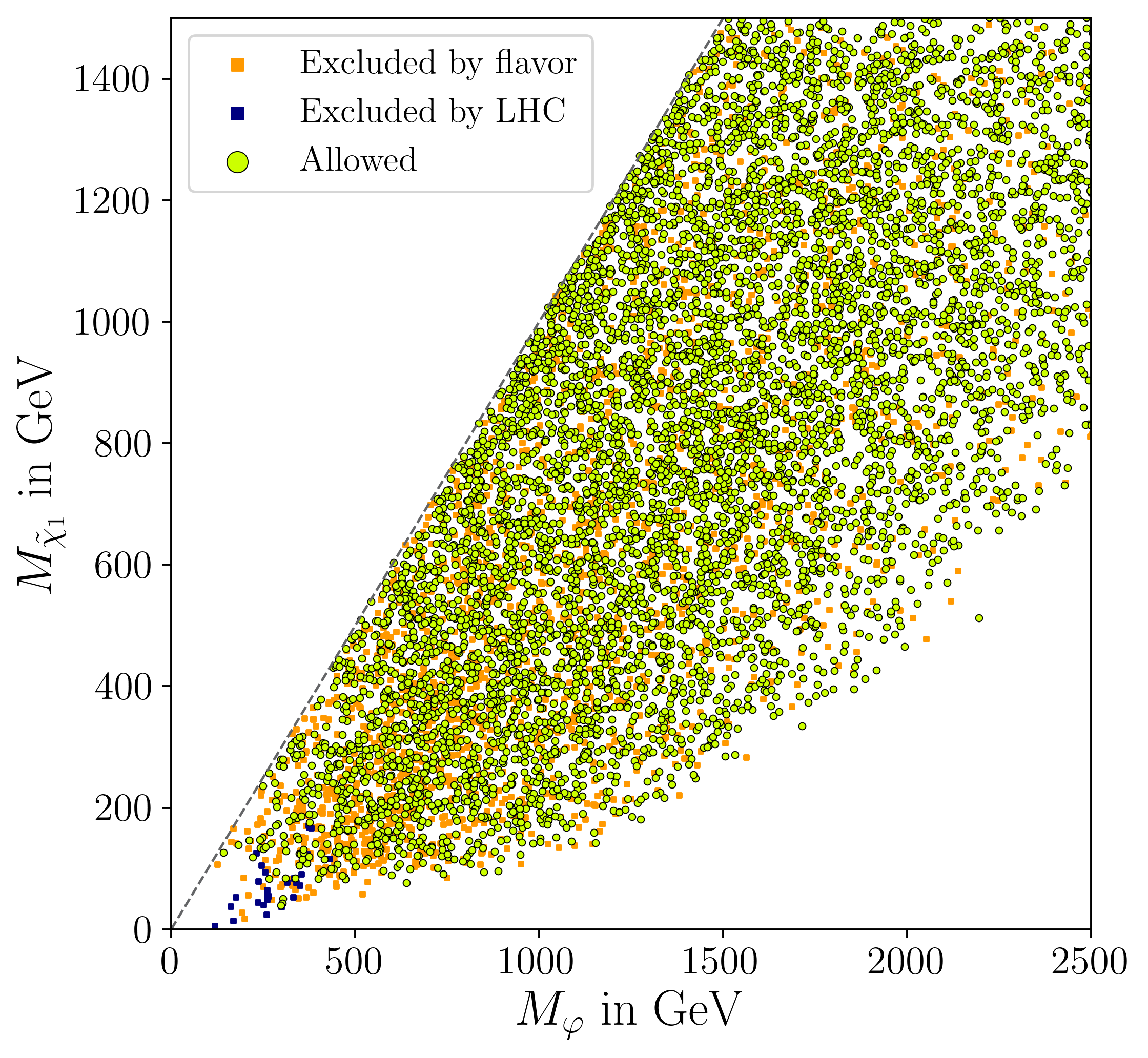}
    \end{subfigure}
    \begin{subfigure}[t]{0.49\textwidth}
        \includegraphics[width=\textwidth]{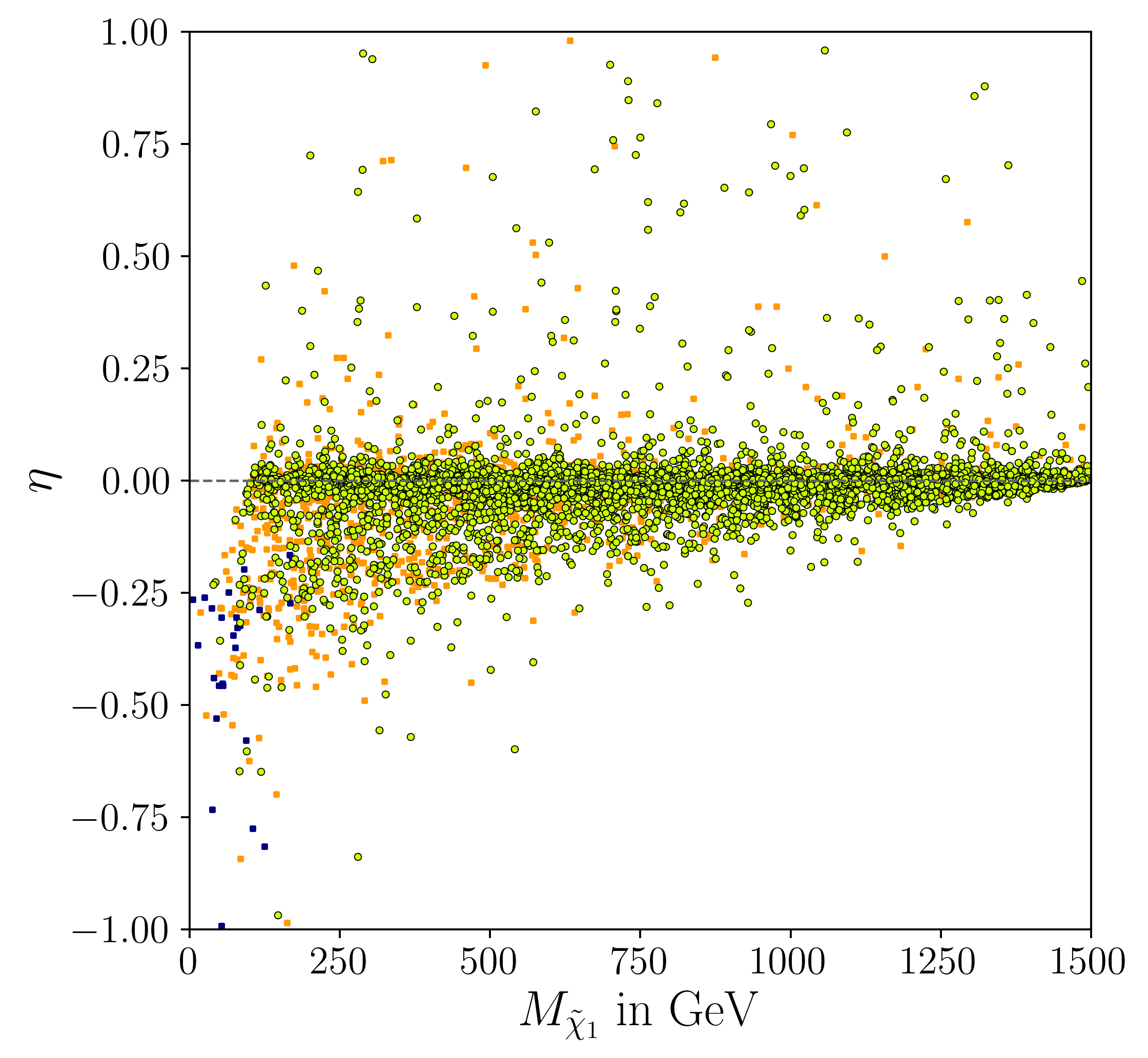}
    \end{subfigure}
    \begin{subfigure}[t]{0.49\textwidth}
        \includegraphics[width=\textwidth]{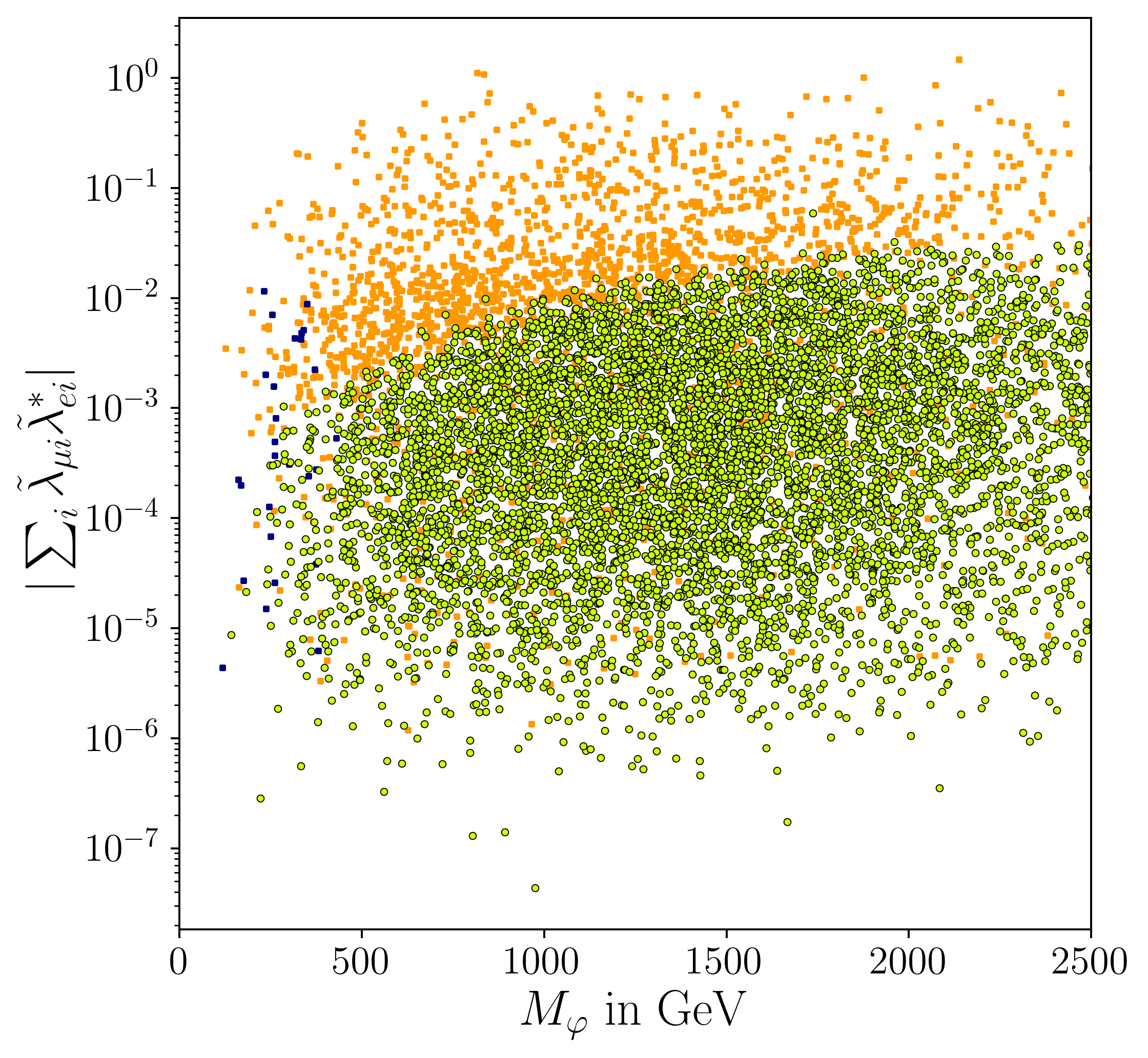}
    \end{subfigure}
    \begin{subfigure}[t]{0.49\textwidth}
        \includegraphics[width=\textwidth]{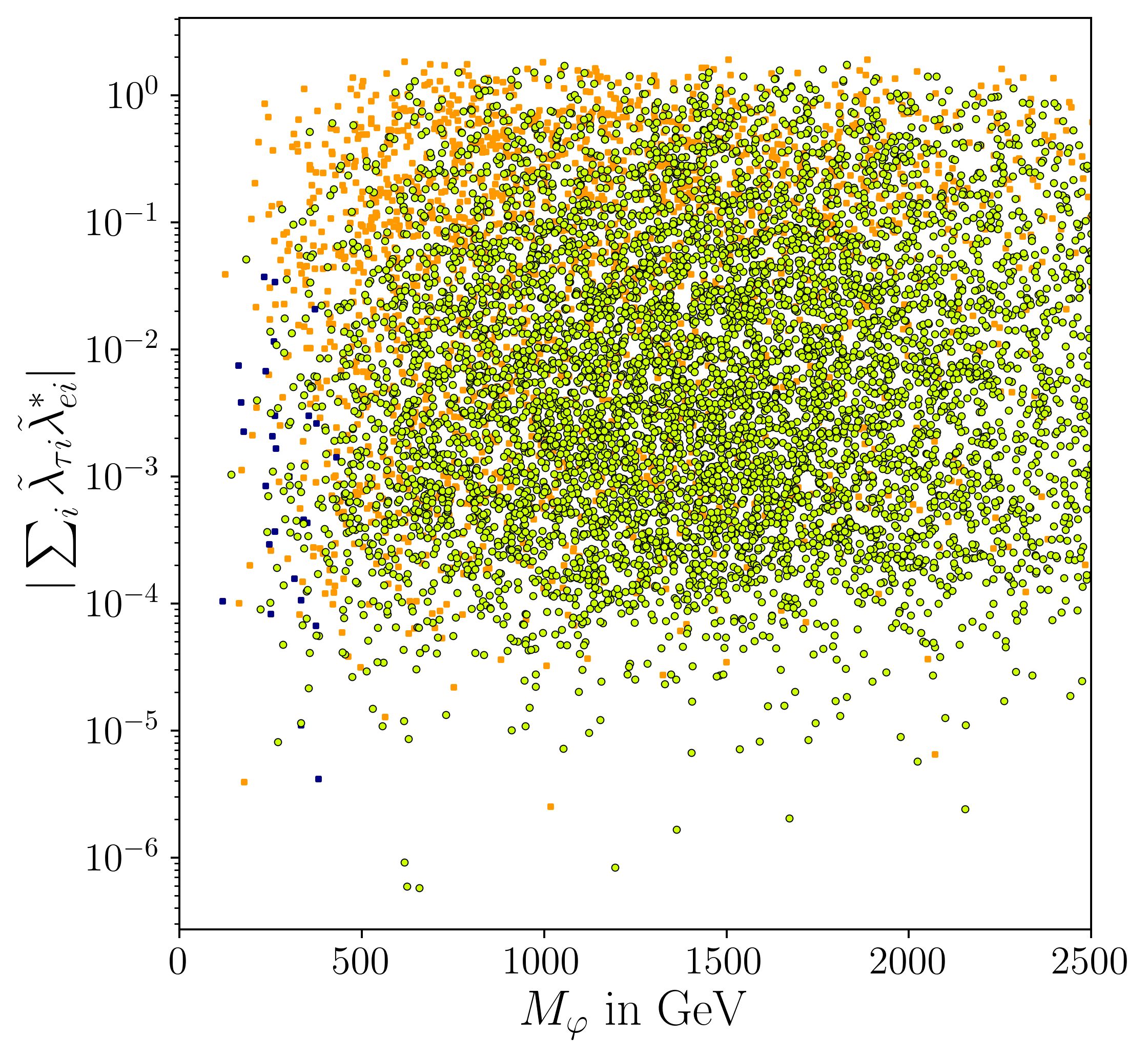}
    \end{subfigure}
    \caption{Constraints on the parameter space for the Majorana DM model coupling to right-handed leptons. Points shown by orange square markers are excluded by flavor observables while still passing the LHC and relic abundance constraints. Blue square markers fulfill only the relic abundance constraints but are excluded by constraints from the LHC. The remaining points that pass all constraints are shown with green circles. The upper left panel shows the parameter space in the $M_{\tilde{\chi}_1}$–$M_\varphi$ plane, with the gray dashed line indicating $M_{\tilde{\chi}_1} = M_\varphi$. The upper right panel displays the mass splitting parameter $\eta$ versus $M_{\tilde{\chi}_1}$. The lower panels show the sum over all DM states of $\tilde{\lambda}_{\mu i} \tilde{\lambda}^*_{e i}$ on the left and $\tilde{\lambda}_{\tau i} \tilde{\lambda}^*_{e i}$ on the right versus $M_\varphi$.}
    \label{fig:allowed_leptons}
\end{figure}

\begin{figure}
    \centering
        \includegraphics[width=0.5\textwidth]{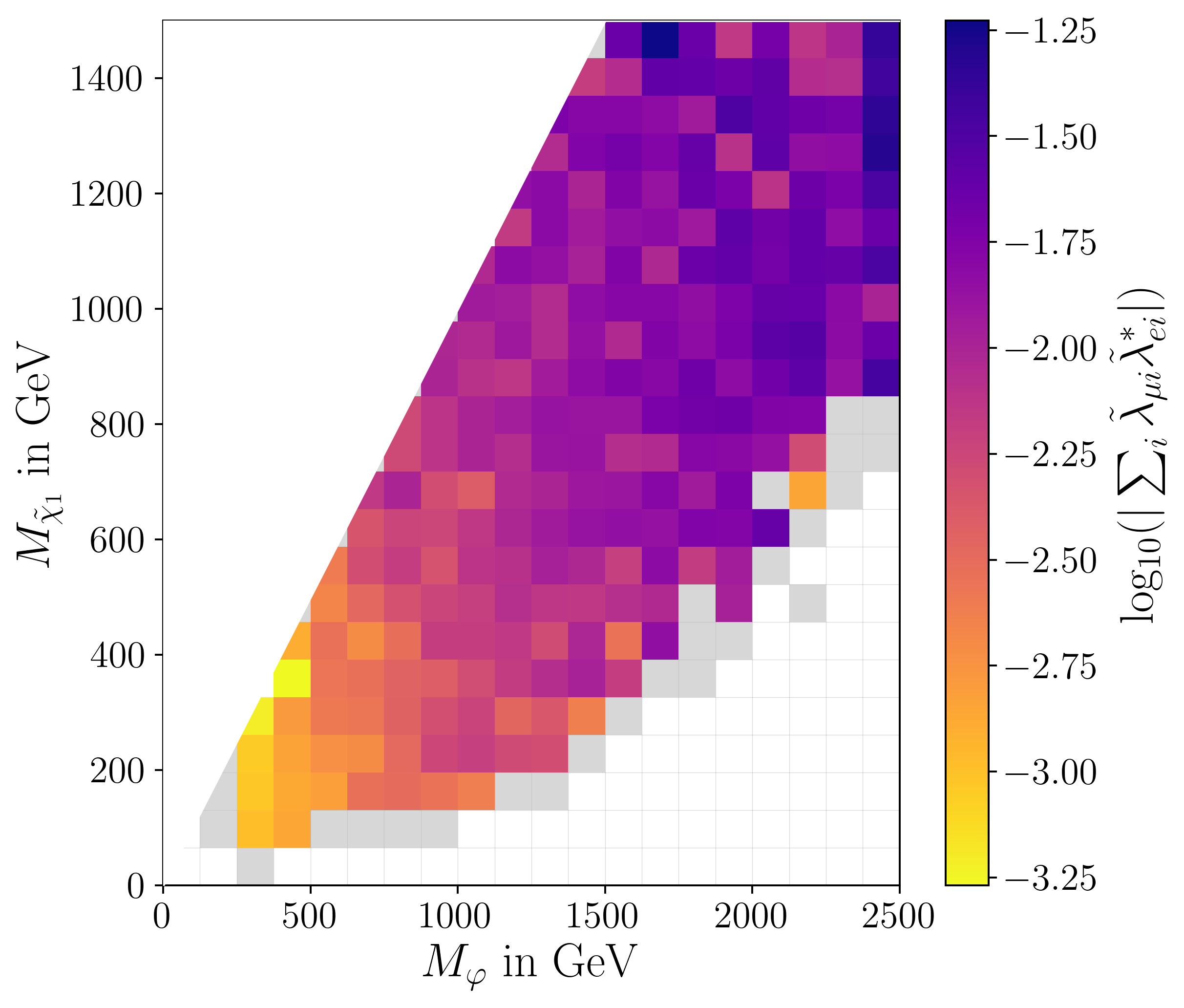}
    \caption{{Largest (absolute) value of the sum over all DM states of $\tilde{\lambda}_{\mu i}\tilde{\lambda}^*_{e i}$ found among the flavor-allowed points in each bin of the $M_{\tilde{\chi}_1}$–$M_\varphi$ plane. This provides a scan-based estimate of the maximally allowed coupling strength in different mass regions. Empty bins are left blank, while bins containing fewer than ten points are shown in gray.}}
    \label{fig:couplings_leptons}
\end{figure}

We generate approximately 350,000 points and sequentially apply constraints from relic density, direct and indirect detection, collider searches (via \textsc{micrOMEGAs} and \textsc{SModelS}), and flavor observables (\textsc{Matchete}, \textsc{smelli}). {The results are presented in Figs.~\ref{fig:relic_abundance_leptons}--\ref{fig:couplings_leptons}.} 
Points surviving all constraints are shown with lime green circles in Fig.~\ref{fig:allowed_leptons}. The top left panel displays allowed values in the $(M_{\tilde{\chi}_1}, M_\varphi)$ plane, with the gray dashed line denoting the minimal mediator mass. The top right panel shows the allowed mass-splitting parameter $\eta$ versus $M_{\tilde{\chi}_1}$. The bottom panels present viable couplings in terms of the absolute value of the sum over all DM states of $\tilde{\lambda}_{\mu i} \tilde{\lambda}^*_{e i}$ on the left and $\tilde{\lambda}_{\tau i} \tilde{\lambda}^*_{e i}$ on the right for a given $M_\varphi$. Points excluded by specific constraints are shown with square-shaped markers in blue (collider) and orange (flavor).

\paragraph{Relic density constraints:}
{In Fig.\ \ref{fig:relic_abundance_leptons} we show the calculated relic density in two different projections of the parameter space, as described in the figure caption. Parameter points that predict overabundant DM have been discarded and are not shown. We observe that the minimum allowed value of the coupling strength $\sqrt{D_1^2+D_2^2+D_3^2}$ increases with increasing mediator mass $M_\varphi$, and that larger couplings tend to favor underabundant DM. Furthermore, the calculated relic abundance depends on the mass splitting between DM and mediator, $M_\varphi/M_{\tilde\chi_1}-1$. For sizable coupling strengths, $\sqrt{D_1^2+D_2^2+D_3^2}\gtrsim 2$, the mass splitting required to reproduce the observed relic density increases.}

{In the following, unless stated otherwise, we focus on the case in which the model reproduces the full observed DM abundance. We return to the case of underabundant DM at the end of this subsection. Requiring the model to reproduce the observed relic density poses a fairly restrictive constraint in our parameter space scan. We find that approximately 97\% of the scanned points fail to yield the measured DM abundance within the assigned theoretical uncertainty.}
Given the large fraction of excluded points, we only show those that satisfy this constraint in {Fig.~\ref{fig:allowed_leptons}}. The dominant contribution to the relic abundance stems from coannihilations between the different $\tilde{\chi}_i$ states into leptons, mediated by $t$- and $u$-channel exchange of the scalar mediator. For small couplings, however, the main process for DM depletion is mediator pair annihilation into photons 
as the DM particles and the mediator are still in thermal equilibrium because they are almost mass degenerate.

A key feature of the viable points is a lower bound on the DM mass~$M_{\tilde{\chi}_1}$ for a given mediator mass. This behavior arises from the scaling of the annihilation cross section, which roughly grows as $M_{\tilde{\chi}_1}^2 / M_\varphi^4$ in the limit of sizable mass splittings and fixed couplings. Since the mediator mass $M_\varphi$ is bounded from below, this scaling implies that for sufficiently small $M_{\tilde{\chi}_1}$, the annihilation cross section becomes too small to yield the correct relic abundance, even when the coupling matrix elements $D_i$ saturate their maximal values. This effect is reflected in the absence of points in the lower right area in the upper left panel of Fig.~\ref{fig:allowed_leptons}.\footnote{Note that the unpopulated region extends beyond the region that is not covered by our parameter scan.}

We also observe a dependence of the allowed points on the mass splitting parameter~$\eta$. Large positive values of~$\eta$ are allowed across the entire mass range of $M_{\tilde{\chi}_1}$, whereas large negative values are only compatible with small DM masses. This is because in our scan we choose the upper value for $M_{\tilde{\chi}}$ to be 1500\,GeV meaning that for $\eta < 0$ the individual values for the masses $M_{\tilde{\chi}_i}$ will be smaller than 1500\,GeV. For~$M_{\tilde{\chi}_1}$ to be close to 1500\,GeV if the absolute value for $\eta < 0$ is large, very small couplings are needed to have very small mass splittings. In this case, the relic abundance becomes too large so that these points are excluded. Note that the region with $M_{\tilde{\chi_1}} < 100$\,GeV is only populated for negative $\eta$ since $M_{\tilde{\chi}}\ge 100$\,GeV. Note also that the absence of points in the lower-left corner in the upper left panel of Fig.~\ref{fig:allowed_leptons} at $M_\varphi\lesssim 200\,$GeV (and relatively small mass splittings) is due to the relic abundance being too low in this region, as efficient coannihilation with the mediator over-depletes the DM density even for small DM couplings; this region could, however, become viable in the regime of conversion-driven freeze-out for very small couplings $\lesssim 10^{-6}$.\footnote{See Refs.~\cite{Herms:2021fql,Heisig:2024mwr} for a respective analysis in the case of two DM flavors.}

\paragraph{Collider constraints:}
We now turn to constraints from LHC searches, evaluated using \textsc{SModelS}. These searches exclude only 0.3\% of the parameter points that have the correct relic abundance, all confined to the low-mass region with $M_{\tilde{\chi}_1} < 200\,\mathrm{GeV}$ and $M_\varphi < 500\,\mathrm{GeV}$ (see upper left panel of Fig.~\ref{fig:allowed_leptons}), while the bulk of the parameter space remains unconstrained. This result is similar to the findings of Ref.~\cite{Acaroglu:2022hrm,Acaroglu:2023cza}, where complex scalar DM coupling to right-handed leptons was studied. 

As seen in the upper right panel of Fig.~\ref{fig:allowed_leptons}, points with negative~$\eta$ and small~$M_{\tilde{\chi}_1}$ are most affected.\footnote{Note that our scan setup does not cover the mass range $M_{\tilde{\chi}_1} < 100\,$GeV for positive~$\eta$.} For negative~$\eta$, $\tilde{\chi}_1$~has the largest coupling, leading to dominant mediator decays into $\tilde{\chi}_1$ and leptons, resulting in two-lepton plus missing energy signatures well covered by \textsc{SModelS}.  For positive~$\eta$, by contrast, the mediator decays dominantly into heavier $\tilde{\chi}_i$ states, which subsequently undergo cascade decays producing multi-lepton final states not covered by current searches implemented in \textsc{SModelS}. As a result, the parameter space with negative $\eta$ is more effectively constrained, although dedicated multilepton searches could significantly enhance sensitivity.

The dominant LHC signal arises from Drell-Yan production of mediator pairs decaying to leptons and missing energy. Although Majorana DM allows for same-sign dileptons, these are suppressed by initial-state statistics~\cite{Buonocore_2020}, and dedicated searches are lacking. Mixed-flavor channels are also possible~\cite{Acaroglu:2022hrm} but to our knowledge are not covered by experimental searches and are not included in \textsc{SModelS}. These channels are also the most frequently identified prompt missing topologies by \textsc{SModelS}, suggesting that these merit further experimental attention.

The most stringent constraint comes from a CMS search for two same-flavor, opposite-sign light leptons plus missing energy~($\slashed{E}_T$)~\cite{CMS-SUS-20-001}, which excludes points with large $\tilde{\chi}_1$--$\mu$ couplings. Other searches~\cite{ATLAS:2014zve, CMS:2018eqb} are less sensitive, excluding only one point with large $\tilde{\chi}_1$--$e$ couplings. Searches for $\tau\bar{\tau} + \slashed{E}_T$ exclude no points, explaining why large $\tilde{\chi}_1$–$\tau$ couplings remain allowed. 

Finally, as the model couples to right-handed leptons, lepton colliders like LEP can constrain it via $e^+e^- \rightarrow \tilde{\chi}_i \tilde{\chi}_j \gamma$ resulting in mono-photon~+~$\slashed{E}_T$ signatures. While not included in \textsc{SModelS}, we expect LEP to set a lower bound of about 100\,GeV on the mediator mass~\cite{DELPHI_2003}, which we find to be respected by all points satisfying the relic abundance constraint in our scan. 

\paragraph{Direct detection constraints:} In lepton-flavored DM models, tree-level interactions with nuclei are absent, making direct detection constraints significantly weaker than in scenarios with quark couplings. The only possible contributions arise from scattering off electrons or from loop-induced couplings to nuclei. The latter proceeds dominantly through photon exchange via the anapole operator. However, following the findings of Ref.~\cite{Kawamura:2020qxo}, we do not expect the current experimental data to impose constraints on the parameter space.\footnote{An interpretation of XENON1T, SuperCDMS and CRESST-III data in terms of the anapole moment can be found in~\cite{Ibarra:2022nzm}.}

\paragraph{Indirect detection constraints:} {Indirect detection experiments impose only weak constraints on our model, with no points satisfying the relic abundance requirement being excluded in our scan. In the leptophilic scenario, the leading two-body annihilation channel $\tilde{\chi}_1 \tilde{\chi}_1 \to \ell^+ \ell^-$ is $p$-wave suppressed and hence inefficient in the present Universe. We therefore include the radiative three-body process $\tilde{\chi}_1 \tilde{\chi}_1 \to \gamma \ell^+ \ell^-$ according to the prescription described in Sec.~\ref{sec
:IDgen}. Even with this contribution included, no parameter point in our scan is excluded by indirect detection.}

\paragraph{Flavor constraints:}
We finally apply flavor constraints to the points that remain after imposing relic density and collider bounds. These observables exclude an additional $\sim25\%$ of the parameter space, as shown in orange in Fig.~\ref{fig:allowed_leptons}. The most stringent bound arises from the LFV decay $\mu \to e\gamma$~\cite{ParticleDataGroup:2016lqr}, with $\mu \to eee$~\cite{ParticleDataGroup:2016lqr} providing a complementary constraint. While the latter can occur via internal photon conversion in $\mu \to e\gamma$, its dominant contribution stems from $Z$-penguin and in particular box diagrams involving $\varphi$ and $\tilde{\chi}_i$. This observable has not been included in previous analyses of Dirac~\cite{Chen:2015jkt} or scalar~\cite{Acaroglu:2022hrm} lepton-flavored DM.

As expected, LFV constraints are stronger at low mediator and DM masses, where loop contributions are less suppressed. 
{This is illustrated in Fig.~\ref{fig:couplings_leptons}, where we show, in each bin of the $M_{\tilde{\chi}_1}$–$M_\varphi$ plane, the largest value of the coupling combination entering $\mu\to e$ flavor transitions found among the flavor-allowed points in our scan. This provides a scan-based estimate of the maximally allowed coupling strength as a function of the DM and mediator masses.}
Moreover, the upper right panel of Fig.~\ref{fig:allowed_leptons} seems to show a bias toward excluding points with small absolute values of $\eta$, though this is a scan artifact due to the logarithmic prior and linear plotting scale. The exclusion fraction is in fact independent of the value of $\eta$.

Flavor bounds primarily constrain the structure of the coupling matrix~$\tilde{\lambda}$, which controls the size of LFV effects. Large couplings to both the electron and muon are strongly disfavored, as shown in the lower left panel of Fig.~\ref{fig:allowed_leptons}, while scenarios with hierarchical couplings -- where only one of the first two rows of $\tilde{\lambda}$  dominates -- can remain viable. The constraint on the relevant coupling parameters weakens with increasing mediator mass as the latter suppresses LFV amplitudes.
Constraints from $\tau$ LFV decays~\cite{ParticleDataGroup:2016lqr,Hayasaka:2010np} are generally weaker, as seen in the lower right panel of Fig.~\ref{fig:allowed_leptons}, and only become relevant for small values of $M_\varphi$. Even in that case larger $\tilde{\lambda}_{\tau i}$ values remain allowed unless accompanied by a sizable coupling to $e$ or $\mu$.

{\paragraph{Underabundant dark matter:} In order to assess the impact of the relic density constraint on our results, we also compared the case of our model reproducing the measured abundance with the cases of only 10\% or 1\% of the observed DM density stemming from our model. As before, we assumed a 10\% theoretical uncertainty in the relic density computation. Compared to about 3\% of the generated points reproducing the full relic density, in the two underabundant scenarios only 0.4\% and 0.03\% of points, respectively, pass the corresponding relic density constraint, leading to significantly less statistics for further analysis. At the same time, the preference of underabundant DM for larger coupling strengths tends to enhance the impact of collider and flavor constraints, while direct and indirect detection still exclude no parameter points.}

\subsection{Coupling to Right-Handed Down-Type Quarks}
\label{sec:down-quarks}

As a second example, we investigated again a Majorana-fermion DM field but coupling to right-handed down-type quarks, which has also not been discussed in the literature so far. 
Because the flavor constraints, particularly meson–anti-meson mixing limits, are even stronger in this case than for the leptophilic model, we perform dedicated scans guided by the insights from Appendix~\ref{app:lambda} to efficiently probe the phenomenologically viable regions of parameter space. 
In the following, we first motivate our scan setup and assumptions, before discussing the results of our analysis.

\subsubsection{Scan Setups}
\label{sec:scan-setup}
\begin{figure}[tbp]
    \centering
    \includegraphics[width=0.85\linewidth]{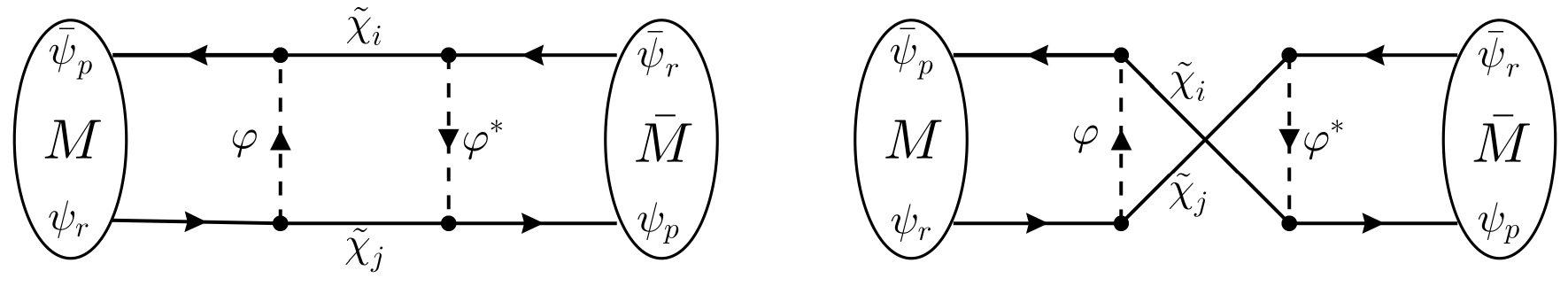}
    \caption{Feynman diagrams contributing to meson--anti-meson mixing in quarkphilic Majorana DM models. While the diagram on the left-hand side is present also in Dirac models, the diagram on the right-hand side only contributes in the {Majorana} scenario.}
    \label{fig:meson-mixing-diag}
\end{figure}

In order to have a sufficient number of parameter points that meet the strong constraints deriving from $\Delta F=2$ processes, we have to design an efficient parameter scan setup with care.
Since we are dealing with Majorana fermions, the common box diagram shown on the left-hand side of Fig.~\ref{fig:meson-mixing-diag}, which contributes to meson--anti-meson mixing, is supplemented by the crossed box diagram shown on the right-hand side of Fig.~\ref{fig:meson-mixing-diag}~\cite{Acaroglu:2021qae}.
The former diagram is proportional to $\tilde{\lambda}_{pi}\tilde{\lambda}^\ast_{ri}\tilde{\lambda}_{pj} \tilde{\lambda}^\ast_{rj}$, whereas the latter scales like ${\tilde{\lambda}_{pi}}^2\,\tilde{\lambda}_{rj}^{\ast\,2}$.
This indicates that large contributions to $\Delta F=2$ processes can be present for Majorana models, even if all off-diagonal entries of~$\tilde{\lambda}$ vanish. 
Already the presence of two non-negligible entries of~$\tilde{\lambda}$ suffices for a parameter point to be in tension with meson-mixing data.

In practice, this means that we want a strong hierarchy between the eigenvalues of the coupling matrix~$\tilde{\lambda}$ in order to sufficiently suppress these flavor-changing processes. 
We implemented this hierarchy in two different ways that we dub \emph{general hierarchical} and \emph{bottom-philic} scenarios.
In the following, we compare both the setups and their results.
We start by describing the details of the two scan setups and qualitatively motivating them. 
A~more detailed and quantitative argument for the exact structure of these scans is provided in Appendix~\ref{app:lambda}.

\paragraph{General hierarchical scenario:}  
In order to efficiently scan the parameters \( D_1 \), \( D_2 \), and \( D_3 \), which determine the magnitude of the eigenvalues of the coupling matrix~$\lambda$ [cf. Eq.~\eqref{eq:lambda-param}], we sample the values such that their product lies within a specified range while allowing for hierarchical structures among them. 
Specifically, we sample:
\begin{itemize}
    \item a total scale $ d_{\text{mul}} = D_1 D_2 D_3 \in [(10^{-4})^3, 2^3]$ logarithmically, which sets the overall coupling strength of the DM multiplet to the SM, allowing efficient exploration of very small as well as large values; 
    \item two independent ratios \( r_{12} = D_1/D_2 \) and \( r_{23} = D_2/D_3 \) logarithmically in the range \( [5\times 10^{-5},\, 2\times 10^{4}] \), which ensures that most parameter points obey a hierarchy among the eigenvalues of~$\lambda$ as discussed before.
\end{itemize}
The individual parameters are obtained as
\begin{align}
    D_1 &= \left( d_{\text{mul}} \, r_{12}^2 \, r_{23} \right)^{1/3}, 
    &
    D_2 &= \left( d_{\text{mul}} \, \frac{r_{23}}{r_{12}} \right)^{1/3}, 
    &
    D_3 &= \left( \frac{d_{\text{mul}}}{r_{12} \, r_{23}^2} \right)^{1/3},
\end{align}
and we retain only those points for which all $D_i$ lie within the range $[10^{-4},2]$. The upper bound of that range ensures that the UV couplings remain perturbative, whereas the lower bound is chosen conservatively to {guarantee} that all dark-sector states remain in thermal equilibrium with each other during freeze-out, consistent with the assumptions underlying the standard \textsc{micrOMEGAs} routines used in the scan. The remaining model parameters are scanned according to the ranges and priors listed in Table~\ref{tab:scan_ranges}. Note that the mixing angles $\theta$ and $\phi$ are scanned with logarithmic priors, ensuring efficient coverage of small angles.\footnote{The structure of the general hierarchical scan is closely related to the first scenario discussed in Appendices~\ref{app:scenarios} and~\ref{app:neutral-mesons}, where small mixing angles $\theta_{ij}$, $\phi_{ij}$ are assumed (i.e.~suppressed off-diagonal entries in~$\lambda$). However, for sizable couplings (as required by the relic-density constraint in the bulk of parameter space) small angles alone are not always sufficient to satisfy the bounds from neutral-meson mixing -- in particular when $M_\varphi \sim M_{\tilde{\chi}}$, where the box diagram on the right side of Fig.~\ref{fig:meson-mixing-diag} is relevant.
In this regime, an additional hierarchy among the diagonal entries of~$\lambda$ is needed, as shown, for instance, in Eq.~\eqref{eq:D1D2D3} (cf.~also the first line in Eq. \eqref{eq:lim-tot-1}, where $\epsilon$ corresponds to the size of the small mixing angles $\theta_{ij}$).
}

This procedure allows for a wide variety of hierarchies among the~$D_i$, including cases where one parameter is significantly larger than the others, while keeping the geometric mean fixed within a physically relevant range. 
In addition, we require either $\sqrt{D_1^2 + D_2^2 + D_3^2} > 0.6$ or $M_\varphi/M_{\tilde{\chi}_1} - 1 < 0.1$, as indicated by the dashed lines in the {lower} left panel of Fig.~\ref{fig:relic_abundance_down}.
{This condition excludes regions that we found {to overshoot} the observed relic abundance.} Specifically, it ensures that either the overall coupling strength of the DM multiplet to the SM is large enough for efficient $\tilde{\chi}_i \tilde{\chi}_j$ annihilation, or that the mass splitting between the DM state~$\tilde{\chi}_1$ and the mediator~$\varphi$ is small enough for efficient coannihilation via the sizable QCD interactions of the mediator.
The resulting set of accepted points constitutes our set of \emph{generated} points.

\begin{table}[htb]
	\centering
	\begin{tabular}{c|c|c} 
    \textbf{Parameter}& \textbf{Range}& \textbf{Prior} \\ 
    \hline
	$D_i$ & $[10^{-4},2]$ & description in text \\ 
    $\theta_{ij}$& $[10^{-4},\pi/4]$& log\\
    $\phi_{ij}$& $[10^{-4},\pi/4]$& log\\
    $\delta_{ij}$& $[0,2\pi)$& lin\\
    $\gamma_i$& $[0,2\pi)$& lin\\
    $\lambda_{\varphi H,1}$ & $[-1,-0.001]\cup [0.001, 1]$& log\\
    \hline
    $\eta$& $[-1,-0.01]\cup [0.01, 1]$& log\\ 
    $M_{\tilde{\chi}}\,[\mathrm{GeV}]$& $[100,1500]$& lin\\
    $M_\varphi\,[\mathrm{GeV}]$& $[M_{\tilde{\chi}_1},M_{\tilde{\chi}_1}+1600]$& lin\\
    \hline
	\end{tabular}
    \caption{Ranges and priors for the parameters in the general hierarchical scan scenario with couplings to right-handed down-type quarks. 
    The priors indicate whether the scan range is sampled linearly~(lin) or logarithmically~(log). 
    The upper part of the table shows the parameters needed for the couplings and the lower part shows the mass parameters.} 
	\label{tab:scan_ranges}
\end{table}

\paragraph{Bottom-philic scenario:}
\begin{table}[htb]
	\centering
	\begin{tabular}{c|c|c} 
    \textbf{Parameter}& \textbf{Range}& \textbf{Prior} \\ 
    \hline
	$D_3$ & $[10^{-1},2]$ & lin \\  
    $\epsilon$ & $[10^{-3},0.5]$ & log 
    \\ 
    $D_2$ & $[0.3,1.3]\times \epsilon\, D_3$ & lin 
    \\ 
    $D_1$ & $[0.3,1.3]\times \epsilon \,D_2$ & lin 
    \\ 
    $\theta_{12}$ & $[0.3,1.3] \times \epsilon$ & lin
    \\
    $\theta_{23}$ & $[0.3,1.3] \times \epsilon$ & lin
    \\
    $\theta_{13}$ & $[0.3,1.3] \times \epsilon^2$ & lin   
    \\  
    $\phi_{ij}$ & $[0,\pi/4]$ & lin 
    \\
    $\delta_{ij}$& $[0,2\pi)$  
    & lin 
    \\
    $\gamma_i$& $[0,2\pi)$ 
    & lin 
    \\
    $\lambda_{\varphi H,1}$ & $[-1,-0.001] \cup [0.001, 1]$ & log 
    \\
    \hline
    $\eta$ & $[-1,-0.01] \cup [0.01, 1]$ & log
    \\ 
    $M_{\tilde{\chi}}\,[\mathrm{GeV}]$ & $[100,1500]$ & lin
    \\
    $M_\varphi\,[\mathrm{GeV}]$ & $[M_{\tilde{\chi}_1},M_{\tilde{\chi}_1}+1600]$ & lin
    \\
    \hline
	\end{tabular}
    \caption{Scan ranges and priors for all parameters in the bottom-philic scenario with a DM flavor multiplet coupled dominantly to the third generation.}
    \label{tab:scan-range_flavorsafe}
\end{table}
In addition to the rather general scan setup discussed above, covering a wide range of possible hierarchies, it is instructive to consider a more targeted scan focusing on regions that largely evade the bounds from flavor-changing observables.

One such scenario follows a flavor hierarchy, with the DM flavor multiplet coupling dominantly to the bottom quark.
This scenario can be interpreted as a toy model representing the large class of NP theories with a dominant coupling to the third generation of SM fermions.
A~bottom-philic coupling structure can be achieved by assuming the hierarchy $D_3 \gg D_2 \gg D_1$ among the eigenvalues of the coupling matrix~$\lambda$.
In addition, we have to require $\theta_{ij} \ll 1$, i.e., small misalignment between the quark mass basis and the basis where $\lambda$ is diagonal, to ensure the dominant third-generation coupling.
In contrast, no special alignment between this matrix and the DM flavor multiplet is required and we typically have $\phi_{ij}=\cO(1)$.
Following from Eq.~\eqref{eq:lambda-param}, we find the texture for the coupling matrix
\begin{align}
    \lambda 
    \sim
    \begin{pmatrix}
        \epsilon^2 & \epsilon^2 & \epsilon^2
        \\
        \epsilon & \epsilon & \epsilon
        \\
        1 & 1 & 1
    \end{pmatrix}
    \,,
    \label{eq:bottom-coupling-pattern}
\end{align}
where $\epsilon \ll 1$ and every entry is, of course, multiplied by an order one number.\footnote{This scenario is described in Appendix \ref{app:scenarios} and \ref{app:neutral-mesons} as \emph{Scenario III}.
Following this scenario, it could also be interesting to consider setups where the dark sector is dominantly coupled to the first or second generation, i.e., interchanging the rows of Eq.~\eqref{eq:bottom-coupling-pattern}. We leave these studies for future work.}

This texture can be implemented in a scan in the following way:
We start by sampling the largest eigenvalue of~$\lambda$ linearly in $D_3 \in [10^{-1},2]$. This choice selects couplings in the ballpark required to obtain the observed relic abundance, thereby improving the efficiency of the scan.
Next, we introduce the hierarchy parameter~$\epsilon$ that we sample logarithmically in $[10^{-3},0.5]$.
Larger values would yield too large contributions to meson-mixing observables  through the box diagrams of Fig.~\ref{fig:meson-mixing-diag}, whereas lower values might lead to too small coupling for some of the dark flavors, making the underlying assumption of thermal equilibrium in the dark sector questionable.\footnote{In our scan, however, we checked that each dark-sector state undergoes at least one efficient inelastic scattering process to maintain chemical equilibrium within the dark sector during freeze-out. For a given state $\tilde\chi_j$, it typically requires one coupling $\tilde \lambda_{ij} \gtrsim 10^{-6}$; see~\cite{Garny:2017rxs} for a detailed discussion. In fact, we specifically observe $\tilde\lambda_{sj}>10^{-5}$ for all cosmologically viable parameter points in the scan.}
The other eigenvalues can then be computed using $D_2 = \cO(1) \times \epsilon D_3$ and $D_1 = \cO(1) \times \epsilon D_2$.
As already mentioned, to maintain that the flavor multiplet is predominantly coupled to the bottom quark, we have to additionally ensure small rotation angles~($\theta_{ij} \ll 1$).
This is achieved requiring $\theta_{12} = \cO(1) \times \epsilon$, $\theta_{23} = \cO(1) \times \epsilon$, and $\theta_{13} = \cO(1) \times \epsilon^2$.

The remaining scan parameters are left unchanged with respect to the general hierarchical scan scenario, except for the angles $\phi_{ij}$, for which the scan range remains the same but they are scanned linearly in the bottom-philic scenario, since we aim at sampling $\phi_{ij} = \mathcal{O}(1)$ mixing angles. The complete list of all scan ranges and priors used in this scenario is summarized in Tab.~\ref{tab:scan-range_flavorsafe}.
More details on the specific values chosen in the scan and a more thorough motivation for this specific setup are provided in Appendix~\ref{app:lambda}.

\subsubsection{Constraints and Results}

\begin{figure}
    \centering
    \begin{subfigure}[t]{0.49\textwidth}
        \centering
        \qquad\quad General hierarchical scenario\vspace*{.3mm}
        
        \includegraphics[width=\textwidth]{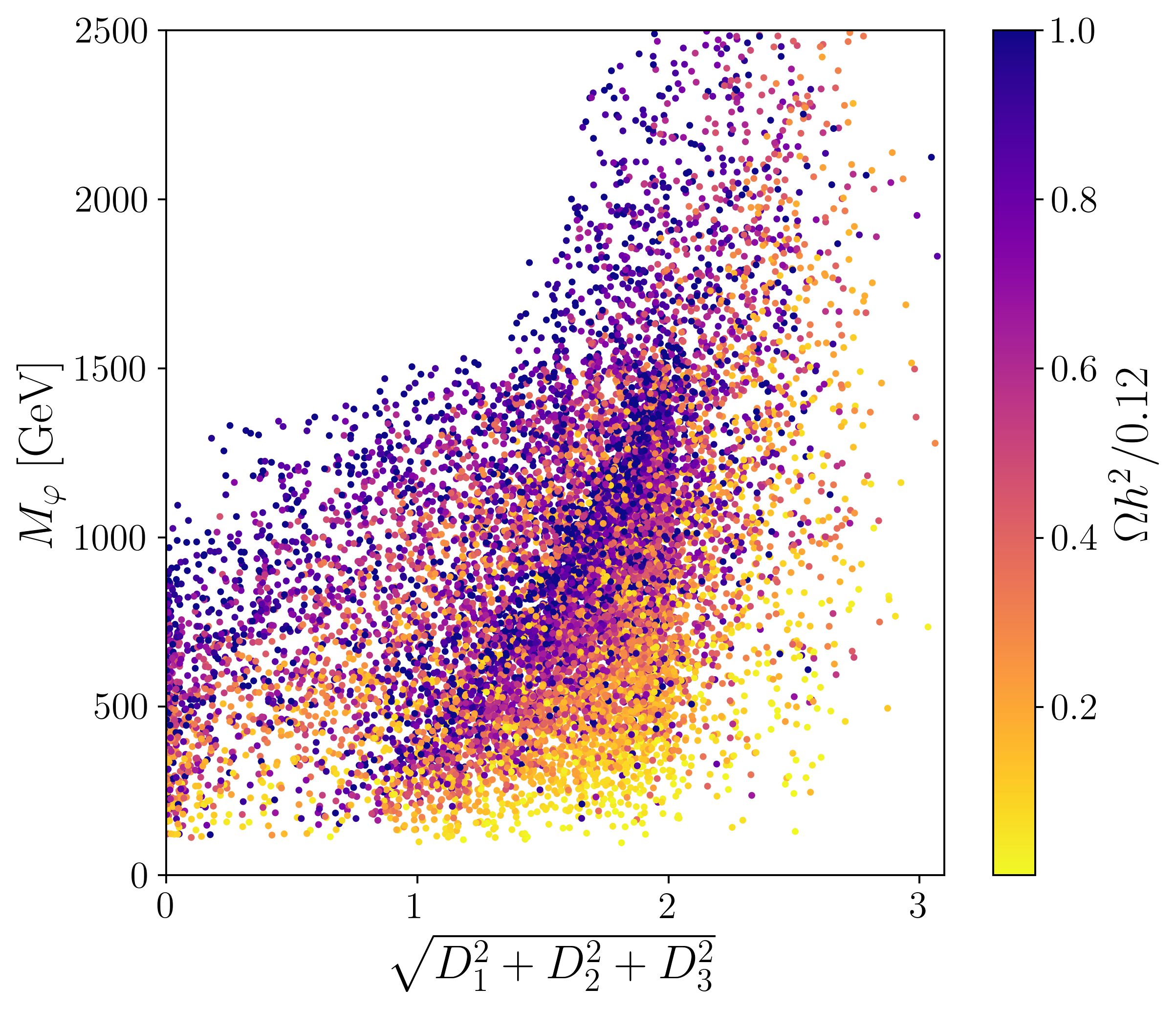}
    \end{subfigure}
    \begin{subfigure}[t]{0.49\textwidth}
        \centering
        \qquad\quad Bottom-philic scenario\vspace*{-.5mm}
        
        \includegraphics[width=\textwidth]{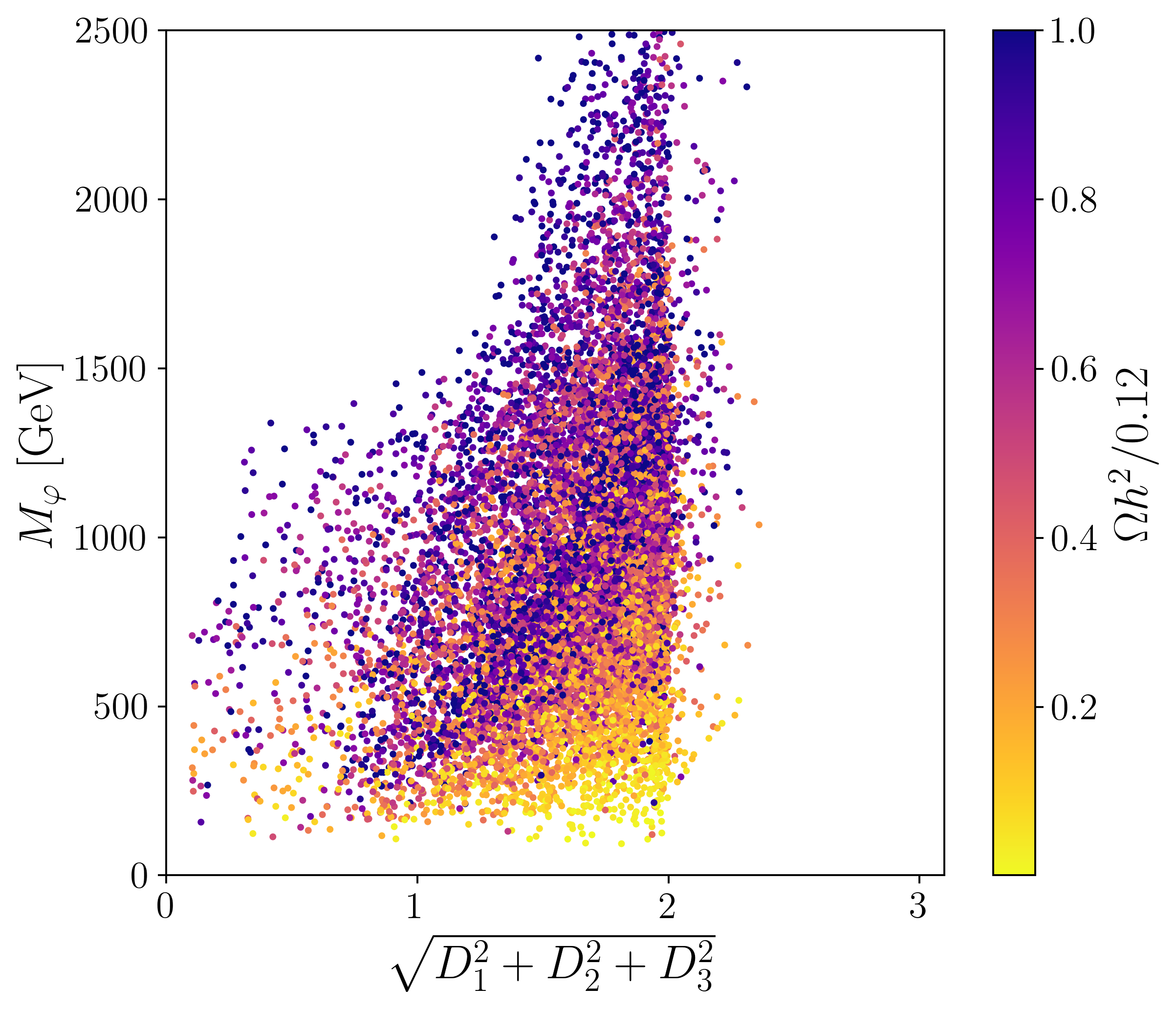}
    \end{subfigure}
    \begin{subfigure}[t]{0.49\textwidth}
        \centering        
        \includegraphics[width=\textwidth]{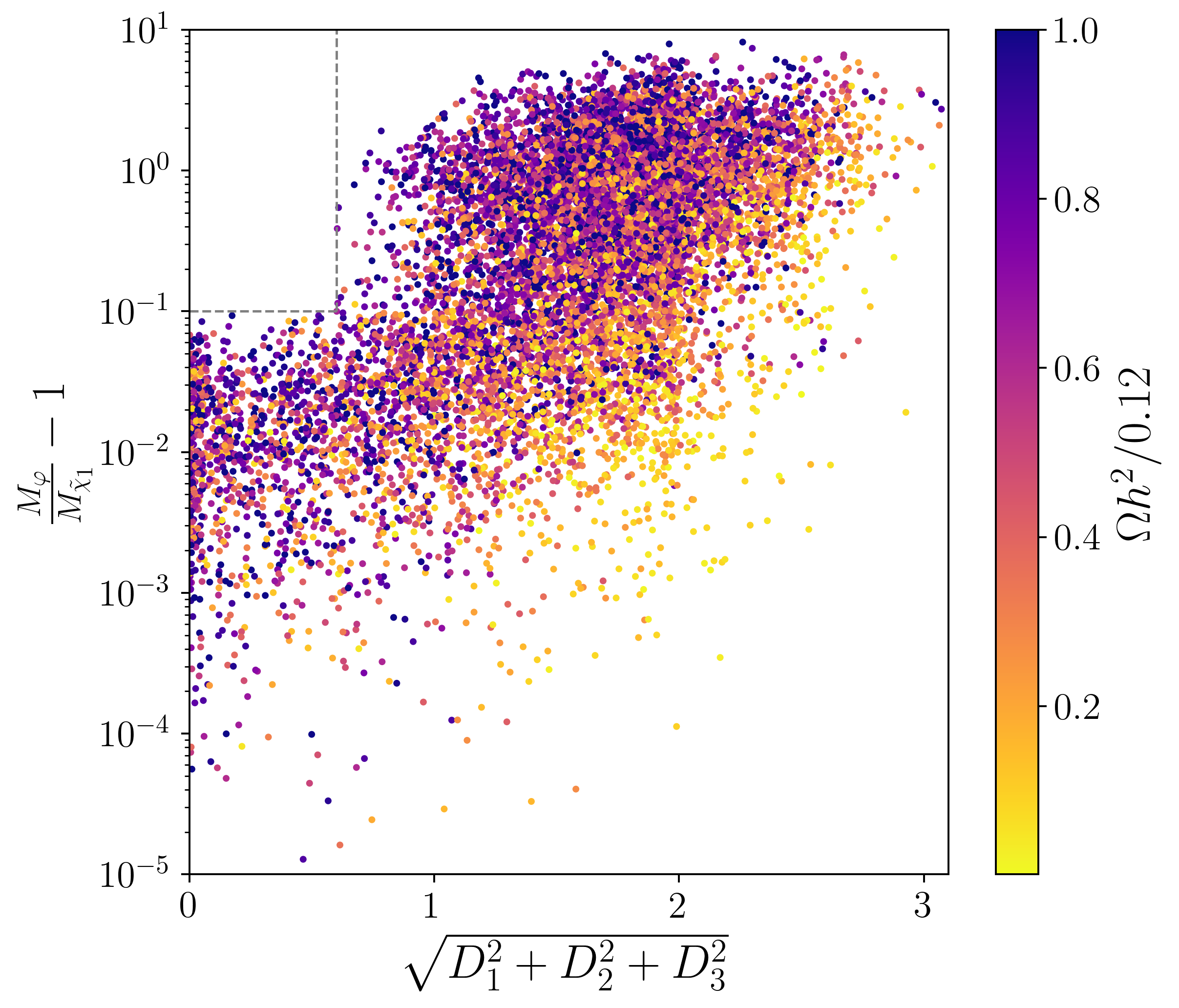}
    \end{subfigure}
    \begin{subfigure}[t]{0.49\textwidth}
        \centering
        \includegraphics[width=\textwidth]{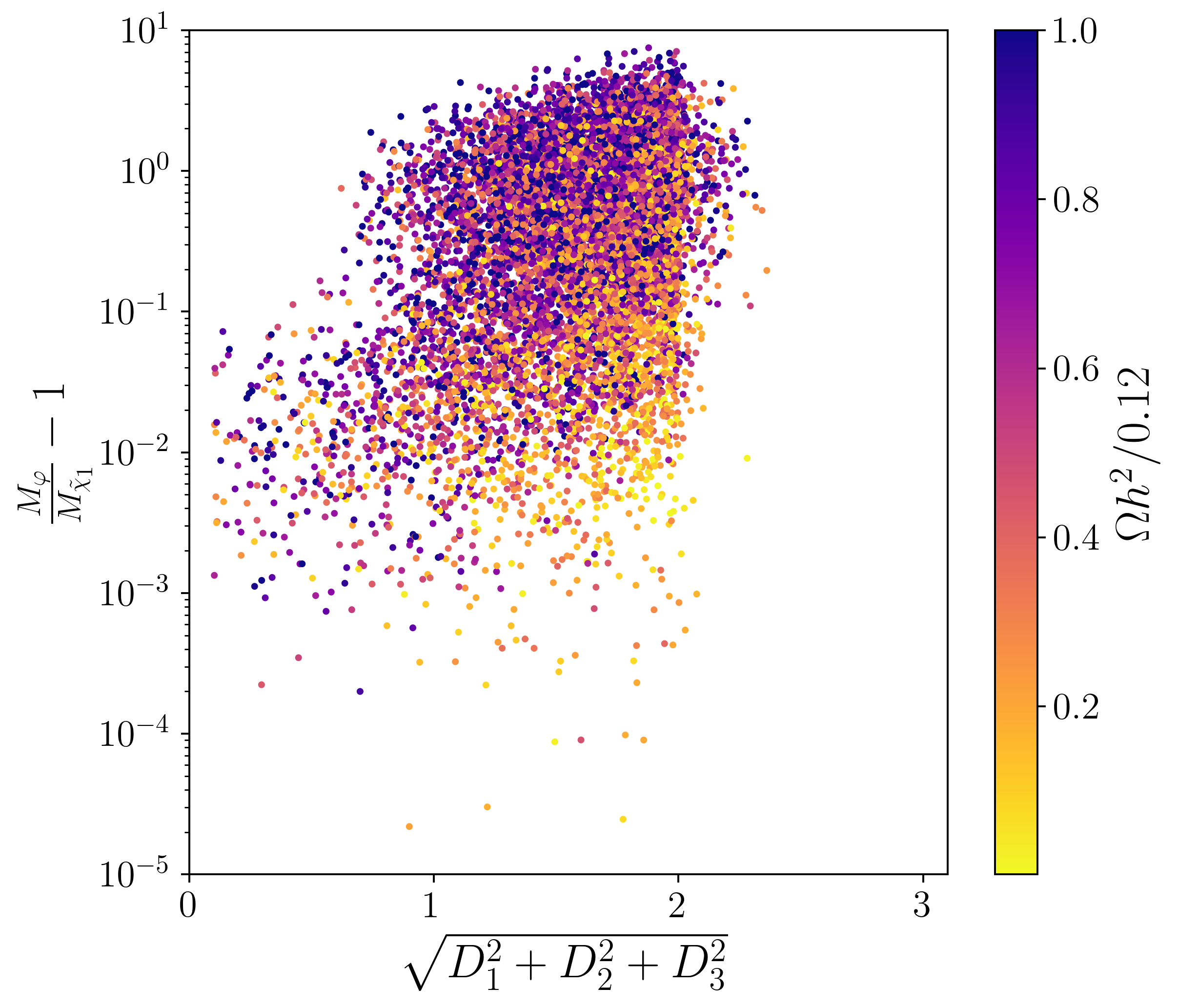}
    \end{subfigure}
    \caption{
{Calculated relic density in the $\sqrt{D_1^2 + D_2^2 + D_3^2}$ - $M_\varphi$ plane (upper panels) and in the $\sqrt{D_1^2 + D_2^2 + D_3^2}$ - $(M_\varphi/M_{\tilde{\chi}_1} - 1)$ plane (lower panels)
    for the Majorana DM model coupling to right-handed down-type quarks. % where 
    The left panel shows the results for the general hierarchical scenario and the right panel shows the ones for the bottom-philic scenario.  The dashed gray lines in the left panel depict the region excluded in the general hierarchical scan by requiring either $\sqrt{D_1^2 + D_2^2 + D_3^2} > 0.6$ or $M_\varphi/M_{\tilde{\chi}_1} - 1 < 0.1$.}}
    \label{fig:relic_abundance_down}
\end{figure}

\begin{figure}
    \centering
    \begin{subfigure}[t]{0.49\textwidth}
        \centering
        \qquad\quad General hierarchical scenario\vspace*{.1mm}
        
        \includegraphics[width=\textwidth]{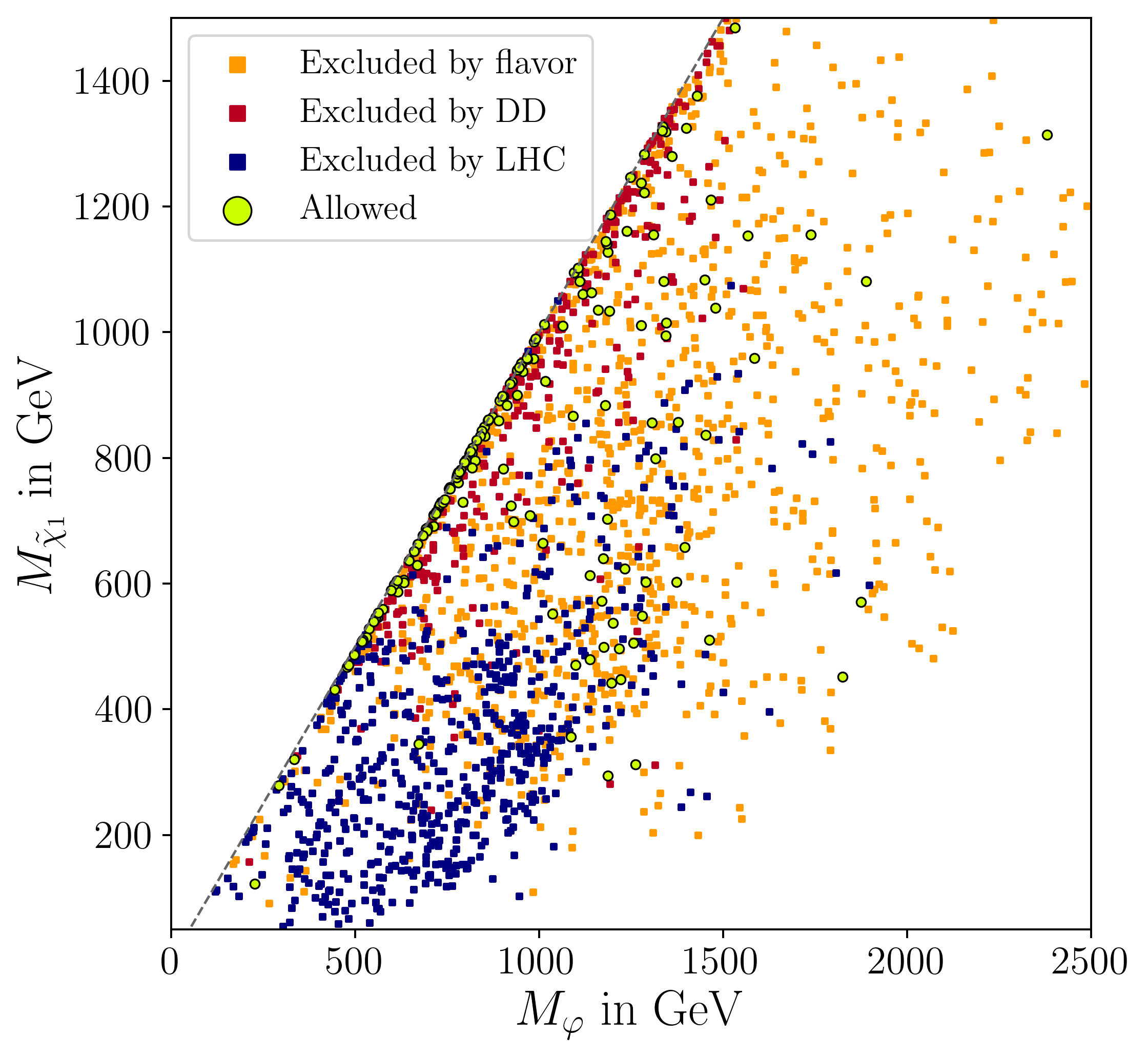}
    \end{subfigure}
    \begin{subfigure}[t]{0.49\textwidth}
        \centering
        \qquad Bottom-philic scenario \vspace*{-.5mm}
        
        \includegraphics[width=\textwidth]{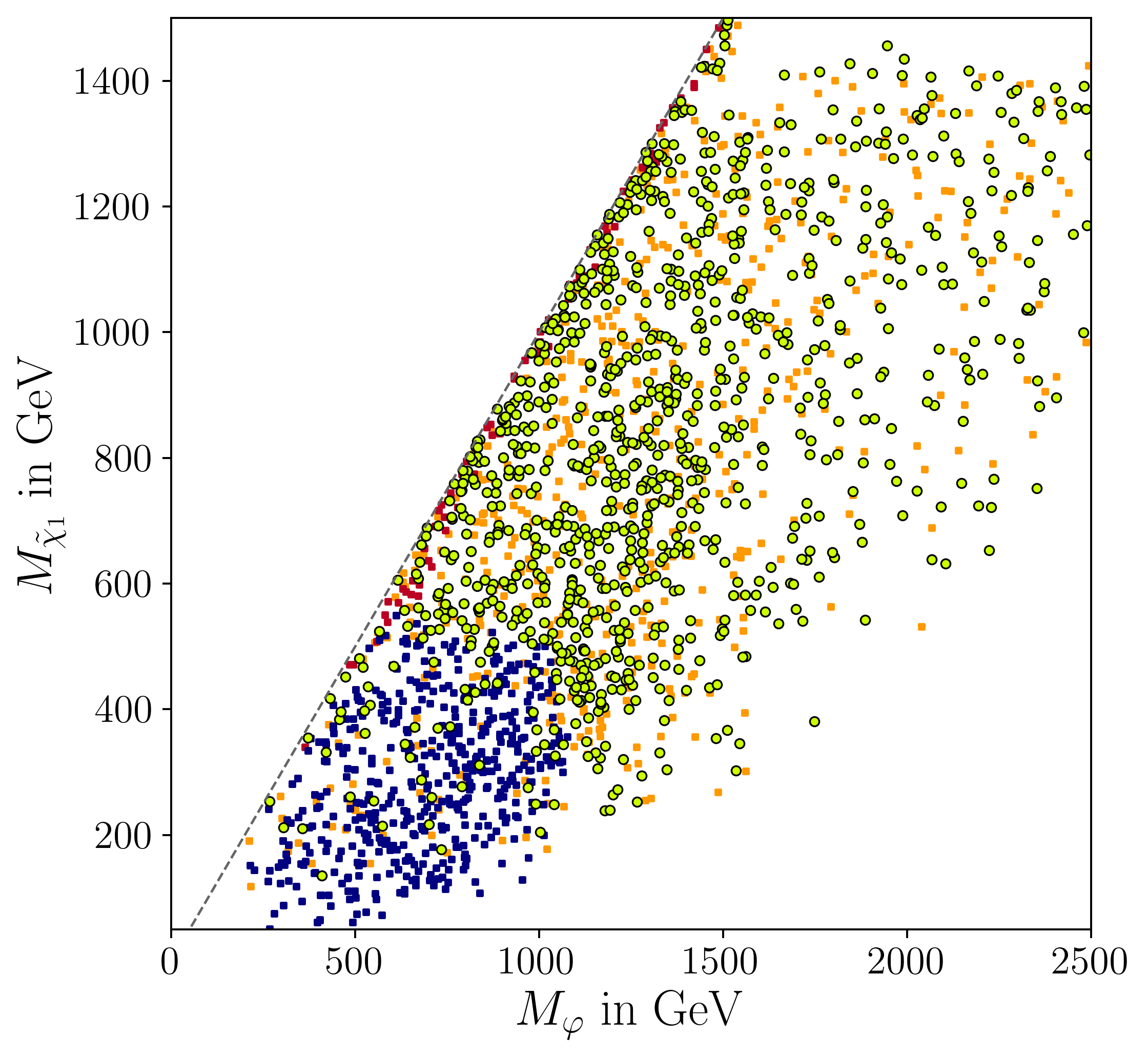}
    \end{subfigure}
    \begin{subfigure}[t]{0.49\textwidth}
        \includegraphics[width=\textwidth]{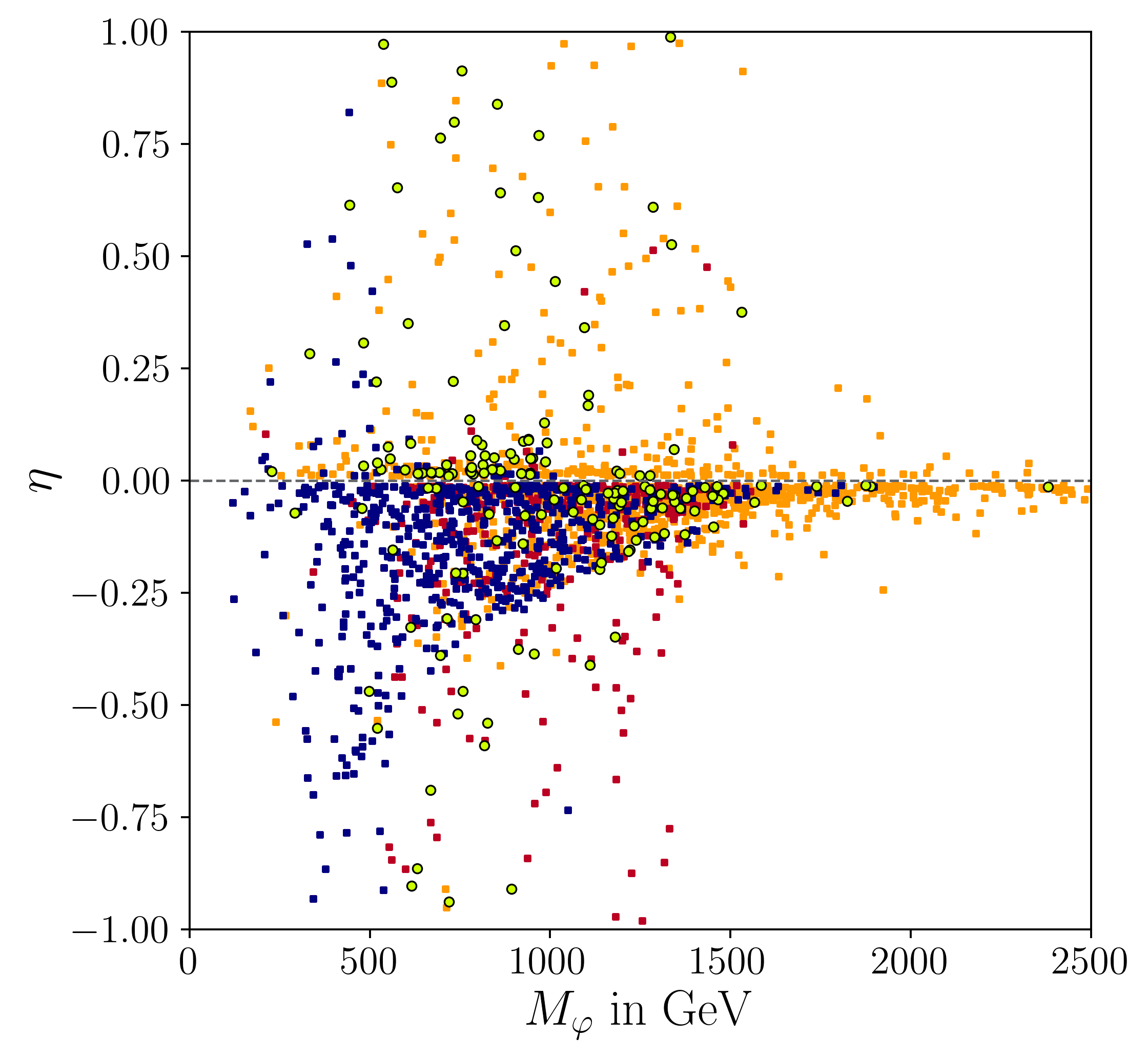}
    \end{subfigure}
    \begin{subfigure}[t]{0.49\textwidth}
        \includegraphics[width=\textwidth]{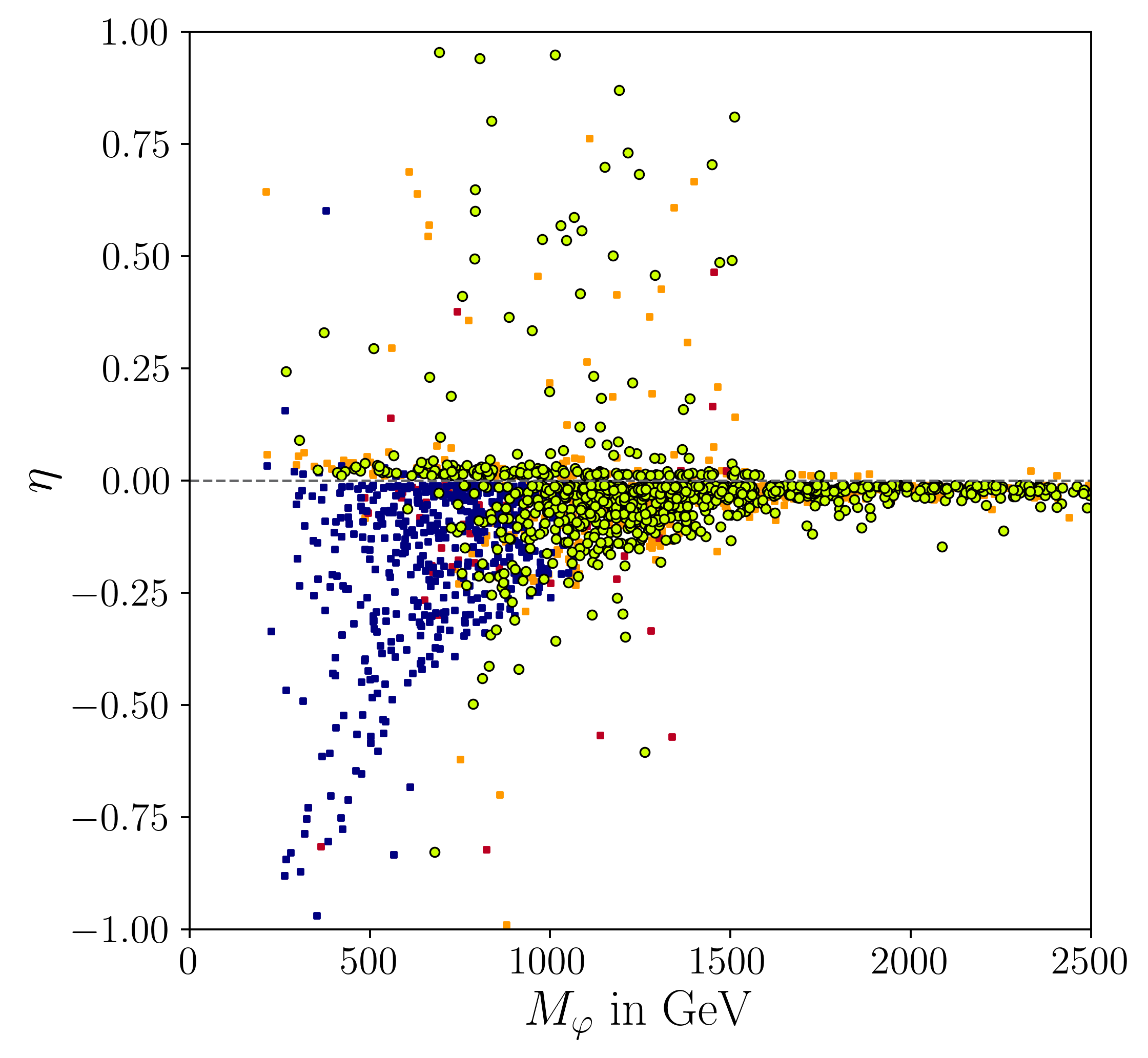}
    \end{subfigure}
    \caption{Comparison of the constraints in the general hierarchical scenario (left) and the bottom-philic scenario (right). The upper panels show the $M_{\tilde{\chi}_1}$–$M_\varphi$ plane and the lower panels show the $\eta$-$M_\varphi$ plane. {The color-code is the same as in Fig. \ref{fig:allowed_leptons} with the addition of direct detection constraints in red.}}
    \label{fig:allowed_quarks_masses}
\end{figure}

\begin{figure}
    \centering
    \begin{subfigure}[t]{0.49\textwidth}
        \centering
        \qquad\quad General hierarchical scenario\vspace*{.3mm}
        
        \includegraphics[width=\textwidth]{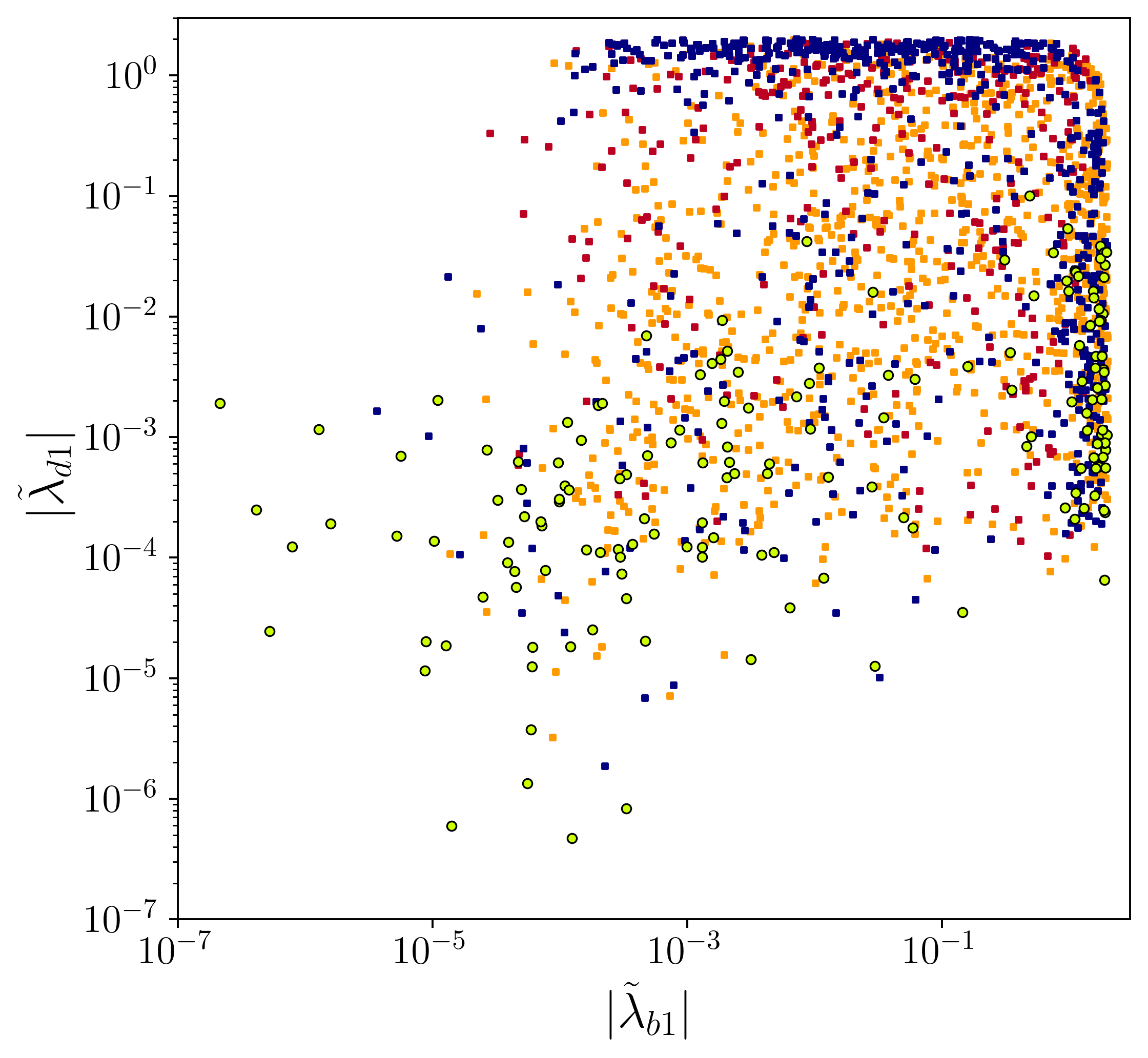}
    \end{subfigure}
    \begin{subfigure}[t]{0.49\textwidth}
        \centering
        \qquad\quad Bottom-philic scenario\vspace*{-.5mm}
        
        \includegraphics[width=\textwidth]{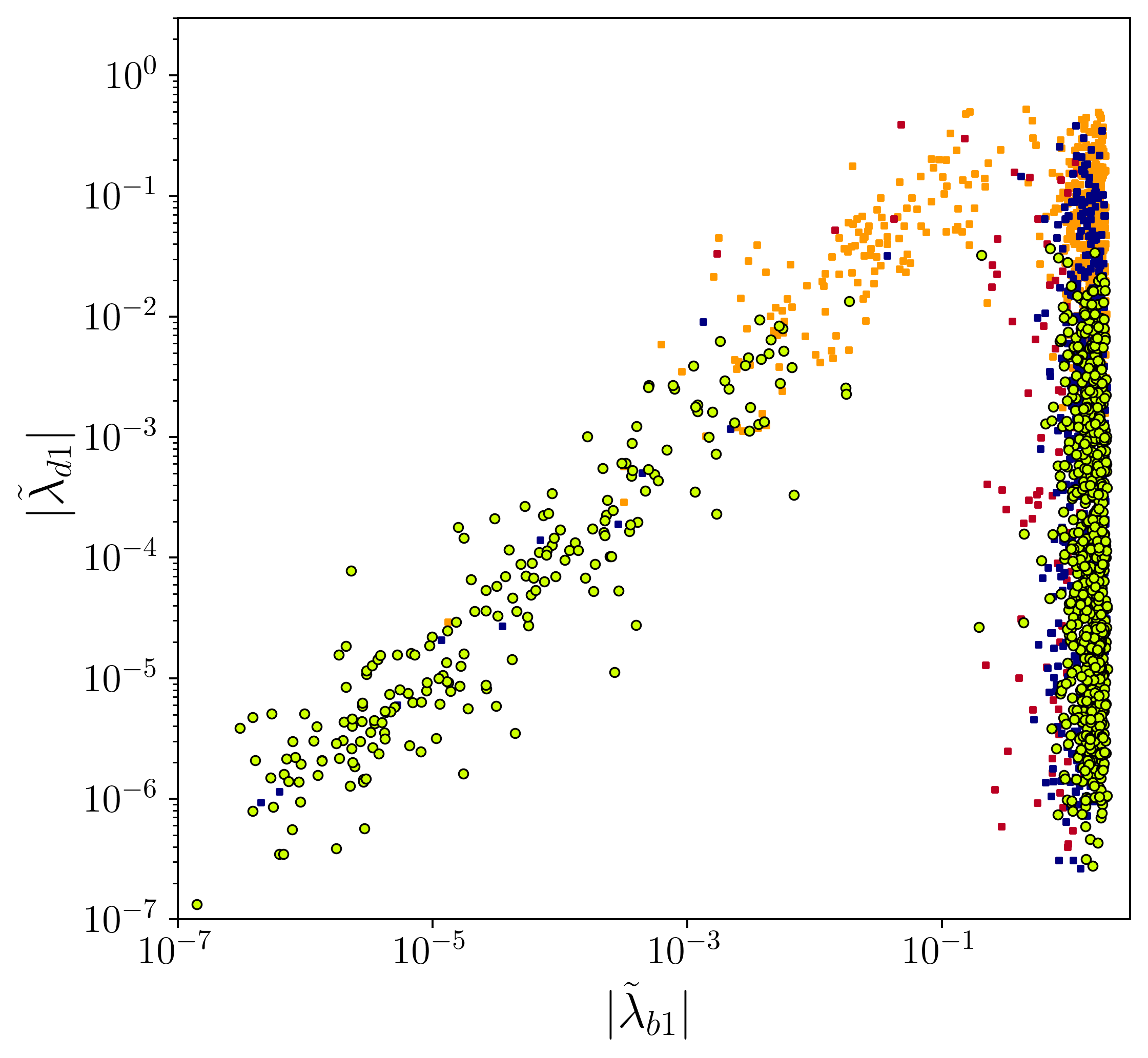}
    \end{subfigure}
    \begin{subfigure}[t]{0.49\textwidth}
        \includegraphics[width=\textwidth]{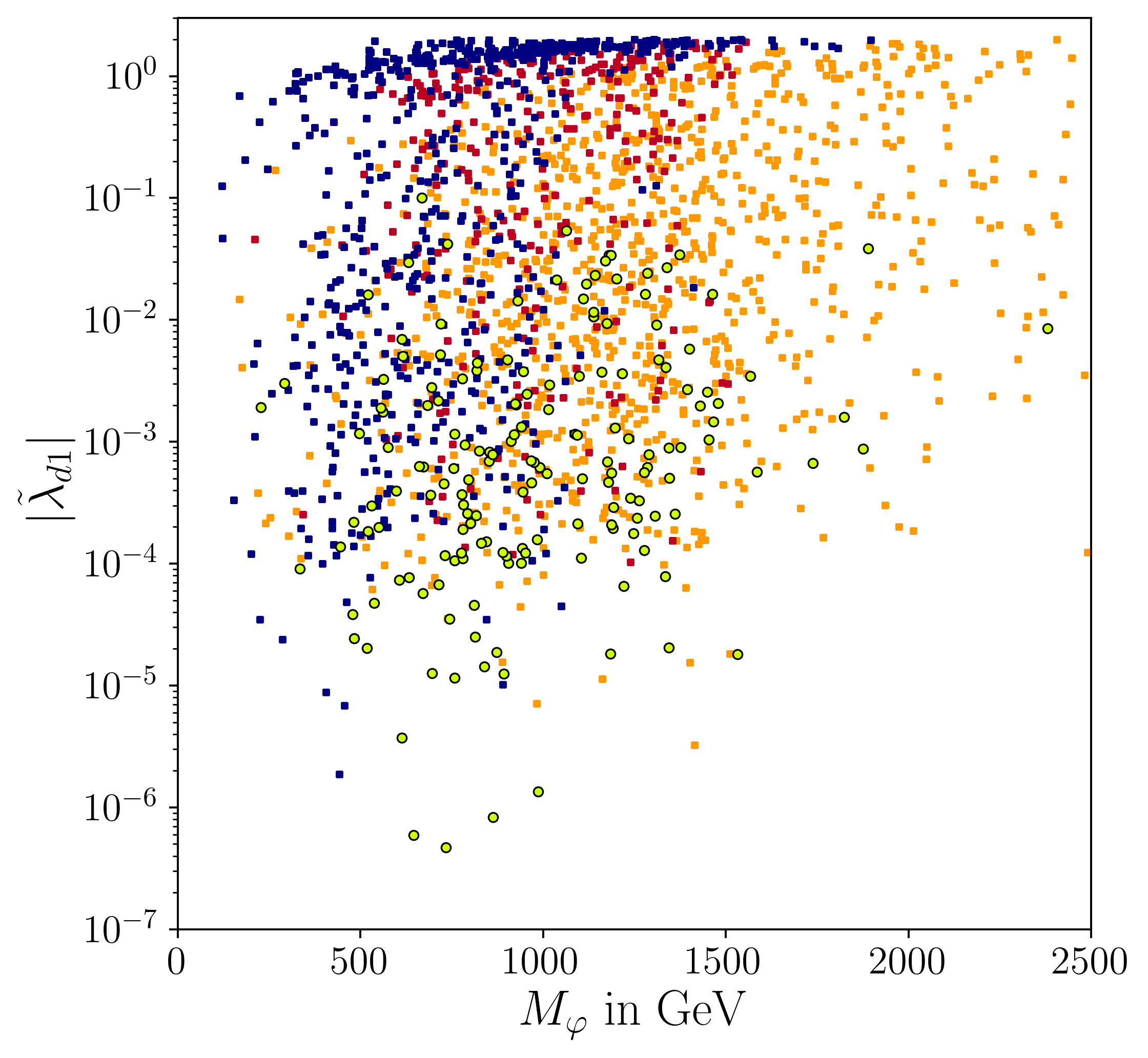}
    \end{subfigure}
    \begin{subfigure}[t]{0.49\textwidth}
        \includegraphics[width=\textwidth]{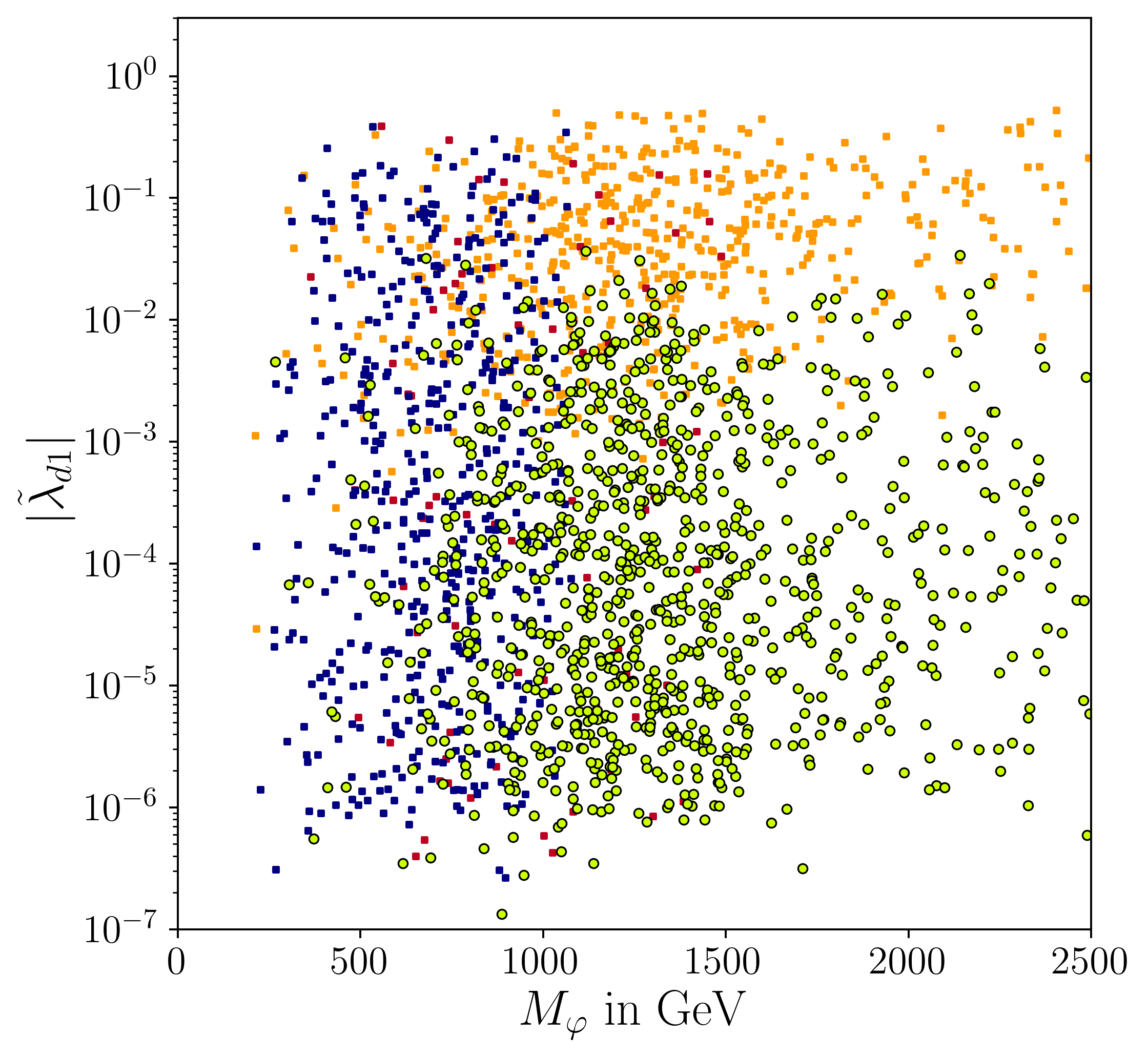}
    \end{subfigure}
    \caption{
Comparison of the constraints in the general hierarchical scenario (left) 
and the bottom-philic scenario (right). 
The upper panels show the $|\tilde{\lambda}_{b1}|$–$|\tilde{\lambda}_{d1}|$ plane 
and the lower panels show the $M_\varphi$–$|\tilde{\lambda}_{d1}|$ plane. 
The color coding for the different constraints is the same as in {Fig.~\ref{fig:allowed_quarks_masses}}.
    }
    \label{fig:allowed_quarks_couplings}
\end{figure}

\begin{figure}
    \centering
    \begin{subfigure}[t]{0.49\textwidth}
        \centering
        \qquad\quad General hierarchical scenario\vspace*{.3mm}
        
        \includegraphics[width=\textwidth]{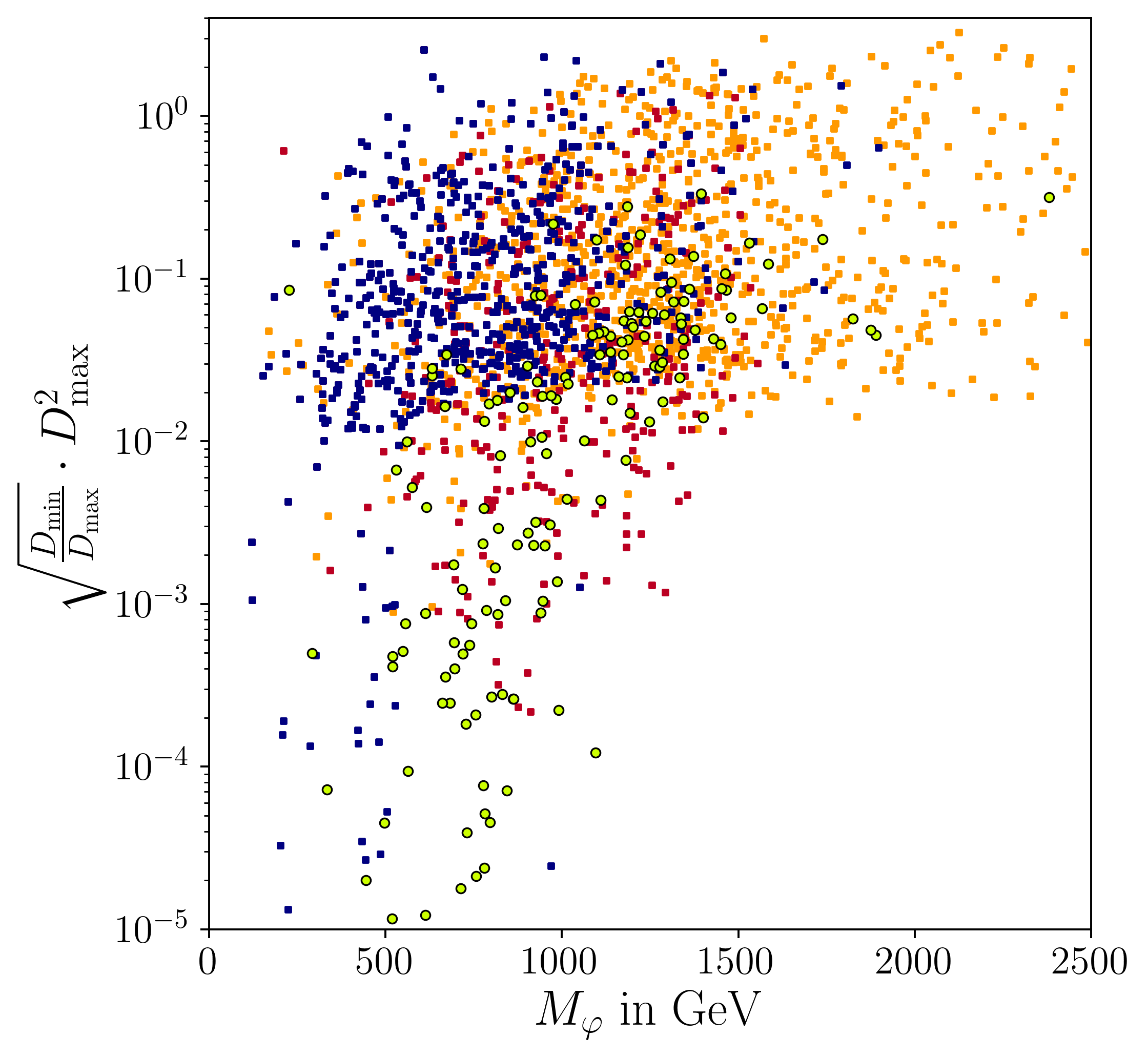}
    \end{subfigure}
    \begin{subfigure}[t]{0.49\textwidth}
        \centering
        \qquad\quad Bottom-philic scenario\vspace*{-.5mm}
        
        \includegraphics[width=\textwidth]{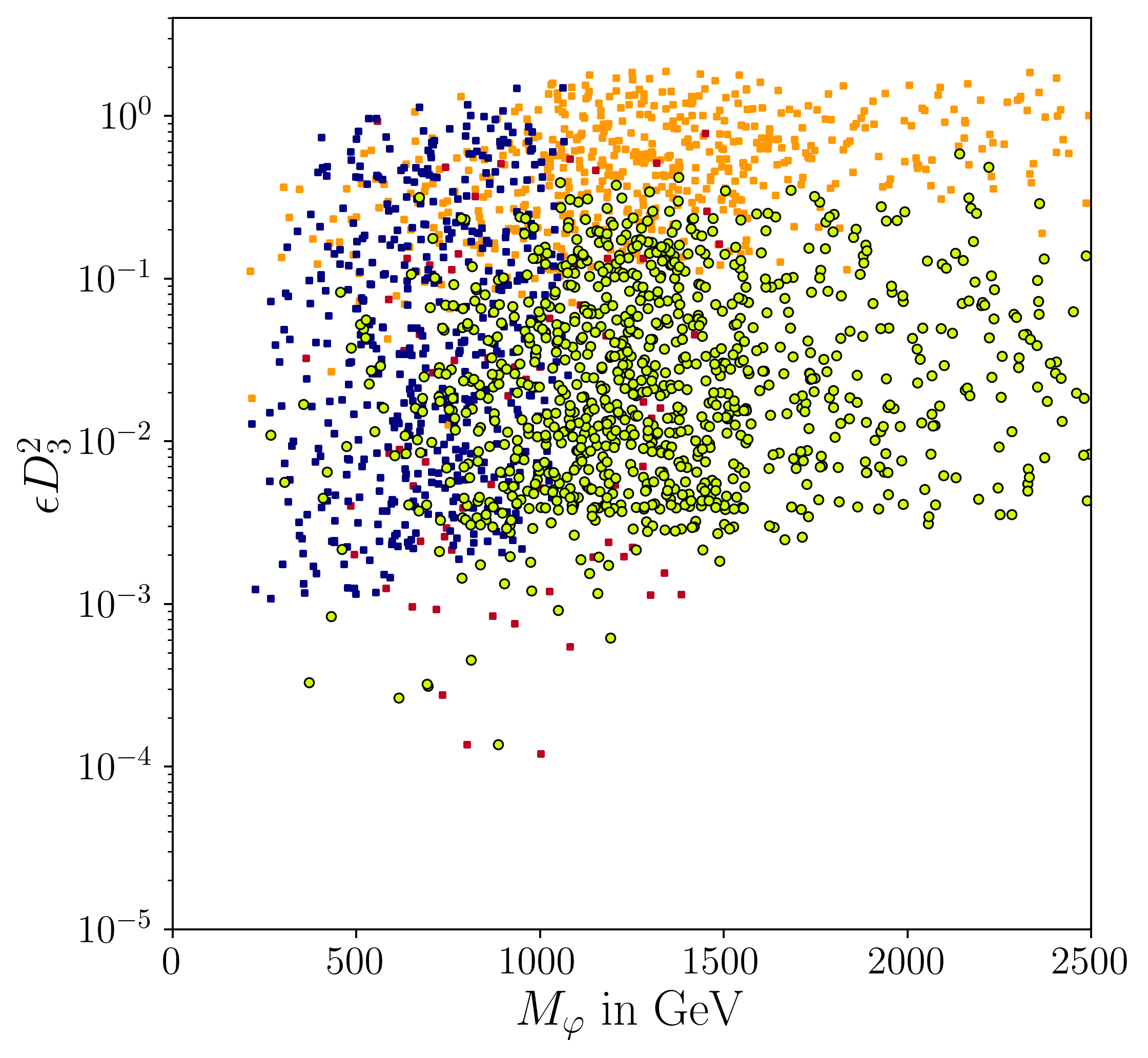}
    \end{subfigure}
    \caption{Comparison of the constraints in the general hierarchical scenario (left) and the bottom-philic scenario (right). Shown are $\sqrt{D_{\rm min}/D_{\rm max}}\,D_{\rm max}^2$ (left) and $\epsilon D_3^2$ (right) versus the mediator mass $M_\varphi$. The two former quantities are related to each other; see the main text for details. The color coding for the different constraints is the same as in {Fig.~\ref{fig:allowed_quarks_masses}}. Note that, to achieve a common y-axis range, the left panel is truncated at the lower end.
    }
\label{fig:epsilon_constraints}
\end{figure}

We generate 250,000 points in both scans and apply the relic density, direct and indirect detection, collider, and flavor constraints discussed in the following. The results for both scans are presented in {Figs.~\ref{fig:relic_abundance_down}--\ref{fig:epsilon_constraints}}, where the general hierarchical scenario is shown on the left and the bottom-philic scenario on the right. The color coding for the constraints is the same as for the leptophilic case, with the addition of direct-detection constraints shown in red.

\paragraph{Relic density constraints:}
{The parametric dependence of the calculated relic density is displayed in Fig.~\ref{fig:relic_abundance_down}, where, as in the leptophilic model, we discarded parameter points predicting overabundant DM. We observe a qualitatively similar picture, with larger coupling strengths as well as smaller masses and mass splittings implying a smaller theoretical value for the relic abundance. However, due to the increased importance of mediator (co)annihilation effects sensitive to the strong coupling constant, the quantitative picture changes.}

{In the following, unless stated otherwise, we focus on the case in which the model reproduces the full observed DM abundance, and we return to the case of underabundant DM at the end of this subsection. Requiring the model to account for the full relic density remains the most stringent selection criterion in the down-type quark scenario, excluding about 99\% of the generated points within our assumed 10\% theoretical uncertainty.\footnote{Note that the nearly identical number of points rejected by the relic density in the two scenarios -- occurring despite the different scan setups described in Sec.~\ref{sec:scan-setup} -- is largely coincidental.}}
Only the surviving points are shown in the plots. The interplay between coupling strength and {masses} in Fig.~\ref{fig:relic_abundance_down}, where $\sqrt{D_1^2 + D_2^2 + D_3^2}$ serves as a proxy for the overall interaction strength of the DM multiplet with the SM, and 
 $M_\varphi/M_{\tilde{\chi}_1} - 1$ quantifies the relative mass splitting that controls the relevance of mediator coannihilations. 
 {Small couplings $\sqrt{D_1^2 + D_2^2 + D_3^2}<1$ can provide the correct relic abundance for mediator masses $M_\varphi\lesssim 1.5 $~TeV, as indicated by the upper panels of the figure.
 The lower panels show}
 that for large splittings, $M_\varphi/M_{\tilde{\chi}_1} - 1 \gtrsim 0.1$, achieving the observed relic abundance requires sizable couplings of order unity, while for smaller splittings efficient coannihilation with the mediator allows for smaller coupling strengths.
 The region with very small values of $\sqrt{D_1^2 + D_2^2 + D_3^2}$ corresponds to regimes where mediator pair annihilations dominate the depletion of the dark sector while DM is kept in chemical equilibrium with the mediator by efficient conversion processes. This region is only populated to a noticeable extent in the general hierarchical scan, owing to the logarithmic prior on the product of the~$D_i$, whereas in the bottom-philic scan the linear prior on $D_3 > 0.1$ prevents access to very small effective couplings. 

Similar to the leptophilic case (see Sec.~\ref{sec:rh-leptons}), the region of large mass splittings $(M_\varphi - M_{\tilde{\chi}_1})$ corresponds to suppressed annihilation cross sections, causing the relic density to exceed the observed value and, hence, leaving a non-accessible region in the lower-right corner of the upper panels of Fig.~\ref{fig:allowed_quarks_masses}.\footnote{We have checked in an exploratory scan allowing for larger $M_\varphi$ values that our choice for the mass range of $M_\varphi$ does not cut into the accessible region.}

The allowed parameter space further shows a preference for negative values of the mass-splitting parameter~$\eta$, see lower panels of Fig.~\ref{fig:allowed_quarks_masses}. Positive~$\eta$ implies that the lightest DM state couples weakest to the SM, such that obtaining the correct relic density would rely on coannihilation among different $\tilde{\chi}_i$ flavors. Because these processes are suppressed in the hierarchical flavor structure, viable points predominantly feature negative~$\eta$, where the lightest state couples most strongly and can efficiently annihilate without relying on coannihilations. Coannihilation with the mediator~$\varphi$ may still occur for small mass splittings but remains somewhat disfavored by the linear mass priors used in our scans. 
Despite structural differences, both scan setups yield very similar viable regions under the relic-density constraint, as the freeze-out dynamics depends mainly on the total coupling strength and masses rather than on the detailed flavor pattern. This is expected since in the bulk of the parameter space the typical freeze-out temperature exceeds the bottom-quark mass, so all quark flavors the dark sector couples to behave effectively alike.

\paragraph{Collider constraints:}
We now turn to collider constraints from LHC searches, evaluated using \textsc{SModelS}, with excluded points shown in blue in all figures.
In the case of the model coupling to down-type quarks, these constraints are significantly more restrictive, excluding 30.1\% of the parameter points that satisfy the relic abundance requirement in the general hierarchical scan and 23.9\% in the bottom-philic scenario. 
The exclusion reaches up to masses of $M_{\tilde{\chi}_1} \approx 1100$\,GeV (600\,GeV) and $M_{\varphi} \approx 1900$\,GeV (1100\,GeV) in the most extreme case for the general hierarchical (bottom-philic) scenario. 
The dominant LHC signal originates from pair production of the scalar mediators, both through QCD interactions and via $t$-channel exchange of $\tilde{\chi}_i$, followed by their decay into a DM particle and a down-type quark. The latter production channel is particularly relevant for large $\tilde{\lambda}_{di}$ couplings as it is enhanced by the down-quark content of the proton.\footnote{For a sizable Majorana mass $M_{\tilde\chi_i}$  the same-sign process $dd\to\varphi\varphi$ becomes dominant in this case \cite{Acaroglu:2021qae,Acaroglu:2023phy}.}
In the general hierarchical scenario all generations of down-type quarks are possible in the final state, depending on the specific hierarchy of a given parameter point.
On the contrary, the bottom-philic scenario is dominated by the $b$-quark final state, which explains the lower energy reach of the collider limits in this case.

In the general hierarchical scenario, the most constraining searches are ATLAS-SUSY-2018-22~\cite{ATLAS:2020syg} searching only for a di-jet signature plus missing transverse energy~($\slashed{E}_T$) and CMS-SUS-19-006~\cite{CMS:2019zmd}, CMS-SUS-16-033~\cite{CMS:2017abv}, and CMS-SUS-16-036~\cite{CMS:2017okm} that also target bottom-quark final states and, to a lesser extent, charm-quark final states.\footnote{%
Note that \textsc{SModelS} also determines that the $c\bar{c}+\slashed{E}_T$ topology becomes relevant for this model due to a peculiarity of the respective CMS searches~\cite{CMS:2017okm,CMS:2021beq}. 
These do not employ a dedicated charm tagger, but exploit the $\sim 10\,\%$ misidentification rate of charm quarks in the $b$-jet tagger, which allows one to gain sensitivity on the $c\bar{c}+\slashed{E}_T$ channel. 
This also means that the signal region of $c\bar{c}+\slashed{E}_T$ is actually identical to the signal region of $b\bar{b}+\slashed{E}_T$ for this search. 
Moreover, the CMS analyses investigate stop~($\tilde{t}$) production with the subsequent decay channels $\tilde{t}\to b \tilde{\chi}^\pm$ or $\tilde{t}\to c \tilde{\chi}^0$, where $\tilde{\chi}^\pm$ is a chargino and $\tilde{\chi}^0$ a neutralino, i.e., the DM candidate. 
Only the latter can lead to the topologies relevant for the DM models we consider here, since the chargino decay would lead to further charged particles in the final state. 
Thus, the $c\bar{c}+\slashed{E}_T$ topology still corresponds physically to a $b\bar{b}+\slashed{E}_T$ final state.
} 
Other searches have a weaker impact on the parameter space, excluding comparatively few points in our scan~\cite{CMS:2017kil, ATLAS:2013lcn, CMS:2021beq, ATLAS:2017avc, ATLAS:2017mjy, ATLAS:2014jxt, CMS-SUS-13-019, ATLAS:2014hqe, CMS:2016ybj, ATLAS:2019gqq}. 
These searches exclude mostly large couplings of $\tilde{\chi}_1$ to either of the quark generations, as can be seen in the upper left panel of Fig.~\ref{fig:allowed_quarks_couplings}. 
In the lower left panel, we also see that the strongest collider constraints apply to parameter points with large first-generation quark couplings leading to high-energetic $\text{di-jet}+\slashed{E}_T$ signals.

For the bottom-philic scenario, the most relevant searches are CMS-SUS-19-006~\cite{CMS:2019zmd} and CMS-SUS-16-036~\cite{CMS:2017okm}, which target the $b\bar{b}+\slashed{E}_T$ channel and exclude the largest number of parameter points. 
Other searches, such as~\cite{CMS:2017kil,CMS:2014nia,ATLAS:2013lcn,CMS:2017abv,CMS:2021beq,ATLAS:2017avc}, exclude only a comparatively small subset of points (where they provide the strongest constraint).
The top-right panel of Fig.~\ref{fig:allowed_quarks_couplings} shows that mainly parameter points with large couplings to $b$~quarks are excluded, as expected.
From the lower-right panel, we see that there is no dependence of the collider limits on the first generation coupling.

As in the leptophilic case, the points with negative values of the mass splitting parameter~$\eta$ and small~$M_{\tilde{\chi}_1}$ are the most affected. 
This is because for negative~$\eta$, $\tilde{\chi}_1$~has the largest coupling, leading to dominant mediator decays directly into~$\tilde{\chi}_1$ and quarks. For light (third-generation) quarks, this results in high-efficiency di-jet ($b\bar{b}$) plus missing transverse  energy  signatures that are well covered by the relevant LHC analyses.
For positive~$\eta$, the mediator decays predominantly into heavier $\tilde{\chi}_i$~states, which undergo cascade decays producing multi-jet final states which are less constraining.
The stronger constraints for negative~$\eta$ can also be observed in the bottom panels of Fig.~\ref{fig:allowed_quarks_masses}.

The missing topologies identified by \textsc{SModelS} in both scenarios correspond primarily to final states containing a single $b$~quark or a $b$~quark accompanied by a light-flavor jet. These signatures are not currently covered by the database, leaving open the possibility that dedicated searches for such final states could further extend the reach of collider constraints in this model.

\paragraph{Direct detection constraints:} 
{In the case of down-type quark couplings, direct detection constraints become substantially more restrictive than in the leptophilic case, because the DM particle couples directly to quarks, giving rise to tree-level SD scattering as well as to a DM–gluon interaction $\bar\chi\chi G^{\mu\nu}G_{\mu\nu}$ at one-loop level~\cite{Drees:1993bu,Gondolo:2013wwa}. The latter constitutes the leading contribution to SI scattering for the considered case of Majorana DM and is implemented in \textsc{micrOMEGAs}, where it is matched onto the gluon content of the nucleon via the trace anomaly in the standard way~\cite{Jungman:1995df}. For points yielding the measured relic abundance, current direct detection searches exclude 32.5\% in the general hierarchical scan and around 5\% in the bottom-philic case (excluded points shown in red). In the general hierarchical scenario, the majority of excluded points are ruled out by SD scattering limits from the LZ experiment~\cite{Aalbers_2023}. These points feature sizable couplings to first-generation quarks, which are the most relevant for tree-level SD scattering due to the significant up- and down-quark content of the nucleons. A much smaller fraction of points is ruled out by loop-induced SI scattering constrained by LZ and XENON1T~\cite{Aprile_2018}. This explains the stronger bounds in the general hierarchical scan compared to the bottom-philic case, where the coupling to the down quark is generally suppressed (cf.~left and right panels of Fig.~\ref{fig:allowed_quarks_couplings}), so that direct detection is dominated by the loop-induced SI contribution.}

In the upper row of Fig.~\ref{fig:allowed_quarks_masses}, it can be seen that the sensitivity of these searches extends up to the upper edge of the scanned DM mass range, with the constraints being stronger when the mediator mass is close to the DM mass, since in the zero-momentum limit relevant for direct detection the scattering amplitude is suppressed by the mediator mass. 

\paragraph{Indirect detection constraints:} {In our analysis, the constraints from indirect detection experiments do not impose any relevant limits on the model, with no parameter points satisfying the relic abundance requirement being excluded in our scan. In the quark-philic scenario, the leading tree-level annihilation process $\tilde{\chi}_1 \tilde{\chi}_1 \to q \bar{q}$ is $p$-wave suppressed and therefore inefficient in the present Universe, which limits the sensitivity of current indirect searches. As in the leptophilic case (see Sec.~\ref{sec:rh-leptons}), the relevant present-day contribution then comes from the radiative three-body processes $\tilde{\chi}_1 \tilde{\chi}_1 \to \gamma q \bar{q}$ and $\tilde{\chi}_1 \tilde{\chi}_1 \to g q \bar{q}$, which are included following the prescription described in Sec.~\ref{sec
:IDgen}. Even with these contributions included, however, no parameter point in our scan is excluded by indirect detection.}

\paragraph{Flavor constraints:} Finally, we apply the flavor constraints to the remaining points that pass the relic-density, direct-detection, and collider bounds, with excluded points shown in orange. {In the down-type quark coupling scenario considered, these observables are particularly powerful, excluding an additional {85.6\%} of the surviving parameter points in the general hierarchical scenario. In the bottom-philic scenario, the constraints are, by construction, less restrictive, with only {35.7\%} of the remaining parameter points being excluded. This leaves us with {165} allowed points in the general hierarchical and {973} points in the bottom-philic scenario.}

\begin{table}[tb]
    \centering
    {
    {%
    \begin{tabular}{r|c|c|c|c|c|c}
        Observable\! &
        $\epsilon_K$~\cite{ParticleDataGroup:2018ovx} & $S_{\psi K_S}$~\cite{HFLAV:2022esi} & $\Delta M_d$~\cite{HFLAV:2014fzu} & $S_{\psi\phi}$~\cite{HFLAV:2022esi} & $\Delta M_s$~\cite{HFLAV:2014fzu} & $\frac{\epsilon^\prime}{\epsilon_K}$~\cite{ParticleDataGroup:2016lqr} \!
        \\\hline
        \!General hier.\! &
        76\% & 47\% & 46\% & 39\% & 37\% & 1\%
        \\
        \!Bottom-phil.\! &
        28\% & 19\% & 18\% & 26\% & 24\% & <1\% 
    \end{tabular}
    }}
    \caption{%
    List of relevant flavor observables and their impact on our scan. 
    The first row lists all flavor observables for which at least one parameter point in our scan has a pull above~2\,$\sigma$ with respect to the SM expectation.
    The second and third rows show the percentage of points in the general hierarchical and bottom-philic scenarios, respectively, that satisfy all other (non-flavor) constraints but 
    have a pull above $2\,\sigma$ with respect to the~SM for the corresponding flavor observable. 
    }
    \label{tab:flavor-observables}
\end{table}

The most stringent bound in both scenarios arises from the $CP$-violating parameter $\epsilon_K$~\cite{ParticleDataGroup:2018ovx} in $K^0$--$\bar{K}^0$ mixing, where the NP contribution originates from the standard and crossed box diagrams involving the scalar mediator and the $\tilde{\chi}_i$ states, shown in Fig.~\ref{fig:meson-mixing-diag}. Further relevant observables are $S_{\psi K_S}$~\cite{HFLAV:2022esi}, 
$\Delta M_d$~\cite{HFLAV:2014fzu}, $S_{\psi\phi}$~\cite{HFLAV:2022esi}, $\Delta M_s$~\cite{HFLAV:2014fzu}, and $\epsilon^\prime/\epsilon_K$~\cite{ParticleDataGroup:2016lqr}. Here, $\Delta M_d$ and $\Delta M_s$ denote the mass differences in the $B_d$ and $B_s$ systems,
while $S_{\psi K_S}$ and $S_{\psi\phi}$ quantify $CP$-violation in the interference of $B^0_q$--$\bar{B}^0_q$ mixing with $b \to c\bar{c}q$ decays ($q=d,s$). The fractions of points that have passed all other constraints but show a pull above $2\sigma$ relative to the SM in these observables are listed in Tab.~\ref{tab:flavor-observables}. They reveal that the general hierarchical scenario is far more restricted by flavor physics. The largest difference between the two scan setups appears, as expected, in $\epsilon_K$, but the other observables also have a significantly reduced impact in the bottom-philic scan. This behavior is reflected in {Figs.~\ref{fig:allowed_quarks_masses}}--\ref{fig:epsilon_constraints}.  It is also visible from the top panels in Fig.~\ref{fig:allowed_quarks_masses} that the flavor constraints can probe high scales and are relevant over the entire energy range covered by our scan.

As shown in Fig.~\ref{fig:allowed_quarks_couplings}, these constraints exclude almost all points with a DM coupling to down quarks, $\tilde\lambda_{d1}$, larger than~0.1, while large couplings to $b$~quarks are less severely constrained. 
The resulting bounds strongly shape the allowed flavor structure of the coupling matrix, favoring hierarchical patterns that avoid simultaneously enhancing multiple flavor-changing amplitudes, as mentioned above. For better performance, this hierarchy is incorporated directly in our two scan setups for the values {of}~$D_{1,2,3}$, as described in Sect.~\ref{sec:scan-setup}.

The pattern of parameter points visible in the top right panel of Fig.~\ref{fig:allowed_quarks_couplings} can be interpreted by inspecting the matrix~$\tilde{\lambda}$ in the mass basis for the bottom-philic scenario.
The couplings to the first and second generation SM quarks are $\tilde{\lambda}_{di}= \mathcal{O}(\epsilon^2)$ and $\tilde{\lambda}_{si}= \mathcal{O}(\epsilon)$, respectively, as given in Eq.\ \eqref{eq:bottom-coupling-pattern}. For the third family, two of the couplings are sizable, $\tilde{\lambda}_{bi}= \mathcal{O}(1)$, however, one turns out to be suppressed, $\tilde{\lambda}_{bj}= \mathcal{O}(\epsilon^2)$.\footnote{For details see the derivation of Eq.~\eqref{eq:tildelambda3} in Appendix~\ref{app:scenarios}.} This small coupling is associated with the
first (lightest) DM generation when $\eta>0$, whereas for
$\eta<0$ one finds $\tilde{\lambda}_{b1}=\mathcal{O}(1)$. 
As discussed above in connection with Fig.~\ref{fig:allowed_quarks_masses}, only a few points with $\eta>0$ satisfy the relic
density constraint. Consequently, the region away from
$\tilde\lambda_{b1}=\mathcal{O}(1)$ in the top-right panel of  Fig.~\ref{fig:allowed_quarks_couplings} is only sparsely populated.

To better understand the constraints from neutral meson mixing in the general hierarchical scenario, it is instructive to compare our results to approximate bounds derived from the leading box-diagram contributions (see the discussion on \emph{Scenario I} in Appendices \ref{app:scenarios} and \ref{app:neutral-mesons} for details).
For example, considering small mixing angles and mass splittings, $M_{\tilde{\chi}}\sim M_\varphi$, the dominant terms that contribute to $\Delta F=2$ processes scale as $D_i^2 D_j^2$, with $i,j$ depending on the meson system.
We obtain the approximate upper bounds
\begin{align}\label{eq:DiDj-1}
& D_1D_2 < \big( 0.2 \mbox{ - } 4.4 \big) \times 10^{-2}\, \Big(\frac{M_\varphi}{\mathrm{TeV}}\Big) \, ,
\quad
D_1D_3 < 5 \times 10^{-2}\, \Big(\frac{M_\varphi}{\mathrm{TeV}}\Big) \, , \quad
D_2D_3 < 0.2\, \Big(\frac{M_\varphi}{\mathrm{TeV}}\Big)\,,
\end{align}
from $K$, $B_d$, and $B_s$~mixing, respectively [cf. Eqs.~\eqref{eq:lim-K-2}, \eqref{eq:lim-Bmesons2}, and \eqref{eq:lim-Bsmesons2}].\footnote{For $M_\chi^2\ll M_\varphi^2$, the relevant contributions become proportional to $(\sin\theta_{ij}\,|D_j^2-D_i^2|)^2$ leading to the bounds in the first lines of Eqs.~\eqref{eq:lim-K-1}, \eqref{eq:lim-B-1}, \eqref{eq:lim-Bs-1}.} The two limits on $D_1D_2$ correspond to the $CP$-violating and $CP$-conserving constraints from the neutral kaon system. The system associated with the two largest~$D_i$ is subject to the strongest constraint. We can estimate the size of this contribution by considering the combination $\sqrt{D_{\min}/D_{\max}}\times D_{\max}^2$, where $D_{\min}$ and $D_{\max}$ denote the smallest and largest of the~$D_i$, respectively.\footnote{Note that we can identify $D_\text{max} \sim D_3$ and $\sqrt{D_\text{min}/D_\text{max}} \sim \epsilon$ when comparing to the bottom-philic scenario. Moreover, we can write $\sqrt{D_\text{min}/D_\text{max}}\times D_\text{max}^2=\sqrt{D_\text{min}D_\text{max}}\times D_\text{max}$, where $\sqrt{D_\text{min}D_\text{max}}$ is the geometric mean of the second largest~$D_i$.}
This quantity can be strongly constrained particularly when the kaon system receives the largest contribution,
as emerges from Eq.~\eqref{eq:DiDj-1}.
The corresponding result for this quantity from our parameter scan 
is shown in Fig.~\ref{fig:epsilon_constraints},
where we see in the left panel that flavor constraints can be satisfied for $\sqrt{D_\text{min}/D_\text{max}}\times D_\text{max}^2 \lesssim 10^{-1}$, in accordance with the estimates in Eq.~\eqref{eq:DiDj-1}. The exact bound on $\sqrt{D_\text{min}/D_\text{max}}\times D_\text{max}^2$ depends on which meson system is affected by the largest couplings. 
Note that the  limits in Eq.\ \eqref{eq:DiDj-1} can also be fulfilled for non-hierarchical couplings provided that their overall scale is suppressed, $D_1,D_2,D_3 \ll 1$.

Similarly, we find that in the bottom-philic scenario, the contribution to the dominant $\Delta F=2$ process is controlled by $\epsilon\, D_3^2$. From $B_s$~mixing, we obtain the approximate constraint
\begin{align}
    &\epsilon\, D_3^2 \lesssim 0.1 \, \Big(\frac{M_\varphi}{\mathrm{TeV}}\Big)\, ,
\end{align}
[cf.~the third lines of Eqs.~\eqref{eq:lim-Bmesons1} and~\eqref{eq:lim-Bsmesons2} with $\lambda_0\approx D_3$] which can also be observed in the right panel of Fig.~\ref{fig:epsilon_constraints}. 
This highlights, again, the clearer distinction between viable and forbidden parameter space in the bottom-philic scenario.

Before concluding, we discuss two possible issues of our flavor analysis: the validity of the underlying EFT approach; and the assumptions about the CKM parameters\@.
Since the experimental energy scale of all these observables is around the $b$-quark mass~$m_b\sim 4\,\mathrm{GeV}$ or below, we expect the EFT to provide a good approximation for the entire range of the DM and mediator masses listed in Tab.~\ref{tab:scan_ranges}, cf. the discussion in Sec.~\ref{sec:DM-EFT}.
In addition, all parameter points with very low DM masses, where the EFT description is potentially questionable, are already excluded by collider constraints, as can be seen in the top panels of Fig.~\ref{fig:allowed_quarks_masses}.
Notice that the physical DM mass~$M_{\tilde{\chi}_1}$ can also reach below the lower bound of $M_{\tilde{\chi}} \geq 100\,\mathrm{GeV}$ for the Lagrangian mass parameter~$M_{\tilde\chi}$ in our scan. 
This is due to the physical mass matrix in Eq.~\eqref{eq:mass-matrix} receiving (possibly negative) corrections from the coupling matrix under the DMFV assumption.

Notice that for this analysis, we chose to fix the CKM elements to their SM values in \textsc{smelli}. 
Determining CKM elements from data in the presence of NP~\cite{Descotes-Genon:2018foz} can be problematic due to the potentially too large effects of NP in the observables used to extract the CKM elements, leading to the program crashing. 
Since CKM elements are extracted from $B$~decays, this problem is particularly severe for NP models coupled to the down-type quarks.
However, we have checked that the points that crash during the CKM extraction are also rejected in our analysis by other observables when fixing the CKM elements to their SM values.
Furthermore, we have also verified that for the parameter points that are not problematic, the results with a CKM matrix extracted under the assumption of the SM or NP are in good agreement.

{\paragraph{Underabundant dark matter:} To assess the impact of the relic density constraint on our results, we also considered the cases in which the model accounts for only 10\% or 1\% of the observed DM density, again assuming a 10\% theoretical uncertainty in the relic density computation. In both the general hierarchical and bottom-philic scenarios, the fraction of generated points passing the corresponding relic density constraint is significantly smaller than in the 100\%-DM scenario. In the general hierarchical case, this fraction decreases from 0.8\% in the 100\%-DM scenario to 0.08\% and 0.003\% in the 10\%-DM and 1\%-DM scenarios, corresponding to 200 and 7 viable points, respectively. In the bottom-philic case, it decreases from 0.8\% to 0.07\% and 0.004\%, corresponding to 183 and 10 viable points. At the same time, the preference of underabundant DM for larger coupling strengths enhances the impact of collider and flavor constraints. Most notably, direct detection becomes significantly more restrictive: in the general hierarchical scenario, the fraction of points excluded by direct detection increases from 31.7\% in the 100\%-DM scenario to 78.5\% in the 10\%-DM scenario and 100\% in the 1\%-DM scenario, while in the bottom-philic scenario it rises from 5.0\% to 71.0\% and 100\%, respectively. This behavior can be understood from the fact that the relic density is often controlled by coannihilation with the mediator, whereas the direct-detection cross section scales more strongly with the DM coupling. As a result, the rescaling of the local DM density in the underabundant scenarios does not compensate the enhancement of the direct-detection rate from the larger couplings. By contrast, indirect detection still excludes no parameter points in either scenario.}

% - - - - CONCLUSION - - - - %
\section{Conclusions}\label{sec:conlc}

In this work, we have developed and applied a comprehensive toolchain for the study of flavored DM models,  combining relic density computations, direct and indirect detection bounds, collider constraints, and a global likelihood analysis of flavor observables based on one-loop matching to the SMEFT and RG evolution to the low-energy scale.

As a proof of principle, we investigated two benchmark scenarios with a Majorana DM flavor triplet coupling either to right-handed charged leptons or to right-handed down-type quarks. In both cases, we performed extensive parameter scans, taking into account the relevant theoretical and experimental constraints.

For the leptophilic scenario, we find that the relic density requirement is the most restrictive constraint, excluding roughly 97\% of the parameter space points before applying other bounds. The surviving points are further shaped by strong limits from charged-lepton flavor-violating decays such as $\mu \to e \gamma$, which remove large regions of parameter space with sizable off-diagonal couplings. Collider searches for missing energy signatures still allow for a substantial viable region, as current limits only exclude a subset of points with $M_{\tilde{\chi}_1} < 200\,\mathrm{GeV}$ and $M_\varphi < 500\,\mathrm{GeV}$, while higher masses remain unconstrained by the LHC\@.

In the quark-coupling scenario, the combination of relic density, direct detection, and flavor constraints leaves only a small fraction of viable points. As flavor data places stringent limits on the parameter space, we study two particular coupling structures -- a general hierarchical scenario and a bottom-philic scenario. While both scenarios behave similarly concerning the relic density constraint, limits from direct detection, collider and flavor observables differ significantly. {Direct detection limits from LZ and XENON1T exclude about one third of the points in the general hierarchical scenario -- predominantly points with sizable couplings to first-generation quarks, for which tree-level SD scattering is most relevant -- while in the bottom-philic scenario those limits rule out only 5\% of the cosmologically allowed parameter points, with the loop-induced SI contribution dominating.} Collider searches using jets$\,+\,\slashed E_T$ and $b$-tagged final states potentially reaching mediator masses up to roughly 1.9~TeV and DM masses up to about 1.1~TeV exclude a subset of points in this region. Due to the suppressed bottom content of colliding protons, in the bottom-philic case, mediator-pair production is dominated by its strong interactions and therefore the maximal LHC reach is reduced in this case, constraining points with mediator masses up to around 1.1~TeV only.
In addition, we find that higher multiplicity final states, obtained when the mediator decays to a heavier dark flavor leading to cascade decays, are suppressed. {For underabundant DM, we find qualitatively similar behavior, with collider and flavor constraints becoming stronger and direct detection gaining additional relevance in the quark-coupling scenario.}

The most stringent flavor constraint in both scenarios arises from $\varepsilon_K$, followed by $\Delta M_{d,s}$ and the $CP$ asymmetries $S_{\psi K_S}$ and $S_{\psi \phi}$.
In the general hierarchical scenario, together these remove close to 86\% of the remaining points, while in the bottom-philic scenario roughly 36\% of the remaining points do not pass the flavor constraints. A general feature of both scenarios is a preference for hierarchical flavor structures in the coupling matrix~$\lambda$, which allow the models to evade simultaneous enhancement of several $\Delta F =2$ processes. 
We have designed the general hierarchical scenario to probe a wide range of different hierarchies, while the bottom-philic scenario singles out the case where the dark sector is dominantly coupled to the bottom quark. As we have shown, the latter case leaves significantly more room for NP, due to the specific choices implemented directly into the scan setup.
This finding quantitatively confirms the expectation that flavored DM models are strongly constrained by the flavor sector and motivates future model-building efforts toward naturally hierarchical couplings.

The framework we have presented is generic and flexible: UFO models for all 20 combinations of DM and mediator flavor representations are provided together with the full analysis toolchain on GitHub~\href{https://github.com/lena-ra/Flavored-Dark-Matter}{\faicon{github}},\textsuperscript{\,\ref{fn:github}\!} enabling straightforward extension of our results to other scenarios. Many of the combinations listed in the summary Table~\ref{tab:overview-of-DM-mediator-combinations} remain unexplored. In particular, scenarios with left-handed mediators and with scalar DM offer a wide range of phenomenological implications -- including richer decay topologies, enhanced electroweak production mechanisms, and potential links to Higgs-sector phenomenology -- which we leave for future work.

% - - - - ACKNOWLEDGMENTS - - - - %
\renewcommand{\headrulewidth}{0pt}
\rhead{}
\phantomsection
\section*{Acknowledgments}

This research was supported by the Deutsche Forschungsgemeinschaft (DFG, German Research Foundation) under grant 396021762 -- TRR~257. MB,~MK,~and FW are grateful to the Mainz Institute for Theoretical Physics (MITP) of the Cluster of Excellence PRISMA+ (Project ID 390831469) for its hospitality and partial support during the final stages of this work.

% - - - - APPENDIX - - - - %
\begin{appendix}

\section{Naming Conventions}\label{app:lagran}

In this appendix, we describe the naming conventions that we use for the parameters in the model files for the different flavored DM models which are published on GitHub~\href{https://github.com/lena-ra/Flavored-Dark-Matter}{\faicon{github}}.\textsuperscript{\,\ref{fn:github}\!} The models are listed in the different columns and specified by the DM particle type (real or complex scalar, Dirac, or Majorana fermion) and the SM fermion to which the particles couple [i.e. right-handed up-type quarks (u), right-handed down-type quarks (d), left-handed quark doublet (q), right-handed leptons (e), and left-handed lepton doublet (l)]. The rows list the parameters of the different models as introduced in Sec. \ref{sec:classific}.

\begin{table}[h]
    \centering
    \begin{tabular}{c|| c | c | c | c | c}
     & Real\_q & Real\_u & Real\_d & Real\_l & Real\_e \\
     \hline
     $\Tilde{S}$ & \multicolumn{5}{c}{\texttt{XS1}, \texttt{XS2}, \texttt{XS3}}\\
     $M_{\Tilde{S}}$ & \multicolumn{5}{c}{\texttt{MXS1}, \texttt{MXS2}, \texttt{MXS3}}\\
     $\psi$ & \texttt{YFq} & \texttt{YFu} & \texttt{YFd} & \texttt{YFl} & \texttt{YFe} \\
     \multirow{2}{*}{$M_\psi$} & \texttt{MYFqu} & \multirow{2}{*}{\texttt{MYFu}} & \multirow{2}{*}{\texttt{MYFd}} & \texttt{MYFle} & \multirow{2}{*}{\texttt{MYFe}}\\
      & \texttt{MYFqd} & & & \texttt{MYFlv} & \\
    $f$ & \texttt{QL} & \texttt{uR} & \texttt{dR} & \texttt{LL} & \texttt{lR} \\
     \multirow{2}{*}{$\lambda_{ij}$} & \texttt{lamSqRe}, & \texttt{lamSuRe}, & \texttt{lamSdRe}, & \texttt{lamSlRe}, & \texttt{lamSeRe}, \\
     & \texttt{lamSqIm} & \texttt{lamSuIm} & \texttt{lamSdIm} & \texttt{lamSlIm} & \texttt{lamSeIm} \\
     $\lambda_{\Tilde{S} H}$ & \multicolumn{5}{c}{\texttt{lamSHRe}, \texttt{lamSHIm}} \\
     $\lambda_{\Tilde{S} \Tilde{S}}$ & \multicolumn{5}{c}{\texttt{lam2S}}
\end{tabular}
    \caption{Parameters in real scalar DM models.}
    \label{tab:my_label}
\end{table}

\begin{table}[h]
    \centering
    \begin{tabular}{c|| c | c | c | c | c}
     & Complex\_q & Complex\_u & Complex\_d & Complex\_l & Complex\_e \\
     \hline
     $S$ & \multicolumn{5}{c}{\texttt{XC1}, \texttt{XC2}, \texttt{XC3}}\\
     $M_S$ & \multicolumn{5}{c}{\texttt{MXC1}, \texttt{MXC2}, \texttt{MXC3}}\\
     $\psi$ & \texttt{YFq} & \texttt{YFu} & \texttt{YFd} & \texttt{YFl} & \texttt{YFe} \\
     \multirow{2}{*}{$M_\psi$} & \texttt{MYFqu} & \multirow{2}{*}{\texttt{MYFu}} & \multirow{2}{*}{\texttt{MYFd}} & \texttt{MYFle} & \multirow{2}{*}{\texttt{MYFe}}\\
      & \texttt{MYFqd} & & & \texttt{MYFlv} & \\
    $f$ & \texttt{QL} & \texttt{uR} & \texttt{dR} & \texttt{LL} & \texttt{lR} \\
     \multirow{2}{*}{$\lambda_{ij}$} & \texttt{lamCqRe}, & \texttt{lamCuRe}, & \texttt{lamCdRe}, & \texttt{lamClRe}, & \texttt{lamCeRe}, \\
     & \texttt{lamCqIm} & \texttt{lamCuIm} & \texttt{lamCdIm} & \texttt{lamClIm} & \texttt{lamCeIm} \\
     $\lambda_{S H}$ & \multicolumn{5}{c}{\texttt{lamCHRe}, \texttt{lamCHIm}} \\
     $\lambda_{S S}$ & \multicolumn{5}{c}{\texttt{lam2C}}
\end{tabular}
    \caption{Parameters in complex scalar DM models.}
    \label{tab:my_label}
\end{table}

\begin{table}[h]
    \centering
    \begin{tabular}{c|| c | c | c | c | c}
     & Majorana\_q & Majorana\_u & Majorana\_d & Majorana\_l & Majorana\_e \\
     \hline
     $\Tilde{\chi}$ & \multicolumn{5}{c}{\texttt{XM1}, \texttt{XM2}, \texttt{XM3}}\\
     $M_{\Tilde{\chi}}$ & \multicolumn{5}{c}{\texttt{MXM1}, \texttt{MXM2}, \texttt{MXM3}}\\
     $\varphi$ & \texttt{YSq} & \texttt{YSu} & \texttt{YSd} & \texttt{YSl} & \texttt{YSe} \\
     \multirow{3}{*}{$M_\varphi$} & \texttt{MYSqu} & \multirow{3}{*}{\texttt{MYSu}} & \multirow{3}{*}{\texttt{MYSd}} & \texttt{MYSle} & \multirow{3}{*}{\texttt{MYSe}}\\
     & \texttt{MYSqd} & & & \texttt{MYSlvR} & \\
     & & & & \texttt{MYSlvI} & \\
    $f$ & \texttt{QL} & \texttt{uR} & \texttt{dR} & \texttt{LL} & \texttt{lR} \\
     \multirow{2}{*}{$\lambda_{ij}$} & \texttt{lamMqRe}, & \texttt{lamMuRe}, & \texttt{lamMdRe}, & \texttt{lamMlRe}, & \texttt{lamMeRe}, \\
     & \texttt{lamMqIm} & \texttt{lamMuIm} & \texttt{lamMdIm} & \texttt{lamMlIm} & \texttt{lamMeIm} \\
     $\lambda_{\varphi H,1}$ & \texttt{lamYHq1} & \texttt{lamYHu} & \texttt{lamYHd} & \texttt{lamYHl1} & \texttt{lamYHe} \\
     $\lambda_{\varphi H,2}$ & \texttt{lamYHq2} & -- & -- & \texttt{lamYHl2} & -- \\
     $\lambda_{\varphi H,3}$ & -- & -- & -- & \texttt{lamYHl3} & -- \\
     $\lambda_{\varphi \varphi}$ & \texttt{lam2Yq} & \texttt{lam2Yu} & \texttt{lam2Yd} & \texttt{lam2Yl} & \texttt{lam2Ye}
\end{tabular}
    \caption{Parameters in Majorana DM models.}
    \label{tab:my_label}
\end{table}

\clearpage

\begin{table}[h]
    \centering
    \begin{tabular}{c|| c | c | c | c | c}
     & Dirac\_q & Dirac\_u & Dirac\_d & Dirac\_l & Dirac\_e \\
     \hline
     $\chi$ & \multicolumn{5}{c}{\texttt{XD1}, \texttt{XD2}, \texttt{XD3}}\\
     $M_\chi$ & \multicolumn{5}{c}{\texttt{MXD1}, \texttt{MXD2}, \texttt{MXD3}}\\
     $\varphi$ & \texttt{YSq} & \texttt{YSu} & \texttt{YSd} & \texttt{YSl} & \texttt{YSe} \\
     \multirow{3}{*}{$M_\varphi$} & \texttt{MYSqu} & \multirow{3}{*}{\texttt{MYSu}} & \multirow{3}{*}{\texttt{MYSd}} & \texttt{MYSle} & \multirow{3}{*}{\texttt{MYSe}}\\
     & \texttt{MYSqd} & & & \texttt{MYSlvR} & \\
     & & & & \texttt{MYSlvI} & \\
     $f$ & \texttt{QL} & \texttt{uR} & \texttt{dR} & \texttt{LL} & \texttt{lR} \\
     \multirow{2}{*}{$\lambda_{ij}$} & \texttt{lamDqRe}, & \texttt{lamDuRe}, & \texttt{lamDdRe}, & \texttt{lamDlRe}, & \texttt{lamDeRe}, \\
     & \texttt{lamDqIm} & \texttt{lamDuIm} & \texttt{lamDdIm} & \texttt{lamDlIm} & \texttt{lamDeIm} \\
     $\lambda_{\varphi H,1}$ & \texttt{lamYHq1} & \texttt{lamYHu} & \texttt{lamYHd} & \texttt{lamYHl1} & \texttt{lamYHe} \\
     $\lambda_{\varphi H,2}$ & \texttt{lamYHq2} & -- & -- & \texttt{lamYHl2} & -- \\
     $\lambda_{\varphi H,3}$ & -- & -- & -- & \texttt{lamYHl3} & -- \\
     $\lambda_{\varphi \varphi}$ & \texttt{lam2Yq} & \texttt{lam2Yu} & \texttt{lam2Yd} & \texttt{lam2Yl} & \texttt{lam2Ye}
\end{tabular}
    \caption{Parameters in Dirac DM models.}
    \label{tab:my_label}
\end{table}

\section{Coupling Matrix and Flavor Constraints}
\label{app:lambda}

In this section we analyze the constraints and relations between the parameters of the matrix~$\lambda$.
We start by summarizing the dark flavor transformations in Appendix~\ref{app:DFBT}.
In Appendix~\ref{app:scenarios} we describe the relations between the parameters in different bases for various coupling scenarios. In Appendix~\ref{app:neutral-mesons} we analyze the relevant experimental constraints on $\lambda$ from flavor changing phenomena.

\subsection{Dark Flavor Basis Transformations}
\label{app:DFBT}

The dark-flavor-breaking terms of the Lagrangian are
\begin{align}
&-\frac{1}{2}\overline{\tilde{\chi}_L}\mathcal{C}\,M_{\tilde{\chi}} \overline{\tilde{\chi}_L}^\intercal-\overline{d_{Ri}}\,\lambda\,\tilde{\chi}\phi+\text{h.c.}
\end{align}
In the model presented in this work, the field $\tilde{\chi}$ transforms as a triplet under a global~$O(3)$ symmetry. Under a flavor transformation we have
\begin{align}\label{eq:basis}
&  \tilde{\chi}\rightarrow \tilde{\chi}'=R_f\, \tilde{\chi} \, , \qquad M_{\tilde{\chi}} \rightarrow 
M'_{\tilde{\chi}}=R_f\, M_{\tilde{\chi}} \, R_f^\intercal \, , \qquad \lambda \rightarrow \lambda'= \lambda \, R_f^\intercal\,,
\end{align}
where $R_f$ is an orthogonal matrix.
Under the DMFV assumption the mass matrix emerges as
\begin{align}\label{eq:mass}
&\tilde{M}_{\tilde{\chi}} = M_{\tilde{\chi}} \left[ \unit + \frac{\eta}{2} (\lambda^\dagger \lambda + \lambda^\intercal \lambda^*) + \mathcal{O}(\lambda^4)\right] \, 
\end{align}  
working to leading order in the spurion~($\lambda$) expansion.
Since $\tilde{M}_{\tilde{\chi}}$ is real we can go to the mass basis using an orthogonal transformation 
\begin{align}\label{eq:mass-basis}
\tilde{\chi}\rightarrow\, \tilde{\tilde{\chi}}=W \, \tilde{\chi} \, , \qquad \tilde{M}_{\tilde{\chi}} \rightarrow \,
\tilde{M}_{\tilde{\chi}}^D=W\, \tilde{M}_{\tilde{\chi}} \, W^\intercal  \, , \qquad \lambda \rightarrow\, \tilde{\lambda}= \lambda\, W^\intercal \,,
\end{align}
where the orthogonal matrix~$W$ diagonalizes the symmetric mass matrix~$\tilde{M}_{\tilde{\chi}}$, and $\tilde{M}_{\tilde{\chi}}^D$ is the diagonal matrix.

The $\lambda $ matrix can be diagonalized by a biunitary transformation:
\begin{align}\label{eq:lambda-diag}
& U^\dag\, \lambda  \, V = D \,,
\end{align}
where $D$ is diagonal.
We can choose to absorb three phases of $U$ into $V$.
Following~\cite{Acaroglu:2021qae}, $V$ can be written in terms of two orthogonal matrices ($O$, $R$) and one diagonal matrix of phases ($d$)~\cite{FUHR201832}
\begin{align}\label{eq:UR-OFR}
& \lambda= U D V^\dag = U\, D\, OdR \,.
\end{align}
A basis with minimal number of parameters can be obtained
by applying the transformation~\eqref{eq:basis} in order to absorb the rotation $R$.
In particular, by rotating the $\tilde{\chi}$ species with $R^\intercal$ one has in the new basis $\tilde{\chi}'$:
\begin{align}
& \lambda'= \lambda \, R^\intercal =U\, D Od \, , \qquad  
\tilde{M}'_{\tilde{\chi}}=R\, \tilde{M}_{\tilde{\chi}} \, R^\intercal \, , \qquad 
\end{align}
The transformation which brings us to the mass basis from the minimal parameterization reads
\begin{align}
 \tilde{\chi}' \rightarrow\, \tilde{\tilde{\chi}}=W^{\,\prime}\, \tilde{\chi}'
 \, , \qquad
\tilde{M}^{\,\prime}_{\tilde{\chi}} \longrightarrow \,  &
\tilde{M}^D_{\tilde{\chi}}=W^{\,\prime}\, \tilde{M}^{\,\prime}_{\tilde{\chi}} \, W^{\,\prime\,\intercal}=  W^{\,\prime}\, R\, \tilde{M}_{\tilde{\chi}} \, R^\intercal \, W^{\,\prime\,\intercal}  \, , \nonumber \\ 
\lambda' \longrightarrow\, & \tilde{\lambda}= \lambda' \, W^{\,\prime\,\intercal} =U\, D\, Od  \, W^{\,\prime\,\intercal}
\end{align}
with
\begin{align}
& W^{\,\prime\,\intercal} =R\, W^\intercal \, .
\end{align}
In particular, in a scenario where the mass matrix emerges as
Eq.~\eqref{eq:mass} we have
\begin{align}
\label{eq:mass-diag}
& W\left( \lambda^\dag\lambda+ \lambda^\intercal\lambda^* \right)W^\intercal=\text{diagonal} \,,\nonumber \\ &
 W\, R^\intercal\left(d^*O^\intercal D^2Od+ dO^\intercal D^2Od^* \right)R\, W^\intercal=\text{diagonal}
\end{align}
or, in the basis of minimal parameterization $\tilde{\chi}'$
\begin{align}
& 
 W^{\,\prime}\left(d^*O^\intercal D^2Od+ dO^\intercal D^2Od^* \right)W^{\,\prime\,\intercal}=\text{diagonal}\,.
\end{align}
Notice that $W^{\,\prime}$ here corresponds to $W$ in Eq.~\eqref{eq:W}.\footnote{In a scenario with $d\propto \mathbf{1}$, we would have that 
$V^\dag=O_R$, $\lambda = U D d O_R$, with $O_R$ an orthogonal transformation.
This implies that $ \lambda^\dag\lambda$ is symmetric and real and diagonalized by an orthogonal matrix.
Then, the mass matrix is diagonalized by the orthogonal matrix $W=O_R$, with eigenvalues given by $D$.
The minimal parameterization is obtained by the transformation $R=O_R=W$:
$\lambda'=\lambda\, O_R^\intercal=U D$, $M'_{\tilde{\chi}}=O_R M_{\tilde{\chi}} O_R^\intercal =M_{\tilde{\chi}}^\text{diag}$, $W^{\,\prime}=\mathbf{1}$.}

\bigskip

\subsection{Indicative Scenarios}
\label{app:scenarios}

The parameters contained in the coupling matrix $\lambda$ determine the relic density as well as the contribution of the new fields to physical processes, such as flavor changing phenomena, which provide potentially stringent constraints. 
Therefore, it is essential to analyze how large the couplings can be.
In particular, it is important to understand the role of the different parameters and the respective experimental limits
in any useful flavor basis.
Hence, it is useful to clarify the relations between the parameters in different bases. 
In this section, we analyze these relations (and in \ref{app:neutral-mesons} the corresponding constraints) starting from some illustrative scenarios characterized by specific textures of the $\lambda$ matrix.

\subsubsection*{Scenario I}

For instance, we can assume that the
diagonal elements are larger than the off-diagonal ones, i.e. the coupling matrix in an initial flavor basis can be written as
\begin{align}\label{eq:lambda1}
&\lambda = \lambda_0 \left(
\begin{array}{ccc}
\hat{\lambda}_{d1} & \epsilon \hat{\lambda}_{d2} & \epsilon \hat{\lambda}_{d3} \\
 \epsilon \hat{\lambda}_{s1} & \hat{\lambda}_{s2} &  \epsilon \hat{\lambda}_{s3}  \\
  \epsilon \hat{\lambda}_{b1} &  \epsilon \hat{\lambda}_{b2} &   \hat{\lambda}_{b3}  
\end{array}
\right)\,,
\end{align}
where $\lambda_0$ is a real parameter of the order of the largest coupling, 
$\epsilon$ is a small real parameter, while $\hat{\lambda}_{\alpha i}$ are parameters which can be of order $\mathcal{O}(1)$.
In this scenario we can compute the rotation matrices and the respective eigenvalues in terms of the Lagrangian parameters as an expansion 
in the small parameter $\epsilon$.
Assuming that the masses of the $\tilde{\chi}$ species are given by Eq.~\eqref{eq:mass}, the orthogonal matrix $W$ [see Eq.~\eqref{eq:mass-basis}] at first order in $\epsilon$ is given by 
\begin{small}
\begin{align}\label{eq:Rchi1}
& W \!= \!
\left(\!
{\footnotesize
\begin{array}{c@{\hspace{1.5\tabcolsep}}c @{\hspace{.5\tabcolsep}}  c}
1 &  
-[W]_{21}
& 
-[W]_{31}
\\
\frac{|\hat{\lambda}_{s1} \hat{\lambda}_{s2}^*|\cos(\delta_{s1}-\delta_{s2})+ |\hat{\lambda}_{d1} \hat{\lambda}_{d2}^*|\cos(\delta_{d1}-\delta_{d2})}{|\hat{\lambda}_{s2}|^2-|\hat{\lambda}_{d1}|^2} \, \epsilon & 
1 &  
-[W]_{32}
\\
 \frac{|\hat{\lambda}_{b1} \hat{\lambda}_{b3}^*|\cos(\delta_{b1}-\delta_{b3})+ |\hat{\lambda}_{d1} \hat{\lambda}_{d3}^*|\cos(\delta_{d1}-\delta_{d3})}{|\hat{\lambda}_{b3}|^2-|\hat{\lambda}_{u1}|^2} \, \epsilon 
& 
\frac{|\hat{\lambda}_{s2} \hat{\lambda}_{s3}^*|\cos(\delta_{s2}-\delta_{s3})+ |\hat{\lambda}_{b2} \hat{\lambda}_{b3}^*|\cos(\delta_{b2}-\delta_{b3})}{|\hat{\lambda}_{b3}|^2-|\hat{\lambda}_{s2}|^2} \, \epsilon
& 1
\end{array}
}
\! \right)\!+\!\mathcal{O}( \epsilon^2 )
\end{align}
\end{small}%
The coupling matrix $\lambda$ can be written in terms of $U$, $V$ and $D$ in Eq.~\eqref{eq:lambda-diag} with the unitary matrices given by
\begin{align}\label{eq:UL1}
& U= 
\left(
\begin{array}{ccc}
1 &  \frac{\hat{\lambda}^*_{s1} \hat{\lambda}_{d1}+ \hat{\lambda}^*_{s2} \hat{\lambda}_{d2}}{|\hat{\lambda}_{s2}|^2-|\hat{\lambda}_{d1}|^2} \, \epsilon
& \frac{\hat{\lambda}_{b1}^* \hat{\lambda}_{d1}+ \hat{\lambda}_{b3}^* \hat{\lambda}_{d3}}{|\hat{\lambda}_{b3}|^2-|\hat{\lambda}_{d1}|^2}\, \epsilon \\
-  \frac{\hat{\lambda}_{s1} \hat{\lambda}_{d1}^*+ \hat{\lambda}_{s2} \hat{\lambda}_{d2}^*}{|\hat{\lambda}_{s2}|^2-|\hat{\lambda}_{d1}|^2} \, \epsilon & 1 & 
-  \frac{\hat{\lambda}_{s2} \hat{\lambda}_{b2}^*+ \hat{\lambda}_{s3} \hat{\lambda}_{b3}^*}{|\hat{\lambda}_{s2}|^2-|\hat{\lambda}_{b3}|^2}
\, \epsilon   \\
- \ \frac{\hat{\lambda}_{b1} \hat{\lambda}_{d1}^*+ \hat{\lambda}_{b3} \hat{\lambda}_{d3}^*}{|\hat{\lambda}_{b3}|^2-|\hat{\lambda}_{d1}|^2} \, \epsilon & 
 \frac{\hat{\lambda}_{s2}^* \hat{\lambda}_{b2}+ \hat{\lambda}_{s3}^* \hat{\lambda}_{b3}}{|\hat{\lambda}_{s2}|^2-|\hat{\lambda}_{b3}|^2}
\, \epsilon  & 1
\end{array}
\right)+\mathcal{O}( \epsilon^2 )
\end{align}
and, with $\text{arg}(\hat{\lambda}_{\alpha i})=\delta_{\alpha i}$
\begin{align}\label{eq:UR1}
& V= 
\left(
\begin{array}{ccc}
e^{-i\delta_{d1}} &  \frac{\hat{\lambda}^*_{s1} \hat{\lambda}_{s2}+ \hat{\lambda}^*_{d1} \hat{\lambda}_{d2}}{|\hat{\lambda}_{s2}|^2-|\hat{\lambda}_{d1}|^2} e^{-i\delta_{s2}} \, \epsilon
& \frac{\hat{\lambda}_{b1}^* \hat{\lambda}_{b3}+ \hat{\lambda}_{d1}^* \hat{\lambda}_{d3}}{|\hat{\lambda}_{b3}|^2-|\hat{\lambda}_{d1}|^2}\,e^{-i\delta_{b3}} \epsilon \\
-  \frac{\hat{\lambda}_{s1} \hat{\lambda}_{s2}^*+ \hat{\lambda}_{d1} \hat{\lambda}_{d2}^*}{|\hat{\lambda}_{s2}|^2-|\hat{\lambda}_{d1}|^2}  e^{-i \delta_{d1}} \, \epsilon & e^{-i\delta_{s2}} & 
-  \frac{\hat{\lambda}_{s2}^* \hat{\lambda}_{s3}+ \hat{\lambda}_{b2}^* \hat{\lambda}_{b3}}{|\hat{\lambda}_{s2}|^2-|\hat{\lambda}_{b3}|^2}
e^{-i\delta_{b3}} \, \epsilon   \\
- \ \frac{\hat{\lambda}_{b1} \hat{\lambda}_{b3}^*+ \hat{\lambda}_{d1} \hat{\lambda}_{d3}^*}{|\hat{\lambda}_{b3}|^2-|\hat{\lambda}_{d1}|^2} e^{-i\delta_{d1}} \, \epsilon & 
 \frac{\hat{\lambda}_{s2} \hat{\lambda}_{s3}^*+ \hat{\lambda}_{b2} \hat{\lambda}_{b3}^*}{|\hat{\lambda}_{s2}|^2-|\hat{\lambda}_{b3}|^2}
e^{-i\delta_{s2}} \, \epsilon  & e^{-i\delta_{b3}}
\end{array}
\right)+\mathcal{O}(\epsilon^2)\,,
\end{align}
where we also transferred three of the phases in $U$ into $V$.
Notice that these expressions are only valid in the small angles approximation, that is, 
the denominators cannot be small, implying for the diagonal couplings that $||\hat{\lambda}_{ii}|^2-|\hat{\lambda}_{jj}|^2|\gg \epsilon$.
As for the eigenvalues $D$, at first order they are given by the diagonal elements of the matrix $\lambda$ in Eq.~\eqref{eq:lambda1}:
\begin{align}
& D=\lambda_0\cdot \text{diag}
\Big( |\hat{\lambda}_{d1}|+\mathcal{O}(\epsilon^2)\, , |\hat{\lambda}_{s2}| +\mathcal{O}(\epsilon^2)\,  ,  |\hat{\lambda}_{b3}| +\mathcal{O}(\epsilon^2) \Big)\,. & 
\end{align}
In order to obtain a minimal parameterization, the unitary matrix $V$ can be written as 
in Eq.~\eqref{eq:UR-OFR} $V^\dag=OFR$, where $O$ is the orthogonal matrix diagonalizing the symmetric
unitary matrix $V^\dag V^*$. At first order in $\epsilon$ we have
\begin{align}\label{eq:O1}
& O \!= \!
\left(\!
{\footnotesize
\begin{array}{c  c @{\hspace{1\tabcolsep}}  c}
1  
& 
-[O]_{21}
&  
-[O]_{31}
\\
-\frac{|\hat{\lambda}_{s1}\hat{\lambda}_{s2}|\sin(\delta_{s1}-\delta_{s2})+|\hat{\lambda}_{d1}\hat{\lambda}_{d2}|\sin(\delta_{d1}-\delta_{d2})}{(|\hat{\lambda}_{s2}|^2-|\hat{\lambda}_{d1}|^2)\, \sin(\delta_{s2}-\delta_{d1})}\, \epsilon
& 1
& 
-[O]_{32}
\\
-\frac{|\hat{\lambda}_{b1}\hat{\lambda}_{b3}|\sin(\delta_{b1}-\delta_{b3})+|\hat{\lambda}_{d1}\hat{\lambda}_{d3}|\sin(\delta_{d1}-\delta_{d3})}{(|\hat{\lambda}_{b3}|^2-|\hat{\lambda}_{d1}|^2)\, \sin(\delta_{b3}-\delta_{d1})}\, \epsilon
&-\frac{|\hat{\lambda}_{s2}\hat{\lambda}_{s3}|\sin(\delta_{s2}-\delta_{s3})+|\hat{\lambda}_{b2}\hat{\lambda}_{b3}|\sin(\delta_{b2}-\delta_{b3})}{(|\hat{\lambda}_{s2}|^2-|\hat{\lambda}_{b3}|^2)\, \sin(\delta_{s2}-\delta_{b3})}\, \epsilon
& 1
\end{array}
}
\!\right)
\!+\!\mathcal{O}(\epsilon^2)\,.
\end{align}
As regards the matrix of phases $d$, 
we have
\begin{align}
& d=
\left(\begin{array}{ccc}
e^{i \delta_{d1}} \big(1+ i\, \theta_{d1}[\hat{\lambda}_{\alpha i}]\epsilon^2 \big) & 0 & 0 \\
0 & e^{i \delta_{s2}} \big(1+ i\, \theta_{s2}[\hat{\lambda}_{\alpha i}]\epsilon^2 \big) & 0 \\
0 & 0 & e^{i \delta_{b3}} \big(1+ i\, \theta_{b3}[\hat{\lambda}_{\alpha i}]\epsilon^2 \big)
\end{array}
\right)\,.
\end{align}

Let us remark that the expression~\eqref{eq:O1} is only valid for small angles. This means that not only the moduli of the diagonal couplings should differ between each other $||\hat{\lambda}_{ii}|^2-|\hat{\lambda}_{jj}|^2|\gg \epsilon$, 
but also the differences between respective phases should be sufficiently large $|\sin(\delta_{ii}-\delta_{jj})|\gg \epsilon$.
In fact, small angles in $V$ do not imply small angles for the orthogonal matrices $O$ and $R$,
which instead can contain large rotations,
depending on the values of the phases contained in the matrix $\lambda$.
Namely, in a situation with small angles in $V$ and large angles in $O$,
also $R$, and thus $W^{\,\prime\,\intercal} =R\, W^\intercal $, would exhibit large rotations, while small phases would appear in
the diagonal phase matrix.

The texture \eqref{eq:lambda1} can be in conflict with experimental limits, since
the magnitude of the diagonal couplings can be in general subject to constraints from flavor changing phenomena (see~\ref{app:neutral-mesons}).
A viable texture would then contain some hierarchy between the eigenvalues, as
\begin{align}
& D \approx 
\text{diag}\left( \epsilon_1 \hat{\lambda}_{d1} \, , \epsilon_2 \hat{\lambda}_{s2} \, , \hat{\lambda}_{b3} 
\right)
\end{align}
together with small mixing angles.

\subsubsection*{Scenario II }
Let us assume a different kind of scenario, in which one species of $\tilde{\chi}$ is more strongly coupled than the others, for instance with couplings of the type:
\begin{align}\label{eq:lambda3}
&\lambda =\lambda_0 \left(
\begin{array}{ccc}
\hat{\lambda}_{d1} \epsilon^2 &  \hat{\lambda}_{d2} \epsilon &  \hat{\lambda}_{d3}  \\
\hat{\lambda}_{s1} \epsilon^2 & \hat{\lambda}_{s2} \epsilon &  \hat{\lambda}_{s3}  \\
 \hat{\lambda}_{b1} \epsilon^2 &   \hat{\lambda}_{b2} \epsilon &   \hat{\lambda}_{b3}  
\end{array}
\right) \,.
\end{align}
In this scenario the mixing angles from the left are generically large while the mixing angles from the right can be small:
\begin{align}\label{eq:UR3}
& U_{\alpha i}\approx \mathcal{O}(1) \, ,
\qquad 
V \approx  \left(
\begin{array}{ccc}
e^{-i\delta_{d1}}&  \mathcal{O}(\epsilon) &  \mathcal{O}(\epsilon^2)  \\
\mathcal{O}(\epsilon) &  e^{-i\delta_{s2}} 
&  \mathcal{O}(\epsilon)  \\
\mathcal{O}(\epsilon^2) &    
\mathcal{O}(\epsilon)
&   e^{-i\delta_{b3}}
\end{array}
\right)\,.
\end{align}
The orthogonal matrix $W$ also shows small mixing angles of order $\epsilon$ and $\epsilon^2$: 
\begin{small}
\begin{align}\label{eq:Rchi3}
&W \! \approx\!  \left(\!\!
{\footnotesize
\begin{array}{c @{\hspace{1\tabcolsep}} c @{\hspace{-16\tabcolsep}} c}
1\!+\! \mathcal{O}(\epsilon^2) &  -\frac{ |\hat{\lambda}_{s1} \hat{\lambda}_{s2}| \cos(\delta_{s1} - \delta_{s2}) +
 |\hat{\lambda}_{b1} \hat{\lambda}_{b2}| \cos(\delta_{b1} - \delta_{b2}) + 
|\hat{\lambda}_{d1} \hat{\lambda}_{d2}| \cos(\delta_{d1} - \delta_{d2}) }{|\hat{\lambda}_{b2}|^2 + |\hat{\lambda}_{s2}|^2+ |\hat{\lambda}_{d2}|^2}  \epsilon & \mathcal{O}(\epsilon^2)  \\
-[W]_{12}
& 1\!+\! \mathcal{O}(\epsilon^2) &   
-\frac{ |\hat{\lambda}_{s2} \hat{\lambda}_{s3}| \cos(\delta_{s2} - \delta_{s3}) +
 |\hat{\lambda}_{b2} \hat{\lambda}_{b3}| \cos(\delta_{b2} - \delta_{b3}) + 
|\hat{\lambda}_{d2} \hat{\lambda}_{d3}| \cos(\delta_{d2} - \delta_{d3}) }{|\hat{\lambda}_{b3}|^2 + |\hat{\lambda}_{s3}|^2+ |\hat{\lambda}_{d3}|^2}   \epsilon
 \\
\mathcal{O}(\epsilon^2) &  -[W]_{23}  &  1\! + \!  \mathcal{O}(\epsilon^2) 
\end{array}
}
\!\!\right).
\end{align}
\end{small}
As for the eigenvalues we have
\begin{align}\label{eq:D3}
&D = \lambda_0 \left(
\begin{array}{ccc}
\mathcal{O}(\epsilon^2) &  0 &  0 \\
0  &  \mathcal{O}(\epsilon)  &  0  \\
0 &   0 &   \sqrt{|\hat{\lambda}_{b3}|^2 + |\hat{\lambda}_{s3}|^2+ |\hat{\lambda}_{d3}|^2}   +   \mathcal{O}(\epsilon^2) 
\end{array}
\right)\,.
\end{align}
Let us note that in a model with the mass matrix given by Eq.~\eqref{eq:mass} (dark minimal flavor violation) the choice of the third species as the more coupled is only for illustration, in fact any column of the $\lambda$ matrix could be the largest since the order of the eigenvalues can be reshuffled and large left rotations are equivalent to a renaming of the DM fields.

Neutral mesons systems can impose stringent constraints on the magnitude of the products of the strongest couplings
(e.g. $\lambda_{d3}\lambda_{s3}$, $\lambda_{d3}\lambda_{b3}$, $\lambda_{s3}\lambda_{b3}$ in the example, respectively from neutral kaons and $B$ mesons).
Then a more realistic texture for the couplings would show additional hierarchy between the three larger couplings.

\subsubsection*{Scenario III}

We can envisage a scenario where the DM couplings to the SM fermions manifest some hierarchy depending on the SM family. Namely, the DM species can be more strongly coupled to one SM family than the others, for example, with the third family:
\begin{align}\label{eq:lambda4}
&\lambda = \lambda_0 \left(
\begin{array}{ccc}
\hat{\lambda}_{d1} \epsilon^2 &  \hat{\lambda}_{d2} \epsilon^2 &  \hat{\lambda}_{d3} \epsilon^2 \\
\hat{\lambda}_{s1} \epsilon & \hat{\lambda}_{s2} \epsilon &  \hat{\lambda}_{s3} \epsilon  \\
 \hat{\lambda}_{b1} &   \hat{\lambda}_{b2} &   \hat{\lambda}_{b3}  
\end{array}
\right)\,.
\end{align}
This corresponds to the \emph{bottom-philic} scenario considered in Sec.~\ref{sec:down-quarks}. In this case, since the matrix $\lambda^\dag\lambda$ contains elements of the same order of magnitude,
the mixing angles from the right are generically large [as well as the angles of the orthogonal matrix which diagonalizes the $\tilde{\chi}$ mass matrix \eqref{eq:mass}]
\begin{align}
    & V_{ ij}\approx \mathcal{O}(1) \, ,
    \qquad W_{ ij}\approx \mathcal{O}(1)
\end{align}
while the mixing angles from the left show hierarchy.
Specifically, we can find the mixing from the left by diagonalizing the matrix $\lambda\lambda^\dag$:
\begin{align}\label{eq:UL3r}
&U =  \left(
\begin{array}{ccc}
1 &  
\frac{\hat{\lambda}_{s1}^*\hat{\lambda}_{d1}+\hat{\lambda}_{s2}^*\hat{\lambda}_{d2}+\hat{\lambda}_{s3}^*\hat{\lambda}_{d3}}{|\hat{\lambda}_{s1}|^2 + |\hat{\lambda}_{s2}|^2+ |\hat{\lambda}_{s3}|^2}\,  \epsilon
&  \mathcal{O}(\epsilon^2)  \\
-\frac{\hat{\lambda}_{s1}\hat{\lambda}_{d1}^*+\hat{\lambda}_{s2}\hat{\lambda}_{d2}^*+\hat{\lambda}_{s3}\hat{\lambda}_{d3}^*}{|\hat{\lambda}_{s1}|^2 + |\hat{\lambda}_{s2}|^2+ |\hat{\lambda}_{s3}|^2}\,  \epsilon
&  1 
&   
\frac{\hat{\lambda}_{s3}\hat{\lambda}_{b3}^*+\hat{\lambda}_{s2}\hat{\lambda}_{b2}^*+\hat{\lambda}_{s1}\hat{\lambda}_{b1}^*}{|\hat{\lambda}_{b1}|^2 + |\hat{\lambda}_{b2}|^2+ |\hat{\lambda}_{b3}|^2}\,  \epsilon  \\
\mathcal{O}(\epsilon^2) &    
-\frac{\hat{\lambda}_{s3}^*\hat{\lambda}_{b3}+\hat{\lambda}_{s2}^*\hat{\lambda}_{b2}+\hat{\lambda}_{s1}^*\hat{\lambda}_{b1}}{|\hat{\lambda}_{b1}|^2 + |\hat{\lambda}_{b2}|^2+ |\hat{\lambda}_{b3}|^2}\,  \epsilon
&   1
\end{array}
\right)+ \mathcal{O}(\epsilon^2)\,.
\end{align}
For the eigenvalues we get
\begin{align}\label{eq:D3r}
&D^2 = \lambda_0^2 \left(
{\footnotesize
\begin{array}{c@{\hspace{-1\tabcolsep}}c@{\hspace{-6\tabcolsep}}c}
\mathcal{O}(\epsilon^4) &  0 &  0 \\
0  &  \Big(|\hat{\lambda}_{s1}|^2 + |\hat{\lambda}_{s2}|^2+ |\hat{\lambda}_{s3}|^2   - 
\frac{\big|\hat{\lambda}_{s1}^*\hat{\lambda}_{b1}+\hat{\lambda}_{s2}^*\hat{\lambda}_{b2}+\hat{\lambda}_{s3}^*\hat{\lambda}_{b3}\big|^2}{|\hat{\lambda}_{b1}|^2 + |\hat{\lambda}_{b2}|^2+ |\hat{\lambda}_{b3}|^2}\Big)\,\epsilon^2
+\mathcal{O}(\epsilon^4)   &  0  \\
0 &   0 &   |\hat{\lambda}_{b1}|^2 + |\hat{\lambda}_{b2}|^2+ |\hat{\lambda}_{b3}|^2   +   \mathcal{O}(\epsilon^2) 
\end{array}
}
\right)\,.
\end{align}

Let us notice that one element of the matrix $\lambda$ can always be made zero by an orthogonal rotation of the $\tilde{\chi}$ fields.
In particular, in the limit $\epsilon\rightarrow 0$,  the matrix $\lambda$ has the form
\begin{align}
& \lambda_{\epsilon=0}= \left(
\begin{array}{ccc}
0 &  0 &  0 \\
0 &  0 &  0   \\
 \lambda_{b1} &   \lambda_{b2} &   \lambda_{b3} 
\end{array}
\right) \quad 
\rightarrow \quad \tilde{\lambda}_{\epsilon=0}= \lambda_{\epsilon=0} W_0^\intercal = \left(
\begin{array}{ccc}
0 &  0 &  0 \\
0 &  0 &  0   \\
0 &   \tilde{\lambda}_{0 b2} &   \tilde{\lambda}_{0 b3}
\end{array}
\right)\,,
\end{align}
i.e.,\,. we can perform an orthogonal rotation $\lambda\rightarrow \lambda W^\intercal$ to set e.g. $[\lambda W^\intercal]_{11}=0$, 
which implies the conditions:
\begin{align}\label{eq:W0}
&\sum_i \lambda_{bi}W_{0\, 1i}=0 \,.
\end{align}
This has to be the same transformation which diagonalizes $\lambda^\dag\lambda$ and its real part $(\lambda^\dag\lambda+\lambda^\intercal\lambda^*)/2$. In fact, 
in the limit $\epsilon\rightarrow 0$ the matrix $\lambda^\dag\lambda$ has two zero eigenvalues, while its real part has one zero eigenvalue, e.g., the first one.
Namely, the diagonalization of $(\lambda^\dag\lambda+\lambda^\intercal\lambda^*)/2$ gives
\begin{align}
&\big[W \, \text{Re}\big(\lambda^\dag\lambda\big) \,W^\intercal   \big]_{\ell n}= \text{Re}\Big(\sum_{i,j}\lambda_i^*W_{\ell i }\lambda_j W_{ n j}\Big) = 0 \,, \qquad \ell\neq n 
\end{align}
and for the eigenvalues we have
\begin{align}
& \Big|  \sum_i \lambda_i W_{0\, 1i} \Big|^2=0 \, , \qquad
 \Big|  \sum_i \lambda_i W_{0\, 3i} \Big|^2
\sim \Big|  \sum_i \lambda_i W_{0\, 2i} \Big|^2 \sim \mathcal{O}(\lambda_0^2)\,.
\end{align}

For $\epsilon\neq 0$, 
we have
\begin{align}
&\big[ \lambda^\dag\lambda \big]_{ij}= \lambda_{bi}^*\lambda_{bj}+ \mathcal{O}(\epsilon^2) \, .
\end{align}
The trace and the determinant of $(\lambda^\dag\lambda+\lambda^\intercal\lambda^*)$ 
are respectively of order $\lambda_0^2$ and $\epsilon^2\lambda_0^6$, with e.g. the first eigenvalue which is zero in the limit $\epsilon\rightarrow 0$.
Then, the diagonalization of the mass matrix reads
\begin{align}
&W \, \text{Re}\big(\lambda^\dag\lambda\big) \,W^\intercal   = 
\text{Re} \big[W_1 W_0 \, \big(\lambda^\dag\lambda\big) \,W_0^\intercal W_1^\intercal   \big]=
W_1
\left(\begin{array}{ccc}
\mathcal{O}(\epsilon^2) & \mathcal{O}(\epsilon^2) & \mathcal{O}(\epsilon^2) \\
\mathcal{O}(\epsilon^2) & \mathcal{O}(\lambda_0^2) & \mathcal{O}(\epsilon^2) \\
\mathcal{O}(\epsilon^2) & \mathcal{O}(\epsilon^2) & \mathcal{O}(\lambda_0^2)
\end{array}
\right)
W_1^\intercal   \, .
\end{align}
Hence, the transformation leading to the mass basis is
\begin{align}\label{eq:W1}
& W^\intercal=W_0^\intercal W_1^\intercal \approx  W_0^\intercal \left(\begin{array}{ccc}
1 & \mathcal{O}(\epsilon^2) & \mathcal{O}(\epsilon^2) \\
\mathcal{O}(\epsilon^2) & 1 & \mathcal{O}(\epsilon^2) \\
\mathcal{O}(\epsilon^2) & \mathcal{O}(\epsilon^2) & 1
\end{array}
\right) \, .
\end{align}
Thus, using \eqref{eq:W0} and \eqref{eq:W1}, in the mass basis we have
\begin{align}\label{eq:tildelambda3}
&\tilde{\lambda}= \lambda W^\intercal = \left(
\begin{array}{ccc}
\mathcal{O}(\epsilon^2) &  \mathcal{O}(\epsilon^2) &  \mathcal{O}(\epsilon^2) \\
\mathcal{O}(\epsilon) &  \mathcal{O}(\epsilon) &  \mathcal{O}(\epsilon)   \\
\mathcal{O}(\epsilon^2) &   \tilde{\lambda}_{b2} &   \tilde{\lambda}_{b3}
\end{array}
\right) \, .
\end{align}

\subsection{Neutral Mesons}
\label{app:neutral-mesons}

In the SM, the short-distance contribution to the transition 
$K^0(d\bar{s})\leftrightarrow \bar{K}^0(\bar{d}s)$
arises from weak box diagrams.
The weak short-distance contribution to the mass splitting $\Delta m_K=m_{K_L}-m_{K_S}$
and the $CP$-violating effects 
are described by
the off-diagonal term $M_{12}$ of the mass matrix of neutral kaons
$M_{12}=- \langle K^0|\mathcal{L}_{\Delta S=2}|\bar{K}^0\rangle /( 2m_{K})$, which in the SM is
\begin{equation} \label{eq:M12sm}
M_{12}^\text{SM}=
-\frac{G^2_F m^2_W f^2_Km_{K}B_K}{12\pi^2}
\left(\lambda^{*2}_c \, S(x_c)+\lambda^{*2}_t \, S(x_t)+2\,\lambda^*_c\lambda^*_t \, S(x_c,x_t)\right)\,,
\end{equation}
where $\lambda_a=V_{as}^*V_{ad}$, $x_a={m_a^2}/{m_W^2}$,
$f_K$ is the kaon decay constant, which can be determined in lattice QCD to be
$f_K=155.7(0.7)$ MeV~\cite{FlavourLatticeAveragingGroupFLAG:2021npn},
 $m_{K}=497.611\pm0.013$~MeV is the neutral kaon mass, 
the factor $B_{K}$ is the correction to the vacuum insertion approximation which is calculated in
lattice QCD,
$B_K=0.7533(91)$~\cite{FlavourLatticeAveragingGroupFLAG:2024oxs},
and $S(x_i)$ are the Inami-Lim
functions~\cite{Inami:1980fz} (corrected by short-distance QCD effects~\cite{Buchalla:1995vs}).

The modulus and the imaginary part of the mixing mass $M^K_{12}$
describe short-distance contributions in the mass splitting and $CP$-violation in $\bar{K}^0\leftrightarrow K^0$ transitions
\cite{Buchalla:1995vs}
\begin{align}
&\Delta m_K \approx 2|M^K_{12}|
+\Delta m_{K,\text{LD}} %\\ &
\, , \quad
 |\epsilon_K| \approx \frac{|\text{Im} M_{12}^K|}{\sqrt{2}\Delta m_K}~,
\end{align}
(in the standard parameterization of $V_\text{CKM}$, i.e. $\lambda_u$ real).
$\Delta m_{K,\text{LD}}$ represents the long-distance contributions which are difficult to evaluate \cite{Bai:2014cva,Bai:2018mdv}.
However the short distance contribution 
gives the dominant contribution to
the experimental determination 
$\Delta m_{K,\text{exp}}=(3.484\pm 0.006) \times 10^{-15}$~GeV~\cite{ParticleDataGroup:2024cfk}. From experimental data one also obtains $|\epsilon_{K}|_{\rm exp}=(2.228 \pm 0.011) \times 10^{-3}$~\cite{ParticleDataGroup:2024cfk}.

The dominant short-distance contribution to the $B^0_{d}$-$\bar{B}_{d}^0$ mixing
in the SM is given by
\begin{align}
\label{eq:dmBB}
&\Delta M_{B_{d},\text{SM}}=2|M_{12,\text{SM}}^B|=
m_{B_{d}}f^2_{B_{d}}B_{B_{d}}\frac{G_F^2m_W^2}{6\pi^2}|(V_{tb}V_{td}^*)^2|S(x_t)~,
\end{align}
where $M_{12,\text{SM}}^{B_d}=- \langle B_d^0|\mathcal{L}_{\Delta B_d=2}|\bar{B}_d^0\rangle /( 2m_{B_d^{0}})$ 
and 
$B_{B_{d}}$ is the correction factor to the vacuum-insertion approximation. In neutral $B$-mesons system long distance contributions are estimated to be small. The analogous expression can be obtained for
$B^0_{s}$-$\bar{B}_{s}^0$ system by substituting $d \to s$.
 Lattice QCD calculations yield
$f_{B_d}\sqrt{B_{B_d}}=210.6(5.5)$~MeV and $f_{B_s}\sqrt{B_{B_s}}=256.1(5.7)$~MeV~\cite{FlavourLatticeAveragingGroupFLAG:2021npn}.
The experimental 
results are 
$\Delta M_{B_d,\text{exp}}=(3.334\pm 0.013) \times 10^{-13} \, \text{GeV}$,
$\Delta M_{B_s,\text{exp}}=(1.1693\pm 0.0004) \times 10^{-11} \, \text{GeV}$~\cite{ParticleDataGroup:2024cfk}.

In the illustrated model with Majorana DM species and the extra scalar field coupled to the SM down quarks,
new contributions to neutral meson mixing are generated through box diagrams (see Fig.~\ref{fig:meson-mixing-diag}).
For instance, the additional contribution to the neutral kaon system reads (see Ref.~\cite{Acaroglu:2021qae}):
\begin{align}
& M_{12,\text{NP}}^K\approx \frac{1}{3} m_K f_K^2\, \frac{1}{128\pi^2M_\varphi^2}
\left( \sum_{ij}
 \tilde{\lambda}_{di}^*\tilde{\lambda}_{si}\tilde{\lambda}_{dj}^*\tilde{\lambda}_{sj} \, F(x_i,x_j) - 
  \sum_{ij} (\tilde{\lambda}_{si}\tilde{\lambda}_{dj}^{*})^2 \, 2\,G(x_i,x_j) \right)\,,
\end{align}
where $x_i=M_{\tilde{\chi}_i}^2/M_\varphi^2$,
\begin{align}
& \tilde{\lambda}= \lambda^{\,\prime}\, W^{\,\prime\, T}= \lambda\, W^\intercal =U\, D \, Od\, W^{\prime\, \intercal}
\end{align}
and $F(x_i,x_j)$ and $G(x_i,x_j)$ are functions with values $1 \geq F(x_i,x_j) \geq 0$ and $0 \leq 2\, G(x_i,x_j)  \leq \frac{1}{3}$.
For $x_{i,j}\leq 1$, $F(x_i,x_j)$ is a decreasing function of $x_i,x_j$ assuming values between $1$ and $1/3$ while
$2\,G(x_i,x_j)$ is an increasing function of $x_i,x_j$ assuming values between $0$ and $1/3$.
In particular, for $x_{i,j}\rightarrow 0$, $F(x_i,x_j)\rightarrow 1$ and $G(x_i,x_j)\rightarrow 0$,
for $x_{j}\rightarrow 0$ and $x_{i}\rightarrow 1$, $F(x_i,x_j)\rightarrow 1/2$ and $G(x_i,x_j)\rightarrow 0$,
while for $x_{j},x_i\rightarrow 1$, $F(x_i,x_j),\, 2\,G(x_i,x_j) \rightarrow 1/3$.
For $x_{i(j)}\rightarrow \infty$, $F(x_i,x_j)\rightarrow 0$ and $G(x_i,x_j)\rightarrow 0$.

Approximate bounds on the new physics contribution can be estimated 
by comparison with the SM contribution,
as $|M_{12,\rm NP}^K|<|M_{12, \text{SM}}^K|\,\Delta_{K} $,
$ |\text{Im}M_{12,NP}^K|<|\text{Im}M_{12, \text{SM}}^K| \, \Delta_{\epsilon_{K}} $, e.g. by setting
$\Delta_{K}=1$ and, approximately using the results in Ref.~\cite{Bona:2022zhn} at $95\%$ CL, 
$\Delta_{\epsilon_{K}}= 0.3$.
By comparing the effective Lagrangians we obtain
\begin{align}\label{eq:limit-K-general}
&\left| \sum_{ij}
 \tilde{\lambda}_{di}^*\tilde{\lambda}_{si}\tilde{\lambda}_{dj}^*\tilde{\lambda}_{sj} \, F(x_i,x_j) - 
  \sum_{ij} (\tilde{\lambda}_{si}\tilde{\lambda}_{dj}^{*})^2 \, 2\,G(x_i,x_j)  \right| <0.6\times 10^{-5}  \, \Big(\frac{M_\varphi}{10^2\,\text{GeV}}\Big)^2 \,,
   \nonumber \\ 
  &\left| \text{Im}\left[\sum_{ij}
 \tilde{\lambda}_{di}^*\tilde{\lambda}_{si}\tilde{\lambda}_{dj}^*\tilde{\lambda}_{sj} \, F(x_i,x_j) - 
  \sum_{ij} (\tilde{\lambda}_{si}\tilde{\lambda}_{dj}^{*})^2 \, 2\,G(x_i,x_j) \right] \right| < 1.4\times 10^{-8} \, \Big(\frac{M_\varphi}{10^2\,\text{GeV}}\Big)^2 \,.
\end{align}

Analogous expressions can be written for the new contributions to neutral $B$-mesons systems:
\begin{align}\label{eq:limit-B-general}
& M_{12,\rm NP}^{B_d}\approx \frac{1}{3} m_{B_{d}} f_{B_{d}}^2\, \frac{1}{128\pi^2M_\varphi^2}\left( \sum_{ij}
 \tilde{\lambda}_{di}^*\tilde{\lambda}_{bi}\tilde{\lambda}_{dj}^*\tilde{\lambda}_{bj} \, F(x_i,x_j) - 2\, \sum_{ij} (\tilde{\lambda}_{bi}\tilde{\lambda}_{dj}^{*})^2 \, G(x_i,x_j) \right)
\end{align}
and similarly for $B_s$ with the substitution $d\rightarrow s$.
The results obtained in Ref.~\cite{Bona:2022zhn} at $95\%$ CL
approximately give a constraint
$\Delta M_{B_{d(s)}}^\text{new}< \Delta M_{B_{d(s)}}^\text{SM} \Delta_{B_{d(s)}}$ 
with $\Delta_{B_d}=0.3$, $\Delta_{B_s}= 0.2$. 
Then, the combination of couplings and Inami-Lim functions in the different scenarios are limited respectively as
\begin{align}\label{eq:limit-B-general}
& \left| \sum_{ij}
 \tilde{\lambda}_{di}^*\tilde{\lambda}_{bi}\tilde{\lambda}_{dj}^*\tilde{\lambda}_{bj} \, F(x_i,x_j) - 2\, \sum_{ij} (\tilde{\lambda}_{bi}\tilde{\lambda}_{dj}^{*})^2 \, G(x_i,x_j) \right|
< 0.8\times 10^{-5} \, \Big(\frac{M_\varphi}{10^2\,\text{GeV}}\Big)^2 \,,
\\ 
& \left|  \sum_{ij}
 \tilde{\lambda}_{si}^*\tilde{\lambda}_{bi}\tilde{\lambda}_{sj}^*\tilde{\lambda}_{bj} \, F(x_i,x_j) - 2\, \sum_{ij} (\tilde{\lambda}_{bi}\tilde{\lambda}_{sj}^{*})^2 \, G(x_i,x_j)  \right|
 < 1.2\times 10^{-4} \, \Big(\frac{M_\varphi}{10^2\,\text{GeV}}\Big)^2\,. 
\end{align}

In the following, we study these constraints in different scenarios. Namely, 
we analyze the constraints in different regimes of mass differences $M_\varphi^2-M_{\tilde{\chi}_i}^2$, considering the scenarios generated by the three textures in Eqs.~\eqref{eq:lambda1},~\eqref{eq:lambda3}, and~\eqref{eq:lambda4}.

\paragraph{1.}
For instance, with $x_{i}\rightarrow 0$ for all species we would have the extra contribution to kaon mixing
\begin{align}\label{eq:KK00}
 M_{12,\text{NP}}^K &
 \approx 
 \frac{1}{3} m_K f_K^2 \frac{1}{128\pi^2M_\varphi^2} \big[\big(\tilde{\lambda}\tilde{\lambda}^\dag\big)_{sd}\big]^2
 \nonumber \\ & =
  \frac{1}{3} m_K f_K^2 \frac{1}{128\pi^2M_\varphi^2} \big[\big(\lambda\lambda^\dag\big)_{sd}\big]^2 
   \nonumber \\ & =
   \frac{1}{3} m_K f_K^2 \frac{1}{128\pi^2M_\varphi^2} \big(\sum_iU_{s i}D_i^2U_{di}^*\big)^2 \,.
\end{align}
The relevance of the bounds \eqref{eq:limit-K-general} also depend on the texture of the coupling matrix.
For instance, for the three textures I, II, III in Eqs.~\eqref{eq:lambda1},~\eqref{eq:lambda3}, and~\eqref{eq:lambda4}
we would have respectively
\begin{align}\label{eq:lim-K-1}
 \Big|\big(\lambda \lambda^\dag\big)_{sd} \Big|  =  
 \Big|\big(U D^2U^\dag\big)_{sd}\Big|
&  \approx 
 \left\{
 \begin{array}{l}
 \text{I:}\quad \epsilon \, \big|  \hat{\lambda}_{d1}^*\hat{\lambda}_{s1} +\hat{\lambda}_{d2}^*\hat{\lambda}_{s2} \big| \lambda_0^2 \approx
 |s_{12}^\theta| \, \big| D_2^2 - D_1^2  \big| \\
\\
\text{II:}\quad |\lambda_{sj(=3)}\lambda_{dj(=3)}^*| \approx |U_{d3}^*U_{s3}|D_3^2 \hspace{1cm} < 2.5\times 10^{-3}\, \Big(\frac{M_\varphi}{10^2\,\text{GeV}}\Big) \\
 \\
\text{III:}\quad \epsilon^{n(=3)}\,\big| \sum_i \hat{\lambda}_{di}^*\hat{\lambda}_{si} \big| \lambda_0^2 
\approx | s^\theta_{13} s^\theta_{23} \, e^{i(\delta_{13}-\delta_{23})} \, D_3^2+s^\theta_{12} \, e^{i\delta_{12}}\, D_2^2|
 \end{array} \right. 
\end{align}
[where $s_{12}^\theta = \sin(\theta_{ij})$ as defined in \eqref{eq:U_parameterization}]. The index $j$ indicates the $\tilde{\chi}$ species with larger couplings [$j=3$ in~Eq.~\eqref{eq:lambda3}] 
and 
$n$ depends on which SM species is mostly coupled to DM [$n=3$ in Eq.~\eqref{eq:lambda3}].
Note that $s_{12}^\theta \sim \epsilon$ in the first case, $|U_{Ld3}^*U_{Ls3}|\sim 1$ generically in the second case, while for the third 
$\lambda$ texture we have $s_{12}^\theta \, , s_{23}^\theta \sim \epsilon$,
$s_{13}^\theta  \sim \epsilon^2$ with $D_2\sim \epsilon\, \lambda_0$. 
From the bound on $CP$-violation 
the respective imaginary parts should be $<2\times 10^{-4}\, m_\phi/10^2\,\text{GeV}$,
e.g. for the first scenario 
\begin{align}\label{eq:cp-limit1}
& \Big|  \text{Im}\big[ \big(\lambda  \lambda^\dag\big)_{sd}^2\big] \Big|^{1/2}
\approx  \big|s_{12}^\theta \,\big( D_1^2 -D_2^2  \big)\big| \big|\sin(2\delta_{12})\big|^{1/2} < 1.2\times 10^{-4}\, \Big(\frac{M_\varphi}{10^2\,\text{GeV}}\Big) \,.
\end{align}
We can infer a limit on $\epsilon$ from the first and third scenario, which
taking into account the $CP$-violating effect (assuming generic phases) reads
\begin{align}\label{eq:lim-kaon-eps}
\text{I:}\quad
 & \epsilon \, \lambda_0^2  \lesssim \big(0.12 \mbox{ - } 2.5\big)\, \times 10^{-3} \, \Big(\frac{M_\varphi}{10^2\,\text{GeV}}\Big) \, , 
 \nonumber \\ 
\text{III:}\quad
& \epsilon\, \lambda_0^{2/3} \lesssim \big(0.5 \mbox{ - } 1.4\big)\,\times 10^{-1}\, \Big(\frac{M_\varphi}{10^2\,\text{GeV}}\Big)^{1/3} \,,
\end{align}
where $\lambda_0$ is of the order of the largest coupling.
In the second scenario instead the contribution is unsuppressed and we get a limit on the coupling
\begin{align}\label{eq:lambda-1}
&
 \text{II:}\quad
\lambda_0\lesssim \big(1.1 \mbox{ - } 5\big)\times10^{-2} \, \Big(\frac{M_\varphi}{10^2\,\text{GeV}}\Big)^{1/2} \,.
\end{align}

Analogously from $B$ and $B_s$-mesons systems we get 
\begin{align}\label{eq:lim-B-1}
 \Big|\sum_{i}
 \lambda_{di}^*\lambda_{bi} \Big|  =  
 \Big| \sum_iU_{b i}D_i^2U_{di}^*\Big|
& \approx 
 \left\{
 \begin{array}{l}
\text{I:}\quad  \epsilon \, \big|  \hat{\lambda}_{d1}^*\hat{\lambda}_{b1} +\hat{\lambda}_{d3}^*\hat{\lambda}_{b3} \big| \lambda_0^2 \approx
 |s_{13}^\theta| \, \big| D_3^2 - D_1^2  \big|    \\
\\
\text{II:}\quad |\lambda_{bj(=3)}\lambda_{dj(=3)}^*|  \hspace{3cm}   < 2.8\times 10^{-3}\, \frac{M_\varphi}{10^2\,\text{GeV}}\\
 \\
\text{III:}\quad \epsilon^{n(=2)}\,\big| \sum_i \hat{\lambda}_{di}^*\hat{\lambda}_{bi} \big| \lambda_0^2 
\approx |s_{13}^\theta| \,D_3^2
 \end{array} \right. 
\end{align}
and
\begin{align}\label{eq:lim-Bs-1}
 \Big|\sum_{i}
 \lambda_{si}^*\lambda_{bi} \Big|  =  
 \Big| \sum_iU_{b i}D_i^2U_{si}^*\Big|
& \approx 
 \left\{
 \begin{array}{l}
\text{I:}\quad  \epsilon \,  \big|  \hat{\lambda}_{s2}^*\hat{\lambda}_{b2} +\hat{\lambda}_{s3}^*\hat{\lambda}_{b3} \big| \lambda_0^2 \approx
 |s_{23}^\theta| \, \big| D_3^2 - D_2^2  \big| \\
\\
\text{II:}\quad |\lambda_{bj(=3)}\lambda_{sj(=3)}^*| \hspace{3cm}  < 1.1\times 10^{-2}\, \frac{M_\varphi}{10^2\,\text{GeV}} \\
 \\
\text{III:}\quad \epsilon^{n(=1)}\,\big| \sum_i \hat{\lambda}_{si}^*\hat{\lambda}_{bi} \big| \lambda_0^2 
\approx |s_{23}^\theta| \,D_3^2
 \end{array} \right. 
\end{align}
providing respectively the approximate bounds in the three scenarios
\begin{align}\label{eq:lim-Bmesons1}
&\text{I:}\quad
 (B)\quad \epsilon \, \lambda_0^2  \lesssim 3\times10^{-3} \, \Big(\frac{M_\varphi}{10^2\,\text{GeV}}\Big)\,,   &&
 (B_s)\quad  \epsilon \, \lambda_0^2  \lesssim 10^{-2} \, \Big(\frac{M_\varphi}{10^2\,\text{GeV}}\Big)\,, 
  \nonumber \\
& \text{II:}\quad
 (B)\quad \lambda_0\lesssim 5.3\times 10^{-2} \, \Big(\frac{M_\varphi}{10^2\,\text{GeV}}\Big)^{1/2}\,,  &&  (B_s)\quad
\lambda_0\lesssim 0.1 \, \Big(\frac{M_\varphi}{10^2\,\text{GeV}}\Big)^{1/2} \,, 
  \nonumber \\
&
\text{III:}\quad
 (B)\quad \epsilon\, \lambda_0 \lesssim 5.3\times 10^{-2}\, \Big(\frac{M_\varphi}{10^2\,\text{GeV}}\Big)^{1/2}\,,  && (B_s)\quad
\epsilon\, \lambda_0^2 \lesssim 10^{-2} \, \Big(\frac{M_\varphi}{10^2\,\text{GeV}}\Big)\,. 
\end{align}

Let us notice that from Eqs.~\eqref{eq:cp-limit1} and \eqref{eq:lambda-1} we have that
for $\lambda_0\sim D_3 \lesssim 0.01$ the hierarchy becomes irrelevant in suppressing the flavor constraints. 
We can also obtain a combined estimate from the three systems 
considering the limits together 
\begin{align}\label{eq:lim-tot-1}
& 
 \begin{array}{l}
\text{I:}\quad \epsilon^{1/2} \, \big(|D_2^2-D_1^2||D_3^2-D_1^2||D_3^2-D_2^2|\big)^{1/6} \sim \epsilon^{1/2} \, \lambda_0 \\
\text{II:}\quad D_3\approx \lambda_0  \\
\text{III:}\quad\epsilon\,D_3\approx \epsilon\,\lambda_0 
 \end{array} 
 \lesssim
\big(0.04 \mbox{ - } 0.07\big)\, \Big(\frac{M_\varphi}{10^2\,\text{GeV}}\Big)^{1/2}\,.
\end{align}

In this setting with $x_i\rightarrow 0$, 
the second scenario with one dark species mainly coupled to standard particles would contribute without suppression 
to the neutral mesons mixing.
Some larger coupling can be allowed in presence of additional hierarchy between the three largest couplings.\footnote{Note that in the first scenario in case of hierarchical values for the $D_i$ we would have the approximate bound $\epsilon^{3/2} \, D_2 D_3^2<\big(0.2 \mbox{ - } 0.9\big)\times 10^{-2}\, \big(M_\varphi/1\,\text{TeV}\big)^{3/2}$.}

\paragraph{2.}

In the limit in which $x_{i}\rightarrow 1$ for all species, $(M_\varphi^2-M_{\tilde{\chi}}^2)/M_\varphi^2 \ll 1$,
(and the mass difference between the $\tilde{\chi}$ species can be neglected)
we have the extra contribution to neutral kaon mixing
\begin{align}
M_{12,\text{NP}}^K & \approx 
\frac{1}{3} m_K f_K^2 \frac{1}{128\pi^2M_\varphi^2}\, \frac{1}{3}
 \left(\big[\big(\tilde{\lambda}\tilde{\lambda}^\dag\big)_{sd}\big]^2
 - \big(\tilde{\lambda}\tilde{\lambda}^\intercal \big)_{ss} \big(\tilde{\lambda}^*\tilde{\lambda}^\dag\big)_{dd}\right) 
\nonumber \\ &
=\frac{1}{3} m_K f_K^2 \frac{1}{128\pi^2M_\varphi^2}\, \frac{1}{3}
\left( \big[\big(\lambda \lambda^\dag \big)_{sd}\big]^2
 - \big(\lambda \lambda^\intercal \big)_{ss} \big(\lambda^* \lambda^\dag \big)_{dd} \right)
\nonumber \\ &
=\frac{1}{3} m_K f_K^2 \frac{1}{128\pi^2M_\varphi^2}\, \frac{1}{3}
\left( \big(U D^2U^\dag \big)_{sd}^2
-\big(U DV^\dag V^*DU_L^\intercal \big)_{ss} \big(U^* D V^\intercal V D U^\dag \big)_{dd}
\right)
\nonumber \\ &
=\frac{1}{3} m_K f_K^2 \frac{1}{128\pi^2M_\varphi^2}\, \frac{1}{3}
\left( \big(U D^2U^\dag \big)_{sd}^2
-\big(U D  O d^2 O^\intercal D U^\intercal \big)_{ss} \big(U^* D O d^{*2} O^\intercal D U^\dag \big)_{dd}
\right)\,.
\end{align}
The combinations of couplings which are relevant depend on the texture of the $\lambda$ matrix.
For example, for the three textures in Eqs.~\eqref{eq:lambda1},~\eqref{eq:lambda3}, and~\eqref{eq:lambda4}
we have respectively
\begin{align}
\Big| \big[\big(\lambda \lambda^\dag \big)_{sd}\big]^2-
\big(\lambda \lambda^\intercal \big)_{ss} \big(\lambda^* \lambda^\dag \big)_{dd} \Big|  &
 \approx 
 \left\{
 \begin{array}{l}
\text{I:}\quad \big|\big(\hat{\lambda}_{s2}\hat{\lambda}_{d1}^*\big)^2  \big| \lambda_0^4
\approx D_2^2 \, D_1^2    \\
\\
\text{II:} \quad \epsilon^2 \, \big| \big( \hat{\lambda}_{s3}\hat{\lambda}_{d2}^*-\hat{\lambda}_{s2}\hat{\lambda}_{d3}^*  \big)^2 \big|
\lambda_0^4
 \\
 \\
\text{III:}\quad \sim \epsilon^{6} \, \lambda_0^4
 \end{array} \right. 
 < 2 \times 10^{-5}\, \Big(\frac{M_\varphi}{10^2\,\text{GeV}}\Big)^2\,.
 \label{eq:lim-K-2}
\end{align}
The limit from 
the $CP$-violating parameter $\epsilon_K$ which apply to the imaginary part of the contributions reads 
\begin{align}
    &\Big| \text{Im}\Big[ \big[\big(\lambda \lambda^\dag \big)_{sd}\big]^2-
\big(\lambda \lambda^\intercal \big)_{ss} \big(\lambda^* \lambda^\dag \big)_{dd} \Big]\Big|  < 4\times 10^{-8}\, \Big(\frac{M_\varphi}{10^2\,\text{GeV}}\Big)^2 \,,
\end{align}
e.g., in the first case we get
\begin{align}
& \big|\hat{\lambda}_{s2}\hat{\lambda}_{d1}  \big|^2 \lambda_0^4
\big|\sin\big[2\,\big(\delta_{s2}-\delta_{d1}\big)\big] \big| \approx
D_1^2D_2^2  \, \big|\sin\big[2\,\big(\gamma_{2}-\gamma_{1}\big)\big] \big|
< 4\times 10^{-8}\, \Big(\frac{M_\varphi}{10^2\,\text{GeV}}\Big)^2 \,.
\end{align}
As regards the B-mesons systems we have in the three scenarios
\begin{align}\label{eq:lim-Bmesons2}
\Big| \big[\big(\lambda \lambda^\dag \big)_{bd}\big]^2-
\big(\lambda \lambda^\intercal \big)_{bb} \big(\lambda^* \lambda^\dag \big)_{dd} \Big|  &
 \approx 
 \left\{
 \begin{array}{l}
\text{I:}\quad D_3^2 \, D_1^2 \\
\\
\text{II:}\quad \epsilon^2 \, \big| \big( \hat{\lambda}_{b3}\hat{\lambda}_{d2}^*-\hat{\lambda}_{b2}\hat{\lambda}_{d3}^*  \big)^2 \big|
\lambda_0^4
\\
 \\
\text{III:}\quad \epsilon^{4} \,\lambda_0^4
 \end{array} \right. 
 < 2.4 \times 10^{-5}\, \Big(\frac{M_\varphi}{10^2\,\text{GeV}}\Big)^2 
\end{align}
and
\begin{align}\label{eq:lim-Bsmesons2}
\Big| \big[\big(\lambda \lambda^\dag \big)_{bs}\big]^2-
\big(\lambda \lambda^\intercal \big)_{bb} \big(\lambda^* \lambda^\dag \big)_{ss} \Big|  &
 \approx 
 \left\{
 \begin{array}{l}
\text{I:}\quad D_3^2 \, D_2^2 \\
\\
\text{II:}\quad \epsilon^2 \, \big| \big( \hat{\lambda}_{b2}\hat{\lambda}_{s3}^*-\hat{\lambda}_{b3}\hat{\lambda}_{s2}^*  \big)^2 \big|
\lambda_0^4 \\
 \\
\text{III:}\quad \epsilon^{2} \, \lambda_0^4
 \end{array} \right. 
 < 3.7 \times 10^{-4}\, \Big(\frac{M_\varphi}{10^2\,\text{GeV}}\Big)^2\,.
\end{align}
From these limits we can extract estimates on the hierarchy parameter $\epsilon$ when $\lambda_0\sim\mathcal{O}(1)$. 
For $\lambda_0\lesssim 0.02$ the hierarchy becomes unnecessary.

In this setting with $M_{\tilde{\chi}}^2\approx M_\varphi^2$, 
the first scenario, which is the one with larger diagonal couplings, gives an unsuppressed contribution
to the neutral mesons mixing. 
Assembling together the constraints from the three neutral mesons systems one obtains
\begin{align}\label{eq:D1D2D3}
&
 |D_1D_2D_3|< \big(0.4 \mbox{ - } 2 \big) \times 10^{-2}\, \Big(\frac{M_\varphi}{\text{TeV}}\Big)^{3/2}
\end{align}
which for values of $D_i$ of the same order would imply
\begin{align}
    &D_{1,2,3} \lesssim \big(0.16 \mbox{ - } 0.3 \big) \, \Big(\frac{M_\varphi}{\text{TeV}}\Big)^{1/2} \,.
\end{align}
In order to achieve a coupling of order $\mathcal{O}(1)$, the $\lambda$ matrix should exhibit some hierarchy 
between the diagonal couplings.

\paragraph{3.}
We can envisage scenarios in which the mass splitting between DM particles is not negligible.
In these scenarios some suppression in the flavor changing phenomena can emerge also in situations in which the DM interacts with similar couplings to all SM families.
For instance, we can consider
one species with mass similar to the mediator mass and two lighter species for which $x\rightarrow 0$.
We can take for example
 $x_{3}\rightarrow 1$ and $x_{1,2}\rightarrow 0$:
\begin{align}\label{eq:KK01}
& M_{12,\text{NP}}^K\approx 
\frac{1}{3} m_K f_K^2 \frac{1}{128\pi^2M_\varphi^2} 
\left[ \Big[\Big(\lambda \lambda^\dag\Big)_{sd}\Big]^2
+ \frac{1}{2} \Big(  \tilde{\lambda}_{d3}^* \tilde{\lambda}_{s3}  \Big)^2
- \frac{3}{2} \Big(  \tilde{\lambda}_{d3}^* \tilde{\lambda}_{s3}  \Big)\Big(\lambda \lambda^\dag\Big)_{sd}
\right]
\nonumber \\ & =
  \frac{1}{3} m_K f_K^2 \frac{1}{128\pi^2M_\varphi^2} 
\left[ \Big[\Big(\lambda \lambda^\dag\Big)_{sd}\Big]^2
+ \frac{1}{2} \Big( \lambda_{sj}W_{ 3j} \lambda_{di}^*W_{3i}   \Big)^2
- \frac{3}{2} \Big( \lambda_{sj}W_{3j} \lambda_{di}^*W_{3i}   \Big)\Big(\lambda \lambda^\dag\Big)_{sd}
 \right].
\end{align}
Here the two contributions $2 G(x_3=1,x_3=1)$ and $F(x_3=1,x_3=1)$ cancel each other.
Notice that in these types of scenarios the mass basis rotations (and the mixing from the right) appear.
In this situation there is a suppression also for the first two scenarios (I and II) with textures in Eqs.~\eqref{eq:lambda1},~\eqref{eq:lambda3}, with $M_{12,\text{NP}}\propto \epsilon^2\lambda_0^4$ for all the three neutral mesons systems (while the third scenario show a suppression similar to the other mass ranges).

We could also have considered one lighter species, e.g. $x_{1}\rightarrow 0$, and two species for which $x_{2,3}\rightarrow 1$.
In this case the scenario I would give an unsuppressed contribution only in one of the three systems ($B_s$).

The new effects are reduced also in scenarios with DM species heavier than the scalar field. For instance
in a scenario with two heavy species, e.g. $x_{2,3}\rightarrow \infty$ and 
one light species $x_{1}\rightarrow 0$ the new contribution would read
\begin{align}
& M_{12,\text{NP}}^K \approx \frac{1}{3} m_K f_K^2 \frac{1}{128\pi^2M_\varphi^2} 
( \tilde{\lambda}_{d1}^* \tilde{\lambda}_{s1})^2 = \frac{1}{3} m_K f_K^2 \hat{B}_K\frac{1}{128\pi^2M_\varphi^2} 
( \lambda_{di}^*W_{1i} \lambda_{sj} W_{1j})^2 \,.
 \end{align}
(In a scenario with two heavy species, e.g., $x_{2,3}\rightarrow \infty$ and 
one lighter species $x_{1}\rightarrow 1$, the contribution from the two box diagrams would even cancel each other.)
Similarly, in scenarios with one heavier DM particle some contributions can be reduced depending on the couplings and masses of the lighter species.
For instance, taking $x_{3}\rightarrow \infty$ only the couplings $\lambda_{d1,2}$, $\tilde{\lambda}_{s12}$ are involved
and some $\epsilon$ suppression emerge unless $x_{1,2}\rightarrow 1$ for which the first scenario does not show suppression.

\medskip
Summarizing, the flavor constraints are more safely avoided when the DM species couple more strongly to one generation of the standard fermions (as one would expect). While the power of the suppression in the different processes depends on which generation is the most strongly coupled, the hierarchy of the couplings always leads to a suppression which is independent of the mass splittings in this kind of scenario.
In cases where the DM fields couple with all the SM families there can be some unsuppressed contributions in flavor changing processes. 
Some different attenuation of these effects in the various scenarios can be gained depending on the mass splittings between the DM particles and the scalar field and among the DM particles themselves.

\end{appendix}

% - - - - REFERENCES - - - - %
\bibliography{references}

@article{Arina:2025zpi,
    author = "Arina, Chiara and others",
    title = "{t-channel dark matter models {\textendash} a whitepaper}",
    eprint = "2504.10597",
    archivePrefix = "arXiv",
    primaryClass = "hep-ph",
    reportNumber = "CERN-LPCC-2025-001, IRMP-CP3-25-07, TTK-25-07",
    doi = "10.1140/epjc/s10052-025-14635-7",
    journal = "Eur. Phys. J. C",
    volume = "85",
    number = "9",
    pages = "975",
    year = "2025"
}

@article{Garny:2017rxs,
    author = {Garny, Mathias and Heisig, Jan and L{\"u}lf, Benedikt and Vogl, Stefan},
    title = "{Coannihilation without chemical equilibrium}",
    eprint = "1705.09292",
    archivePrefix = "arXiv",
    primaryClass = "hep-ph",
    reportNumber = "TUM-HEP-1085-17, TTK-17-18",
    doi = "10.1103/PhysRevD.96.103521",
    journal = "Phys. Rev. D",
    volume = "96",
    number = "10",
    pages = "103521",
    year = "2017"
}

@article{DAgnolo:2017dbv,
    author = "D'Agnolo, Raffaele Tito and Pappadopulo, Duccio and Ruderman, Joshua T.",
    title = "{Fourth Exception in the Calculation of Relic Abundances}",
    eprint = "1705.08450",
    archivePrefix = "arXiv",
    primaryClass = "hep-ph",
    doi = "10.1103/PhysRevLett.119.061102",
    journal = "Phys. Rev. Lett.",
    volume = "119",
    number = "6",
    pages = "061102",
    year = "2017"
}

@article{Bona:2022zhn,
    author = "Bona, Marcella and others",
    title = "{Unitarity Triangle global fits beyond the Standard Model: UTfit 2021 NP update}",
    doi = "10.22323/1.398.0500",
    journal = "PoS",
    volume = "EPS-HEP2021",
    pages = "500",
    year = "2022"
}

@article{Bai:2013iqa,
    author = "Bai, Yang and Berger, Joshua",
    title = "{Fermion Portal Dark Matter}",
    eprint = "1308.0612",
    archivePrefix = "arXiv",
    primaryClass = "hep-ph",
    reportNumber = "SLAC-PUB-15704",
    doi = "10.1007/JHEP11(2013)171",
    journal = "JHEP",
    volume = "11",
    pages = "171",
    year = "2013"
}

@article{DiFranzo:2013vra,
    author = "DiFranzo, Anthony and Nagao, Keiko I. and Rajaraman, Arvind and Tait, Tim M. P.",
    title = "{Simplified Models for Dark Matter Interacting with Quarks}",
    eprint = "1308.2679",
    archivePrefix = "arXiv",
    primaryClass = "hep-ph",
    reportNumber = "UCI-HEP-TR-2013-17, KEK-TH-1659",
    doi = "10.1007/JHEP11(2013)014",
    journal = "JHEP",
    volume = "11",
    pages = "014",
    year = "2013",
    note = "[Erratum: JHEP 01, 162 (2014)]"
}

@article{Ambrogi:2018jqj,
    author = "Ambrogi, Federico and Arina, Chiara and Backovic, Mihailo and Heisig, Jan and Maltoni, Fabio and Mantani, Luca and Mattelaer, Olivier and Mohlabeng, Gopolang",
    title = "{MadDM v.3.0: a Comprehensive Tool for Dark Matter Studies}",
    eprint = "1804.00044",
    archivePrefix = "arXiv",
    primaryClass = "hep-ph",
    reportNumber = "CP3-18-26, MCnet-18-07, MCNET-18-07",
    doi = "10.1016/j.dark.2018.11.009",
    journal = "Phys. Dark Univ.",
    volume = "24",
    pages = "100249",
    year = "2019"
}

@article{Arina:2021gfn,
    author = "Arina, Chiara and Heisig, Jan and Maltoni, Fabio and Massaro, Daniele and Mattelaer, Olivier",
    title = "{Indirect dark-matter detection with MadDM v3.2 {\textendash} Lines and Loops}",
    eprint = "2107.04598",
    archivePrefix = "arXiv",
    primaryClass = "hep-ph",
    reportNumber = "CP3-21-46, MCNET-21-09, TTK-21-25",
    doi = "10.1140/epjc/s10052-023-11377-2",
    journal = "Eur. Phys. J. C",
    volume = "83",
    number = "3",
    pages = "241",
    year = "2023"
}

@article{Heisig:2024mwr,
    author = "Heisig, Jan",
    title = "{Conversion-Driven Leptogenesis: A Testable Theory of Dark Matter and Baryogenesis at the Electroweak Scale}",
    eprint = "2404.12428",
    archivePrefix = "arXiv",
    primaryClass = "hep-ph",
    reportNumber = "TTK-24-12",
    doi = "10.1103/PhysRevLett.133.191803",
    journal = "Phys. Rev. Lett.",
    volume = "133",
    number = "19",
    pages = "191803",
    year = "2024"
}

@article{Acaroglu:2023phy,
    author = {Acaro\u{g}lu, Harun and Blanke, Monika and Heisig, Jan and Kr\"amer, Michael and Rathmann, Lena},
    title = "{Flavoured Majorana Dark Matter then and now: from freeze-out scenarios to LHC signatures}",
    eprint = "2312.09274",
    archivePrefix = "arXiv",
    primaryClass = "hep-ph",
    reportNumber = "TTP23-059, P3H-23-102, TTK-23-35",
    doi = "10.1007/JHEP06(2024)179",
    journal = "JHEP",
    volume = "06",
    pages = "179",
    year = "2024"
}

@article{Ma:2006km,
    author = "Ma, Ernest",
    title = "{Verifiable radiative seesaw mechanism of neutrino mass and dark matter}",
    eprint = "hep-ph/0601225",
    archivePrefix = "arXiv",
    reportNumber = "UCRHEP-T403",
    doi = "10.1103/PhysRevD.73.077301",
    journal = "Phys. Rev. D",
    volume = "73",
    pages = "077301",
    year = "2006"
}

@article{Herms:2021fql,
    author = "Herms, Johannes and Ibarra, Alejandro",
    title = "{Production and signatures of multi-flavour dark matter scenarios with t-channel mediators}",
    eprint = "2103.10392",
    archivePrefix = "arXiv",
    primaryClass = "hep-ph",
    reportNumber = "TUM-HEP 1321/21",
    doi = "10.1088/1475-7516/2021/10/026",
    journal = "JCAP",
    volume = "10",
    pages = "026",
    year = "2021"
}

@article{Acaroglu:2022hrm,
    author = "Acaro\u{g}lu, Harun and Agrawal, Prateek and Blanke, Monika",
    title = "{Lepton-flavoured scalar dark matter in Dark Minimal Flavour Violation}",
    eprint = "2211.03809",
    archivePrefix = "arXiv",
    primaryClass = "hep-ph",
    reportNumber = "TTP22-063; P3H-22-104",
    doi = "10.1007/JHEP05(2023)106",
    journal = "JHEP",
    volume = "05",
    pages = "106",
    year = "2023"
}

@article{Chen:2015jkt,
    author = "Chen, Mu-Chun and Huang, Jinrui and Takhistov, Volodymyr",
    title = "{Beyond Minimal Lepton Flavored Dark Matter}",
    eprint = "1510.04694",
    archivePrefix = "arXiv",
    primaryClass = "hep-ph",
    reportNumber = "UCI-TR-2015-17, LA-UR-15-27938",
    doi = "10.1007/JHEP02(2016)060",
    journal = "JHEP",
    volume = "02",
    pages = "060",
    year = "2016"
}

@article{Agrawal:2014aoa,
    author = "Agrawal, Prateek and Blanke, Monika and Gemmler, Katrin",
    title = "{Flavored dark matter beyond Minimal Flavor Violation}",
    eprint = "1405.6709",
    archivePrefix = "arXiv",
    primaryClass = "hep-ph",
    reportNumber = "CERN-PH-TH-2014-098, FERMILAB-PUB-14-141-T",
    doi = "10.1007/JHEP10(2014)072",
    journal = "JHEP",
    volume = "10",
    pages = "072",
    year = "2014"
}

@article{Acaroglu:2022boc,
    author = "Acaro\u{g}lu, Harun and Agrawal, Prateek and Blanke, Monika",
    title = "{Flavoured $(g-2)_\mu$ with dark lepton seasoning}",
    eprint = "2212.08142",
    archivePrefix = "arXiv",
    primaryClass = "hep-ph",
    reportNumber = "TTP22-071; P3H-22-124",
    doi = "10.21468/SciPostPhys.15.4.176",
    journal = "SciPost Phys.",
    volume = "15",
    number = "4",
    pages = "176",
    year = "2023"
}

@article{Blanke:2017tnb,
    author = "Blanke, Monika and Kast, Simon",
    title = "{Top-Flavoured Dark Matter in Dark Minimal Flavour Violation}",
    eprint = "1702.08457",
    archivePrefix = "arXiv",
    primaryClass = "hep-ph",
    reportNumber = "TTP17-008",
    doi = "10.1007/JHEP05(2017)162",
    journal = "JHEP",
    volume = "05",
    pages = "162",
    year = "2017"
}

@article{Blanke:2017fum,
    author = "Blanke, Monika and Das, Satrajit and Kast, Simon",
    title = "{Flavoured Dark Matter Moving Left}",
    eprint = "1711.10493",
    archivePrefix = "arXiv",
    primaryClass = "hep-ph",
    reportNumber = "TTP17-049",
    doi = "10.1007/JHEP02(2018)105",
    journal = "JHEP",
    volume = "02",
    pages = "105",
    year = "2018"
}

@article{Acaroglu:2021qae,
    author = "Acaro\u{g}lu, Harun and Blanke, Monika",
    title = "{Tasting flavoured Majorana dark matter}",
    eprint = "2109.10357",
    archivePrefix = "arXiv",
    primaryClass = "hep-ph",
    reportNumber = "TTP21-031; P3H-21-065, TTP21-031, P3H-21-065",
    doi = "10.1007/JHEP05(2022)086",
    journal = "JHEP",
    volume = "05",
    pages = "086",
    year = "2022"
}

@article{Jubb:2017rhm,
    author = "Jubb, Thomas and Kirk, Matthew and Lenz, Alexander",
    title = "{Charming Dark Matter}",
    eprint = "1709.01930",
    archivePrefix = "arXiv",
    primaryClass = "hep-ph",
    reportNumber = "IPPP-17-64",
    doi = "10.1007/JHEP12(2017)010",
    journal = "JHEP",
    volume = "12",
    pages = "010",
    year = "2017"
}

@article{Kilic:2015vka,
    author = "Kilic, Can and Klimek, Matthew D. and Yu, Jiang-Hao",
    title = "{Signatures of Top Flavored Dark Matter}",
    eprint = "1501.02202",
    archivePrefix = "arXiv",
    primaryClass = "hep-ph",
    reportNumber = "UTTG-28-14, TCC-027-14",
    doi = "10.1103/PhysRevD.91.054036",
    journal = "Phys. Rev. D",
    volume = "91",
    number = "5",
    pages = "054036",
    year = "2015"
}

@article{Acaroglu:2023cza,
    author = "Acaro\u{g}lu, Harun and Blanke, Monika and Tabet, Mustafa",
    title = "{Opening the Higgs portal to lepton-flavoured dark matter}",
    eprint = "2309.10700",
    archivePrefix = "arXiv",
    primaryClass = "hep-ph",
    reportNumber = "TTP23-028, P3H-23-050, DO-TH 23/11",
    doi = "10.1007/JHEP11(2023)079",
    journal = "JHEP",
    volume = "11",
    pages = "079",
    year = "2023"
}

@article{Agrawal:2015kje,
    author = "Agrawal, Prateek and Chacko, Zackaria and Fortes, Elaine C. F. S. and Kilic, Can",
    title = "{Skew-Flavored Dark Matter}",
    eprint = "1511.06293",
    archivePrefix = "arXiv",
    primaryClass = "hep-ph",
    reportNumber = "UTTG-20-15-, TCC-009-15, FERMILAB-PUB-15-622-T",
    doi = "10.1103/PhysRevD.93.103510",
    journal = "Phys. Rev. D",
    volume = "93",
    number = "10",
    pages = "103510",
    year = "2016"
}

@article{Bensalem:2021qtj,
    author = "Bensalem, Wafia and Stolarski, Daniel",
    title = "{Flavor and CP violation from a QCD-like hidden sector}",
    eprint = "2111.05515",
    archivePrefix = "arXiv",
    primaryClass = "hep-ph",
    doi = "10.1007/JHEP02(2022)011",
    journal = "JHEP",
    volume = "02",
    pages = "011",
    year = "2022"
}

@article{Blanke:2020bsf,
    author = "Blanke, Monika and Pani, Priscilla and Polesello, Giacomo and Rovelli, Giulia",
    title = "{Single-top final states as a probe of top-flavoured dark matter models at the LHC}",
    eprint = "2010.10530",
    archivePrefix = "arXiv",
    primaryClass = "hep-ph",
    reportNumber = "TTP20-034, P3H-20-059, DESY-21-041",
    doi = "10.1007/JHEP01(2021)194",
    journal = "JHEP",
    volume = "01",
    pages = "194",
    year = "2021"
}

@article{Batell:2011tc,
    author = "Batell, Brian and Pradler, Josef and Spannowsky, Michael",
    title = "{Dark Matter from Minimal Flavor Violation}",
    eprint = "1105.1781",
    archivePrefix = "arXiv",
    primaryClass = "hep-ph",
    doi = "10.1007/JHEP08(2011)038",
    journal = "JHEP",
    volume = "08",
    pages = "038",
    year = "2011"
}

@article{Agrawal:2014una,
    author = "Agrawal, Prateek and Batell, Brian and Hooper, Dan and Lin, Tongyan",
    title = "{Flavored Dark Matter and the Galactic Center Gamma-Ray Excess}",
    eprint = "1404.1373",
    archivePrefix = "arXiv",
    primaryClass = "hep-ph",
    reportNumber = "FERMILAB-PUB-14-069-A-T",
    doi = "10.1103/PhysRevD.90.063512",
    journal = "Phys. Rev. D",
    volume = "90",
    number = "6",
    pages = "063512",
    year = "2014"
}

@article{Batell:2013zwa,
    author = "Batell, Brian and Lin, Tongyan and Wang, Lian-Tao",
    title = "{Flavored Dark Matter and R-Parity Violation}",
    eprint = "1309.4462",
    archivePrefix = "arXiv",
    primaryClass = "hep-ph",
    doi = "10.1007/JHEP01(2014)075",
    journal = "JHEP",
    volume = "01",
    pages = "075",
    year = "2014"
}

@article{Lopez-Honorez:2013wla,
    author = "Lopez-Honorez, Laura and Merlo, Luca",
    title = "{Dark matter within the minimal flavour violation ansatz}",
    eprint = "1303.1087",
    archivePrefix = "arXiv",
    primaryClass = "hep-ph",
    reportNumber = "FTUAM-13-129, IFT-UAM-CSIC-13-018, CERN-PH-TH-2013-034",
    doi = "10.1016/j.physletb.2013.04.015",
    journal = "Phys. Lett. B",
    volume = "722",
    pages = "135--143",
    year = "2013"
}

@article{Kumar:2013hfa,
    author = "Kumar, Abhishek and Tulin, Sean",
    title = "{Top-flavored dark matter and the forward-backward asymmetry}",
    eprint = "1303.0332",
    archivePrefix = "arXiv",
    primaryClass = "hep-ph",
    reportNumber = "MCTP-13-04",
    doi = "10.1103/PhysRevD.87.095006",
    journal = "Phys. Rev. D",
    volume = "87",
    number = "9",
    pages = "095006",
    year = "2013"
}

@article{Agrawal:2011ze,
    author = "Agrawal, Prateek and Blanchet, Steve and Chacko, Zackaria and Kilic, Can",
    title = "{Flavored Dark Matter, and Its Implications for Direct Detection and Colliders}",
    eprint = "1109.3516",
    archivePrefix = "arXiv",
    primaryClass = "hep-ph",
    reportNumber = "UMD-PP-011-014, RUNHETC-2011-17, UTTG-19-11, TCC-020-11",
    doi = "10.1103/PhysRevD.86.055002",
    journal = "Phys. Rev. D",
    volume = "86",
    pages = "055002",
    year = "2012"
}

@article{Desai:2020rwz,
    author = "Desai, Niral and Kilic, Can and Yang, Yuan-Pao and Youn, Taewook",
    title = "{Suppressed flavor violation in Lepton Flavored Dark Matter from an extra dimension}",
    eprint = "2001.00720",
    archivePrefix = "arXiv",
    primaryClass = "hep-ph",
    reportNumber = "UTTG 14-2019",
    doi = "10.1103/PhysRevD.101.075043",
    journal = "Phys. Rev. D",
    volume = "101",
    pages = "075043",
    year = "2020"
}

@article{Zurek:2008qg,
    author = "Zurek, Kathryn M.",
    title = "{Multi-Component Dark Matter}",
    eprint = "0811.4429",
    archivePrefix = "arXiv",
    primaryClass = "hep-ph",
    reportNumber = "FERMILAB-PUB-08-542-A",
    doi = "10.1103/PhysRevD.79.115002",
    journal = "Phys. Rev. D",
    volume = "79",
    pages = "115002",
    year = "2009"
}

@article{Agrawal:2015tfa,
    author = "Agrawal, Prateek and Chacko, Zackaria and Kilic, Can and Verhaaren, Christopher B.",
    title = "{A Couplet from Flavored Dark Matter}",
    eprint = "1503.03057",
    archivePrefix = "arXiv",
    primaryClass = "hep-ph",
    reportNumber = "UTTG-06-15, TCC-001-15, FERMILAB-PUB-15-059-T",
    doi = "10.1007/JHEP08(2015)072",
    journal = "JHEP",
    volume = "08",
    pages = "072",
    year = "2015"
}

@article{Lee:2014rba,
    author = "Lee, Chao-Jung and Tandean, Jusak",
    title = "{Lepton-Flavored Scalar Dark Matter with Minimal Flavor Violation}",
    eprint = "1410.6803",
    archivePrefix = "arXiv",
    primaryClass = "hep-ph",
    doi = "10.1007/JHEP04(2015)174",
    journal = "JHEP",
    volume = "04",
    pages = "174",
    year = "2015"
}

@article{Hamze:2014wca,
    author = "Hamze, Ali and Kilic, Can and Koeller, Jason and Trendafilova, Cynthia and Yu, Jiang-Hao",
    title = "{Lepton-Flavored Asymmetric Dark Matter and Interference in Direct Detection}",
    eprint = "1410.3030",
    archivePrefix = "arXiv",
    primaryClass = "hep-ph",
    reportNumber = "UTTG-20-14, TCC-022-14",
    doi = "10.1103/PhysRevD.91.035009",
    journal = "Phys. Rev. D",
    volume = "91",
    number = "3",
    pages = "035009",
    year = "2015"
}

@article{Alloul:2013bka,
    author = "Alloul, Adam and Christensen, Neil D. and Degrande, C\'eline and Duhr, Claude and Fuks, Benjamin",
    title = "{FeynRules  2.0 - A complete toolbox for tree-level phenomenology}",
    eprint = "1310.1921",
    archivePrefix = "arXiv",
    primaryClass = "hep-ph",
    reportNumber = "CERN-PH-TH-2013-239, MCNET-13-14, IPPP-13-71, DCPT-13-142, PITT-PACC-1308",
    doi = "10.1016/j.cpc.2014.04.012",
    journal = "Comput. Phys. Commun.",
    volume = "185",
    pages = "2250--2300",
    year = "2014"
}

@article{Belyaev_2013,
   title={CalcHEP 3.4 for collider physics within and beyond the Standard Model},
   volume={184},
   ISSN={0010-4655},
   url={http://dx.doi.org/10.1016/j.cpc.2013.01.014},
   DOI={10.1016/j.cpc.2013.01.014},
   number={7},
   journal={Computer Physics Communications},
   publisher={Elsevier BV},
   author={Belyaev, Alexander and Christensen, Neil D. and Pukhov, Alexander},
   year={2013},
   month=jul, pages={1729–1769} }

@article{Ibarra:2022nzm,
    author = "Ibarra, Alejandro and Reichard, Merlin and Nagai, Ryo",
    title = "{Anapole moment of Majorana fermions and implications for direct detection of neutralino dark matter}",
    eprint = "2207.01014",
    archivePrefix = "arXiv",
    primaryClass = "hep-ph",
    doi = "10.1007/JHEP01(2023)086",
    journal = "JHEP",
    volume = "01",
    pages = "086",
    year = "2023"
}

@article{Kawamura:2020qxo,
author = "Kawamura, Junichiro and Okawa, Shohei and Omura, Yuji",
title = "{Current status and muon $g-2$ explanation of lepton portal dark matter}",
eprint = "2002.12534",
archivePrefix = "arXiv",
primaryClass = "hep-ph",
doi = "10.1007/JHEP08(2020)042",
journal = "JHEP",
volume = "08",
pages = "042",
year = "2020"
}

@article{Belanger:2020gnr,
    author = "Belanger, Genevieve and Mjallal, Ali and Pukhov, Alexander",
    title = "{Recasting direct detection limits within micrOMEGAs and implication for non-standard Dark Matter scenarios}",
    eprint = "2003.08621",
    archivePrefix = "arXiv",
    primaryClass = "hep-ph",
    doi = "10.1140/epjc/s10052-021-09012-z",
    journal = "Eur. Phys. J. C",
    volume = "81",
    number = "3",
    pages = "239",
    year = "2021"
}

@article{Belanger:2010gh,
    author = "Belanger, G. and Boudjema, F. and Brun, P. and Pukhov, A. and Rosier-Lees, S. and Salati, P. and Semenov, A.",
    title = "{Indirect search for dark matter with micrOMEGAs2.4}",
    eprint = "1004.1092",
    archivePrefix = "arXiv",
    primaryClass = "hep-ph",
    reportNumber = "IRFU-10-24, LAPTH-012-10.",
    doi = "10.1016/j.cpc.2010.11.033",
    journal = "Comput. Phys. Commun.",
    volume = "182",
    pages = "842--856",
    year = "2011"
}

@article{Barducci:2016pcb,
    author = "Barducci, D. and Belanger, G. and Bernon, J. and Boudjema, F. and Da Silva, J. and Kraml, S. and Laa, U. and Pukhov, A.",
    title = "{Collider limits on new physics within micrOMEGAs$\_$4.3}",
    eprint = "1606.03834",
    archivePrefix = "arXiv",
    primaryClass = "hep-ph",
    doi = "10.1016/j.cpc.2017.08.028",
    journal = "Comput. Phys. Commun.",
    volume = "222",
    pages = "327--338",
    year = "2018"
}

@article{Alguero:2023zol,
    author = "Alguero, G. and Belanger, G. and Boudjema, F. and Chakraborti, S. and Goudelis, A. and Kraml, S. and Mjallal, A. and Pukhov, A.",
    title = "{micrOMEGAs 6.0: N-component dark matter}",
    eprint = "2312.14894",
    archivePrefix = "arXiv",
    primaryClass = "hep-ph",
    doi = "10.1016/j.cpc.2024.109133",
    journal = "Comput. Phys. Commun.",
    volume = "299",
    pages = "109133",
    year = "2024"
}

@article{Alguero:2021dig,
    author = {Alguero, Ga\"el and Heisig, Jan and Khosa, Charanjit K. and Kraml, Sabine and Kulkarni, Suchita and Lessa, Andre and Reyes-Gonz\'alez, Humberto and Waltenberger, Wolfgang and Wongel, Alicia},
    title = "{Constraining new physics with SModelS version 2}",
    eprint = "2112.00769",
    archivePrefix = "arXiv",
    primaryClass = "hep-ph",
    reportNumber = "TTK-21-50",
    doi = "10.1007/JHEP08(2022)068",
    journal = "JHEP",
    volume = "08",
    pages = "068",
    year = "2022"
}

@article{MahdiAltakach:2023bdn,
    author = "Altakach, Mohammad Mahdi and Kraml, Sabine and Lessa, Andre and Narasimha, Sahana and Pascal, Timoth\'ee and Waltenberger, Wolfgang",
    title = "{SModelS v2.3: Enabling global likelihood analyses}",
    eprint = "2306.17676",
    archivePrefix = "arXiv",
    primaryClass = "hep-ph",
    doi = "10.21468/SciPostPhys.15.5.185",
    journal = "SciPost Phys.",
    volume = "15",
    number = "5",
    pages = "185",
    year = "2023"
}

@article{Altakach:2024jwk,
    author = "Altakach, Mohammad Mahdi and Kraml, Sabine and Lessa, Andre and Narasimha, Sahana and Pascal, Timoth{\'e}e and Ramos, Camila and Villamizar, Yoxara and Waltenberger, Wolfgang",
    title = "{SModelS v3: going beyond $ \mathcal{Z} _{2}$ topologies}",
    eprint = "2409.12942",
    archivePrefix = "arXiv",
    primaryClass = "hep-ph",
    doi = "10.1007/JHEP11(2024)074",
    journal = "JHEP",
    volume = "11",
    pages = "074",
    year = "2024"
}

@article{Planck:2018,
    author = "Aghanim, N. and others",
    collaboration = "Planck",
    title = "{Planck 2018 results. VI. Cosmological parameters}",
    eprint = "1807.06209",
    archivePrefix = "arXiv",
    primaryClass = "astro-ph.CO",
    doi = "10.1051/0004-6361/201833910",
    journal = "Astron. Astrophys.",
    volume = "641",
    pages = "A6",
    year = "2020",
    note = "[Erratum: Astron.Astrophys. 652, C4 (2021)]"
}

@article{Aalbers_2023,
   title={First Dark Matter Search Results from the LUX-ZEPLIN (LZ) Experiment},
   volume={131},
   ISSN={1079-7114},
   url={http://dx.doi.org/10.1103/PhysRevLett.131.041002},
   DOI={10.1103/physrevlett.131.041002},
   number={4},
   journal={Physical Review Letters},
   publisher={American Physical Society (APS)},
   author={Aalbers, J. and Akerib, D. S. and Akerlof, C. W. and Al Musalhi, A. K. and Alder, F. and Alqahtani, A. and Alsum, S. K. and Amarasinghe, C. S. and Ames, A. and Anderson, T. J. and Angelides, N. and Araújo, H. M. and Armstrong, J. E. and Arthurs, M. and Azadi, S. and Bailey, A. J. and Baker, A. and Balajthy, J. and Balashov, S. and Bang, J. and Bargemann, J. W. and Barry, M. J. and Barthel, J. and Bauer, D. and Baxter, A. and Beattie, K. and Belle, J. and Beltrame, P. and Bensinger, J. and Benson, T. and Bernard, E. P. and Bhatti, A. and Biekert, A. and Biesiadzinski, T. P. and Birch, H. J. and Birrittella, B. and Blockinger, G. M. and Boast, K. E. and Boxer, B. and Bramante, R. and Brew, C. A. J. and Brás, P. and Buckley, J. H. and Bugaev, V. V. and Burdin, S. and Busenitz, J. K. and Buuck, M. and Cabrita, R. and Carels, C. and Carlsmith, D. L. and Carlson, B. and Carmona-Benitez, M. C. and Cascella, M. and Chan, C. and Chawla, A. and Chen, H. and Cherwinka, J. J. and Chott, N. I. and Cole, A. and Coleman, J. and Converse, M. V. and Cottle, A. and Cox, G. and Craddock, W. W. and Creaner, O. and Curran, D. and Currie, A. and Cutter, J. E. and Dahl, C. E. and David, A. and Davis, J. and Davison, T. J. R. and Delgaudio, J. and Dey, S. and de Viveiros, L. and Dobi, A. and Dobson, J. E. Y. and Druszkiewicz, E. and Dushkin, A. and Edberg, T. K. and Edwards, W. R. and Elnimr, M. M. and Emmet, W. T. and Eriksen, S. R. and Faham, C. H. and Fan, A. and Fayer, S. and Fearon, N. M. and Fiorucci, S. and Flaecher, H. and Ford, P. and Francis, V. B. and Fraser, E. D. and Fruth, T. and Gaitskell, R. J. and Gantos, N. J. and Garcia, D. and Geffre, A. and Gehman, V. M. and Genovesi, J. and Ghag, C. and Gibbons, R. and Gibson, E. and Gilchriese, M. G. D. and Gokhale, S. and Gomber, B. and Green, J. and Greenall, A. and Greenwood, S. and van der Grinten, M. G. D. and Gwilliam, C. B. and Hall, C. R. and Hans, S. and Hanzel, K. and Harrison, A. and Hartigan-O’Connor, E. and Haselschwardt, S. J. and Hernandez, M. A. and Hertel, S. A. and Heuermann, G. and Hjemfelt, C. and Hoff, M. D. and Holtom, E. and Hor, J. Y-K. and Horn, M. and Huang, D. Q. and Hunt, D. and Ignarra, C. M. and Jacobsen, R. G. and Jahangir, O. and James, R. S. and Jeffery, S. N. and Ji, W. and Johnson, J. and Kaboth, A. C. and Kamaha, A. C. and Kamdin, K. and Kasey, V. and Kazkaz, K. and Keefner, J. and Khaitan, D. and Khaleeq, M. and Khazov, A. and Khurana, I. and Kim, Y. D. and Kocher, C. D. and Kodroff, D. and Korley, L. and Korolkova, E. V. and Kras, J. and Kraus, H. and Kravitz, S. and Krebs, H. J. and Kreczko, L. and Krikler, B. and Kudryavtsev, V. A. and Kyre, S. and Landerud, B. and Leason, E. A. and Lee, C. and Lee, J. and Leonard, D. S. and Leonard, R. and Lesko, K. T. and Levy, C. and Li, J. and Liao, F.-T. and Liao, J. and Lin, J. and Lindote, A. and Linehan, R. and Lippincott, W. H. and Liu, R. and Liu, X. and Liu, Y. and Loniewski, C. and Lopes, M. I. and Lopez Asamar, E. and López Paredes, B. and Lorenzon, W. and Lucero, D. and Luitz, S. and Lyle, J. M. and Majewski, P. A. and Makkinje, J. and Malling, D. C. and Manalaysay, A. and Manenti, L. and Mannino, R. L. and Marangou, N. and Marzioni, M. F. and Maupin, C. and McCarthy, M. E. and McConnell, C. T. and McKinsey, D. N. and McLaughlin, J. and Meng, Y. and Migneault, J. and Miller, E. H. and Mizrachi, E. and Mock, J. A. and Monte, A. and Monzani, M. E. and Morad, J. A. and Morales Mendoza, J. D. and Morrison, E. and Mount, B. J. and Murdy, M. and Murphy, A. St. J. and Naim, D. and Naylor, A. and Nedlik, C. and Nehrkorn, C. and Neves, F. and Nguyen, A. and Nikoleyczik, J. A. and Nilima, A. and O’Dell, J. and O’Neill, F. G. and O’Sullivan, K. and Olcina, I. and Olevitch, M. A. and Oliver-Mallory, K. C. and Orpwood, J. and Pagenkopf, D. and Pal, S. and Palladino, K. J. and Palmer, J. and Pangilinan, M. and Parveen, N. and Patton, S. J. and Pease, E. K. and Penning, B. and Pereira, C. and Pereira, G. and Perry, E. and Pershing, T. and Peterson, I. B. and Piepke, A. and Podczerwinski, J. and Porzio, D. and Powell, S. and Preece, R. M. and Pushkin, K. and Qie, Y. and Ratcliff, B. N. and Reichenbacher, J. and Reichhart, L. and Rhyne, C. A. and Richards, A. and Riffard, Q. and Rischbieter, G. R. C. and Rodrigues, J. P. and Rodriguez, A. and Rose, H. J. and Rosero, R. and Rossiter, P. and Rushton, T. and Rutherford, G. and Rynders, D. and Saba, J. S. and Santone, D. and Sazzad, A. B. M. R. and Schnee, R. W. and Scovell, P. R. and Seymour, D. and Shaw, S. and Shutt, T. and Silk, J. J. and Silva, C. and Sinev, G. and Skarpaas, K. and Skulski, W. and Smith, R. and Solmaz, M. and Solovov, V. N. and Sorensen, P. and Soria, J. and Stancu, I. and Stark, M. R. and Stevens, A. and Stiegler, T. M. and Stifter, K. and Studley, R. and Suerfu, B. and Sumner, T. J. and Sutcliffe, P. and Swanson, N. and Szydagis, M. and Tan, M. and Taylor, D. J. and Taylor, R. and Taylor, W. C. and Temples, D. J. and Tennyson, B. P. and Terman, P. A. and Thomas, K. J. and Tiedt, D. R. and Timalsina, M. and To, W. H. and Tomás, A. and Tong, Z. and Tovey, D. R. and Tranter, J. and Trask, M. and Tripathi, M. and Tronstad, D. R. and Tull, C. E. and Turner, W. and Tvrznikova, L. and Utku, U. and Va’vra, J. and Vacheret, A. and Vaitkus, A. C. and Verbus, J. R. and Voirin, E. and Waldron, W. L. and Wang, A. and Wang, B. and Wang, J. J. and Wang, W. and Wang, Y. and Watson, J. R. and Webb, R. C. and White, A. and White, D. T. and White, J. T. and White, R. G. and Whitis, T. J. and Williams, M. and Wisniewski, W. J. and Witherell, M. S. and Wolfs, F. L. H. and Wolfs, J. D. and Woodford, S. and Woodward, D. and Worm, S. D. and Wright, C. J. and Xia, Q. and Xiang, X. and Xiao, Q. and Xu, J. and Yeh, M. and Yin, J. and Young, I. and Zarzhitsky, P. and Zuckerman, A. and Zweig, E. A.},
   year={2023},
   month=jul }

@article{Aprile_2018,
   title={Dark Matter Search Results from a One Ton-Year Exposure of XENON1T},
   volume={121},
   ISSN={1079-7114},
   url={http://dx.doi.org/10.1103/PhysRevLett.121.111302},
   DOI={10.1103/physrevlett.121.111302},
   number={11},
   journal={Physical Review Letters},
   publisher={American Physical Society (APS)},
   author={Aprile, E. and Aalbers, J. and Agostini, F. and Alfonsi, M. and Althueser, L. and Amaro, F. D. and Anthony, M. and Arneodo, F. and Baudis, L. and Bauermeister, B. and Benabderrahmane, M. L. and Berger, T. and Breur, P. A. and Brown, A. and Brown, A. and Brown, E. and Bruenner, S. and Bruno, G. and Budnik, R. and Capelli, C. and Cardoso, J. M. R. and Cichon, D. and Coderre, D. and Colijn, A. P. and Conrad, J. and Cussonneau, J. P. and Decowski, M. P. and de Perio, P. and Di Gangi, P. and Di Giovanni, A. and Diglio, S. and Elykov, A. and Eurin, G. and Fei, J. and Ferella, A. D. and Fieguth, A. and Fulgione, W. and Gallo Rosso, A. and Galloway, M. and Gao, F. and Garbini, M. and Geis, C. and Grandi, L. and Greene, Z. and Qiu, H. and Hasterok, C. and Hogenbirk, E. and Howlett, J. and Itay, R. and Joerg, F. and Kaminsky, B. and Kazama, S. and Kish, A. and Koltman, G. and Landsman, H. and Lang, R. F. and Levinson, L. and Lin, Q. and Lindemann, S. and Lindner, M. and Lombardi, F. and Lopes, J. A. M. and Mahlstedt, J. and Manfredini, A. and Marrodán Undagoitia, T. and Masbou, J. and Masson, D. and Messina, M. and Micheneau, K. and Miller, K. and Molinario, A. and Morå, K. and Murra, M. and Naganoma, J. and Ni, K. and Oberlack, U. and Pelssers, B. and Piastra, F. and Pienaar, J. and Pizzella, V. and Plante, G. and Podviianiuk, R. and Priel, N. and Ramírez García, D. and Rauch, L. and Reichard, S. and Reuter, C. and Riedel, B. and Rizzo, A. and Rocchetti, A. and Rupp, N. and dos Santos, J. M. F. and Sartorelli, G. and Scheibelhut, M. and Schindler, S. and Schreiner, J. and Schulte, D. and Schumann, M. and Scotto Lavina, L. and Selvi, M. and Shagin, P. and Shockley, E. and Silva, M. and Simgen, H. and Thers, D. and Toschi, F. and Trinchero, G. and Tunnell, C. and Upole, N. and Vargas, M. and Wack, O. and Wang, H. and Wang, Z. and Wei, Y. and Weinheimer, C. and Wittweg, C. and Wulf, J. and Ye, J. and Zhang, Y. and Zhu, T.},
   year={2018},
   month=sep }

@article{Aprile_2019,
   title={Constraining the Spin-Dependent WIMP-Nucleon Cross Sections with XENON1T},
   volume={122},
   ISSN={1079-7114},
   url={http://dx.doi.org/10.1103/PhysRevLett.122.141301},
   DOI={10.1103/physrevlett.122.141301},
   number={14},
   journal={Physical Review Letters},
   publisher={American Physical Society (APS)},
   author={Aprile, E. and Aalbers, J. and Agostini, F. and Alfonsi, M. and Althueser, L. and Amaro, F. D. and Anthony, M. and Antochi, V. C. and Arneodo, F. and Baudis, L. and Bauermeister, B. and Benabderrahmane, M. L. and Berger, T. and Breur, P. A. and Brown, A. and Brown, A. and Brown, E. and Bruenner, S. and Bruno, G. and Budnik, R. and Capelli, C. and Cardoso, J. M. R. and Cichon, D. and Coderre, D. and Colijn, A. P. and Conrad, J. and Cussonneau, J. P. and Decowski, M. P. and de Perio, P. and Di Gangi, P. and Di Giovanni, A. and Diglio, S. and Elykov, A. and Eurin, G. and Fei, J. and Ferella, A. D. and Fieguth, A. and Fulgione, W. and Gallo Rosso, A. and Galloway, M. and Gao, F. and Garbini, M. and Grandi, L. and Greene, Z. and Hasterok, C. and Hogenbirk, E. and Howlett, J. and Iacovacci, M. and Itay, R. and Joerg, F. and Kazama, S. and Kish, A. and Koltman, G. and Kopec, A. and Landsman, H. and Lang, R. F. and Levinson, L. and Lin, Q. and Lindemann, S. and Lindner, M. and Lombardi, F. and Lopes, J. A. M. and López Fune, E. and Macolino, C. and Mahlstedt, J. and Manfredini, A. and Marignetti, F. and Marrodán Undagoitia, T. and Masbou, J. and Masson, D. and Mastroianni, S. and Messina, M. and Micheneau, K. and Miller, K. and Molinario, A. and Morå, K. and Mosbacher, Y. and Murra, M. and Naganoma, J. and Ni, K. and Oberlack, U. and Odgers, K. and Pelssers, B. and Piastra, F. and Pienaar, J. and Pizzella, V. and Plante, G. and Podviianiuk, R. and Priel, N. and Qiu, H. and Ramírez García, D. and Reichard, S. and Riedel, B. and Rizzo, A. and Rocchetti, A. and Rupp, N. and dos Santos, J. M. F. and Sartorelli, G. and Šarčević, N. and Scheibelhut, M. and Schindler, S. and Schreiner, J. and Schulte, D. and Schumann, M. and Scotto Lavina, L. and Selvi, M. and Shagin, P. and Shockley, E. and Silva, M. and Simgen, H. and Therreau, C. and Thers, D. and Toschi, F. and Trinchero, G. and Tunnell, C. and Upole, N. and Vargas, M. and Wack, O. and Wang, H. and Wang, Z. and Wei, Y. and Weinheimer, C. and Wenz, D. and Wittweg, C. and Wulf, J. and Xu, Z. and Ye, J. and Zhang, Y. and Zhu, T. and Zopounidis, J. P.},
   year={2019},
   month=apr }

@article{Agnes_2018,
   title={Low-Mass Dark Matter Search with the DarkSide-50 Experiment},
   volume={121},
   ISSN={1079-7114},
   url={http://dx.doi.org/10.1103/PhysRevLett.121.081307},
   DOI={10.1103/physrevlett.121.081307},
   number={8},
   journal={Physical Review Letters},
   publisher={American Physical Society (APS)},
   author={Agnes, P. and Albuquerque, I. F. M. and Alexander, T. and Alton, A. K. and Araujo, G. R. and Asner, D. M. and Ave, M. and Back, H. O. and Baldin, B. and Batignani, G. and Biery, K. and Bocci, V. and Bonfini, G. and Bonivento, W. and Bottino, B. and Budano, F. and Bussino, S. and Cadeddu, M. and Cadoni, M. and Calaprice, F. and Caminata, A. and Canci, N. and Candela, A. and Caravati, M. and Cariello, M. and Carlini, M. and Carpinelli, M. and Catalanotti, S. and Cataudella, V. and Cavalcante, P. and Cavuoti, S. and Cereseto, R. and Chepurnov, A. and Cicalò, C. and Cifarelli, L. and Cocco, A. G. and Covone, G. and D’Angelo, D. and D’Incecco, M. and D’Urso, D. and Davini, S. and De Candia, A. and De Cecco, S. and De Deo, M. and De Filippis, G. and De Rosa, G. and De Vincenzi, M. and Demontis, P. and Derbin, A. V. and Devoto, A. and Di Eusanio, F. and Di Pietro, G. and Dionisi, C. and Downing, M. and Edkins, E. and Empl, A. and Fan, A. and Fiorillo, G. and Fomenko, K. and Franco, D. and Gabriele, F. and Gabrieli, A. and Galbiati, C. and Garcia Abia, P. and Ghiano, Chiara and Giagu, S. and Giganti, C. and Giovanetti, G. K. and Gorchakov, O. and Goretti, A. M. and Granato, F. and Gromov, M. and Guan, M. and Guardincerri, Y. and Gulino, M. and Hackett, B. R. and Hassanshahi, M. H. and Herner, K. and Hosseini, B. and Hughes, D. and Humble, P. and Hungerford, E. V. and Ianni, Al. and Ianni, An. and Ippolito, V. and James, I. and Johnson, T. N. and Kahn, Y. and Keeter, K. and Kendziora, C. L. and Kochanek, I. and Koh, G. and Korablev, D. and Korga, G. and Kubankin, A. and Kuss, M. and La Commara, M. and Lai, M. and Li, X. and Lisanti, M. and Lissia, M. and Loer, B. and Longo, G. and Ma, Y. and Machado, A. A. and Machulin, I. N. and Mandarano, A. and Mapelli, L. and Mari, S. M. and Maricic, J. and Martoff, C. J. and Messina, A. and Meyers, P. D. and Milincic, R. and Mishra-Sharma, S. and Monte, A. and Morrocchi, M. and Mount, B. J. and Muratova, V. N. and Musico, P. and Nania, R. and Navrer Agasson, A. and Nozdrina, A. O. and Oleinik, A. and Orsini, M. and Ortica, F. and Pagani, L. and Pallavicini, M. and Pandola, L. and Pantic, E. and Paoloni, E. and Pazzona, F. and Pelczar, K. and Pelliccia, N. and Pesudo, V. and Pocar, A. and Pordes, S. and Poudel, S. S. and Pugachev, D. A. and Qian, H. and Ragusa, F. and Razeti, M. and Razeto, A. and Reinhold, B. and Renshaw, A. L. and Rescigno, M. and Riffard, Q. and Romani, A. and Rossi, B. and Rossi, N. and Sablone, D. and Samoylov, O. and Sands, W. and Sanfilippo, S. and Sant, M. and Santorelli, R. and Savarese, C. and Scapparone, E. and Schlitzer, B. and Segreto, E. and Semenov, D. A. and Shchagin, A. and Sheshukov, A. and Singh, P. N. and Skorokhvatov, M. D. and Smirnov, O. and Sotnikov, A. and Stanford, C. and Stracka, S. and Suffritti, G. B. and Suvorov, Y. and Tartaglia, R. and Testera, G. and Tonazzo, A. and Trinchese, P. and Unzhakov, E. V. and Verducci, M. and Vishneva, A. and Vogelaar, B. and Wada, M. and Waldrop, T. J. and Wang, H. and Wang, Y. and Watson, A. W. and Westerdale, S. and Wojcik, M. M. and Wojcik, M. and Xiang, X. and Xiao, X. and Yang, C. and Ye, Z. and Zhu, C. and Zichichi, A. and Zuzel, G.},
   year={2018},
   month=aug }

@article{Amole_2019,
   title={Dark matter search results from the complete exposure of the PICO-60 C$_3$F$_8$ bubble chamber},
   volume={100},
   ISSN={2470-0029},
   url={http://dx.doi.org/10.1103/PhysRevD.100.022001},
   DOI={10.1103/physrevd.100.022001},
   number={2},
   journal={Physical Review D},
   publisher={American Physical Society (APS)},
   author={Amole, C. and Ardid, M. and Arnquist, I. J. and Asner, D. M. and Baxter, D. and Behnke, E. and Bressler, M. and Broerman, B. and Cao, G. and Chen, C. J. and Chowdhury, U. and Clark, K. and Collar, J. I. and Cooper, P. S. and Coutu, C. B. and Cowles, C. and Crisler, M. and Crowder, G. and Cruz-Venegas, N. A. and Dahl, C. E. and Das, M. and Fallows, S. and Farine, J. and Felis, I. and Filgas, R. and Girard, F. and Giroux, G. and Hall, J. and Hardy, C. and Harris, O. and Hillier, T. and Hoppe, E. W. and Jackson, C. M. and Jin, M. and Klopfenstein, L. and Kozynets, T. and Krauss, C. B. and Laurin, M. and Lawson, I. and Leblanc, A. and Levine, I. and Licciardi, C. and Lippincott, W. H. and Loer, B. and Mamedov, F. and Mitra, P. and Moore, C. and Nania, T. and Neilson, R. and Noble, A. J. and Oedekerk, P. and Ortega, A. and Piro, M.-C. and Plante, A. and Podviyanuk, R. and Priya, S. and Robinson, A. E. and Sahoo, S. and Scallon, O. and Seth, S. and Sonnenschein, A. and Starinski, N. and Štekl, I. and Sullivan, T. and Tardif, F. and Vázquez-Jáuregui, E. and Walkowski, N. and Weima, E. and Wichoski, U. and Wierman, K. and Yan, Y. and Zacek, V. and Zhang, J.},
   year={2019},
   month=jul }

@article{Abdelhameed_2019,
   title={First results from the CRESST-III low-mass dark matter program},
   volume={100},
   ISSN={2470-0029},
   url={http://dx.doi.org/10.1103/PhysRevD.100.102002},
   DOI={10.1103/physrevd.100.102002},
   number={10},
   journal={Physical Review D},
   publisher={American Physical Society (APS)},
   author={Abdelhameed, A. H. and Angloher, G. and Bauer, P. and Bento, A. and Bertoldo, E. and Bucci, C. and Canonica, L. and D’Addabbo, A. and Defay, X. and Di Lorenzo, S. and Erb, A. and Feilitzsch, F. v. and Fichtinger, S. and Ferreiro Iachellini, N. and Fuss, A. and Gorla, P. and Hauff, D. and Jochum, J. and Kinast, A. and Kluck, H. and Kraus, H. and Langenkämper, A. and Mancuso, M. and Mokina, V. and Mondragon, E. and Münster, A. and Olmi, M. and Ortmann, T. and Pagliarone, C. and Pattavina, L. and Petricca, F. and Potzel, W. and Pröbst, F. and Reindl, F. and Rothe, J. and Schäffner, K. and Schieck, J. and Schipperges, V. and Schmiedmayer, D. and Schönert, S. and Schwertner, C. and Stahlberg, M. and Stodolsky, L. and Strandhagen, C. and Strauss, R. and Türkoǧlu, C. and Usherov, I. and Willers, M. and Zema, V.},
   year={2019},
   month=nov }

@article{Calore:2022stf,
    author = "Calore, Francesca and Cirelli, Marco and Derome, Laurent and Genolini, Yoann and Maurin, David and Salati, Pierre and Serpico, Pasquale Dario",
    title = "{AMS-02 antiprotons and dark matter: Trimmed hints and robust bounds}",
    eprint = "2202.03076",
    archivePrefix = "arXiv",
    primaryClass = "hep-ph",
    reportNumber = "LAPTH-003/22",
    doi = "10.21468/SciPostPhys.12.5.163",
    journal = "SciPost Phys.",
    volume = "12",
    number = "5",
    pages = "163",
    year = "2022"
}

@article{Cuoco:2017iax,
    author = {Cuoco, Alessandro and Heisig, Jan and Korsmeier, Michael and Kr{\"a}mer, Michael},
    title = "{Constraining heavy dark matter with cosmic-ray antiprotons}",
    eprint = "1711.05274",
    archivePrefix = "arXiv",
    primaryClass = "hep-ph",
    reportNumber = "TTK-17-34",
    doi = "10.1088/1475-7516/2018/04/004",
    journal = "JCAP",
    volume = "04",
    pages = "004",
    year = "2018"
}

@misc{armand2021,
      title={Combined dark matter searches towards dwarf spheroidal galaxies with Fermi-LAT, HAWC, H.E.S.S., MAGIC, and VERITAS}, 
      author={Celine Armand and Eric Charles and Mattia di Mauro and Chiara Giuri and J. Patrick Harding and Daniel Kerszberg and Tjark Miener and Emmanuel Moulin and Louise Oakes and Vincent Poireau and Elisa Pueschel and Javier Rico and Lucia Rinchiuso and Daniel Salazar-Gallegos and Kirsten Tollefson and Benjamin Zitzer},
      year={2021},
      eprint={2108.13646},
      archivePrefix={arXiv},
      primaryClass={hep-ex},
      url={https://arxiv.org/abs/2108.13646}, 
}

@article{Archambault_2017,
   title={Dark matter constraints from a joint analysis of dwarf Spheroidal galaxy observations with VERITAS},
   volume={95},
   ISSN={2470-0029},
   url={http://dx.doi.org/10.1103/PhysRevD.95.082001},
   DOI={10.1103/physrevd.95.082001},
   number={8},
   journal={Physical Review D},
   publisher={American Physical Society (APS)},
   author={Archambault, S. and Archer, A. and Benbow, W. and Bird, R. and Bourbeau, E. and Brantseg, T. and Buchovecky, M. and Buckley, J. H. and Bugaev, V. and Byrum, K. and Cerruti, M. and Christiansen, J. L. and Connolly, M. P. and Cui, W. and Daniel, M. K. and Feng, Q. and Finley, J. P. and Fleischhack, H. and Fortson, L. and Furniss, A. and Geringer-Sameth, A. and Griffin, S. and Grube, J. and Hütten, M. and Håkansson, N. and Hanna, D. and Hervet, O. and Holder, J. and Hughes, G. and Hummensky, B. and Johnson, C. A. and Kaaret, P. and Kar, P. and Kelley-Hoskins, N. and Kertzman, M. and Kieda, D. and Koushiappas, S. and Krause, M. and Krennrich, F. and Lang, M. J. and Lin, T. T. Y. and McArthur, S. and Moriarty, P. and Mukherjee, R. and Nieto, D. and O’Brien, S. and Ong, R. A. and Otte, A. N. and Park, N. and Pohl, M. and Popkow, A. and Pueschel, E. and Quinn, J. and Ragan, K. and Reynolds, P. T. and Richards, G. T. and Roache, E. and Rulten, C. and Sadeh, I. and Santander, M. and Sembroski, G. H. and Shahinyan, K. and Smith, A. W. and Staszak, D. and Telezhinsky, I. and Trepanier, S. and Tucci, J. V. and Tyler, J. and Wakely, S. P. and Weinstein, A. and Wilcox, P. and Williams, D. A. and Zitzer, B.},
   year={2017},
   month=apr }

@misc{acharyya2024,
      title={An indirect search for dark matter with a combined analysis of dwarf spheroidal galaxies from VERITAS}, 
      author={A. Acharyya and C. B. Adams and P. Bangale and J. T. Bartkoske and P. Batista and W. Benbow and J. L. Christiansen and A. J. Chromey and A. Duerr and M. Errando and A. Falcone and Q. Feng and G. M. Foote and L. Fortson and A. Furniss and W. Hanlon and D. Hanna and O. Hervet and C. E. Hinrichs and J. Holder and T. B. Humensky and W. Jin and M. N. Johnson and P. Kaaret and M. Kertzman and D. Kieda and T. K. Kleiner and N. Korzoun and S. Kumar and M. J. Lang and M. Lundy and G. Maier and Conor E. McGrath and M. J. Millard and C. L. Mooney and P. Moriarty and R. Mukherjee and W. Ning and S. O'Brien and R. A. Ong and N. Park and M. Pohl and E. Pueschel and J. Quinn and P. L. Rabinowitz and K. Ragan and P. T. Reynolds and D. Ribeiro and E. Roache and J. L. Ryan and I. Sadeh and L. Saha and G. H. Sembroski and R. Shang and M. Splettstoesser and Donggeun Tak and A. K. Talluri and J. V. Tucci and V. V. Vassiliev and A. Weinstein and D. A. Williams and S. L. Wong},
      year={2024},
      eprint={2407.16518},
      archivePrefix={arXiv},
      primaryClass={astro-ph.HE},
      url={https://arxiv.org/abs/2407.16518}, 
}

@article{John_2021,
   title={Cosmic-ray positrons strongly constrain leptophilic dark matter},
   volume={2021},
   ISSN={1475-7516},
   url={http://dx.doi.org/10.1088/1475-7516/2021/12/007},
   DOI={10.1088/1475-7516/2021/12/007},
   number={12},
   journal={Journal of Cosmology and Astroparticle Physics},
   publisher={IOP Publishing},
   author={John, Isabelle and Linden, Tim},
   year={2021},
   month=dec, pages={007} 
}

@article{Cirelli:2024ssz,
    author = "Cirelli, Marco and Strumia, Alessandro and Zupan, Jure",
    title = "{Dark Matter}",
    eprint = "2406.01705",
    archivePrefix = "arXiv",
    primaryClass = "hep-ph",
    month = "6",
    year = "2024"
}

@misc{2023IceCubeDM,
      title={Search for neutrino lines from dark matter annihilation and decay with IceCube}, 
      author={The IceCube Collaboration and R. Abbasi and M. Ackermann and J. Adams and S. K. Agarwalla and J. A. Aguilar and M. Ahlers and J. M. Alameddine and N. M. Amin and K. Andeen and G. Anton and C. Argüelles and Y. Ashida and S. Athanasiadou and S. N. Axani and X. Bai and A. Balagopal V. and M. Baricevic and S. W. Barwick and V. Basu and R. Bay and J. J. Beatty and K. -H. Becker and J. Becker Tjus and J. Beise and C. Bellenghi and S. BenZvi and D. Berley and E. Bernardini and D. Z. Besson and G. Binder and D. Bindig and E. Blaufuss and S. Blot and F. Bontempo and J. Y. Book and C. Boscolo Meneguolo and S. Böser and O. Botner and J. Böttcher and E. Bourbeau and J. Braun and B. Brinson and J. Brostean-Kaiser and R. T. Burley and R. S. Busse and D. Butterfield and M. A. Campana and K. Carloni and E. G. Carnie-Bronca and S. Chattopadhyay and C. Chen and Z. Chen and D. Chirkin and S. Choi and B. A. Clark and L. Classen and A. Coleman and G. H. Collin and A. Connolly and J. M. Conrad and P. Coppin and P. Correa and S. Countryman and D. F. Cowen and P. Dave and C. De Clercq and J. J. DeLaunay and D. Delgado López and H. Dembinski and S. Deng and K. Deoskar and A. Desai and P. Desiati and K. D. de Vries and G. de Wasseige and T. DeYoung and A. Diaz and J. C. Díaz-Vélez and M. Dittmer and A. Domi and H. Dujmovic and M. A. DuVernois and T. Ehrhardt and C. El Aisati and P. Eller and R. Engel and H. Erpenbeck and J. Evans and P. A. Evenson and K. L. Fan and K. Fang and A. R. Fazely and A. Fedynitch and N. Feigl and S. Fiedlschuster and C. Finley and L. Fischer and D. Fox and A. Franckowiak and E. Friedman and A. Fritz and P. Fürst and T. K. Gaisser and J. Gallagher and E. Ganster and A. Garcia and S. Garrappa and L. Gerhardt and A. Ghadimi and C. Glaser and T. Glauch and T. Glüsenkamp and M. Gustafsson and N. Goehlke and J. G. Gonzalez and S. Goswami and D. Grant and S. J. Gray and S. Griffin and S. Griswold and C. Günther and P. Gutjahr and C. Haack and A. Hallgren and R. Halliday and L. Halve and F. Halzen and T. Hambye and H. Hamdaoui and M. Ha Minh and K. Hanson and J. Hardin and A. A. Harnisch and P. Hatch and A. Haungs and S. Hauser and K. Helbing and J. Hellrung and F. Henningsen and L. Heuermann and S. Hickford and A. Hidvegi and C. Hill and G. C. Hill and K. D. Hoffman and K. Hoshina and W. Hou and T. Huber and K. Hultqvist and M. Hünnefeld and R. Hussain and K. Hymon and S. In and N. Iovine and A. Ishihara and M. Jacquart and M. Jansson and G. S. Japaridze and K. Jayakumar and M. Jeong and M. Jin and B. J. P. Jones and D. Kang and W. Kang and X. Kang and A. Kappes and D. Kappesser and L. Kardum and T. Karg and M. Karl and A. Karle and U. Katz and M. Kauer and J. L. Kelley and A. Khatee Zathul and A. Kheirandish and K. Kin and J. Kiryluk and S. R. Klein and A. Kochocki and R. Koirala and H. Kolanoski and T. Kontrimas and L. Köpke and C. Kopper and D. J. Koskinen and P. Koundal and M. Kovacevich and M. Kowalski and T. Kozynets and K. Kruiswijk and E. Krupczak and A. Kumar and E. Kun and N. Kurahashi and N. Lad and C. Lagunas Gualda and M. Lamoureux and M. J. Larson and F. Lauber and J. P. Lazar and J. W. Lee and K. Leonard DeHolton and A. Leszczyńska and M. Lincetto and Q. R. Liu and M. Liubarska and E. Lohfink and C. Love and C. J. Lozano Mariscal and L. Lu and F. Lucarelli and A. Ludwig and W. Luszczak and Y. Lyu and W. Y. Ma and J. Madsen and K. B. M. Mahn and Y. Makino and S. Mancina and W. Marie Sainte and I. C. Mari{ş} and S. Marka and Z. Marka and M. Marsee and I. Martinez-Soler and R. Maruyama and F. Mayhew and T. McElroy and F. McNally and J. V. Mead and K. Meagher and S. Mechbal and A. Medina and M. Meier and S. Meighen-Berger and Y. Merckx and L. Merten and J. Micallef and D. Mockler and T. Montaruli and R. W. Moore and Y. Morii and R. Morse and M. Moulai and T. Mukherjee and R. Naab and R. Nagai and M. Nakos and U. Naumann and J. Necker and M. Neumann and H. Niederhausen and M. U. Nisa and A. Noell and S. C. Nowicki and A. Obertacke Pollmann and M. Oehler and B. Oeyen and A. Olivas and R. Orsoe and J. Osborn and E. O'Sullivan and H. Pandya and N. Park and G. K. Parker and E. N. Paudel and L. Paul and C. Pérez de los Heros and J. Peterson and S. Philippen and S. Pieper and A. Pizzuto and M. Plum and Y. Popovych and M. Prado Rodriguez and B. Pries and R. Procter-Murphy and G. T. Przybylski and C. Raab and J. Rack-Helleis and K. Rawlins and Z. Rechav and A. Rehman and P. Reichherzer and G. Renzi and E. Resconi and S. Reusch and W. Rhode and M. Richman and B. Riedel and E. J. Roberts and S. Robertson and S. Rodan and G. Roellinghoff and M. Rongen and C. Rott and T. Ruhe and L. Ruohan and D. Ryckbosch and S. Athanasiadou and I. Safa and J. Saffer and D. Salazar-Gallegos and P. Sampathkumar and S. E. Sanchez Herrera and A. Sandrock and M. Santander and S. Sarkar and S. Sarkar and J. Savelberg and P. Savina and M. Schaufel and H. Schieler and S. Schindler and B. Schlüter and T. Schmidt and J. Schneider and F. G. Schröder and L. Schumacher and G. Schwefer and S. Sclafani and D. Seckel and S. Seunarine and A. Sharma and S. Shefali and N. Shimizu and M. Silva and B. Skrzypek and B. Smithers and R. Snihur and J. Soedingrekso and A. Søgaard and D. Soldin and G. Sommani and C. Spannfellner and G. M. Spiczak and C. Spiering and M. Stamatikos and T. Stanev and R. Stein and T. Stezelberger and T. Stürwald and T. Stuttard and G. W. Sullivan and I. Taboada and S. Ter-Antonyan and W. G. Thompson and J. Thwaites and S. Tilav and K. Tollefson and C. Tönnis and S. Toscano and D. Tosi and A. Trettin and C. F. Tung and R. Turcotte and J. P. Twagirayezu and B. Ty and M. A. Unland Elorrieta and A. K. Upadhyay and K. Upshaw and N. Valtonen-Mattila and J. Vandenbroucke and N. van Eijndhoven and D. Vannerom and J. van Santen and J. Vara and J. Veitch-Michaelis and M. Venugopal and S. Verpoest and D. Veske and C. Walck and T. B. Watson and C. Weaver and P. Weigel and A. Weindl and J. Weldert and C. Wendt and J. Werthebach and M. Weyrauch and N. Whitehorn and C. H. Wiebusch and N. Willey and D. R. Williams and M. Wolf and G. Wrede and J. Wulff and X. W. Xu and J. P. Yanez and E. Yildizci and S. Yoshida and F. Yu and S. Yu and T. Yuan and Z. Zhang and P. Zhelnin},
      year={2023},
      eprint={2303.13663},
      archivePrefix={arXiv},
      primaryClass={astro-ph.HE},
      url={https://arxiv.org/abs/2303.13663}, 
}

@article{ANTARES_DM,
   title={Search for dark matter towards the Galactic Centre with 11 years of ANTARES data},
   volume={805},
   ISSN={0370-2693},
   url={http://dx.doi.org/10.1016/j.physletb.2020.135439},
   DOI={10.1016/j.physletb.2020.135439},
   journal={Physics Letters B},
   publisher={Elsevier BV},
   author={Albert, A. and André, M. and Anghinolfi, M. and Anton, G. and Ardid, M. and Aubert, J.-J. and Aublin, J. and Baret, B. and Basa, S. and Belhorma, B. and Bertin, V. and Biagi, S. and Bissinger, M. and Boumaaza, J. and Bourret, S. and Bouta, M. and Bouwhuis, M.C. and Brânzaş, H. and Bruijn, R. and Brunner, J. and Busto, J. and Capone, A. and Caramete, L. and Carr, J. and Celli, S. and Chabab, M. and Chau, T.N. and Cherkaoui El Moursli, R. and Chiarusi, T. and Circella, M. and Coleiro, A. and Colomer, M. and Coniglione, R. and Costantini, H. and Coyle, P. and Creusot, A. and Díaz, A.F. and de Wasseige, G. and Deschamps, A. and Distefano, C. and Di Palma, I. and Domi, A. and Donzaud, C. and Dornic, D. and Drouhin, D. and Eberl, T. and El Bojaddaini, I. and El Khayati, N. and Elsässer, D. and Enzenhöfer, A. and Ettahiri, A. and Fassi, F. and Fermani, P. and Ferrara, G. and Filippini, F. and Fusco, L. and Gay, P. and Glotin, H. and Gozzini, R. and Gracia Ruiz, R. and Graf, K. and Guidi, C. and Hallmann, S. and van Haren, H. and Heijboer, A.J. and Hello, Y. and Hernández-Rey, J.J. and Hößl, J. and Hofestädt, J. and Illuminati, G. and James, C.W. and de Jong, M. and de Jong, P. and Jongen, M. and Kadler, M. and Kalekin, O. and Katz, U. and Khan-Chowdhury, N.R. and Kouchner, A. and Kreter, M. and Kreykenbohm, I. and Kulikovskiy, V. and Lahmann, R. and Le Breton, R. and Lefèvre, D. and Leonora, E. and Levi, G. and Lincetto, M. and Lopez-Coto, D. and Loucatos, S. and Maggi, G. and Manczak, J. and Marcelin, M. and Margiotta, A. and Marinelli, A. and Martínez-Mora, J.A. and Mele, R. and Melis, K. and Migliozzi, P. and Moser, M. and Moussa, A. and Muller, R. and Nauta, L. and Navas, S. and Nezri, E. and Nielsen, C. and Nuñez-Castiñeyra, A. and O’Fearraigh, B. and Organokov, M. and Păvălaş, G.E. and Pellegrino, C. and Perrin-Terrin, M. and Piattelli, P. and Poirè, C. and Popa, V. and Pradier, T. and Quinn, L. and Randazzo, N. and Riccobene, G. and Sánchez-Losa, A. and Salah-Eddine, A. and Samtleben, D.F.E. and Sanguineti, M. and Sapienza, P. and Schüssler, F. and Spurio, M. and Stolarczyk, Th. and Strandberg, B. and Taiuti, M. and Tayalati, Y. and Thakore, T. and Tingay, S.J. and Trovato, A. and Vallage, B. and Van Elewyck, V. and Versari, F. and Viola, S. and Vivolo, D. and Wilms, J. and Zaborov, D. and Zegarelli, A. and Zornoza, J.D. and Zúñiga, J.},
   year={2020},
   month=jun, pages={135439} 
}

@article{Abdallah_2016,
   title={Search for Dark Matter Annihilations towards the Inner Galactic Halo from 10 Years of Observations with H.E.S.S.},
   volume={117},
   ISSN={1079-7114},
   url={http://dx.doi.org/10.1103/PhysRevLett.117.111301},
   DOI={10.1103/physrevlett.117.111301},
   number={11},
   journal={Physical Review Letters},
   publisher={American Physical Society (APS)},
   author={Abdallah, H. and Abramowski, A. and Aharonian, F. and Ait Benkhali, F. and Akhperjanian, A. G. and Angüner, E. and Arrieta, M. and Aubert, P. and Backes, M. and Balzer, A. and Barnard, M. and Becherini, Y. and Becker Tjus, J. and Berge, D. and Bernhard, S. and Bernlöhr, K. and Birsin, E. and Blackwell, R. and Böttcher, M. and Boisson, C. and Bolmont, J. and Bordas, P. and Bregeon, J. and Brun, F. and Brun, P. and Bryan, M. and Bulik, T. and Capasso, M. and Carr, J. and Casanova, S. and Chakraborty, N. and Chalme-Calvet, R. and Chaves, R. C. G. and Chen, A. and Chevalier, J. and Chrétien, M. and Colafrancesco, S. and Cologna, G. and Condon, B. and Conrad, J. and Couturier, C. and Cui, Y. and Davids, I. D. and Degrange, B. and Deil, C. and deWilt, P. and Djannati-Ataï, A. and Domainko, W. and Donath, A. and Drury, L. O’C. and Dubus, G. and Dutson, K. and Dyks, J. and Dyrda, M. and Edwards, T. and Egberts, K. and Eger, P. and Ernenwein, J.-P. and Eschbach, S. and Farnier, C. and Fegan, S. and Fernandes, M. V. and Fiasson, A. and Fontaine, G. and Förster, A. and Funk, S. and Füßling, M. and Gabici, S. and Gajdus, M. and Gallant, Y. A. and Garrigoux, T. and Giavitto, G. and Giebels, B. and Glicenstein, J. F. and Gottschall, D. and Goyal, A. and Grondin, M.-H. and Grudzińska, M. and Hadasch, D. and Hahn, J. and Hawkes, J. and Heinzelmann, G. and Henri, G. and Hermann, G. and Hervet, O. and Hillert, A. and Hinton, J. A. and Hofmann, W. and Hoischen, C. and Holler, M. and Horns, D. and Ivascenko, A. and Jacholkowska, A. and Jamrozy, M. and Janiak, M. and Jankowsky, D. and Jankowsky, F. and Jingo, M. and Jogler, T. and Jouvin, L. and Jung-Richardt, I. and Kastendieck, M. A. and Katarzyński, K. and Katz, U. and Kerszberg, D. and Khélifi, B. and Kieffer, M. and King, J. and Klepser, S. and Klochkov, D. and Kluźniak, W. and Kolitzus, D. and Komin, Nu. and Kosack, K. and Krakau, S. and Kraus, M. and Krayzel, F. and Krüger, P. P. and Laffon, H. and Lamanna, G. and Lau, J. and Lees, J.-P. and Lefaucheur, J. and Lefranc, V. and Lemière, A. and Lemoine-Goumard, M. and Lenain, J.-P. and Leser, E. and Lohse, T. and Lorentz, M. and Lui, R. and Lypova, I. and Marandon, V. and Marcowith, A. and Mariaud, C. and Marx, R. and Maurin, G. and Maxted, N. and Mayer, M. and Meintjes, P. J. and Menzler, U. and Meyer, M. and Mitchell, A. M. W. and Moderski, R. and Mohamed, M. and Morå, K. and Moulin, E. and Murach, T. and de Naurois, M. and Niederwanger, F. and Niemiec, J. and Oakes, L. and Odaka, H. and Ohm, S. and Öttl, S. and Ostrowski, M. and Oya, I. and Padovani, M. and Panter, M. and Parsons, R. D. and Paz Arribas, M. and Pekeur, N. W. and Pelletier, G. and Petrucci, P.-O. and Peyaud, B. and Pita, S. and Poon, H. and Prokhorov, D. and Prokoph, H. and Pühlhofer, G. and Punch, M. and Quirrenbach, A. and Raab, S. and Reimer, A. and Reimer, O. and Renaud, M. and de los Reyes, R. and Rieger, F. and Romoli, C. and Rosier-Lees, S. and Rowell, G. and Rudak, B. and Rulten, C. B. and Sahakian, V. and Salek, D. and Sanchez, D. A. and Santangelo, A. and Sasaki, M. and Schlickeiser, R. and Schüssler, F. and Schulz, A. and Schwanke, U. and Schwemmer, S. and Seyffert, A. S. and Shafi, N. and Simoni, R. and Sol, H. and Spanier, F. and Spengler, G. and Spieß, F. and Stawarz, L. and Steenkamp, R. and Stegmann, C. and Stinzing, F. and Stycz, K. and Sushch, I. and Tavernet, J.-P. and Tavernier, T. and Taylor, A. M. and Terrier, R. and Tluczykont, M. and Trichard, C. and Tuffs, R. and van der Walt, J. and van Eldik, C. and van Soelen, B. and Vasileiadis, G. and Veh, J. and Venter, C. and Viana, A. and Vincent, P. and Vink, J. and Voisin, F. and Völk, H. J. and Vuillaume, T. and Wadiasingh, Z. and Wagner, S. J. and Wagner, P. and Wagner, R. M. and White, R. and Wierzcholska, A. and Willmann, P. and Wörnlein, A. and Wouters, D. and Yang, R. and Zabalza, V. and Zaborov, D. and Zacharias, M. and Zdziarski, A. A. and Zech, A. and Zefi, F. and Ziegler, A. and Żywucka, N.},
   year={2016},
   month=sep 
}

@article{Albert_2017,
   title={Searching for Dark Matter Annihilation in Recently Discovered Milky Way Satellites with Fermi-LAT},
   volume={834},
   ISSN={1538-4357},
   url={http://dx.doi.org/10.3847/1538-4357/834/2/110},
   DOI={10.3847/1538-4357/834/2/110},
   number={2},
   journal={The Astrophysical Journal},
   publisher={American Astronomical Society},
   author={Albert, A. and Anderson, B. and Bechtol, K. and Drlica-Wagner, A. and Meyer, M. and Sánchez-Conde, M. and Strigari, L. and Wood, M. and Abbott, T. M. C. and Abdalla, F. B. and Benoit-Lévy, A. and Bernstein, G. M. and Bernstein, R. A. and Bertin, E. and Brooks, D. and Burke, D. L. and Rosell, A. Carnero and Kind, M. Carrasco and Carretero, J. and Crocce, M. and Cunha, C. E. and D’Andrea, C. B. and da Costa, L. N. and Desai, S. and Diehl, H. T. and Dietrich, J. P. and Doel, P. and Eifler, T. F. and Evrard, A. E. and Neto, A. Fausti and Finley, D. A. and Flaugher, B. and Fosalba, P. and Frieman, J. and Gerdes, D. W. and Goldstein, D. A. and Gruen, D. and Gruendl, R. A. and Honscheid, K. and James, D. J. and Kent, S. and Kuehn, K. and Kuropatkin, N. and Lahav, O. and Li, T. S. and Maia, M. A. G. and March, M. and Marshall, J. L. and Martini, P. and Miller, C. J. and Miquel, R. and Neilsen, E. and Nord, B. and Ogando, R. and Plazas, A. A. and Reil, K. and Romer, A. K. and Rykoff, E. S. and Sanchez, E. and Santiago, B. and Schubnell, M. and Sevilla-Noarbe, I. and Smith, R. C. and Soares-Santos, M. and Sobreira, F. and Suchyta, E. and Swanson, M. E. C. and Tarle, G. and Vikram, V. and Walker, A. R. and Wechsler, R. H.},
   year={2017},
   month=jan, pages={110} }

@article{Magic_2016,
   title={Limits to dark matter annihilation cross-section from a combined analysis of MAGIC and Fermi-LAT observations of dwarf satellite galaxies},
   author = {MAGIC collaboration},
   volume={2016},
   ISSN={1475-7516},
   url={http://dx.doi.org/10.1088/1475-7516/2016/02/039},
   DOI={10.1088/1475-7516/2016/02/039},
   number={02},
   journal={Journal of Cosmology and Astroparticle Physics},
   publisher={IOP Publishing},
   year={2016},
   month=feb, pages={039–039} 
}

@article{Ackermann_2015,
   title={Searching for Dark Matter Annihilation from Milky Way Dwarf Spheroidal Galaxies with Six Years of Fermi Large Area Telescope Data},
   volume={115},
   ISSN={1079-7114},
   url={http://dx.doi.org/10.1103/PhysRevLett.115.231301},
   DOI={10.1103/physrevlett.115.231301},
   number={23},
   journal={Physical Review Letters},
   publisher={American Physical Society (APS)},
   author={Ackermann, M. and Albert, A. and Anderson, B. and Atwood, W. B. and Baldini, L. and Barbiellini, G. and Bastieri, D. and Bechtol, K. and Bellazzini, R. and Bissaldi, E. and Blandford, R. D. and Bloom, E. D. and Bonino, R. and Bottacini, E. and Brandt, T. J. and Bregeon, J. and Bruel, P. and Buehler, R. and Caliandro, G. A. and Cameron, R. A. and Caputo, R. and Caragiulo, M. and Caraveo, P. A. and Cecchi, C. and Charles, E. and Chekhtman, A. and Chiang, J. and Chiaro, G. and Ciprini, S. and Claus, R. and Cohen-Tanugi, J. and Conrad, J. and Cuoco, A. and Cutini, S. and D’Ammando, F. and de Angelis, A. and de Palma, F. and Desiante, R. and Digel, S. W. and Di Venere, L. and Drell, P. S. and Drlica-Wagner, A. and Essig, R. and Favuzzi, C. and Fegan, S. J. and Ferrara, E. C. and Focke, W. B. and Franckowiak, A. and Fukazawa, Y. and Funk, S. and Fusco, P. and Gargano, F. and Gasparrini, D. and Giglietto, N. and Giordano, F. and Giroletti, M. and Glanzman, T. and Godfrey, G. and Gomez-Vargas, G. A. and Grenier, I. A. and Guiriec, S. and Gustafsson, M. and Hays, E. and Hewitt, J. W. and Horan, D. and Jogler, T. and Jóhannesson, G. and Kuss, M. and Larsson, S. and Latronico, L. and Li, J. and Li, L. and Llena Garde, M. and Longo, F. and Loparco, F. and Lubrano, P. and Malyshev, D. and Mayer, M. and Mazziotta, M. N. and McEnery, J. E. and Meyer, M. and Michelson, P. F. and Mizuno, T. and Moiseev, A. A. and Monzani, M. E. and Morselli, A. and Murgia, S. and Nuss, E. and Ohsugi, T. and Orienti, M. and Orlando, E. and Ormes, J. F. and Paneque, D. and Perkins, J. S. and Pesce-Rollins, M. and Piron, F. and Pivato, G. and Porter, T. A. and Rainò, S. and Rando, R. and Razzano, M. and Reimer, A. and Reimer, O. and Ritz, S. and Sánchez-Conde, M. and Schulz, A. and Sehgal, N. and Sgrò, C. and Siskind, E. J. and Spada, F. and Spandre, G. and Spinelli, P. and Strigari, L. and Tajima, H. and Takahashi, H. and Thayer, J. B. and Tibaldo, L. and Torres, D. F. and Troja, E. and Vianello, G. and Werner, M. and Winer, B. L. and Wood, K. S. and Wood, M. and Zaharijas, G. and Zimmer, S.},
   year={2015},
   month=nov }

@article{Aleksi__2014,
   title={Optimized dark matter searches in deep observations of Segue 1 with MAGIC},
   volume={2014},
   ISSN={1475-7516},
   url={http://dx.doi.org/10.1088/1475-7516/2014/02/008},
   DOI={10.1088/1475-7516/2014/02/008},
   number={02},
   journal={Journal of Cosmology and Astroparticle Physics},
   publisher={IOP Publishing},
   author={Aleksić, J. and Ansoldi, S. and Antonelli, L.A. and Antoranz, P. and Babic, A. and Bangale, P. and de Almeida, U. Barres and Barrio, J.A. and González, J. Becerra and Bednarek, W. and Berger, K. and Bernardini, E. and Biland, A. and Blanch, O. and Bock, R.K. and Bonnefoy, S. and Bonnoli, G. and Borracci, F. and Bretz, T. and Carmona, E. and Carosi, A. and Fidalgo, D. Carreto and Colin, P. and Colombo, E. and Contreras, J.L. and Cortina, J. and Covino, S. and Da Vela, P. and Dazzi, F. and De Angelis, A. and De Caneva, G. and De Lotto, B. and Mendez, C. Delgado and Doert, M. and Domínguez, A. and Prester, D. Dominis and Dorner, D. and Doro, M. and Einecke, S. and Eisenacher, D. and Elsaesser, D. and Farina, E. and Ferenc, D. and Fonseca, M.V. and Font, L. and Frantzen, K. and Fruck, C. and López, R.J. García and Garczarczyk, M. and Terrats, D. Garrido and Gaug, M. and Giavitto, G. and Godinović, N. and Muñoz, A. González and Gozzini, S.R. and Hadasch, D. and Hayashida, M. and Herrero, A. and Hildebrand, D. and Hose, J. and Hrupec, D. and Idec, W. and Kadenius, V. and Kellermann, H. and Kodani, K. and Konno, Y. and Krause, J. and Kubo, H. and Kushida, J. and La Barbera, A. and Lelas, D. and Lewandowska, N. and Lindfors, E. and Lombardi, S. and López, M. and López-Coto, R. and López-Oramas, A. and Lorenz, E. and Lozano, I. and Makariev, M. and Mallot, K. and Maneva, G. and Mankuzhiyil, N. and Mannheim, K. and Maraschi, L. and Marcote, B. and Mariotti, M. and Martínez, M. and Mazin, D. and Menzel, U. and Meucci, M. and Miranda, J.M. and Mirzoyan, R. and Moralejo, A. and Munar-Adrover, P. and Nakajima, D. and Niedzwiecki, A. and Nilsson, K. and Nishijima, K. and Nowak, N. and Orito, R. and Overkemping, A. and Paiano, S. and Palatiello, M. and Paneque, D. and Paoletti, R. and Paredes, J.M. and Paredes-Fortuny, X. and Partini, S. and Persic, M. and Prada, F. and Moroni, P. G. Prada and Prandini, E. and Preziuso, S. and Puljak, I. and Reinthal, R. and Rhode, W. and Ribó, M. and Rico, J. and Garcia, J. Rodriguez and Rügamer, S. and Saggion, A. and Saito, T. and Saito, K. and Salvati, M. and Satalecka, K. and Scalzotto, V. and Scapin, V. and Schultz, C. and Schweizer, T. and Sillanpää, A. and Sitarek, J. and Snidaric, I. and Sobczynska, D. and Spanier, F. and Stamatescu, V. and Stamerra, A. and Steinbring, T. and Storz, J. and Sun, S. and Surić, T. and Takalo, L. and Takami, H. and Tavecchio, F. and Temnikov, P. and Terzić, T. and Tescaro, D. and Teshima, M. and Thaele, J. and Tibolla, O. and Torres, D.F. and Toyama, T. and Treves, A. and Uellenbeck, M. and Vogler, P. and Wagner, R.M. and Zandanel, F. and Zanin, R. and Ibarra, A.},
   year={2014},
   month=feb, pages={008–008} }

@article{Aebischer:2018iyb,
    author = "Aebischer, Jason and Kumar, Jacky and Stangl, Peter and Straub, David M.",
    title = "{A Global Likelihood for Precision Constraints and Flavour Anomalies}",
    eprint = "1810.07698",
    archivePrefix = "arXiv",
    primaryClass = "hep-ph",
    doi = "10.1140/epjc/s10052-019-6977-z",
    journal = "Eur. Phys. J. C",
    volume = "79",
    number = "6",
    pages = "509",
    year = "2019"
}

@article{Straub:2018kue,
    author = "Straub, David M.",
    title = "{flavio: a Python package for flavour and precision phenomenology in the Standard Model and beyond}",
    eprint = "1810.08132",
    archivePrefix = "arXiv",
    primaryClass = "hep-ph",
    month = "10",
    year = "2018"
}

@article{Aebischer:2018bkb,
    author = "Aebischer, Jason and Kumar, Jacky and Straub, David M.",
    title = "{Wilson: a Python package for the running and matching of Wilson coefficients above and below the electroweak scale}",
    eprint = "1804.05033",
    archivePrefix = "arXiv",
    primaryClass = "hep-ph",
    doi = "10.1140/epjc/s10052-018-6492-7",
    journal = "Eur. Phys. J. C",
    volume = "78",
    number = "12",
    pages = "1026",
    year = "2018"
}

@article{Aebischer:2017ugx,
    author = "Aebischer, Jason and others",
    title = "{WCxf: an exchange format for Wilson coefficients beyond the Standard Model}",
    eprint = "1712.05298",
    archivePrefix = "arXiv",
    primaryClass = "hep-ph",
    reportNumber = "IFIC-17-61, TUM-HEP-1117-17, LMU-ASC-74-17, IFIC/17-61, KA-TP-38-2017, TUM-HEP-1117/17, LMU-ASC 74/17",
    doi = "10.1016/j.cpc.2018.05.022",
    journal = "Comput. Phys. Commun.",
    volume = "232",
    pages = "71--83",
    year = "2018"
}

@article{Descotes-Genon:2018foz,
    author = "Descotes-Genon, S\'ebastien and Falkowski, Adam and Fedele, Marco and Gonz\'alez-Alonso, Mart\'\i{}n and Virto, Javier",
    title = "{The CKM parameters in the SMEFT}",
    eprint = "1812.08163",
    archivePrefix = "arXiv",
    primaryClass = "hep-ph",
    reportNumber = "LPT Orsay 18-92, CERN-TH-2018-276, TUM-HEP-1178/18, MIT-CTP/5081,
  NIOBE-2018-01",
    doi = "10.1007/JHEP05(2019)172",
    journal = "JHEP",
    volume = "05",
    pages = "172",
    year = "2019"
}

@article{Fuentes-Martin:2022jrf,
    author = {Fuentes-Mart\'\i{}n, Javier and K\"onig, Matthias and Pag\`es, Julie and Thomsen, Anders Eller and Wilsch, Felix},
    title = "{A proof of concept for matchete: an automated tool for matching effective theories}",
    eprint = "2212.04510",
    archivePrefix = "arXiv",
    primaryClass = "hep-ph",
    reportNumber = "MITP-22-105, TUM-HEP-1443/22, ZU-TH-58/22",
    doi = "10.1140/epjc/s10052-023-11726-1",
    journal = "Eur. Phys. J. C",
    volume = "83",
    number = "7",
    pages = "662",
    year = "2023"
}

@article{Buchmuller:1985jz,
    author = "Buchmuller, W. and Wyler, D.",
    title = "{Effective Lagrangian Analysis of New Interactions and Flavor Conservation}",
    reportNumber = "CERN-TH-4254/85",
    doi = "10.1016/0550-3213(86)90262-2",
    journal = "Nucl. Phys. B",
    volume = "268",
    pages = "621--653",
    year = "1986"
}

@article{Jenkins:2013zja,
    author = "Jenkins, Elizabeth E. and Manohar, Aneesh V. and Trott, Michael",
    title = "{Renormalization Group Evolution of the Standard Model Dimension Six Operators I: Formalism and lambda Dependence}",
    eprint = "1308.2627",
    archivePrefix = "arXiv",
    primaryClass = "hep-ph",
    doi = "10.1007/JHEP10(2013)087",
    journal = "JHEP",
    volume = "10",
    pages = "087",
    year = "2013"
}

@article{Jenkins:2013wua,
    author = "Jenkins, Elizabeth E. and Manohar, Aneesh V. and Trott, Michael",
    title = "{Renormalization Group Evolution of the Standard Model Dimension Six Operators II: Yukawa Dependence}",
    eprint = "1310.4838",
    archivePrefix = "arXiv",
    primaryClass = "hep-ph",
    reportNumber = "CERN-PH-TH/2015-247",
    doi = "10.1007/JHEP01(2014)035",
    journal = "JHEP",
    volume = "01",
    pages = "035",
    year = "2014"
}

@article{Alonso:2013hga,
    author = "Alonso, Rodrigo and Jenkins, Elizabeth E. and Manohar, Aneesh V. and Trott, Michael",
    title = "{Renormalization Group Evolution of the Standard Model Dimension Six Operators III: Gauge Coupling Dependence and Phenomenology}",
    eprint = "1312.2014",
    archivePrefix = "arXiv",
    primaryClass = "hep-ph",
    reportNumber = "CERN-PH-TH-2013-305, CERN-PH-TH/2013-305",
    doi = "10.1007/JHEP04(2014)159",
    journal = "JHEP",
    volume = "04",
    pages = "159",
    year = "2014"
}

@article{Grzadkowski:2010es,
    author = "Grzadkowski, B. and Iskrzynski, M. and Misiak, M. and Rosiek, J.",
    title = "{Dimension-Six Terms in the Standard Model Lagrangian}",
    eprint = "1008.4884",
    archivePrefix = "arXiv",
    primaryClass = "hep-ph",
    reportNumber = "IFT-9-2010, TTP10-35",
    doi = "10.1007/JHEP10(2010)085",
    journal = "JHEP",
    volume = "10",
    pages = "085",
    year = "2010"
}

@article{Jenkins:2017jig,
    author = "Jenkins, Elizabeth E. and Manohar, Aneesh V. and Stoffer, Peter",
    title = "{Low-Energy Effective Field Theory below the Electroweak Scale: Operators and Matching}",
    eprint = "1709.04486",
    archivePrefix = "arXiv",
    primaryClass = "hep-ph",
    doi = "10.1007/JHEP03(2018)016",
    journal = "JHEP",
    volume = "03",
    pages = "016",
    year = "2018",
    note = "[Erratum: JHEP 12, 043 (2023)]"
}

@article{Dekens:2019ept,
    author = "Dekens, Wouter and Stoffer, Peter",
    title = "{Low-energy effective field theory below the electroweak scale: matching at one loop}",
    eprint = "1908.05295",
    archivePrefix = "arXiv",
    primaryClass = "hep-ph",
    doi = "10.1007/JHEP10(2019)197",
    journal = "JHEP",
    volume = "10",
    pages = "197",
    year = "2019",
    note = "[Erratum: JHEP 11, 148 (2022)]"
}

@article{Jenkins:2017dyc,
    author = "Jenkins, Elizabeth E. and Manohar, Aneesh V. and Stoffer, Peter",
    title = "{Low-Energy Effective Field Theory below the Electroweak Scale: Anomalous Dimensions}",
    eprint = "1711.05270",
    archivePrefix = "arXiv",
    primaryClass = "hep-ph",
    doi = "10.1007/JHEP01(2018)084",
    journal = "JHEP",
    volume = "01",
    pages = "084",
    year = "2018",
    note = "[Erratum: JHEP 12, 042 (2023)]"
}

@article{Buonocore_2020,
   title={Leptons in the proton},
   volume={2020},
   ISSN={1029-8479},
   url={http://dx.doi.org/10.1007/JHEP08(2020)019},
   DOI={10.1007/jhep08(2020)019},
   number={8},
   journal={Journal of High Energy Physics},
   publisher={Springer Science and Business Media LLC},
   author={Buonocore, Luca and Nason, Paolo and Tramontano, Francesco and Zanderighi, Giulia},
   year={2020},
   month=aug }

@article{CMS-SUS-20-001,
   title={Search for supersymmetry in final states with two oppositely charged same-flavor leptons and missing transverse momentum in proton-proton collisions at $\sqrt{s}$ = 13 TeV},
   volume={2021},
   ISSN={1029-8479},
   url={http://dx.doi.org/10.1007/JHEP04(2021)123},
   DOI={10.1007/jhep04(2021)123},
   number={4},
   journal={Journal of High Energy Physics},
   publisher={Springer Science and Business Media LLC},
   author={Sirunyan, A. M. and Tumasyan, A. and Adam, W. and Bergauer, T. and Dragicevic, M. and Escalante Del Valle, A. and Frühwirth, R. and Jeitler, M. and Krammer, N. and Lechner, L. and Liko, D. and Mikulec, I. and Pitters, F. M. and Rad, N. and Schieck, J. and Schöfbeck, R. and Spanring, M. and Templ, S. and Waltenberger, W. and Wulz, C.-E. and Zarucki, M. and Chekhovsky, V. and Litomin, A. and Makarenko, V. and Suarez Gonzalez, J. and Darwish, M. R. and De Wolf, E. A. and Di Croce, D. and Janssen, X. and Kello, T. and Lelek, A. and Pieters, M. and Rejeb Sfar, H. and Van Haevermaet, H. and Van Mechelen, P. and Van Putte, S. and Van Remortel, N. and Blekman, F. and Bols, E. S. and Chhibra, S. S. and D’Hondt, J. and De Clercq, J. and Lontkovskyi, D. and Lowette, S. and Marchesini, I. and Moortgat, S. and Morton, A. and Müller, D. and Python, Q. and Tavernier, S. and Van Doninck, W. and Van Mulders, P. and Beghin, D. and Bilin, B. and Clerbaux, B. and De Lentdecker, G. and Dorney, B. and Favart, L. and Grebenyuk, A. and Kalsi, A. K. and Makarenko, I. and Moureaux, L. and Pétré, L. and Popov, A. and Postiau, N. and Starling, E. and Thomas, L. and Vander Velde, C. and Vanlaer, P. and Vannerom, D. and Wezenbeek, L. and Cornelis, T. and Dobur, D. and Gruchala, M. and Khvastunov, I. and Niedziela, M. and Roskas, C. and Skovpen, K. and Tytgat, M. and Verbeke, W. and Vermassen, B. and Vit, M. and Bruno, G. and Bury, F. and Caputo, C. and David, P. and Delaere, C. and Delcourt, M. and Donertas, I. S. and Giammanco, A. and Lemaitre, V. and Mondal, K. and Prisciandaro, J. and Taliercio, A. and Teklishyn, M. and Vischia, P. and Wertz, S. and Wuyckens, S. and Alves, G. A. and Hensel, C. and Moraes, A. and Aldá Júnior, W. L. and Belchior Batista Das Chagas, E. and BRANDAO MALBOUISSON, H. and Carvalho, W. and Chinellato, J. and Coelho, E. and Da Costa, E. M. and Da Silveira, G. G. and De Jesus Damiao, D. and Fonseca De Souza, S. and Martins, J. and Matos Figueiredo, D. and Medina Jaime, M. and Mora Herrera, C. and Mundim, L. and Nogima, H. and Rebello Teles, P. and Sanchez Rosas, L. J. and Santoro, A. and Silva Do Amaral, S. M. and Sznajder, A. and Thiel, M. and Torres Da Silva De Araujo, F. and Vilela Pereira, A. and Bernardes, C. A. and Calligaris, L. and Fernandez Perez Tomei, T. R. and Gregores, E. M. and Lemos, D. S. and Mercadante, P. G. and Novaes, S. F. and Padula, Sandra S. and Aleksandrov, A. and Antchev, G. and Atanasov, I. and Hadjiiska, R. and Iaydjiev, P. and Misheva, M. and Rodozov, M. and Shopova, M. and Sultanov, G. and Dimitrov, A. and Ivanov, T. and Litov, L. and Pavlov, B. and Petkov, P. and Petrov, A. and Cheng, T. and Fang, W. and Guo, Q. and Wang, H. and Yuan, L. and Ahmad, M. and Bauer, G. and Hu, Z. and Wang, Y. and Yi, K. and Chapon, E. and Chen, G. M. and Chen, H. S. and Chen, M. and Javaid, T. and Kapoor, A. and Leggat, D. and Liao, H. and LIU, Z.-A. and Sharma, R. and Spiezia, A. and Tao, J. and Thomas-wilsker, J. and Wang, J. and Zhang, H. and Zhang, S. and Zhao, J. and Agapitos, A. and Ban, Y. and Chen, C. and Huang, Q. and Levin, A. and Li, Q. and Lu, M. and Lyu, X. and Mao, Y. and Qian, S. J. and Wang, D. and Wang, Q. and Xiao, J. and You, Z. and Gao, X. and Xiao, M. and Avila, C. and Cabrera, A. and Florez, C. and Fraga, J. and Sarkar, A. and Segura Delgado, M. A. and Jaramillo, J. and Mejia Guisao, J. and Ramirez, F. and Ruiz Alvarez, J. D. and Salazar González, C. A. and Vanegas Arbelaez, N. and Giljanovic, D. and Godinovic, N. and Lelas, D. and Puljak, I. and Antunovic, Z. and Kovac, M. and Sculac, T. and Brigljevic, V. and Ferencek, D. and Majumder, D. and Roguljic, M. and Starodumov, A. and Susa, T. and Ather, M. W. and Attikis, A. and Erodotou, E. and Ioannou, A. and Kole, G. and Kolosova, M. and Konstantinou, S. and Mousa, J. and Nicolaou, C. and Ptochos, F. and Razis, P. A. and Rykaczewski, H. and Saka, H. and Tsiakkouri, D. and Finger, M. and Finger, M. and Kveton, A. and Tomsa, J. and Ayala, E. and Carrera Jarrin, E. and Abdalla, H. and Elgammal, S. and Khalil, S. and Mahmoud, M. A. and Mohammed, Y. and Bhowmik, S. and Carvalho Antunes De Oliveira, A. and Dewanjee, R. K. and Ehataht, K. and Kadastik, M. and Raidal, M. and Veelken, C. and Eerola, P. and Forthomme, L. and Kirschenmann, H. and Osterberg, K. and Voutilainen, M. and Brücken, E. and Garcia, F. and Havukainen, J. and Karimäki, V. and Kim, M. S. and Kinnunen, R. and Lampén, T. and Lassila-Perini, K. and Lehti, S. and Lindén, T. and Siikonen, H. and Tuominen, E. and Tuominiemi, J. and Luukka, P. and Tuuva, T. and Amendola, C. and Besancon, M. and Couderc, F. and Dejardin, M. and Denegri, D. and Faure, J. L. and Ferri, F. and Ganjour, S. and Givernaud, A. and Gras, P. and Hamel de Monchenault, G. and Jarry, P. and Lenzi, B. and Locci, E. and Malcles, J. and Rander, J. and Rosowsky, A. and Sahin, M. Ö. and Savoy-Navarro, A. and Titov, M. and Yu, G. B. and Ahuja, S. and Beaudette, F. and Bonanomi, M. and Buchot Perraguin, A. and Busson, P. and Charlot, C. and Davignon, O. and Diab, B. and Falmagne, G. and Granier de Cassagnac, R. and Hakimi, A. and Kucher, I. and Lobanov, A. and Martin Perez, C. and Nguyen, M. and Ochando, C. and Paganini, P. and Rembser, J. and Salerno, R. and Sauvan, J. B. and Sirois, Y. and Zabi, A. and Zghiche, A. and Agram, J.-L. and Andrea, J. and Bloch, D. and Bourgatte, G. and Brom, J.-M. and Chabert, E. C. and Collard, C. and Fontaine, J.-C. and Gelé, D. and Goerlach, U. and Grimault, C. and Le Bihan, A.-C. and Van Hove, P. and Asilar, E. and Beauceron, S. and Bernet, C. and Boudoul, G. and Camen, C. and Carle, A. and Chanon, N. and Contardo, D. and Depasse, P. and El Mamouni, H. and Fay, J. and Gascon, S. and Gouzevitch, M. and Ille, B. and Jain, Sa. and Laktineh, I. B. and Lattaud, H. and Lesauvage, A. and Lethuillier, M. and Mirabito, L. and Shchablo, K. and Torterotot, L. and Touquet, G. and Vander Donckt, M. and Viret, S. and Khvedelidze, A. and Tsamalaidze, Z. and Feld, L. and Klein, K. and Lipinski, M. and Meuser, D. and Pauls, A. and Rauch, M. P. and Schulz, J. and Teroerde, M. and Eliseev, D. and Erdmann, M. and Fackeldey, P. and Fischer, B. and Ghosh, S. and Hebbeker, T. and Hoepfner, K. and Keller, H. and Mastrolorenzo, L. and Merschmeyer, M. and Meyer, A. and Mocellin, G. and Mondal, S. and Mukherjee, S. and Noll, D. and Novak, A. and Pook, T. and Pozdnyakov, A. and Rath, Y. and Reithler, H. and Roemer, J. and Schmidt, A. and Schuler, S. C. and Sharma, A. and Wiedenbeck, S. and Zaleski, S. and Dziwok, C. and Flügge, G. and Haj Ahmad, W. and Hlushchenko, O. and Kress, T. and Nowack, A. and Pistone, C. and Pooth, O. and Roy, D. and Sert, H. and Stahl, A. and Ziemons, T. and Aarup Petersen, H. and Aldaya Martin, M. and Asmuss, P. and Babounikau, I. and Baxter, S. and Behnke, O. and Bermúdez Martínez, A. and Bin Anuar, A. A. and Borras, K. and Botta, V. and Brunner, D. and Campbell, A. and Cardini, A. and Connor, P. and Consuegra Rodríguez, S. and Danilov, V. and De Wit, A. and Defranchis, M. M. and Didukh, L. and Domínguez Damiani, D. and Eckerlin, G. and Eckstein, D. and Estevez Banos, L. I. and Gallo, E. and Geiser, A. and Giraldi, A. and Grohsjean, A. and Guthoff, M. and Harb, A. and Jafari, A. and Jomhari, N. Z. and Jung, H. and Kasem, A. and Kasemann, M. and Kaveh, H. and Kleinwort, C. and Knolle, J. and Krücker, D. and Lange, W. and Lenz, T. and Lidrych, J. and Lipka, K. and Lohmann, W. and Madlener, T. and Mankel, R. and Melzer-Pellmann, I.-A. and Metwally, J. and Meyer, A. B. and Meyer, M. and Missiroli, M. and Mnich, J. and Mussgiller, A. and Myronenko, V. and Otarid, Y. and Pérez Adán, D. and Pflitsch, S. K. and Pitzl, D. and Raspereza, A. and Saggio, A. and Saibel, A. and Savitskyi, M. and Scheurer, V. and Schwanenberger, C. and Singh, A. and Sosa Ricardo, R. E. and Tonon, N. and Turkot, O. and Vagnerini, A. and Van De Klundert, M. and Walsh, R. and Walter, D. and Wen, Y. and Wichmann, K. and Wissing, C. and Wuchterl, S. and Zenaiev, O. and Zlebcik, R. and Aggleton, R. and Bein, S. and Benato, L. and Benecke, A. and De Leo, K. and Dreyer, T. and Ebrahimi, A. and Eich, M. and Feindt, F. and Fröhlich, A. and Garbers, C. and Garutti, E. and Gunnellini, P. and Haller, J. and Hinzmann, A. and Karavdina, A. and Kasieczka, G. and Klanner, R. and Kogler, R. and Kutzner, V. and Lange, J. and Lange, T. and Malara, A. and Niemeyer, C. E. N. and Nigamova, A. and Pena Rodriguez, K. J. and Rieger, O. and Schleper, P. and Schumann, S. and Schwandt, J. and Schwarz, D. and Sonneveld, J. and Stadie, H. and Steinbrück, G. and Tews, A. and Vormwald, B. and Zoi, I. and Bechtel, J. and Berger, T. and Butz, E. and Caspart, R. and Chwalek, T. and De Boer, W. and Dierlamm, A. and Droll, A. and El Morabit, K. and Faltermann, N. and Flöh, K. and Giffels, M. and Gottmann, A. and Hartmann, F. and Heidecker, C. and Husemann, U. and Katkov, I. and Keicher, P. and Koppenhöfer, R. and Maier, S. and Metzler, M. and Mitra, S. and Müller, Th. and Musich, M. and Quast, G. and Rabbertz, K. and Rauser, J. and Savoiu, D. and Schäfer, D. and Schnepf, M. and Schröder, M. and Seith, D. and Shvetsov, I. and Simonis, H. J. and Ulrich, R. and Wassmer, M. and Weber, M. and Wolf, R. and Wozniewski, S. and Anagnostou, G. and Asenov, P. and Daskalakis, G. and Geralis, T. and Kyriakis, A. and Loukas, D. and Paspalaki, G. and Stakia, A. and Diamantopoulou, M. and Karasavvas, D. and Karathanasis, G. and Kontaxakis, P. and Koraka, C. K. and Manousakis-katsikakis, A. and Panagiotou, A. and Papavergou, I. and Saoulidou, N. and Theofilatos, K. and Tziaferi, E. and Vellidis, K. and Vourliotis, E. and Bakas, G. and Kousouris, K. and Papakrivopoulos, I. and Tsipolitis, G. and Zacharopoulou, A. and Evangelou, I. and Foudas, C. and Gianneios, P. and Katsoulis, P. and Kokkas, P. and Manitara, K. and Manthos, N. and Papadopoulos, I. and Strologas, J. and Bartók, M. and Csanad, M. and Gadallah, M. M. A. and Lökös, S. and Major, P. and Mandal, K. and Mehta, A. and Pasztor, G. and Surányi, O. and Veres, G. I. and Bencze, G. and Hajdu, C. and Horvath, D. and Sikler, F. and Veszpremi, V. and Vesztergombi, G. and Czellar, S. and Karancsi, J. and Molnar, J. and Szillasi, Z. and Teyssier, D. and Raics, P. and Trocsanyi, Z. L. and Ujvari, B. and Csorgo, T. and Nemes, F. and Novak, T. and Choudhury, S. and Komaragiri, J. R. and Kumar, D. and Panwar, L. and Tiwari, P. C. and Bahinipati, S. and Dash, D. and Kar, C. and Mal, P. and Mishra, T. and Muraleedharan Nair Bindhu, V. K. and Nayak, A. and Sur, N. and Swain, S. K. and Bansal, S. and Beri, S. B. and Bhatnagar, V. and Chaudhary, G. and Chauhan, S. and Dhingra, N. and Gupta, R. and Kaur, A. and Kaur, S. and Kumari, P. and Meena, M. and Sandeep, K. and Sharma, S. and Singh, J. B. and Virdi, A. K. and Ahmed, A. and Bhardwaj, A. and Choudhary, B. C. and Garg, R. B. and Gola, M. and Keshri, S. and Kumar, A. and Naimuddin, M. and Priyanka, P. and Ranjan, K. and Shah, A. and Bharti, M. and Bhattacharya, R. and Bhattacharya, S. and Bhowmik, D. and Dutta, S. and Ghosh, S. and Gomber, B. and Maity, M. and Nandan, S. and Palit, P. and Rout, P. K. and Saha, G. and Sahu, B. and Sarkar, S. and Sharan, M. and Singh, B. and Thakur, S. and Behera, P. K. and Behera, S. C. and Kalbhor, P. and Muhammad, A. and Pradhan, R. and Pujahari, P. R. and Sharma, A. and Sikdar, A. K. and Dutta, D. and Kumar, V. and Naskar, K. and Netrakanti, P. K. and Pant, L. M. and Shukla, P. and Aziz, T. and Bhat, M. A. and Dugad, S. and Kumar Verma, R. and Mohanty, G. B. and Sarkar, U. and Banerjee, S. and Bhattacharya, S. and Chatterjee, S. and Chudasama, R. and Guchait, M. and Karmakar, S. and Kumar, S. and Majumder, G. and Mazumdar, K. and Mukherjee, S. and Roy, D. and Dube, S. and Kansal, B. and Pandey, S. and Rane, A. and Rastogi, A. and Sharma, S. and Bakhshiansohi, H. and Zeinali, M. and Chenarani, S. and Etesami, S. M. and Khakzad, M. and Mohammadi Najafabadi, M. and Felcini, M. and Grunewald, M. and Abbrescia, M. and Aly, R. and Aruta, C. and Colaleo, A. and Creanza, D. and De Filippis, N. and De Palma, M. and Di Florio, A. and Di Pilato, A. and Elmetenawee, W. and Fiore, L. and Gelmi, A. and Gul, M. and Iaselli, G. and Ince, M. and Lezki, S. and Maggi, G. and Maggi, M. and Margjeka, I. and Mastrapasqua, V. and Merlin, J. A. and My, S. and Nuzzo, S. and Pompili, A. and Pugliese, G. and Ranieri, A. and Selvaggi, G. and Silvestris, L. and Simone, F. M. and Venditti, R. and Verwilligen, P. and Abbiendi, G. and Battilana, C. and Bonacorsi, D. and Borgonovi, L. and Braibant-Giacomelli, S. and Campanini, R. and Capiluppi, P. and Castro, A. and Cavallo, F. R. and Ciocca, C. and Cuffiani, M. and Dallavalle, G. M. and Diotalevi, T. and Fabbri, F. and Fanfani, A. and Fontanesi, E. and Giacomelli, P. and Giommi, L. and Grandi, C. and Guiducci, L. and Iemmi, F. and Lo Meo, S. and Marcellini, S. and Masetti, G. and Navarria, F. L. and Perrotta, A. and Primavera, F. and Rossi, A. M. and Rovelli, T. and Siroli, G. P. and Tosi, N. and Albergo, S. and Costa, S. and Di Mattia, A. and Potenza, R. and Tricomi, A. and Tuve, C. and Barbagli, G. and Cassese, A. and Ceccarelli, R. and Ciulli, V. and Civinini, C. and D’Alessandro, R. and Fiori, F. and Focardi, E. and Latino, G. and Lenzi, P. and Lizzo, M. and Meschini, M. and Paoletti, S. and Seidita, R. and Sguazzoni, G. and Viliani, L. and Benussi, L. and Bianco, S. and Piccolo, D. and Bozzo, M. and Ferro, F. and Mulargia, R. and Robutti, E. and Tosi, S. and Benaglia, A. and Beschi, A. and Brivio, F. and Cetorelli, F. and Ciriolo, V. and De Guio, F. and Dinardo, M. E. and Dini, P. and Gennai, S. and Ghezzi, A. and Govoni, P. and Guzzi, L. and Malberti, M. and Malvezzi, S. and Massironi, A. and Menasce, D. and Monti, F. and Moroni, L. and Paganoni, M. and Pedrini, D. and Ragazzi, S. and Tabarelli de Fatis, T. and Valsecchi, D. and Zuolo, D. and Buontempo, S. and Cavallo, N. and De Iorio, A. and Fabozzi, F. and Fienga, F. and Iorio, A. O. M. and Lista, L. and Meola, S. and Paolucci, P. and Rossi, B. and Sciacca, C. and Azzi, P. and Bacchetta, N. and Bisello, D. and Bortignon, P. and Bragagnolo, A. and Carlin, R. and Checchia, P. and De Castro Manzano, P. and Dorigo, T. and Gasparini, F. and Gasparini, U. and Hoh, S. Y. and Layer, L. and Margoni, M. and Meneguzzo, A. T. and Presilla, M. and Ronchese, P. and Rossin, R. and Simonetto, F. and Strong, G. and Tosi, M. and YARAR, H. and Zanetti, M. and Zotto, P. and Zucchetta, A. and Zumerle, G. and Aime‘, C. and Braghieri, A. and Calzaferri, S. and Fiorina, D. and Montagna, P. and Ratti, S. P. and Re, V. and Ressegotti, M. and Riccardi, C. and Salvini, P. and Vai, I. and Vitulo, P. and Biasini, M. and Bilei, G. M. and Ciangottini, D. and Fanò, L. and Lariccia, P. and Mantovani, G. and Mariani, V. and Menichelli, M. and Moscatelli, F. and Piccinelli, A. and Rossi, A. and Santocchia, A. and Spiga, D. and Tedeschi, T. and Androsov, K. and Azzurri, P. and Bagliesi, G. and Bertacchi, V. and Bianchini, L. and Boccali, T. and Castaldi, R. and Ciocci, M. A. and Dell’Orso, R. and Di Domenico, M. R. and Donato, S. and Giannini, L. and Giassi, A. and Grippo, M. T. and Ligabue, F. and Manca, E. and Mandorli, G. and Messineo, A. and Palla, F. and Ramirez-Sanchez, G. and Rizzi, A. and Rolandi, G. and Roy Chowdhury, S. and Scribano, A. and Shafiei, N. and Spagnolo, P. and Tenchini, R. and Tonelli, G. and Turini, N. and Venturi, A. and Verdini, P. G. and Cavallari, F. and Cipriani, M. and Del Re, D. and Di Marco, E. and Diemoz, M. and Longo, E. and Meridiani, P. and Organtini, G. and Pandolfi, F. and Paramatti, R. and Quaranta, C. and Rahatlou, S. and Rovelli, C. and Santanastasio, F. and Soffi, L. and Tramontano, R. and Amapane, N. and Arcidiacono, R. and Argiro, S. and Arneodo, M. and Bartosik, N. and Bellan, R. and Bellora, A. and Berenguer Antequera, J. and Biino, C. and Cappati, A. and Cartiglia, N. and Cometti, S. and Costa, M. and Covarelli, R. and Demaria, N. and Kiani, B. and Legger, F. and Mariotti, C. and Maselli, S. and Migliore, E. and Monaco, V. and Monteil, E. and Monteno, M. and Obertino, M. M. and Ortona, G. and Pacher, L. and Pastrone, N. and Pelliccioni, M. and Pinna Angioni, G. L. and Ruspa, M. and Salvatico, R. and Siviero, F. and Sola, V. and Solano, A. and Soldi, D. and Staiano, A. and Tornago, M. and Trocino, D. and Belforte, S. and Candelise, V. and Casarsa, M. and Cossutti, F. and Da Rold, A. and Della Ricca, G. and Vazzoler, F. and Dogra, S. and Huh, C. and Kim, B. and Kim, D. H. and Kim, G. N. and Lee, J. and Lee, S. W. and Moon, C. S. and Oh, Y. D. and Pak, S. I. and Radburn-Smith, B. C. and Sekmen, S. and Yang, Y. C. and Kim, H. and Moon, D. H. and Francois, B. and Kim, T. J. and Park, J. and Cho, S. and Choi, S. and Go, Y. and Ha, S. and Hong, B. and Lee, K. and Lee, K. S. and Lim, J. and Park, J. and Park, S. K. and Yoo, J. and Goh, J. and Gurtu, A. and Kim, H. S. and Kim, Y. and Almond, J. and Bhyun, J. H. and Choi, J. and Jeon, S. and Kim, J. and Kim, J. S. and Ko, S. and Kwon, H. and Lee, H. and Lee, K. and Lee, S. and Nam, K. and Oh, B. H. and Oh, M. and Oh, S. B. and Seo, H. and Yang, U. K. and Yoon, I. and Jeon, D. and Kim, J. H. and Ko, B. and Lee, J. S. H. and Park, I. C. and Roh, Y. and Song, D. and Watson, I. J. and Yoo, H. D. and Choi, Y. and Hwang, C. and Jeong, Y. and Lee, H. and Lee, Y. and Yu, I. and Maghrbi, Y. and Veckalns, V. and Juodagalvis, A. and Rinkevicius, A. and Tamulaitis, G. and Vaitkevicius, A. and Wan Abdullah, W. A. T. and Yusli, M. N. and Zolkapli, Z. and Benitez, J. F. and Castaneda Hernandez, A. and Murillo Quijada, J. A. and Valencia Palomo, L. and Ayala, G. and Castilla-Valdez, H. and De La Cruz-Burelo, E. and Heredia-De La Cruz, I. and Lopez-Fernandez, R. and Mondragon Herrera, C. A. and Perez Navarro, D. A. and Sanchez-Hernandez, A. and Carrillo Moreno, S. and Oropeza Barrera, C. and Ramirez-Garcia, M. and Vazquez Valencia, F. and Eysermans, J. and Pedraza, I. and Salazar Ibarguen, H. A. and Uribe Estrada, C. and Morelos Pineda, A. and Mijuskovic, J. and Raicevic, N. and Krofcheck, D. and Bheesette, S. and Butler, P. H. and Ahmad, A. and Asghar, M. I. and Awais, A. and Awan, M. I. M. and Hoorani, H. R. and Khan, W. A. and Shah, M. A. and Shoaib, M. and Waqas, M. and Avati, V. and Grzanka, L. and Malawski, M. and Bialkowska, H. and Bluj, M. and Boimska, B. and Frueboes, T. and Górski, M. and Kazana, M. and Szleper, M. and Traczyk, P. and Zalewski, P. and Bunkowski, K. and Doroba, K. and Kalinowski, A. and Konecki, M. and Krolikowski, J. and Walczak, M. and Araujo, M. and Bargassa, P. and Bastos, D. and Boletti, A. and Faccioli, P. and Gallinaro, M. and Hollar, J. and Leonardo, N. and Niknejad, T. and Seixas, J. and Shchelina, K. and Toldaiev, O. and Varela, J. and Bunin, P. and Ershov, Y. and Gavrilenko, M. and Golunov, A. and Golutvin, I. and Gorbounov, N. and Gorbunov, I. and Kamenev, A. and Karjavine, V. and Lanev, A. and Malakhov, A. and Matveev, V. and Palichik, V. and Perelygin, V. and Savina, M. and Shmatov, S. and Shulha, S. and Smirnov, V. and Teryaev, O. and Trofimov, V. and Yuldashev, B. S. and Zarubin, A. and Gavrilov, G. and Golovtcov, V. and Ivanov, Y. and Kim, V. and Kuznetsova, E. and Murzin, V. and Oreshkin, V. and Smirnov, I. and Sosnov, D. and Sulimov, V. and Uvarov, L. and Volkov, S. and Vorobyev, A. and Andreev, Yu. and Dermenev, A. and Gninenko, S. and Golubev, N. and Karneyeu, A. and Kirsanov, M. and Krasnikov, N. and Pashenkov, A. and Pivovarov, G. and Tlisov, D. and Toropin, A. and Epshteyn, V. and Gavrilov, V. and Lychkovskaya, N. and Nikitenko, A. and Popov, V. and Safronov, G. and Spiridonov, A. and Stepennov, A. and Toms, M. and Vlasov, E. and Zhokin, A. and Aushev, T. and Bychkova, O. and Chadeeva, M. and Philippov, D. and Popova, E. and Rusinov, V. and Andreev, V. and Azarkin, M. and Dremin, I. and Kirakosyan, M. and Terkulov, A. and Belyaev, A. and Boos, E. and Bunichev, V. and Dubinin, M. and Dudko, L. and Klyukhin, V. and Kodolova, O. and Lokhtin, I. and Obraztsov, S. and Perfilov, M. and Petrushanko, S. and Savrin, V. and Snigirev, A. and Blinov, V. and Dimova, T. and Kardapoltsev, L. and Ovtin, I. and Skovpen, Y. and Azhgirey, I. and Bayshev, I. and Kachanov, V. and Kalinin, A. and Konstantinov, D. and Petrov, V. and Ryutin, R. and Sobol, A. and Troshin, S. and Tyurin, N. and Uzunian, A. and Volkov, A. and Babaev, A. and Iuzhakov, A. and Okhotnikov, V. and Sukhikh, L. and Borchsh, V. and Ivanchenko, V. and Tcherniaev, E. and Adzic, P. and Dordevic, M. and Milenovic, P. and Milosevic, J. and Aguilar-Benitez, M. and Alcaraz Maestre, J. and Álvarez Fernández, A. and Bachiller, I. and Barrio Luna, M. and Bedoya, Cristina F. and Carrillo Montoya, C. A. and Cepeda, M. and Cerrada, M. and Colino, N. and De La Cruz, B. and Delgado Peris, A. and Fernández Ramos, J. P. and Flix, J. and Fouz, M. C. and Gonzalez Lopez, O. and Goy Lopez, S. and Hernandez, J. M. and Josa, M. I. and León Holgado, J. and Moran, D. and Navarro Tobar, Á. and Pérez-Calero Yzquierdo, A. and Puerta Pelayo, J. and Redondo, I. and Romero, L. and Sánchez Navas, S. and Soares, M. S. and Urda Gómez, L. and Willmott, C. and Albajar, C. and de Trocóniz, J. F. and Reyes-Almanza, R. and Alvarez Gonzalez, B. and Cuevas, J. and Erice, C. and Fernandez Menendez, J. and Folgueras, S. and Gonzalez Caballero, I. and Palencia Cortezon, E. and Ramón Álvarez, C. and Ripoll Sau, J. and Rodríguez Bouza, V. and Sanchez Cruz, S. and Trapote, A. and Brochero Cifuentes, J. A. and Cabrillo, I. J. and Calderon, A. and Chazin Quero, B. and Duarte Campderros, J. and Fernandez, M. and Fernández Manteca, P. J. and García Alonso, A. and Gomez, G. and Martinez Rivero, C. and Martinez Ruiz del Arbol, P. and Matorras, F. and Piedra Gomez, J. and Prieels, C. and Ricci-Tam, F. and Rodrigo, T. and Ruiz-Jimeno, A. and Scodellaro, L. and Vila, I. and Vizan Garcia, J. M. and Jayananda, M K and Kailasapathy, B. and Sonnadara, D. U. J. and Wickramarathna, D D C and Dharmaratna, W. G. D. and Liyanage, K. and Perera, N. and Wickramage, N. and Aarrestad, T. K. and Abbaneo, D. and Auffray, E. and Auzinger, G. and Baechler, J. and Baillon, P. and Ball, A. H. and Barney, D. and Bendavid, J. and Beni, N. and Bianco, M. and Bocci, A. and Bossini, E. and Brondolin, E. and Camporesi, T. and Capeans Garrido, M. and Cerminara, G. and Cristella, L. and d’Enterria, D. and Dabrowski, A. and Daci, N. and David, A. and De Roeck, A. and Deile, M. and Di Maria, R. and Dobson, M. and Dünser, M. and Dupont, N. and Elliott-Peisert, A. and Emriskova, N. and Fallavollita, F. and Fasanella, D. and Fiorendi, S. and Florent, A. and Franzoni, G. and Fulcher, J. and Funk, W. and Giani, S. and Gigi, D. and Gill, K. and Glege, F. and Gouskos, L. and Guilbaud, M. and Haranko, M. and Hegeman, J. and Iiyama, Y. and Innocente, V. and James, T. and Janot, P. and Kaspar, J. and Kieseler, J. and Komm, M. and Kratochwil, N. and Lange, C. and Laurila, S. and Lecoq, P. and Long, K. and Lourenço, C. and Malgeri, L. and Mallios, S. and Mannelli, M. and Meijers, F. and Mersi, S. and Meschi, E. and Moortgat, F. and Mulders, M. and Orfanelli, S. and Orsini, L. and Pantaleo, F. and Pape, L. and Perez, E. and Peruzzi, M. and Petrilli, A. and Petrucciani, G. and Pfeiffer, A. and Pierini, M. and Quast, T. and Rabady, D. and Racz, A. and Rieger, M. and Rovere, M. and Sakulin, H. and Salfeld-Nebgen, J. and Scarfi, S. and Schäfer, C. and Schwick, C. and Selvaggi, M. and Sharma, A. and Silva, P. and Snoeys, W. and Sphicas, P. and Summers, S. and Tavolaro, V. R. and Treille, D. and Tsirou, A. and Van Onsem, G. P. and Vartak, A. and Verzetti, M. and Wozniak, K. A. and Zeuner, W. D. and Caminada, L. and Erdmann, W. and Horisberger, R. and Ingram, Q. and Kaestli, H. C. and Kotlinski, D. and Langenegger, U. and Rohe, T. and Backhaus, M. and Berger, P. and Calandri, A. and Chernyavskaya, N. and De Cosa, A. and Dissertori, G. and Dittmar, M. and Donegà, M. and Dorfer, C. and Gadek, T. and Gómez Espinosa, T. A. and Grab, C. and Hits, D. and Lustermann, W. and Lyon, A.-M. and Manzoni, R. A. and Meinhard, M. T. and Micheli, F. and Nessi-Tedaldi, F. and Niedziela, J. and Pauss, F. and Perovic, V. and Perrin, G. and Pigazzini, S. and Ratti, M. G. and Reichmann, M. and Reissel, C. and Reitenspiess, T. and Ristic, B. and Ruini, D. and Sanz Becerra, D. A. and Schönenberger, M. and Stampf, V. and Steggemann, J. and Wallny, R. and Zhu, D. H. and Amsler, C. and Botta, C. and Brzhechko, D. and Canelli, M. F. and Del Burgo, R. and Heikkilä, J. K. and Huwiler, M. and Jofrehei, A. and Kilminster, B. and Leontsinis, S. and Macchiolo, A. and Meiring, P. and Mikuni, V. M. and Molinatti, U. and Neutelings, I. and Rauco, G. and Reimers, A. and Robmann, P. and Schweiger, K. and Takahashi, Y. and Adloff, C. and Kuo, C. M. and Lin, W. and Roy, A. and Sarkar, T. and Yu, S. S. and Ceard, L. and Chang, P. and Chao, Y. and Chen, K. F. and Chen, P. H. and Hou, W.-S. and Li, Y. y. and Lu, R.-S. and Paganis, E. and Psallidas, A. and Steen, A. and Yazgan, E. and Asavapibhop, B. and Asawatangtrakuldee, C. and Srimanobhas, N. and Boran, F. and Damarseckin, S. and Demiroglu, Z. S. and Dolek, F. and Dozen, C. and Dumanoglu, I. and Eskut, E. and Gokbulut, G. and Guler, Y. and Gurpinar Guler, E. and Hos, I. and Isik, C. and Kangal, E. E. and Kara, O. and Kayis Topaksu, A. and Kiminsu, U. and Onengut, G. and Ozdemir, K. and Polatoz, A. and Simsek, A. E. and Tali, B. and Tok, U. G. and Turkcapar, S. and Zorbakir, I. S. and Zorbilmez, C. and Isildak, B. and Karapinar, G. and Ocalan, K. and Yalvac, M. and Akgun, B. and Atakisi, I. O. and Gülmez, E. and Kaya, M. and Kaya, O. and Özçelik, Ö. and Tekten, S. and Yetkin, E. A. and Cakir, A. and Cankocak, K. and Komurcu, Y. and Sen, S. and Aydogmus Sen, F. and Cerci, S. and Kaynak, B. and Ozkorucuklu, S. and Sunar Cerci, D. and Grynyov, B. and Levchuk, L. and Bhal, E. and Bologna, S. and Brooke, J. J. and Clement, E. and Cussans, D. and Flacher, H. and Goldstein, J. and Heath, G. P. and Heath, H. F. and Kreczko, L. and Krikler, B. and Paramesvaran, S. and Sakuma, T. and Seif El Nasr-Storey, S. and Smith, V. J. and Stylianou, N. and Taylor, J. and Titterton, A. and Bell, K. W. and Belyaev, A. and Brew, C. and Brown, R. M. and Cockerill, D. J. A. and Ellis, K. V. and Harder, K. and Harper, S. and Linacre, J. and Manolopoulos, K. and Newbold, D. M. and Olaiya, E. and Petyt, D. and Reis, T. and Schuh, T. and Shepherd-Themistocleous, C. H. and Thea, A. and Tomalin, I. R. and Williams, T. and Bainbridge, R. and Bloch, P. and Bonomally, S. and Borg, J. and Breeze, S. and Buchmuller, O. and Bundock, A. and Cepaitis, V. and Chahal, G. S. and Colling, D. and Dauncey, P. and Davies, G. and Della Negra, M. and Fedi, G. and Hall, G. and Iles, G. and Langford, J. and Lyons, L. and Magnan, A.-M. and Malik, S. and Martelli, A. and Milosevic, V. and Nash, J. and Palladino, V. and Pesaresi, M. and Raymond, D. M. and Richards, A. and Rose, A. and Scott, E. and Seez, C. and Shtipliyski, A. and Stoye, M. and Tapper, A. and Uchida, K. and Virdee, T. and Wardle, N. and Webb, S. N. and Winterbottom, D. and Zecchinelli, A. G. and Cole, J. E. and Hobson, P. R. and Khan, A. and Kyberd, P. and Mackay, C. K. and Reid, I. D. and Teodorescu, L. and Zahid, S. and Abdullin, S. and Brinkerhoff, A. and Call, K. and Caraway, B. and Dittmann, J. and Hatakeyama, K. and Kanuganti, A. R. and Madrid, C. and McMaster, B. and Pastika, N. and Sawant, S. and Smith, C. and Wilson, J. and Bartek, R. and Dominguez, A. and Uniyal, R. and Vargas Hernandez, A. M. and Buccilli, A. and Charaf, O. and Cooper, S. I. and Gleyzer, S. V. and Henderson, C. and Perez, C. U. and Rumerio, P. and West, C. and Akpinar, A. and Albert, A. and Arcaro, D. and Cosby, C. and Demiragli, Z. and Gastler, D. and Rohlf, J. and Salyer, K. and Sperka, D. and Spitzbart, D. and Suarez, I. and Yuan, S. and Zou, D. and Benelli, G. and Burkle, B. and Coubez, X. and Cutts, D. and Duh, Y. t. and Hadley, M. and Heintz, U. and Hogan, J. M. and Kwok, K. H. M. and Laird, E. and Landsberg, G. and Lau, K. T. and Lee, J. and Narain, M. and Sagir, S. and Syarif, R. and Usai, E. and Wong, W. Y. and Yu, D. and Zhang, W. and Band, R. and Brainerd, C. and Breedon, R. and Calderon De La Barca Sanchez, M. and Chertok, M. and Conway, J. and Conway, R. and Cox, P. T. and Erbacher, R. and Flores, C. and Funk, G. and Jensen, F. and Ko, W. and Kukral, O. and Lander, R. and Mulhearn, M. and Pellett, D. and Pilot, J. and Shi, M. and Taylor, D. and Tos, K. and Tripathi, M. and Yao, Y. and Zhang, F. and Bachtis, M. and Cousins, R. and Dasgupta, A. and Hamilton, D. and Hauser, J. and Ignatenko, M. and Iqbal, M. A. and Lam, T. and Mccoll, N. and Nash, W. A. and Regnard, S. and Saltzberg, D. and Schnaible, C. and Stone, B. and Valuev, V. and Burt, K. and Chen, Y. and Clare, R. and Gary, J. W. and Hanson, G. and Karapostoli, G. and Long, O. R. and Manganelli, N. and Olmedo Negrete, M. and Si, W. and Wimpenny, S. and Zhang, Y. and Branson, J. G. and Chang, P. and Cittolin, S. and Cooperstein, S. and Deelen, N. and Duarte, J. and Gerosa, R. and Gilbert, D. and Krutelyov, V. and Letts, J. and Masciovecchio, M. and May, S. and Padhi, S. and Pieri, M. and Sathia Narayanan, B. V. and Sharma, V. and Tadel, M. and Würthwein, F. and Yagil, A. and Amin, N. and Campagnari, C. and Citron, M. and Dorsett, A. and Dutta, V. and Incandela, J. and Kilpatrick, M. and Marsh, B. and Mei, H. and Ovcharova, A. and Qu, H. and Quinnan, M. and Richman, J. and Sarica, U. and Stuart, D. and Wang, S. and Bornheim, A. and Cerri, O. and Dutta, I. and Lawhorn, J. M. and Lu, N. and Mao, J. and Newman, H. B. and Ngadiuba, J. and Nguyen, T. Q. and Pata, J. and Spiropulu, M. and Vlimant, J. R. and Wang, C. and Xie, S. and Zhang, Z. and Zhu, R. Y. and Alison, J. and Andrews, M. B. and Ferguson, T. and Mudholkar, T. and Paulini, M. and Vorobiev, I. and Cumalat, J. P. and Ford, W. T. and MacDonald, E. and Patel, R. and Perloff, A. and Stenson, K. and Ulmer, K. A. and Wagner, S. R. and Alexander, J. and Cheng, Y. and Chu, J. and Cranshaw, D. J. and Datta, A. and Frankenthal, A. and Mcdermott, K. and Monroy, J. and Patterson, J. R. and Quach, D. and Ryd, A. and Sun, W. and Tan, S. M. and Tao, Z. and Thom, J. and Wittich, P. and Zientek, M. and Albrow, M. and Alyari, M. and Apollinari, G. and Apresyan, A. and Apyan, A. and Banerjee, S. and Bauerdick, L. A. T. and Beretvas, A. and Berry, D. and Berryhill, J. and Bhat, P. C. and Burkett, K. and Butler, J. N. and Canepa, A. and Cerati, G. B. and Cheung, H. W. K. and Chlebana, F. and Cremonesi, M. and Elvira, V. D. and Freeman, J. and Gecse, Z. and Gray, L. and Green, D. and Grünendahl, S. and Gutsche, O. and Harris, R. M. and Hasegawa, S. and Heller, R. and Herwig, T. C. and Hirschauer, J. and Jayatilaka, B. and Jindariani, S. and Johnson, M. and Joshi, U. and Klabbers, P. and Klijnsma, T. and Klima, B. and Kortelainen, M. J. and Lammel, S. and Lincoln, D. and Lipton, R. and Liu, M. and Liu, T. and Lykken, J. and Maeshima, K. and Mason, D. and McBride, P. and Merkel, P. and Mrenna, S. and Nahn, S. and O’Dell, V. and Papadimitriou, V. and Pedro, K. and Pena, C. and Prokofyev, O. and Ravera, F. and Reinsvold Hall, A. and Ristori, L. and Schneider, B. and Sexton-Kennedy, E. and Smith, N. and Soha, A. and Spalding, W. J. and Spiegel, L. and Stoynev, S. and Strait, J. and Taylor, L. and Tkaczyk, S. and Tran, N. V. and Uplegger, L. and Vaandering, E. W. and Weber, H. A. and Woodard, A. and Acosta, D. and Avery, P. and Bourilkov, D. and Cadamuro, L. and Cherepanov, V. and Errico, F. and Field, R. D. and Guerrero, D. and Joshi, B. M. and Kim, M. and Konigsberg, J. and Korytov, A. and Lo, K. H. and Matchev, K. and Menendez, N. and Mitselmakher, G. and Rosenzweig, D. and Shi, K. and Sturdy, J. and Wang, J. and Zuo, X. and Adams, T. and Askew, A. and Diaz, D. and Habibullah, R. and Hagopian, S. and Hagopian, V. and Johnson, K. F. and Khurana, R. and Kolberg, T. and Martinez, G. and Prosper, H. and Schiber, C. and Yohay, R. and Zhang, J. and Baarmand, M. M. and Butalla, S. and Elkafrawy, T. and Hohlmann, M. and Noonan, D. and Rahmani, M. and Saunders, M. and Yumiceva, F. and Adams, M. R. and Apanasevich, L. and Becerril Gonzalez, H. and Cavanaugh, R. and Chen, X. and Dittmer, S. and Evdokimov, O. and Gerber, C. E. and Hangal, D. A. and Hofman, D. J. and Mills, C. and Oh, G. and Roy, T. and Tonjes, M. B. and Varelas, N. and Viinikainen, J. and Wang, X. and Wu, Z. and Ye, Z. and Alhusseini, M. and Dilsiz, K. and Durgut, S. and Gandrajula, R. P. and Haytmyradov, M. and Khristenko, V. and Köseyan, O. K. and Merlo, J.-P. and Mestvirishvili, A. and Moeller, A. and Nachtman, J. and Ogul, H. and Onel, Y. and Ozok, F. and Penzo, A. and Snyder, C. and Tiras, E. and Wetzel, J. and Amram, O. and Blumenfeld, B. and Corcodilos, L. and Eminizer, M. and Gritsan, A. V. and Kyriacou, S. and Maksimovic, P. and Mantilla, C. and Roskes, J. and Swartz, M. and Vámi, T. Á. and Baldenegro Barrera, C. and Baringer, P. and Bean, A. and Bylinkin, A. and Isidori, T. and Khalil, S. and King, J. and Krintiras, G. and Kropivnitskaya, A. and Lindsey, C. and Minafra, N. and Murray, M. and Rogan, C. and Royon, C. and Sanders, S. and Schmitz, E. and Tapia Takaki, J. D. and Wang, Q. and Williams, J. and Wilson, G. and Duric, S. and Ivanov, A. and Kaadze, K. and Kim, D. and Maravin, Y. and Mitchell, T. and Modak, A. and Mohammadi, A. and Rebassoo, F. and Wright, D. and Adams, E. and Baden, A. and Baron, O. and Belloni, A. and Eno, S. C. and Feng, Y. and Hadley, N. J. and Jabeen, S. and Jeng, G. Y. and Kellogg, R. G. and Koeth, T. and Mignerey, A. C. and Nabili, S. and Seidel, M. and Skuja, A. and Tonwar, S. C. and Wang, L. and Wong, K. and Abercrombie, D. and Allen, B. and Bi, R. and Brandt, S. and Busza, W. and Cali, I. A. and Chen, Y. and D’Alfonso, M. and Gomez Ceballos, G. and Goncharov, M. and Harris, P. and Hsu, D. and Hu, M. and Klute, M. and Kovalskyi, D. and Krupa, J. and Lee, Y.-J. and Luckey, P. D. and Maier, B. and Marini, A. C. and Mironov, C. and Narayanan, S. and Niu, X. and Paus, C. and Rankin, D. and Roland, C. and Roland, G. and Shi, Z. and Stephans, G. S. F. and Sumorok, K. and Tatar, K. and Velicanu, D. and Wang, J. and Wang, T. W. and Wang, Z. and Wyslouch, B. and Chatterjee, R. M. and Evans, A. and Hansen, P. and Hiltbrand, J. and Jain, Sh. and Krohn, M. and Kubota, Y. and Lesko, Z. and Mans, J. and Revering, M. and Rusack, R. and Saradhy, R. and Schroeder, N. and Strobbe, N. and Wadud, M. A. and Acosta, J. G. and Oliveros, S. and Bloom, K. and Chauhan, S. and Claes, D. R. and Fangmeier, C. and Finco, L. and Golf, F. and González Fernández, J. R. and Joo, C. and Kravchenko, I. and Siado, J. E. and Snow, G. R. and Tabb, W. and Yan, F. and Agarwal, G. and Bandyopadhyay, H. and Hay, L. and Iashvili, I. and Kharchilava, A. and McLean, C. and Nguyen, D. and Pekkanen, J. and Rappoccio, S. and Alverson, G. and Barberis, E. and Freer, C. and Haddad, Y. and Hortiangtham, A. and Li, J. and Madigan, G. and Marzocchi, B. and Morse, D. M. and Nguyen, V. and Orimoto, T. and Parker, A. and Skinnari, L. and Tishelman-Charny, A. and Wamorkar, T. and Wang, B. and Wisecarver, A. and Wood, D. and Bhattacharya, S. and Bueghly, J. and Chen, Z. and Gilbert, A. and Gunter, T. and Hahn, K. A. and Odell, N. and Schmitt, M. H. and Sung, K. and Velasco, M. and Bucci, R. and Dev, N. and Goldouzian, R. and Hildreth, M. and Hurtado Anampa, K. and Jessop, C. and Lannon, K. and Loukas, N. and Marinelli, N. and Mcalister, I. and Meng, F. and Mohrman, K. and Musienko, Y. and Ruchti, R. and Siddireddy, P. and Wayne, M. and Wightman, A. and Wolf, M. and Zygala, L. and Alimena, J. and Bylsma, B. and Cardwell, B. and Durkin, L. S. and Francis, B. and Hill, C. and Lefeld, A. and Winer, B. L. and Yates, B. R. and Bonham, B. and Das, P. and Dezoort, G. and Elmer, P. and Greenberg, B. and Haubrich, N. and Higginbotham, S. and Kalogeropoulos, A. and Kopp, G. and Kwan, S. and Lange, D. and Lucchini, M. T. and Luo, J. and Marlow, D. and Mei, K. and Ojalvo, I. and Olsen, J. and Palmer, C. and Piroué, P. and Stickland, D. and Tully, C. and Malik, S. and Norberg, S. and Barnes, V. E. and Chawla, R. and Das, S. and Gutay, L. and Jones, M. and Jung, A. W. and Negro, G. and Neumeister, N. and Peng, C. C. and Piperov, S. and Purohit, A. and Schulte, J. F. and Stojanovic, M. and Trevisani, N. and Wang, F. and Wildridge, A. and Xiao, R. and Xie, W. and Dolen, J. and Parashar, N. and Baty, A. and Dildick, S. and Ecklund, K. M. and Freed, S. and Geurts, F. J. M. and Kumar, A. and Li, W. and Padley, B. P. and Redjimi, R. and Roberts, J. and Rorie, J. and Shi, W. and Stahl Leiton, A. G. and Bodek, A. and de Barbaro, P. and Demina, R. and Dulemba, J. L. and Fallon, C. and Ferbel, T. and Galanti, M. and Garcia-Bellido, A. and Hindrichs, O. and Khukhunaishvili, A. and Ranken, E. and Taus, R. and Chiarito, B. and Chou, J. P. and Gandrakota, A. and Gershtein, Y. and Halkiadakis, E. and Hart, A. and Heindl, M. and Hughes, E. and Kaplan, S. and Karacheban, O. and Laflotte, I. and Lath, A. and Montalvo, R. and Nash, K. and Osherson, M. and Salur, S. and Schnetzer, S. and Somalwar, S. and Stone, R. and Thayil, S. A. and Thomas, S. and Wang, H. and Acharya, H. and Delannoy, A. G. and Spanier, S. and Bouhali, O. and Dalchenko, M. and Delgado, A. and Eusebi, R. and Gilmore, J. and Huang, T. and Kamon, T. and Kim, H. and Luo, S. and Malhotra, S. and Mueller, R. and Overton, D. and Perniè, L. and Rathjens, D. and Safonov, A. and Akchurin, N. and Damgov, J. and Hegde, V. and Kunori, S. and Lamichhane, K. and Lee, S. W. and Mengke, T. and Muthumuni, S. and Peltola, T. and Undleeb, S. and Volobouev, I. and Wang, Z. and Whitbeck, A. and Appelt, E. and Greene, S. and Gurrola, A. and Janjam, R. and Johns, W. and Maguire, C. and Melo, A. and Ni, H. and Padeken, K. and Romeo, F. and Sheldon, P. and Tuo, S. and Velkovska, J. and Arenton, M. W. and Cox, B. and Cummings, G. and Hakala, J. and Hirosky, R. and Joyce, M. and Ledovskoy, A. and Li, A. and Neu, C. and Tannenwald, B. and Wolfe, E. and Karchin, P. E. and Poudyal, N. and Thapa, P. and Black, K. and Bose, T. and Buchanan, J. and Caillol, C. and Dasu, S. and De Bruyn, I. and Everaerts, P. and Galloni, C. and He, H. and Herndon, M. and Hervé, A. and Hussain, U. and Lanaro, A. and Loeliger, A. and Loveless, R. and Madhusudanan Sreekala, J. and Mallampalli, A. and Pinna, D. and Savin, A. and Shang, V. and Sharma, V. and Smith, W. H. and Teague, D. and Trembath-reichert, S. and Vetens, W.},
   year={2021},
   month=apr }

@article{ATLAS:2014zve,
    author = "Aad, Georges and others",
    collaboration = "ATLAS",
    title = "{Search for direct production of charginos, neutralinos and sleptons in final states with two leptons and missing transverse momentum in $pp$ collisions at $\sqrt{s} =$ 8 TeV with the ATLAS detector}",
    eprint = "1403.5294",
    archivePrefix = "arXiv",
    primaryClass = "hep-ex",
    reportNumber = "CERN-PH-EP-2014-037",
    doi = "10.1007/JHEP05(2014)071",
    journal = "JHEP",
    volume = "05",
    pages = "071",
    year = "2014"
}

@article{CMS:2018eqb,
    author = "Sirunyan, Albert M. and others",
    collaboration = "CMS",
    title = "{Search for supersymmetric partners of electrons and muons in proton-proton collisions at $\sqrt{s}=$ 13 TeV}",
    eprint = "1806.05264",
    archivePrefix = "arXiv",
    primaryClass = "hep-ex",
    reportNumber = "CMS-SUS-17-009, CERN-EP-2018-132",
    doi = "10.1016/j.physletb.2019.01.005",
    journal = "Phys. Lett. B",
    volume = "790",
    pages = "140--166",
    year = "2019"
}

@article{DELPHI_2003,
   title={Searches for supersymmetric particles in e + e- collisions up to 208 GeV and interpretation of the results within the MSSM},
    author = "Abdallah, J. and others",
   volume={31},
   ISSN={1434-6052},
   url={http://dx.doi.org/10.1140/epjc/s2003-01355-5},
   DOI={10.1140/epjc/s2003-01355-5},
   number={4},
   journal={The European Physical Journal C},
   publisher={Springer Science and Business Media LLC},
   year={2003},
   month=dec, pages={421–479} }

@article{Darm__2023,
   title={UFO 2.0: the ‘Universal Feynman Output’ format},
   volume={83},
   ISSN={1434-6052},
   url={http://dx.doi.org/10.1140/epjc/s10052-023-11780-9},
   DOI={10.1140/epjc/s10052-023-11780-9},
   number={7},
   journal={The European Physical Journal C},
   publisher={Springer Science and Business Media LLC},
   author={Darmé, Luc and Degrande, Céline and Duhr, Claude and Fuks, Benjamin and Goodsell, Mark and Heinrich, Gudrun and Hirschi, Valentin and Höche, Stefan and Höfer, Marius and Isaacson, Joshua and Mattelaer, Olivier and Ohl, Thorsten and Pagani, Davide and Reuter, Jürgen and Richardson, Peter and Schumann, Steffen and Shao, Hua-Sheng and Siegert, Frank and Zaro, Marco},
   year={2023},
   month=jul }

@article{Mescia:2024rki,
    author = "Mescia, Federico and Okawa, Shohei and Wu, Keyun",
    title = "{Multi-component dark matter from Minimal Flavor Violation}",
    eprint = "2408.16812",
    archivePrefix = "arXiv",
    primaryClass = "hep-ph",
    reportNumber = "KEK-TH-2649",
    doi = "10.1007/JHEP11(2024)114",
    journal = "JHEP",
    volume = "11",
    pages = "114",
    year = "2024"
}

@article{Buras:2000dm,
    author = "Buras, A. J. and Gambino, P. and Gorbahn, M. and Jager, S. and Silvestrini, L.",
    title = "{Universal unitarity triangle and physics beyond the standard model}",
    eprint = "hep-ph/0007085",
    archivePrefix = "arXiv",
    reportNumber = "TUM-HEP-379-00, CERN-TH-2000-190",
    doi = "10.1016/S0370-2693(01)00061-2",
    journal = "Phys. Lett. B",
    volume = "500",
    pages = "161--167",
    year = "2001"
}

@article{Chivukula:1987py,
    author = "Chivukula, R. Sekhar and Georgi, Howard",
    title = "{Composite Technicolor Standard Model}",
    reportNumber = "BUHEP-87-2, HUTP-87/A003",
    doi = "10.1016/0370-2693(87)90713-1",
    journal = "Phys. Lett. B",
    volume = "188",
    pages = "99--104",
    year = "1987"
}

@article{Hall:1990ac,
    author = "Hall, L. J. and Randall, Lisa",
    title = "{Weak scale effective supersymmetry}",
    reportNumber = "UCB-PTH-90/13, LBL-28879",
    doi = "10.1103/PhysRevLett.65.2939",
    journal = "Phys. Rev. Lett.",
    volume = "65",
    pages = "2939--2942",
    year = "1990"
}

@article{DAmbrosio:2002vsn,
    author = "D'Ambrosio, G. and Giudice, G. F. and Isidori, G. and Strumia, A.",
    title = "{Minimal flavor violation: An Effective field theory approach}",
    eprint = "hep-ph/0207036",
    archivePrefix = "arXiv",
    reportNumber = "CERN-TH-2002-147, IFUP-TH-2002-17",
    doi = "10.1016/S0550-3213(02)00836-2",
    journal = "Nucl. Phys. B",
    volume = "645",
    pages = "155--187",
    year = "2002"
}

@article{Kile:2011mn,
    author = "Kile, Jennifer and Soni, Amarjit",
    title = "{Flavored Dark Matter in Direct Detection Experiments and at LHC}",
    eprint = "1104.5239",
    archivePrefix = "arXiv",
    primaryClass = "hep-ph",
    reportNumber = "NUHEP-TH-11-01",
    doi = "10.1103/PhysRevD.84.035016",
    journal = "Phys. Rev. D",
    volume = "84",
    pages = "035016",
    year = "2011"
}

@article{Carmona:2021xtq,
    author = "Carmona, Adrian and Lazopoulos, Achilleas and Olgoso, Pablo and Santiago, Jose",
    title = "{Matchmakereft: automated tree-level and one-loop matching}",
    eprint = "2112.10787",
    archivePrefix = "arXiv",
    primaryClass = "hep-ph",
    doi = "10.21468/SciPostPhys.12.6.198",
    journal = "SciPost Phys.",
    volume = "12",
    number = "6",
    pages = "198",
    year = "2022"
}

@article{DasBakshi:2018vni,
    author = "Das Bakshi, Supratim and Chakrabortty, Joydeep and Patra, Sunando Kumar",
    title = "{CoDEx: Wilson coefficient calculator connecting SMEFT to UV theory}",
    eprint = "1808.04403",
    archivePrefix = "arXiv",
    primaryClass = "hep-ph",
    doi = "10.1140/epjc/s10052-018-6444-2",
    journal = "Eur. Phys. J. C",
    volume = "79",
    number = "1",
    pages = "21",
    year = "2019"
}

@article{Biondini:2025gpg,
    author = "Biondini, Simone and Tiberi, Lorenzo and Panella, Orlando",
    title = "{Connecting t-channel dark matter models to the Standard Model Effective Field Theory}",
    eprint = "2507.00925",
    archivePrefix = "arXiv",
    primaryClass = "hep-ph",
    doi = "10.1007/JHEP10(2025)060",
    journal = "JHEP",
    volume = "10",
    pages = "060",
    year = "2025"
}

@article{ATLAS:2020syg,
    author = "Aad, Georges and others",
    collaboration = "ATLAS",
    title = "{Search for squarks and gluinos in final states with jets and missing transverse momentum using 139 fb$^{-1}$ of $\sqrt{s}$ =13 TeV $pp$ collision data with the ATLAS detector}",
    eprint = "2010.14293",
    archivePrefix = "arXiv",
    primaryClass = "hep-ex",
    reportNumber = "CERN-EP-2020-166",
    doi = "10.1007/JHEP02(2021)143",
    journal = "JHEP",
    volume = "02",
    pages = "143",
    year = "2021"
}

@article{CMS:2019zmd,
    author = "Collaboration, The Cms and others",
    collaboration = "CMS",
    title = "{Search for supersymmetry in proton-proton collisions at 13 TeV in final states with jets and missing transverse momentum}",
    eprint = "1908.04722",
    archivePrefix = "arXiv",
    primaryClass = "hep-ex",
    reportNumber = "CMS-SUS-19-006, CERN-EP-2019-152",
    doi = "10.1007/JHEP10(2019)244",
    journal = "JHEP",
    volume = "10",
    pages = "244",
    year = "2019"
}

@article{CMS:2017abv,
    author = "Sirunyan, Albert M and others",
    collaboration = "CMS",
    title = "{Search for supersymmetry in multijet events with missing transverse momentum in proton-proton collisions at 13 TeV}",
    eprint = "1704.07781",
    archivePrefix = "arXiv",
    primaryClass = "hep-ex",
    reportNumber = "CMS-SUS-16-033, CERN-EP-2017-072",
    doi = "10.1103/PhysRevD.96.032003",
    journal = "Phys. Rev. D",
    volume = "96",
    number = "3",
    pages = "032003",
    year = "2017"
}

@article{CMS:2017okm,
    author = "Sirunyan, Albert M and others",
    collaboration = "CMS",
    title = "{Search for new phenomena with the $M_{\mathrm {T2}}$ variable in the all-hadronic final state produced in proton{\textendash}proton collisions at $\sqrt{s} = 13$ $\,\text {TeV}$}",
    eprint = "1705.04650",
    archivePrefix = "arXiv",
    primaryClass = "hep-ex",
    reportNumber = "CMS-SUS-16-036, CERN-EP-2017-084",
    doi = "10.1140/epjc/s10052-017-5267-x",
    journal = "Eur. Phys. J. C",
    volume = "77",
    number = "10",
    pages = "710",
    year = "2017"
}

@article{HFLAV:2022esi,
    author = "Amhis, Yasmine Sara and others",
    collaboration = "HFLAV",
    title = "{Averages of b-hadron, c-hadron, and {\ensuremath{\tau}}-lepton properties as of 2021}",
    eprint = "2206.07501",
    archivePrefix = "arXiv",
    primaryClass = "hep-ex",
    doi = "10.1103/PhysRevD.107.052008",
    journal = "Phys. Rev. D",
    volume = "107",
    number = "5",
    pages = "052008",
    year = "2023"
}

@article{HFLAV:2014fzu,
    author = "Amhis, Y. and others",
    collaboration = "HFLAV",
    title = "{Averages of $b$-hadron, $c$-hadron, and $\tau$-lepton properties as of summer 2014}",
    eprint = "1412.7515",
    archivePrefix = "arXiv",
    primaryClass = "hep-ex",
    reportNumber = "FERMILAB-PUB-15-004-PPD",
    month = "12",
    year = "2014"
}

@article{ParticleDataGroup:2018ovx,
    author = "Tanabashi, M. and others",
    collaboration = "Particle Data Group",
    title = "{Review of Particle Physics}",
    doi = "10.1103/PhysRevD.98.030001",
    journal = "Phys. Rev. D",
    volume = "98",
    number = "3",
    pages = "030001",
    year = "2018"
}

@article{ParticleDataGroup:2016lqr,
    author = "Patrignani, C. and others",
    collaboration = "Particle Data Group",
    title = "{Review of Particle Physics}",
    doi = "10.1088/1674-1137/40/10/100001",
    journal = "Chin. Phys. C",
    volume = "40",
    number = "10",
    pages = "100001",
    year = "2016"
}

@article{Hayasaka:2010np,
    author = "Hayasaka, K. and others",
    title = "{Search for Lepton Flavor Violating Tau Decays into Three Leptons with 719 Million Produced Tau+Tau- Pairs}",
    eprint = "1001.3221",
    archivePrefix = "arXiv",
    primaryClass = "hep-ex",
    doi = "10.1016/j.physletb.2010.03.037",
    journal = "Phys. Lett. B",
    volume = "687",
    pages = "139--143",
    year = "2010"
}

@article{CMS:2021beq,
    author = "Sirunyan, Albert M and others",
    collaboration = "CMS",
    title = "{Search for top squark production in fully-hadronic final states in proton-proton collisions at $\sqrt{s} =$ 13 TeV}",
    eprint = "2103.01290",
    archivePrefix = "arXiv",
    primaryClass = "hep-ex",
    reportNumber = "CMS-SUS-19-010, CERN-EP-2021-022",
    doi = "10.1103/PhysRevD.104.052001",
    journal = "Phys. Rev. D",
    volume = "104",
    number = "5",
    pages = "052001",
    year = "2021"
}

@article{Brivio:2017vri,
    author = "Brivio, Ilaria and Trott, Michael",
    title = "{The Standard Model as an Effective Field Theory}",
    eprint = "1706.08945",
    archivePrefix = "arXiv",
    primaryClass = "hep-ph",
    doi = "10.1016/j.physrep.2018.11.002",
    journal = "Phys. Rept.",
    volume = "793",
    pages = "1--98",
    year = "2019"
}

@article{Isidori:2023pyp,
    author = "Isidori, Gino and Wilsch, Felix and Wyler, Daniel",
    title = "{The standard model effective field theory at work}",
    eprint = "2303.16922",
    archivePrefix = "arXiv",
    primaryClass = "hep-ph",
    reportNumber = "ZU-TH 14/23",
    doi = "10.1103/RevModPhys.96.015006",
    journal = "Rev. Mod. Phys.",
    volume = "96",
    number = "1",
    pages = "015006",
    year = "2024"
}

@article{Aebischer:2025qhh,
    author = "Aebischer, Jason and Buras, Andrzej J. and Kumar, Jacky",
    title = "{SMEFT ATLAS: The Landscape Beyond the Standard Model}",
    eprint = "2507.05926",
    archivePrefix = "arXiv",
    primaryClass = "hep-ph",
    reportNumber = "AJB-25-1, CERN-TH-2025-129, LA-UR-24-24665",
    month = "7",
    year = "2025"
}

@article{Celis:2017hod,
    author = "Celis, Alejandro and Fuentes-Martin, Javier and Vicente, Avelino and Virto, Javier",
    title = "{DsixTools: The Standard Model Effective Field Theory Toolkit}",
    eprint = "1704.04504",
    archivePrefix = "arXiv",
    primaryClass = "hep-ph",
    reportNumber = "LMU-ASC-24-17, IFIC-17-18",
    doi = "10.1140/epjc/s10052-017-4967-6",
    journal = "Eur. Phys. J. C",
    volume = "77",
    number = "6",
    pages = "405",
    year = "2017"
}

@article{CMS:2017kil,
    author = "Sirunyan, Albert M and others",
    collaboration = "CMS",
    title = "{Search for the pair production of third-generation squarks with two-body decays to a bottom or charm quark and a neutralino in proton{\textendash}proton collisions at $\sqrt{s}$ = 13 TeV}",
    eprint = "1707.07274",
    archivePrefix = "arXiv",
    primaryClass = "hep-ex",
    reportNumber = "CMS-SUS-16-032, CERN-EP-2017-144",
    doi = "10.1016/j.physletb.2018.01.012",
    journal = "Phys. Lett. B",
    volume = "778",
    pages = "263--291",
    year = "2018"
}

@article{ATLAS:2013lcn,
    author = "Aad, Georges and others",
    collaboration = "ATLAS",
    title = "{Search for direct third-generation squark pair production in final states with missing transverse momentum and two $b$-jets in $\sqrt{s} =$ 8 TeV $pp$ collisions with the ATLAS detector}",
    eprint = "1308.2631",
    archivePrefix = "arXiv",
    primaryClass = "hep-ex",
    reportNumber = "CERN-PH-EP-2013-119",
    doi = "10.1007/JHEP10(2013)189",
    journal = "JHEP",
    volume = "10",
    pages = "189",
    year = "2013"
}

@article{ATLAS:2017avc,
    author = "Aaboud, Morad and others",
    collaboration = "ATLAS",
    title = "{Search for supersymmetry in events with $b$-tagged jets and missing transverse momentum in $pp$ collisions at $\sqrt{s}=13$ TeV with the ATLAS detector}",
    eprint = "1708.09266",
    archivePrefix = "arXiv",
    primaryClass = "hep-ex",
    reportNumber = "CERN-EP-2017-154",
    doi = "10.1007/JHEP11(2017)195",
    journal = "JHEP",
    volume = "11",
    pages = "195",
    year = "2017"
}

@article{ATLAS:2017mjy,
    author = "Aaboud, Morad and others",
    collaboration = "ATLAS",
    title = "{Search for squarks and gluinos in final states with jets and missing transverse momentum using 36  fb$^{-1}$ of $\sqrt{s}=13$  TeV pp collision data with the ATLAS detector}",
    eprint = "1712.02332",
    archivePrefix = "arXiv",
    primaryClass = "hep-ex",
    reportNumber = "CERN-EP-2017-136",
    doi = "10.1103/PhysRevD.97.112001",
    journal = "Phys. Rev. D",
    volume = "97",
    number = "11",
    pages = "112001",
    year = "2018"
}

@article{ATLAS:2014jxt,
    author = "Aad, Georges and others",
    collaboration = "ATLAS",
    title = "{Search for squarks and gluinos with the ATLAS detector in final states with jets and missing transverse momentum using $\sqrt{s}=8$ TeV proton--proton collision data}",
    eprint = "1405.7875",
    archivePrefix = "arXiv",
    primaryClass = "hep-ex",
    reportNumber = "CERN-PH-EP-2014-093",
    doi = "10.1007/JHEP09(2014)176",
    journal = "JHEP",
    volume = "09",
    pages = "176",
    year = "2014"
}

@article{CMS-SUS-13-019,
   title={Searches for supersymmetry using the M T2 variable in hadronic events produced in pp collisions at 8 TeV},
   volume={2015},
   ISSN={1029-8479},
   url={http://dx.doi.org/10.1007/JHEP05(2015)078},
   DOI={10.1007/jhep05(2015)078},
   number={5},
   journal={Journal of High Energy Physics},
   publisher={Springer Science and Business Media LLC},
   author={Khachatryan, V. and Sirunyan, A. M. and Tumasyan, A. and Adam, W. and Bergauer, T. and Dragicevic, M. and Erö, J. and Friedl, M. and Frühwirth, R. and Ghete, V. M. and Hartl, C. and Hörmann, N. and Hrubec, J. and Jeitler, M. and Kiesenhofer, W. and Knünz, V. and Krammer, M. and Krätschmer, I. and Liko, D. and Mikulec, I. and Rabady, D. and Rahbaran, B. and Rohringer, H. and Schöfbeck, R. and Strauss, J. and Treberer-Treberspurg, W. and Waltenberger, W. and Wulz, C.-E. and Mossolov, V. and Shumeiko, N. and Suarez Gonzalez, J. and Alderweireldt, S. and Bansal, S. and Cornelis, T. and De Wolf, E. A. and Janssen, X. and Knutsson, A. and Lauwers, J. and Luyckx, S. and Ochesanu, S. and Rougny, R. and Van De Klundert, M. and Van Haevermaet, H. and Van Mechelen, P. and Van Remortel, N. and Van Spilbeeck, A. and Blekman, F. and Blyweert, S. and D’Hondt, J. and Daci, N. and Heracleous, N. and Keaveney, J. and Lowette, S. and Maes, M. and Olbrechts, A. and Python, Q. and Strom, D. and Tavernier, S. and Van Doninck, W. and Van Mulders, P. and Van Onsem, G. P. and Villella, I. and Caillol, C. and Clerbaux, B. and De Lentdecker, G. and Dobur, D. and Favart, L. and Gay, A. P. R. and Grebenyuk, A. and Léonard, A. and Mohammadi, A. and Perniè, L. and Randle-conde, A. and Reis, T. and Seva, T. and Thomas, L. and Vander Velde, C. and Vanlaer, P. and Wang, J. and Zenoni, F. and Adler, V. and Beernaert, K. and Benucci, L. and Cimmino, A. and Costantini, S. and Crucy, S. and Dildick, S. and Fagot, A. and Garcia, G. and Mccartin, J. and Ocampo Rios, A. A. and Ryckbosch, D. and Salva Diblen, S. and Sigamani, M. and Strobbe, N. and Thyssen, F. and Tytgat, M. and Yazgan, E. and Zaganidis, N. and Basegmez, S. and Beluffi, C. and Bruno, G. and Castello, R. and Caudron, A. and Ceard, L. and Da Silveira, G. G. and Delaere, C. and du Pree, T. and Favart, D. and Forthomme, L. and Giammanco, A. and Hollar, J. and Jafari, A. and Jez, P. and Komm, M. and Lemaitre, V. and Nuttens, C. and Perrini, L. and Pin, A. and Piotrzkowski, K. and Popov, A. and Quertenmont, L. and Selvaggi, M. and Vidal Marono, M. and Vizan Garcia, J. M. and Beliy, N. and Caebergs, T. and Daubie, E. and Hammad, G. H. and Aldá Júnior, W. L. and Alves, G. A. and Brito, L. and Correa Martins Junior, M. and Dos Reis Martins, T. and Molina, J. and Mora Herrera, C. and Pol, M. E. and Rebello Teles, P. and Carvalho, W. and Chinellato, J. and Custódio, A. and Da Costa, E. M. and De Jesus Damiao, D. and De Oliveira Martins, C. and Fonseca De Souza, S. and Malbouisson, H. and Matos Figueiredo, D. and Mundim, L. and Nogima, H. and Prado Da Silva, W. L. and Santaolalla, J. and Santoro, A. and Sznajder, A. and Tonelli Manganote, E. J. and Vilela Pereira, A. and Bernardes, C. A. and Dogra, S. and Fernandez Perez Tomei, T. R. and Gregores, E. M. and Mercadante, P. G. and Novaes, S. F. and Padula, Sandra S. and Aleksandrov, A. and Genchev, V. and Hadjiiska, R. and Iaydjiev, P. and Marinov, A. and Piperov, S. and Rodozov, M. and Stoykova, S. and Sultanov, G. and Vutova, M. and Dimitrov, A. and Glushkov, I. and Litov, L. and Pavlov, B. and Petkov, P. and Bian, J. G. and Chen, G. M. and Chen, H. S. and Chen, M. and Cheng, T. and Du, R. and Jiang, C. H. and Plestina, R. and Romeo, F. and Tao, J. and Wang, Z. and Asawatangtrakuldee, C. and Ban, Y. and Li, Q. and Liu, S. and Mao, Y. and Qian, S. J. and Wang, D. and Xu, Z. and Zou, W. and Avila, C. and Cabrera, A. and Chaparro Sierra, L. F. and Florez, C. and Gomez, J. P. and Gomez Moreno, B. and Sanabria, J. C. and Godinovic, N. and Lelas, D. and Polic, D. and Puljak, I. and Antunovic, Z. and Kovac, M. and Brigljevic, V. and Kadija, K. and Luetic, J. and Mekterovic, D. and Sudic, L. and Attikis, A. and Mavromanolakis, G. and Mousa, J. and Nicolaou, C. and Ptochos, F. and Razis, P. A. and Bodlak, M. and Finger, M. and Finger, M. and Assran, Y. and Ellithi Kamel, A. and Mahmoud, M. A. and Radi, A. and Kadastik, M. and Murumaa, M. and Raidal, M. and Tiko, A. and Eerola, P. and Fedi, G. and Voutilainen, M. and Härkönen, J. and Karimäki, V. and Kinnunen, R. and Kortelainen, M. J. and Lampén, T. and Lassila-Perini, K. and Lehti, S. and Lindén, T. and Luukka, P. and Mäenpää, T. and Peltola, T. and Tuominen, E. and Tuominiemi, J. and Tuovinen, E. and Wendland, L. and Talvitie, J. and Tuuva, T. and Besancon, M. and Couderc, F. and Dejardin, M. and Denegri, D. and Fabbro, B. and Faure, J. L. and Favaro, C. and Ferri, F. and Ganjour, S. and Givernaud, A. and Gras, P. and Hamel de Monchenault, G. and Jarry, P. and Locci, E. and Malcles, J. and Rander, J. and Rosowsky, A. and Titov, M. and Baffioni, S. and Beaudette, F. and Busson, P. and Charlot, C. and Dahms, T. and Dalchenko, M. and Dobrzynski, L. and Filipovic, N. and Florent, A. and Granier de Cassagnac, R. and Mastrolorenzo, L. and Miné, P. and Naranjo, I. N. and Nguyen, M. and Ochando, C. and Ortona, G. and Paganini, P. and Regnard, S. and Salerno, R. and Sauvan, J. B. and Sirois, Y. and Veelken, C. and Yilmaz, Y. and Zabi, A. and Agram, J.-L. and Andrea, J. and Aubin, A. and Bloch, D. and Brom, J.-M. and Chabert, E. C. and Collard, C. and Conte, E. and Fontaine, J.-C. and Gelé, D. and Goerlach, U. and Goetzmann, C. and Le Bihan, A.-C. and Skovpen, K. and Van Hove, P. and Gadrat, S. and Beauceron, S. and Beaupere, N. and Bernet, C. and Boudoul, G. and Bouvier, E. and Brochet, S. and Carrillo Montoya, C. A. and Chasserat, J. and Chierici, R. and Contardo, D. and Depasse, P. and El Mamouni, H. and Fan, J. and Fay, J. and Gascon, S. and Gouzevitch, M. and Ille, B. and Kurca, T. and Lethuillier, M. and Mirabito, L. and Perries, S. and Ruiz Alvarez, J. D. and Sabes, D. and Sgandurra, L. and Sordini, V. and Vander Donckt, M. and Verdier, P. and Viret, S. and Xiao, H. and Tsamalaidze, Z. and Autermann, C. and Beranek, S. and Bontenackels, M. and Edelhoff, M. and Feld, L. and Heister, A. and Klein, K. and Ostapchuk, A. and Preuten, M. and Raupach, F. and Sammet, J. and Schael, S. and Schulte, J. F. and Weber, H. and Wittmer, B. and Zhukov, V. and Ata, M. and Brodski, M. and Dietz-Laursonn, E. and Duchardt, D. and Erdmann, M. and Fischer, R. and Güth, A. and Hebbeker, T. and Heidemann, C. and Hoepfner, K. and Klingebiel, D. and Knutzen, S. and Kreuzer, P. and Merschmeyer, M. and Meyer, A. and Millet, P. and Olschewski, M. and Padeken, K. and Papacz, P. and Reithler, H. and Schmitz, S. A. and Sonnenschein, L. and Teyssier, D. and Thüer, S. and Weber, M. and Cherepanov, V. and Erdogan, Y. and Flügge, G. and Geenen, H. and Geisler, M. and Haj Ahmad, W. and Hoehle, F. and Kargoll, B. and Kress, T. and Kuessel, Y. and Künsken, A. and Lingemann, J. and Nowack, A. and Nugent, I. M. and Pooth, O. and Stahl, A. and Aldaya Martin, M. and Asin, I. and Bartosik, N. and Behr, J. and Behrens, U. and Bell, A. J. and Bethani, A. and Borras, K. and Burgmeier, A. and Cakir, A. and Calligaris, L. and Campbell, A. and Choudhury, S. and Costanza, F. and Diez Pardos, C. and Dolinska, G. and Dooling, S. and Dorland, T. and Eckerlin, G. and Eckstein, D. and Eichhorn, T. and Flucke, G. and Garay Garcia, J. and Geiser, A. and Gunnellini, P. and Hauk, J. and Hempel, M. and Jung, H. and Kalogeropoulos, A. and Kasemann, M. and Katsas, P. and Kieseler, J. and Kleinwort, C. and Korol, I. and Krücker, D. and Lange, W. and Leonard, J. and Lipka, K. and Lobanov, A. and Lohmann, W. and Lutz, B. and Mankel, R. and Marfin, I. and Melzer-Pellmann, I.-A. and Meyer, A. B. and Mittag, G. and Mnich, J. and Mussgiller, A. and Naumann-Emme, S. and Nayak, A. and Ntomari, E. and Perrey, H. and Pitzl, D. and Placakyte, R. and Raspereza, A. and Ribeiro Cipriano, P. M. and Roland, B. and Ron, E. and Sahin, M. Ö. and Salfeld-Nebgen, J. and Saxena, P. and Schoerner-Sadenius, T. and Schröder, M. and Seitz, C. and Spannagel, S. and Vargas Trevino, A. D. R. and Walsh, R. and Wissing, C. and Blobel, V. and Centis Vignali, M. and Draeger, A. R. and Erfle, J. and Garutti, E. and Goebel, K. and Görner, M. and Haller, J. and Hoffmann, M. and Höing, R. S. and Junkes, A. and Kirschenmann, H. and Klanner, R. and Kogler, R. and Lange, J. and Lapsien, T. and Lenz, T. and Marchesini, I. and Ott, J. and Peiffer, T. and Perieanu, A. and Pietsch, N. and Poehlsen, J. and Poehlsen, T. and Rathjens, D. and Sander, C. and Schettler, H. and Schleper, P. and Schlieckau, E. and Schmidt, A. and Seidel, M. and Sola, V. and Stadie, H. and Steinbrück, G. and Troendle, D. and Usai, E. and Vanelderen, L. and Vanhoefer, A. and Barth, C. and Baus, C. and Berger, J. and Böser, C. and Butz, E. and Chwalek, T. and De Boer, W. and Descroix, A. and Dierlamm, A. and Feindt, M. and Frensch, F. and Giffels, M. and Gilbert, A. and Hartmann, F. and Hauth, T. and Husemann, U. and Katkov, I. and Kornmayer, A. and Lobelle Pardo, P. and Mozer, M. U. and Müller, T. and Müller, Th. and Nürnberg, A. and Quast, G. and Rabbertz, K. and Röcker, S. and Simonis, H. J. and Stober, F. M. and Ulrich, R. and Wagner-Kuhr, J. and Wayand, S. and Weiler, T. and Wolf, R. and Anagnostou, G. and Daskalakis, G. and Geralis, T. and Giakoumopoulou, V. A. and Kyriakis, A. and Loukas, D. and Markou, A. and Markou, C. and Psallidas, A. and Topsis-Giotis, I. and Agapitos, A. and Kesisoglou, S. and Panagiotou, A. and Saoulidou, N. and Stiliaris, E. and Aslanoglou, X. and Evangelou, I. and Flouris, G. and Foudas, C. and Kokkas, P. and Manthos, N. and Papadopoulos, I. and Paradas, E. and Strologas, J. and Bencze, G. and Hajdu, C. and Hidas, P. and Horvath, D. and Sikler, F. and Veszpremi, V. and Vesztergombi, G. and Zsigmond, A. J. and Beni, N. and Czellar, S. and Karancsi, J. and Molnar, J. and Palinkas, J. and Szillasi, Z. and Makovec, A. and Raics, P. and Trocsanyi, Z. L. and Ujvari, B. and Swain, S. K. and Beri, S. B. and Bhatnagar, V. and Gupta, R. and Bhawandeep, U. and Kalsi, A. K. and Kaur, M. and Kumar, R. and Mittal, M. and Nishu, N. and Singh, J. B. and Kumar, Ashok and Kumar, Arun and Ahuja, S. and Bhardwaj, A. and Choudhary, B. C. and Kumar, A. and Malhotra, S. and Naimuddin, M. and Ranjan, K. and Sharma, V. and Banerjee, S. and Bhattacharya, S. and Chatterjee, K. and Dutta, S. and Gomber, B. and Jain, Sa. and Jain, Sh. and Khurana, R. and Modak, A. and Mukherjee, S. and Roy, D. and Sarkar, S. and Sharan, M. and Abdulsalam, A. and Dutta, D. and Kumar, V. and Mohanty, A. K. and Pant, L. M. and Shukla, P. and Topkar, A. and Aziz, T. and Banerjee, S. and Bhowmik, S. and Chatterjee, R. M. and Dewanjee, R. K. and Dugad, S. and Ganguly, S. and Ghosh, S. and Guchait, M. and Gurtu, A. and Kole, G. and Kumar, S. and Maity, M. and Majumder, G. and Mazumdar, K. and Mohanty, G. B. and Parida, B. and Sudhakar, K. and Wickramage, N. and Bakhshiansohi, H. and Behnamian, H. and Etesami, S. M. and Fahim, A. and Goldouzian, R. and Khakzad, M. and Mohammadi Najafabadi, M. and Naseri, M. and Mehdiabadi, S. Paktinat and Rezaei Hosseinabadi, F. and Safarzadeh, B. and Zeinali, M. and Felcini, M. and Grunewald, M. and Abbrescia, M. and Calabria, C. and Chhibra, S. S. and Colaleo, A. and Creanza, D. and De Filippis, N. and De Palma, M. and Fiore, L. and Iaselli, G. and Maggi, G. and Maggi, M. and My, S. and Nuzzo, S. and Pompili, A. and Pugliese, G. and Radogna, R. and Selvaggi, G. and Sharma, A. and Silvestris, L. and Venditti, R. and Verwilligen, P. and Abbiendi, G. and Benvenuti, A. C. and Bonacorsi, D. and Braibant-Giacomelli, S. and Brigliadori, L. and Campanini, R. and Capiluppi, P. and Castro, A. and Cavallo, F. R. and Codispoti, G. and Cuffiani, M. and Dallavalle, G. M. and Fabbri, F. and Fanfani, A. and Fasanella, D. and Giacomelli, P. and Grandi, C. and Guiducci, L. and Marcellini, S. and Masetti, G. and Montanari, A. and Navarria, F. L. and Perrotta, A. and Primavera, F. and Rossi, A. M. and Rovelli, T. and Siroli, G. P. and Tosi, N. and Travaglini, R. and Albergo, S. and Cappello, G. and Chiorboli, M. and Costa, S. and Giordano, F. and Potenza, R. and Tricomi, A. and Tuve, C. and Barbagli, G. and Ciulli, V. and Civinini, C. and D’Alessandro, R. and Focardi, E. and Gallo, E. and Gonzi, S. and Gori, V. and Lenzi, P. and Meschini, M. and Paoletti, S. and Sguazzoni, G. and Tropiano, A. and Benussi, L. and Bianco, S. and Fabbri, F. and Piccolo, D. and Ferretti, R. and Ferro, F. and Lo Vetere, M. and Robutti, E. and Tosi, S. and Dinardo, M. E. and Fiorendi, S. and Gennai, S. and Gerosa, R. and Ghezzi, A. and Govoni, P. and Lucchini, M. T. and Malvezzi, S. and Manzoni, R. A. and Martelli, A. and Marzocchi, B. and Menasce, D. and Moroni, L. and Paganoni, M. and Pedrini, D. and Ragazzi, S. and Redaelli, N. and Tabarelli de Fatis, T. and Buontempo, S. and Cavallo, N. and Di Guida, S. and Fabozzi, F. and Iorio, A. O. M. and Lista, L. and Meola, S. and Merola, M. and Paolucci, P. and Bacchetta, N. and Bellato, M. and Biasotto, M. and Bisello, D. and Branca, A. and Carlin, R. and Checchia, P. and Dall’Osso, M. and Galanti, M. and Gasparini, F. and Gasparini, U. and Gonella, F. and Gozzelino, A. and Margoni, M. and Meneguzzo, A. T. and Pazzini, J. and Pozzobon, N. and Ronchese, P. and Simonetto, F. and Torassa, E. and Tosi, M. and Vanini, S. and Ventura, S. and Zotto, P. and Zucchetta, A. and Zumerle, G. and Gabusi, M. and Ratti, S. P. and Re, V. and Riccardi, C. and Salvini, P. and Vitulo, P. and Biasini, M. and Bilei, G. M. and Ciangottini, D. and Fanò, L. and Lariccia, P. and Mantovani, G. and Menichelli, M. and Saha, A. and Santocchia, A. and Spiezia, A. and Androsov, K. and Azzurri, P. and Bagliesi, G. and Bernardini, J. and Boccali, T. and Broccolo, G. and Castaldi, R. and Ciocci, M. A. and Dell’Orso, R. and Donato, S. and Fiori, F. and Foà, L. and Giassi, A. and Grippo, M. T. and Ligabue, F. and Lomtadze, T. and Martini, L. and Messineo, A. and Moon, C. S. and Palla, F. and Rizzi, A. and Savoy-Navarro, A. and Serban, A. T. and Spagnolo, P. and Squillacioti, P. and Tenchini, R. and Tonelli, G. and Venturi, A. and Verdini, P. G. and Vernieri, C. and Barone, L. and Cavallari, F. and D’imperio, G. and Del Re, D. and Diemoz, M. and Jorda, C. and Longo, E. and Margaroli, F. and Meridiani, P. and Micheli, F. and Organtini, G. and Paramatti, R. and Rahatlou, S. and Rovelli, C. and Santanastasio, F. and Soffi, L. and Traczyk, P. and Amapane, N. and Arcidiacono, R. and Argiro, S. and Arneodo, M. and Bellan, R. and Biino, C. and Cartiglia, N. and Casasso, S. and Costa, M. and Degano, A. and Demaria, N. and Finco, L. and Mariotti, C. and Maselli, S. and Migliore, E. and Monaco, V. and Musich, M. and Obertino, M. M. and Pacher, L. and Pastrone, N. and Pelliccioni, M. and Pinna Angioni, G. L. and Potenza, A. and Romero, A. and Ruspa, M. and Sacchi, R. and Solano, A. and Staiano, A. and Tamponi, U. and Belforte, S. and Candelise, V. and Casarsa, M. and Cossutti, F. and Della Ricca, G. and Gobbo, B. and La Licata, C. and Marone, M. and Schizzi, A. and Umer, T. and Zanetti, A. and Chang, S. and Kropivnitskaya, A. and Nam, S. K. and Kim, D. H. and Kim, G. N. and Kim, M. S. and Kong, D. J. and Lee, S. and Oh, Y. D. and Park, H. and Sakharov, A. and Son, D. C. and Kim, T. J. and Ryu, M. S. and Kim, J. Y. and Moon, D. H. and Song, S. and Choi, S. and Gyun, D. and Hong, B. and Jo, M. and Kim, H. and Kim, Y. and Lee, B. and Lee, K. S. and Park, S. K. and Roh, Y. and Yoo, H. D. and Choi, M. and Kim, J. H. and Park, I. C. and Ryu, G. and Choi, Y. and Choi, Y. K. and Goh, J. and Kim, D. and Kwon, E. and Lee, J. and Yu, I. and Juodagalvis, A. and Komaragiri, J. R. and Ali, M. A. B. Md and Casimiro Linares, E. and Castilla-Valdez, H. and De La Cruz-Burelo, E. and Heredia-de La Cruz, I. and Hernandez-Almada, A. and Lopez-Fernandez, R. and Sanchez-Hernandez, A. and Carrillo Moreno, S. and Vazquez Valencia, F. and Pedraza, I. and Salazar Ibarguen, H. A. and Morelos Pineda, A. and Krofcheck, D. and Butler, P. H. and Reucroft, S. and Ahmad, A. and Ahmad, M. and Hassan, Q. and Hoorani, H. R. and Khan, W. A. and Khurshid, T. and Shoaib, M. and Bialkowska, H. and Bluj, M. and Boimska, B. and Frueboes, T. and Górski, M. and Kazana, M. and Nawrocki, K. and Romanowska-Rybinska, K. and Szleper, M. and Zalewski, P. and Brona, G. and Bunkowski, K. and Cwiok, M. and Dominik, W. and Doroba, K. and Kalinowski, A. and Konecki, M. and Krolikowski, J. and Misiura, M. and Olszewski, M. and Bargassa, P. and Beirão Da Cruz E Silva, C. and Faccioli, P. and Ferreira Parracho, P. G. and Gallinaro, M. and Lloret Iglesias, L. and Nguyen, F. and Rodrigues Antunes, J. and Seixas, J. and Varela, J. and Vischia, P. and Bunin, P. and Golutvin, I. and Gorbunov, I. and Karjavin, V. and Konoplyanikov, V. and Kozlov, G. and Lanev, A. and Malakhov, A. and Matveev, V. and Moisenz, P. and Palichik, V. and Perelygin, V. and Savina, M. and Shmatov, S. and Shulha, S. and Skatchkov, N. and Smirnov, V. and Zarubin, A. and Golovtsov, V. and Ivanov, Y. and Kim, V. and Kuznetsova, E. and Levchenko, P. and Murzin, V. and Oreshkin, V. and Smirnov, I. and Sulimov, V. and Uvarov, L. and Vavilov, S. and Vorobyev, A. and Vorobyev, An. and Andreev, Yu. and Dermenev, A. and Gninenko, S. and Golubev, N. and Kirsanov, M. and Krasnikov, N. and Pashenkov, A. and Tlisov, D. and Toropin, A. and Epshteyn, V. and Gavrilov, V. and Lychkovskaya, N. and Popov, V. and Pozdnyakov, I. and Safronov, G. and Semenov, S. and Spiridonov, A. and Stolin, V. and Vlasov, E. and Zhokin, A. and Andreev, V. and Azarkin, M. and Dremin, I. and Kirakosyan, M. and Leonidov, A. and Mesyats, G. and Rusakov, S. V. and Vinogradov, A. and Belyaev, A. and Boos, E. and Dubinin, M. and Dudko, L. and Ershov, A. and Gribushin, A. and Klyukhin, V. and Kodolova, O. and Lokhtin, I. and Obraztsov, S. and Petrushanko, S. and Savrin, V. and Snigirev, A. and Azhgirey, I. and Bayshev, I. and Bitioukov, S. and Kachanov, V. and Kalinin, A. and Konstantinov, D. and Krychkine, V. and Petrov, V. and Ryutin, R. and Sobol, A. and Tourtchanovitch, L. and Troshin, S. and Tyurin, N. and Uzunian, A. and Volkov, A. and Adzic, P. and Ekmedzic, M. and Milosevic, J. and Rekovic, V. and Alcaraz Maestre, J. and Battilana, C. and Calvo, E. and Cerrada, M. and Chamizo Llatas, M. and Colino, N. and De La Cruz, B. and Delgado Peris, A. and Domínguez Vázquez, D. and Escalante Del Valle, A. and Fernandez Bedoya, C. and Fernández Ramos, J. P. and Flix, J. and Fouz, M. C. and Garcia-Abia, P. and Gonzalez Lopez, O. and Goy Lopez, S. and Hernandez, J. M. and Josa, M. I. and Navarro De Martino, E. and Pérez-Calero Yzquierdo, A. and Puerta Pelayo, J. and Quintario Olmeda, A. and Redondo, I. and Romero, L. and Soares, M. S. and Albajar, C. and de Trocóniz, J. F. and Missiroli, M. and Moran, D. and Brun, H. and Cuevas, J. and Fernandez Menendez, J. and Folgueras, S. and Gonzalez Caballero, I. and Brochero Cifuentes, J. A. and Cabrillo, I. J. and Calderon, A. and Duarte Campderros, J. and Fernandez, M. and Gomez, G. and Graziano, A. and Lopez Virto, A. and Marco, J. and Marco, R. and Martinez Rivero, C. and Matorras, F. and Munoz Sanchez, F. J. and Piedra Gomez, J. and Rodrigo, T. and Rodríguez-Marrero, A. Y. and Ruiz-Jimeno, A. and Scodellaro, L. and Vila, I. and Vilar Cortabitarte, R. and Abbaneo, D. and Auffray, E. and Auzinger, G. and Bachtis, M. and Baillon, P. and Ball, A. H. and Barney, D. and Benaglia, A. and Bendavid, J. and Benhabib, L. and Benitez, J. F. and Bloch, P. and Bocci, A. and Bonato, A. and Bondu, O. and Botta, C. and Breuker, H. and Camporesi, T. and Cerminara, G. and Colafranceschi, S. and D’Alfonso, M. and d’Enterria, D. and Dabrowski, A. and David, A. and De Guio, F. and De Roeck, A. and De Visscher, S. and Di Marco, E. and Dobson, M. and Dordevic, M. and Dorney, B. and Dupont-Sagorin, N. and Elliott-Peisert, A. and Franzoni, G. and Funk, W. and Gigi, D. and Gill, K. and Giordano, D. and Girone, M. and Glege, F. and Guida, R. and Gundacker, S. and Guthoff, M. and Hammer, J. and Hansen, M. and Harris, P. and Hegeman, J. and Innocente, V. and Janot, P. and Kousouris, K. and Krajczar, K. and Lecoq, P. and Lourenço, C. and Magini, N. and Malgeri, L. and Mannelli, M. and Marrouche, J. and Masetti, L. and Meijers, F. and Mersi, S. and Meschi, E. and Moortgat, F. and Morovic, S. and Mulders, M. and Orsini, L. and Pape, L. and Perez, E. and Petrilli, A. and Petrucciani, G. and Pfeiffer, A. and Pimiä, M. and Piparo, D. and Plagge, M. and Racz, A. and Rolandi, G. and Rovere, M. and Sakulin, H. and Schäfer, C. and Schwick, C. and Sharma, A. and Siegrist, P. and Silva, P. and Simon, M. and Sphicas, P. and Spiga, D. and Steggemann, J. and Stieger, B. and Stoye, M. and Takahashi, Y. and Treille, D. and Tsirou, A. and Veres, G. I. and Wardle, N. and Wöhri, H. K. and Wollny, H. and Zeuner, W. D. and Bertl, W. and Deiters, K. and Erdmann, W. and Horisberger, R. and Ingram, Q. and Kaestli, H. C. and Kotlinski, D. and Langenegger, U. and Renker, D. and Rohe, T. and Bachmair, F. and Bäni, L. and Bianchini, L. and Buchmann, M. A. and Casal, B. and Chanon, N. and Dissertori, G. and Dittmar, M. and Donegà, M. and Dünser, M. and Eller, P. and Grab, C. and Hits, D. and Hoss, J. and Lustermann, W. and Mangano, B. and Marini, A. C. and Marionneau, M. and Martinez Ruiz del Arbol, P. and Masciovecchio, M. and Meister, D. and Mohr, N. and Musella, P. and Nägeli, C. and Nessi-Tedaldi, F. and Pandolfi, F. and Pauss, F. and Perrozzi, L. and Peruzzi, M. and Quittnat, M. and Rebane, L. and Rossini, M. and Starodumov, A. and Takahashi, M. and Theofilatos, K. and Wallny, R. and Weber, H. A. and Amsler, C. and Canelli, M. F. and Chiochia, V. and De Cosa, A. and Hinzmann, A. and Hreus, T. and Kilminster, B. and Lange, C. and Millan Mejias, B. and Ngadiuba, J. and Pinna, D. and Robmann, P. and Ronga, F. J. and Taroni, S. and Verzetti, M. and Yang, Y. and Cardaci, M. and Chen, K. H. and Ferro, C. and Kuo, C. M. and Lin, W. and Lu, Y. J. and Volpe, R. and Yu, S. S. and Chang, P. and Chang, Y. H. and Chao, Y. and Chen, K. F. and Chen, P. H. and Dietz, C. and Grundler, U. and Hou, W.-S. and Liu, Y. F. and Lu, R.-S. and Petrakou, E. and Tzeng, Y. M. and Wilken, R. and Asavapibhop, B. and Singh, G. and Srimanobhas, N. and Suwonjandee, N. and Adiguzel, A. and Bakirci, M. N. and Cerci, S. and Dozen, C. and Dumanoglu, I. and Eskut, E. and Girgis, S. and Gokbulut, G. and Guler, Y. and Gurpinar, E. and Hos, I. and Kangal, E. E. and Kayis Topaksu, A. and Onengut, G. and Ozdemir, K. and Ozturk, S. and Polatoz, A. and Sunar Cerci, D. and Tali, B. and Topakli, H. and Vergili, M. and Zorbilmez, C. and Akin, I. V. and Bilin, B. and Bilmis, S. and Gamsizkan, H. and Isildak, B. and Karapinar, G. and Ocalan, K. and Sekmen, S. and Surat, U. E. and Yalvac, M. and Zeyrek, M. and Albayrak, E. A. and Gülmez, E. and Kaya, M. and Kaya, O. and Yetkin, T. and Cankocak, K. and Vardarlı, F. I. and Levchuk, L. and Sorokin, P. and Brooke, J. J. and Clement, E. and Cussans, D. and Flacher, H. and Goldstein, J. and Grimes, M. and Heath, G. P. and Heath, H. F. and Jacob, J. and Kreczko, L. and Lucas, C. and Meng, Z. and Newbold, D. M. and Paramesvaran, S. and Poll, A. and Sakuma, T. and Seif El Nasr-storey, S. and Senkin, S. and Smith, V. J. and Bell, K. W. and Belyaev, A. and Brew, C. and Brown, R. M. and Cockerill, D. J. A. and Coughlan, J. A. and Harder, K. and Harper, S. and Olaiya, E. and Petyt, D. and Shepherd-Themistocleous, C. H. and Thea, A. and Tomalin, I. R. and Williams, T. and Womersley, W. J. and Worm, S. D. and Baber, M. and Bainbridge, R. and Buchmuller, O. and Burton, D. and Colling, D. and Cripps, N. and Dauncey, P. and Davies, G. and Della Negra, M. and Dunne, P. and Ferguson, W. and Fulcher, J. and Futyan, D. and Hall, G. and Iles, G. and Jarvis, M. and Karapostoli, G. and Kenzie, M. and Lane, R. and Lucas, R. and Lyons, L. and Magnan, A.-M. and Malik, S. and Mathias, B. and Nash, J. and Nikitenko, A. and Pela, J. and Pesaresi, M. and Petridis, K. and Raymond, D. M. and Rogerson, S. and Rose, A. and Seez, C. and Sharp, P. and Tapper, A. and Vazquez Acosta, M. and Virdee, T. and Zenz, S. C. and Cole, J. E. and Hobson, P. R. and Khan, A. and Kyberd, P. and Leggat, D. and Leslie, D. and Reid, I. D. and Symonds, P. and Teodorescu, L. and Turner, M. and Dittmann, J. and Hatakeyama, K. and Kasmi, A. and Liu, H. and Scarborough, T. and Wu, Z. and Charaf, O. and Cooper, S. I. and Henderson, C. and Rumerio, P. and Avetisyan, A. and Bose, T. and Fantasia, C. and Lawson, P. and Richardson, C. and Rohlf, J. and John, J. St. and Sulak, L. and Alimena, J. and Berry, E. and Bhattacharya, S. and Christopher, G. and Cutts, D. and Demiragli, Z. and Dhingra, N. and Ferapontov, A. and Garabedian, A. and Heintz, U. and Kukartsev, G. and Laird, E. and Landsberg, G. and Luk, M. and Narain, M. and Segala, M. and Sinthuprasith, T. and Speer, T. and Swanson, J. and Breedon, R. and Breto, G. and Calderon De La Barca Sanchez, M. and Chauhan, S. and Chertok, M. and Conway, J. and Conway, R. and Cox, P. T. and Erbacher, R. and Gardner, M. and Ko, W. and Lander, R. and Mulhearn, M. and Pellett, D. and Pilot, J. and Ricci-Tam, F. and Shalhout, S. and Smith, J. and Squires, M. and Stolp, D. and Tripathi, M. and Wilbur, S. and Yohay, R. and Cousins, R. and Everaerts, P. and Farrell, C. and Hauser, J. and Ignatenko, M. and Rakness, G. and Takasugi, E. and Valuev, V. and Weber, M. and Burt, K. and Clare, R. and Ellison, J. and Gary, J. W. and Hanson, G. and Heilman, J. and Ivova Rikova, M. and Jandir, P. and Kennedy, E. and Lacroix, F. and Long, O. R. and Luthra, A. and Malberti, M. and Olmedo Negrete, M. and Shrinivas, A. and Sumowidagdo, S. and Wimpenny, S. and Branson, J. G. and Cerati, G. B. and Cittolin, S. and D’Agnolo, R. T. and Holzner, A. and Kelley, R. and Klein, D. and Letts, J. and Macneill, I. and Olivito, D. and Padhi, S. and Palmer, C. and Pieri, M. and Sani, M. and Sharma, V. and Simon, S. and Tadel, M. and Tu, Y. and Vartak, A. and Welke, C. and Würthwein, F. and Yagil, A. and Barge, D. and Bradmiller-Feld, J. and Campagnari, C. and Danielson, T. and Dishaw, A. and Dutta, V. and Flowers, K. and Franco Sevilla, M. and Geffert, P. and George, C. and Golf, F. and Gouskos, L. and Incandela, J. and Justus, C. and Mccoll, N. and Richman, J. and Stuart, D. and To, W. and West, C. and Yoo, J. and Apresyan, A. and Bornheim, A. and Bunn, J. and Chen, Y. and Duarte, J. and Mott, A. and Newman, H. B. and Pena, C. and Pierini, M. and Spiropulu, M. and Vlimant, J. R. and Wilkinson, R. and Xie, S. and Zhu, R. Y. and Azzolini, V. and Calamba, A. and Carlson, B. and Ferguson, T. and Iiyama, Y. and Paulini, M. and Russ, J. and Vogel, H. and Vorobiev, I. and Cumalat, J. P. and Ford, W. T. and Gaz, A. and Krohn, M. and Luiggi Lopez, E. and Nauenberg, U. and Smith, J. G. and Stenson, K. and Wagner, S. R. and Alexander, J. and Chatterjee, A. and Chaves, J. and Chu, J. and Dittmer, S. and Eggert, N. and Mirman, N. and Nicolas Kaufman, G. and Patterson, J. R. and Ryd, A. and Salvati, E. and Skinnari, L. and Sun, W. and Teo, W. D. and Thom, J. and Thompson, J. and Tucker, J. and Weng, Y. and Winstrom, L. and Wittich, P. and Winn, D. and Abdullin, S. and Albrow, M. and Anderson, J. and Apollinari, G. and Bauerdick, L. A. T. and Beretvas, A. and Berryhill, J. and Bhat, P. C. and Bolla, G. and Burkett, K. and Butler, J. N. and Cheung, H. W. K. and Chlebana, F. and Cihangir, S. and Elvira, V. D. and Fisk, I. and Freeman, J. and Gao, Y. and Gottschalk, E. and Gray, L. and Green, D. and Grünendahl, S. and Gutsche, O. and Hanlon, J. and Hare, D. and Harris, R. M. and Hirschauer, J. and Hooberman, B. and Jindariani, S. and Johnson, M. and Joshi, U. and Klima, B. and Kreis, B. and Kwan, S. and Linacre, J. and Lincoln, D. and Lipton, R. and Liu, T. and Lykken, J. and Maeshima, K. and Marraffino, J. M. and Martinez Outschoorn, V. I. and Maruyama, S. and Mason, D. and McBride, P. and Merkel, P. and Mishra, K. and Mrenna, S. and Nahn, S. and Newman-Holmes, C. and O’Dell, V. and Prokofyev, O. and Sexton-Kennedy, E. and Sharma, S. and Soha, A. and Spalding, W. J. and Spiegel, L. and Taylor, L. and Tkaczyk, S. and Tran, N. V. and Uplegger, L. and Vaandering, E. W. and Vidal, R. and Whitbeck, A. and Whitmore, J. and Yang, F. and Acosta, D. and Avery, P. and Bortignon, P. and Bourilkov, D. and Carver, M. and Curry, D. and Das, S. and De Gruttola, M. and Di Giovanni, G. P. and Field, R. D. and Fisher, M. and Furic, I. K. and Hugon, J. and Konigsberg, J. and Korytov, A. and Kypreos, T. and Low, J. F. and Matchev, K. and Mei, H. and Milenovic, P. and Mitselmakher, G. and Muniz, L. and Rinkevicius, A. and Shchutska, L. and Snowball, M. and Sperka, D. and Yelton, J. and Zakaria, M. and Hewamanage, S. and Linn, S. and Markowitz, P. and Martinez, G. and Rodriguez, J. L. and Adams, T. and Askew, A. and Bochenek, J. and Diamond, B. and Haas, J. and Hagopian, S. and Hagopian, V. and Johnson, K. F. and Prosper, H. and Veeraraghavan, V. and Weinberg, M. and Baarmand, M. M. and Hohlmann, M. and Kalakhety, H. and Yumiceva, F. and Adams, M. R. and Apanasevich, L. and Berry, D. and Betts, R. R. and Bucinskaite, I. and Cavanaugh, R. and Evdokimov, O. and Gauthier, L. and Gerber, C. E. and Hofman, D. J. and Kurt, P. and O’Brien, C. and Sandoval Gonzalez, I. D. and Silkworth, C. and Turner, P. and Varelas, N. and Bilki, B. and Clarida, W. and Dilsiz, K. and Haytmyradov, M. and Merlo, J.-P. and Mermerkaya, H. and Mestvirishvili, A. and Moeller, A. and Nachtman, J. and Ogul, H. and Onel, Y. and Ozok, F. and Penzo, A. and Rahmat, R. and Sen, S. and Tan, P. and Tiras, E. and Wetzel, J. and Yi, K. and Barnett, B. A. and Blumenfeld, B. and Bolognesi, S. and Fehling, D. and Gritsan, A. V. and Maksimovic, P. and Martin, C. and Swartz, M. and Baringer, P. and Bean, A. and Benelli, G. and Bruner, C. and Gray, J. and Kenny, R. P. and Majumder, D. and Malek, M. and Murray, M. and Noonan, D. and Sanders, S. and Sekaric, J. and Stringer, R. and Wang, Q. and Wood, J. S. and Chakaberia, I. and Ivanov, A. and Kaadze, K. and Khalil, S. and Makouski, M. and Maravin, Y. and Saini, L. K. and Skhirtladze, N. and Svintradze, I. and Gronberg, J. and Lange, D. and Rebassoo, F. and Wright, D. and Baden, A. and Belloni, A. and Calvert, B. and Eno, S. C. and Gomez, J. A. and Hadley, N. J. and Kellogg, R. G. and Kolberg, T. and Lu, Y. and Mignerey, A. C. and Pedro, K. and Skuja, A. and Tonjes, M. B. and Tonwar, S. C. and Apyan, A. and Barbieri, R. and Busza, W. and Cali, I. A. and Chan, M. and Di Matteo, L. and Gomez Ceballos, G. and Goncharov, M. and Gulhan, D. and Klute, M. and Lai, Y. S. and Lee, Y.-J. and Levin, A. and Luckey, P. D. and Paus, C. and Ralph, D. and Roland, C. and Roland, G. and Stephans, G. S. F. and Sumorok, K. and Velicanu, D. and Veverka, J. and Wyslouch, B. and Yang, M. and Zanetti, M. and Zhukova, V. and Dahmes, B. and Gude, A. and Kao, S. C. and Klapoetke, K. and Kubota, Y. and Mans, J. and Nourbakhsh, S. and Pastika, N. and Rusack, R. and Singovsky, A. and Tambe, N. and Turkewitz, J. and Acosta, J. G. and Oliveros, S. and Avdeeva, E. and Bloom, K. and Bose, S. and Claes, D. R. and Dominguez, A. and Gonzalez Suarez, R. and Keller, J. and Knowlton, D. and Kravchenko, I. and Lazo-Flores, J. and Meier, F. and Ratnikov, F. and Snow, G. R. and Zvada, M. and Dolen, J. and Godshalk, A. and Iashvili, I. and Kharchilava, A. and Kumar, A. and Rappoccio, S. and Alverson, G. and Barberis, E. and Baumgartel, D. and Chasco, M. and Massironi, A. and Morse, D. M. and Nash, D. and Orimoto, T. and Trocino, D. and Wang, R.-J. and Wood, D. and Zhang, J. and Hahn, K. A. and Kubik, A. and Mucia, N. and Odell, N. and Pollack, B. and Pozdnyakov, A. and Schmitt, M. and Stoynev, S. and Sung, K. and Velasco, M. and Won, S. and Brinkerhoff, A. and Chan, K. M. and Drozdetskiy, A. and Hildreth, M. and Jessop, C. and Karmgard, D. J. and Kellams, N. and Lannon, K. and Lynch, S. and Marinelli, N. and Musienko, Y. and Pearson, T. and Planer, M. and Ruchti, R. and Smith, G. and Valls, N. and Wayne, M. and Wolf, M. and Woodard, A. and Antonelli, L. and Brinson, J. and Bylsma, B. and Durkin, L. S. and Flowers, S. and Hart, A. and Hill, C. and Hughes, R. and Kotov, K. and Ling, T. Y. and Luo, W. and Puigh, D. and Rodenburg, M. and Winer, B. L. and Wolfe, H. and Wulsin, H. W. and Driga, O. and Elmer, P. and Hardenbrook, J. and Hebda, P. and Koay, S. A. and Lujan, P. and Marlow, D. and Medvedeva, T. and Mooney, M. and Olsen, J. and Piroué, P. and Quan, X. and Saka, H. and Stickland, D. and Tully, C. and Werner, J. S. and Zuranski, A. and Brownson, E. and Malik, S. and Mendez, H. and Ramirez Vargas, J. E. and Barnes, V. E. and Benedetti, D. and Bortoletto, D. and De Mattia, M. and Gutay, L. and Hu, Z. and Jha, M. K. and Jones, M. and Jung, K. and Kress, M. and Leonardo, N. and Miller, D. H. and Neumeister, N. and Radburn-Smith, B. C. and Shi, X. and Shipsey, I. and Silvers, D. and Svyatkovskiy, A. and Wang, F. and Xie, W. and Xu, L. and Zablocki, J. and Parashar, N. and Stupak, J. and Adair, A. and Akgun, B. and Ecklund, K. M. and Geurts, F. J. M. and Li, W. and Michlin, B. and Padley, B. P. and Redjimi, R. and Roberts, J. and Zabel, J. and Betchart, B. and Bodek, A. and Covarelli, R. and de Barbaro, P. and Demina, R. and Eshaq, Y. and Ferbel, T. and Garcia-Bellido, A. and Goldenzweig, P. and Han, J. and Harel, A. and Hindrichs, O. and Khukhunaishvili, A. and Korjenevski, S. and Petrillo, G. and Vishnevskiy, D. and Ciesielski, R. and Demortier, L. and Goulianos, K. and Mesropian, C. and Arora, S. and Barker, A. and Chou, J. P. and Contreras-Campana, C. and Contreras-Campana, E. and Duggan, D. and Ferencek, D. and Gershtein, Y. and Gray, R. and Halkiadakis, E. and Hidas, D. and Kaplan, S. and Lath, A. and Panwalkar, S. and Park, M. and Patel, R. and Salur, S. and Schnetzer, S. and Sheffield, D. and Somalwar, S. and Stone, R. and Thomas, S. and Thomassen, P. and Walker, M. and Rose, K. and Spanier, S. and York, A. and Bouhali, O. and Castaneda Hernandez, A. and Eusebi, R. and Flanagan, W. and Gilmore, J. and Kamon, T. and Khotilovich, V. and Krutelyov, V. and Montalvo, R. and Osipenkov, I. and Pakhotin, Y. and Perloff, A. and Roe, J. and Rose, A. and Safonov, A. and Suarez, I. and Tatarinov, A. and Ulmer, K. A. and Akchurin, N. and Cowden, C. and Damgov, J. and Dragoiu, C. and Dudero, P. R. and Faulkner, J. and Kovitanggoon, K. and Kunori, S. and Lee, S. W. and Libeiro, T. and Volobouev, I. and Appelt, E. and Delannoy, A. G. and Greene, S. and Gurrola, A. and Johns, W. and Maguire, C. and Mao, Y. and Melo, A. and Sharma, M. and Sheldon, P. and Snook, B. and Tuo, S. and Velkovska, J. and Arenton, M. W. and Boutle, S. and Cox, B. and Francis, B. and Goodell, J. and Hirosky, R. and Ledovskoy, A. and Li, H. and Lin, C. and Neu, C. and Wood, J. and Clarke, C. and Harr, R. and Karchin, P. E. and Kottachchi Kankanamge Don, C. and Lamichhane, P. and Sturdy, J. and Belknap, D. A. and Carlsmith, D. and Cepeda, M. and Dasu, S. and Dodd, L. and Duric, S. and Friis, E. and Hall-Wilton, R. and Herndon, M. and Hervé, A. and Klabbers, P. and Lanaro, A. and Lazaridis, C. and Levine, A. and Loveless, R. and Mohapatra, A. and Ojalvo, I. and Perry, T. and Pierro, G. A. and Polese, G. and Ross, I. and Sarangi, T. and Savin, A. and Smith, W. H. and Taylor, D. and Vuosalo, C. and Woods, N.},
   year={2015},
   month=may }

@article{ATLAS:2014hqe,
    author = "Aad, Georges and others",
    collaboration = "ATLAS",
    title = "{Search for pair-produced third-generation squarks decaying via charm quarks or in compressed supersymmetric scenarios in $pp$ collisions at $\sqrt{s}=8~$TeV with the ATLAS detector}",
    eprint = "1407.0608",
    archivePrefix = "arXiv",
    primaryClass = "hep-ex",
    reportNumber = "CERN-PH-EP-2014-141",
    doi = "10.1103/PhysRevD.90.052008",
    journal = "Phys. Rev. D",
    volume = "90",
    number = "5",
    pages = "052008",
    year = "2014"
}

@article{CMS:2016ybj,
    collaboration = "CMS",
    title = "{Search for heavy stable charged particles with $12.9~\mathrm{fb}^{-1}$ of 2016 data}",
    reportNumber = "CMS-PAS-EXO-16-036",
    year = "2016"
}

@article{ATLAS:2019gqq,
    author = "Aaboud, Morad and others",
    collaboration = "ATLAS",
    title = "{Search for heavy charged long-lived particles in the ATLAS detector in 36.1 fb$^{-1}$ of proton-proton collision data at $\sqrt{s} = 13$ TeV}",
    eprint = "1902.01636",
    archivePrefix = "arXiv",
    primaryClass = "hep-ex",
    reportNumber = "CERN-EP-2018-339",
    doi = "10.1103/PhysRevD.99.092007",
    journal = "Phys. Rev. D",
    volume = "99",
    number = "9",
    pages = "092007",
    year = "2019"
}

@article{CMS:2014nia,
    collaboration = "CMS",
    title = "{Search for direct production of bottom squark pairs}",
    reportNumber = "CMS-PAS-SUS-13-018",
    year = "2014",
    url = {https://cds.cern.ch/record/1693164}
}

@article{Bringmann:2007nk,
    author = "Bringmann, Torsten and Bergstrom, Lars and Edsjo, Joakim",
    title = "{New Gamma-Ray Contributions to Supersymmetric Dark Matter Annihilation}",
    eprint = "0710.3169",
    archivePrefix = "arXiv",
    primaryClass = "hep-ph",
    doi = "10.1088/1126-6708/2008/01/049",
    journal = "JHEP",
    volume = "01",
    pages = "049",
    year = "2008"
}

@article{Garny:2011ii,
    author = "Garny, Mathias and Ibarra, Alejandro and Vogl, Stefan",
    title = "{Dark matter annihilations into two light fermions and one gauge boson: General analysis and antiproton constraints}",
    eprint = "1112.5155",
    archivePrefix = "arXiv",
    primaryClass = "hep-ph",
    reportNumber = "DESY-11-257, TUM-HEP-823-11",
    doi = "10.1088/1475-7516/2012/04/033",
    journal = "JCAP",
    volume = "04",
    pages = "033",
    year = "2012"
}

@article{FlavourLatticeAveragingGroupFLAG:2021npn,
    author = "Aoki, Y. and others",
    collaboration = "Flavour Lattice Averaging Group (FLAG)",
    title = "{FLAG Review 2021}",
    eprint = "2111.09849",
    archivePrefix = "arXiv",
    primaryClass = "hep-lat",
    reportNumber = "CERN-TH-2021-191, JLAB-THY-21-3528, FERMILAB-PUB-21-620-SCD-T",
    doi = "10.1140/epjc/s10052-022-10536-1",
    journal = "Eur. Phys. J. C",
    volume = "82",
    number = "10",
    pages = "869",
    year = "2022"
}

@article{FlavourLatticeAveragingGroupFLAG:2024oxs,
    author = "Aoki, Y. and others",
    collaboration = "Flavour Lattice Averaging Group (FLAG)",
    title = "{FLAG Review 2024}",
    eprint = "2411.04268",
    archivePrefix = "arXiv",
    primaryClass = "hep-lat",
    reportNumber = "CERN-TH-2024-192, FERMILAB-PUB-24-0785-T",
    month = "11",
    year = "2024"
}

@article{Inami:1980fz,
    author = "Inami, T. and Lim, C. S.",
    title = "{Effects of Superheavy Quarks and Leptons in Low-Energy Weak Processes k(L) ---{\ensuremath{>}} mu anti-mu, K+ ---{\ensuremath{>}} pi+ Neutrino anti-neutrino and K0 {\ensuremath{<}}---{\ensuremath{>}} anti-K0}",
    reportNumber = "UT-KOMABA-80-8",
    doi = "10.1143/PTP.65.297",
    journal = "Prog. Theor. Phys.",
    volume = "65",
    pages = "297",
    year = "1981",
    note = "[Erratum: Prog.Theor.Phys. 65, 1772 (1981)]"
}

@article{Buchalla:1995vs,
    author = "Buchalla, Gerhard and Buras, Andrzej J. and Lautenbacher, Markus E.",
    title = "{Weak Decays beyond Leading Logarithms}",
    eprint = "hep-ph/9512380",
    archivePrefix = "arXiv",
    reportNumber = "SLAC-PUB-7009, SLAC-PUB-95-7009, MPI-PH-95-104, TUM-T31-100-95, FERMILAB-PUB-95-305-T",
    doi = "10.1103/RevModPhys.68.1125",
    journal = "Rev. Mod. Phys.",
    volume = "68",
    pages = "1125--1144",
    year = "1996"
}

@article{Bai:2014cva,
    author = "Bai, Z. and Christ, N. H. and Izubuchi, T. and Sachrajda, C. T. and Soni, A. and Yu, J.",
    title = "{$K_L-K_S$ Mass Difference from Lattice QCD}",
    eprint = "1406.0916",
    archivePrefix = "arXiv",
    primaryClass = "hep-lat",
    doi = "10.1103/PhysRevLett.113.112003",
    journal = "Phys. Rev. Lett.",
    volume = "113",
    pages = "112003",
    year = "2014"
}

@article{Bai:2018mdv,
    author = "Bai, Ziyuan and Christ, Norman H. and Sachrajda, Christopher T.",
    editor = "Della Morte, M. and Fritzsch, P. and G{\'a}miz S{\'a}nchez, E. and Pena Ruano, C.",
    title = "{The $K_L$ - $K_S$ Mass Difference}",
    doi = "10.1051/epjconf/201817513017",
    journal = "EPJ Web Conf.",
    volume = "175",
    pages = "13017",
    year = "2018"
}

@article{ParticleDataGroup:2024cfk,
    author = "Navas, S. and others",
    collaboration = "Particle Data Group",
    title = "{Review of particle physics}",
    doi = "10.1103/PhysRevD.110.030001",
    journal = "Phys. Rev. D",
    volume = "110",
    number = "3",
    pages = "030001",
    year = "2024"
}

@article{FUHR201832,
author = "Hartmut Führ and Ziemowit Rzeszotnik",
title = "{A note on factoring unitary matrices}",
doi = "https://doi.org/10.1016/j.laa.2018.02.017",
journal = "Linear Algebra and its Applications",
volume = "547",
pages = "32-44",
year = "2018",
issn = "0024-3795"
}

@article{Drees:1993bu,
    author = "Drees, Manuel and Nojiri, Mihoko",
    title = "{Neutralino - nucleon scattering revisited}",
    eprint = "hep-ph/9307208",
    archivePrefix = "arXiv",
    reportNumber = "MAD-PH-768",
    doi = "10.1103/PhysRevD.48.3483",
    journal = "Phys. Rev. D",
    volume = "48",
    pages = "3483--3501",
    year = "1993"
}

@article{Gondolo:2013wwa,
    author = "Gondolo, Paolo and Scopel, Stefano",
    title = "{On the sbottom resonance in dark matter scattering}",
    eprint = "1307.4481",
    archivePrefix = "arXiv",
    primaryClass = "hep-ph",
    reportNumber = "CETUP2013-008",
    doi = "10.1088/1475-7516/2013/10/032",
    journal = "JCAP",
    volume = "10",
    pages = "032",
    year = "2013"
}

@article{Arina:2023msd,
    author = {Arina, Chiara and Fuks, Benjamin and Heisig, Jan and Kr{\"a}mer, Michael and Mantani, Luca and Panizzi, Luca},
    title = "{Comprehensive exploration of t-channel simplified models of dark matter}",
    eprint = "2307.10367",
    archivePrefix = "arXiv",
    primaryClass = "hep-ph",
    reportNumber = "TTK-23-19",
    doi = "10.1103/PhysRevD.108.115007",
    journal = "Phys. Rev. D",
    volume = "108",
    number = "11",
    pages = "115007",
    year = "2023"
}

@article{Jungman:1995df,
    author = "Jungman, Gerard and Kamionkowski, Marc and Griest, Kim",
    title = "{Supersymmetric dark matter}",
    eprint = "hep-ph/9506380",
    archivePrefix = "arXiv",
    reportNumber = "SU-4240-605, UCSD-PTH-95-02, IASSNS-HEP-95-14, CU-TP-677",
    doi = "10.1016/0370-1573(95)00058-5",
    journal = "Phys. Rept.",
    volume = "267",
    pages = "195--373",
    year = "1996"
}

\end{document}